\documentclass{article}
\usepackage{amsfonts}
\usepackage{amsmath}
\usepackage{makeidx}
\usepackage{eurosym}
\usepackage{amssymb}
\usepackage{graphicx}
\usepackage{caption}
\usepackage{wrapfig}
\usepackage{rotating}
\usepackage{natbib}
\usepackage{appendix}
\usepackage{subfigure}

\numberwithin{equation}{section}
\newtheorem{theorem}{Theorem}
\newtheorem{acknowledgement}[theorem]{Acknowledgement}

\input{tcilatex}
\begin{document}

\title{Hydrodynamics of Markets: \linebreak Hidden Links Between Physics and
Finance}
\author{A. Lipton \\
Advisory Board Member\\
ADIA\ Lab, Abu\ Dhabi, UAE }
\maketitle

\begin{abstract}
An intriguing link between a wide range of problems occurring in physics and
financial engineering is presented. These problems include the evolution of
small perturbations of linear flows in hydrodynamics, the movements of
particles in random fields described by the Kolmogorov and Klein-Kramers
equations, the Ornstein-Uhlenbeck and Feller processes, and their
generalizations. They are reduced to affine differential and
pseudo-differential equations and solved in a unified way by using Kelvin
waves and developing a comprehensive math framework for calculating
transition probabilities and expectations. Kelvin waves are instrumental for
studying the well-known Black-Scholes, Heston, and Stein-Stein models and
more complex path-dependent volatility models, as well as the pricing of
Asian options, volatility and variance swaps, bonds, and bond options.
Kelvin waves help to solve several cutting-edge problems, including hedging
the impermanent loss of Automated Market Makers for cryptocurrency trading.
This title is also available as Open Access on Cambridge Core.


\end{abstract}

\tableofcontents

\begin{flushright}
.\newpage

6accdae13eff7i3l9n4o4qrr4s8t12ux

Letter from Isaac Newton to Henry Oldenburg, 24 October 1676\footnote{%
\textit{"Data aequatione quotcunque fluentes quantitae involvente fluxiones
invenire et vice versa."} -- \textquotedblleft Given an equation involving
any number of fluent quantities to find the fluxions, and vice
versa.\textquotedblright\ V.I.\ Arnold paraphrased the statement as follows:
\textquotedblleft It is useful to solve differential
equations.\textquotedblright\ Cambridge University Library, Department of
Manuscripts and University Archives. ItemReference Code: GBR/0012/MS
Add.9597/2/18/56.}

------------------------------------------------------------------------------------------------

Bob Montagnet: Yeah, good choice, Vlad.

Get back to the security system. How does it work?

Vlad: The way everything works. Mathematics.

\textquotedblleft The Good Thief,\textquotedblright\ screenplay by Neil
Jordan, 2002
\end{flushright}

\section{Introduction\label{Chap1}}

\subsection{Background\label{Sec11}}

Newton's discovery of differential equations and calculus was crucial in
developing classical mechanics because it allowed for the mathematical
description of the motion of objects. This discovery made a groundbreaking
step in unifying mathematics with physics, enabling the prediction of
planetary orbits, the motion of objects under various forces, and much more,
and marked the beginning of a new era in mathematics and science, laying the
cornerstone for over three centuries of advancements.

Newton understood the immediate impact of his discoveries and their
potential to transform the understanding of the natural world. To establish
and protect his intellectual property rights at the same time, he concealed
his discovery in the fundamental anagram of calculus, which he included in
his 1676 letter to Oldenburg. This anagram contained a Latin statement
describing the method of fluxions (his term for calculus) when decoded. The
need for an anagram reflected that Newton was competitive and cautious in
equal measure by balancing the desire for recognition with the fear of
disclosure. The number of occurrences of each Latin character in Newton's
sentence agrees with his anagram, thus proving that the actual sentence was
written in 1676.\footnote{%
The Newton's anagram is an early example of a one-way hash function. An
anagram is easy to calculate, provided the message is known, but not vice
versa. Hash functions are indispensable in modern cryptography, including
its applications to cryptocurrencies such as Bitcoin; see, e.g., \cite%
{Lipton21b}.}

The fact that differential equations are instrumental in mathematics and
physics alike was firmly established in the late seventeenth century.
However, methods for solving these equations remained ad-hoc for more than a
century until the work by Lagrange, Laplace, Fourier, and many other
mathematicians and physicists. In particular, the Fourier transform stands
out as the most potent tool in an applied mathematician's toolkit, enabling
the solving of linear partial differential equations (PDEs) and partial
pseudo-differential equations (PPDEs) with spatially constant coefficients;
it is also invaluable for analyzing time series and tackling other critical
tasks; \cite{Fourier22, Morse53}.

At the heart of the $n-$dimensional Fourier method are wave functions,
expressed as:

\begin{equation}
\begin{array}{c}
\mathcal{F}\left( t,\mathbf{x,k}\right) =\mathsf{a}\left( t\right) \exp
\left( i\mathbf{k}\cdot \mathbf{x}\right) ,%
\end{array}
\label{Eq1.1}
\end{equation}%
where $\mathbf{x}$\ and $\mathbf{k}$ are $n$-dimensional vectors, $\cdot $
denote the scalar product, $\mathsf{a}\left( t\right) $ is the amplitude,
and $\mathbf{k}\cdot \mathbf{x}$ is the phase. Depending on the particular
problem at hand, the amplitude $\mathsf{a}\left( t\right) $ can be a scalar
or a vector, hence the notation. Substituting $\mathcal{F}$ into a PDE with
spatially constant coefficients, one reduces the problem of interest to a
system of ordinary differential equations (ODEs) or a single ODE when $%
\mathsf{a}\left( t\right) $ is scalar. Of course, this system parametrically
depends on $\mathbf{k}$.

This book studies PDEs and PPDEs with coefficients linearly dependent on $%
\mathbf{x}$, which are called affine. Hence, one must use a more general
approach and consider wave functions with time-dependent wave vectors:

\begin{equation}
\begin{array}{c}
\mathcal{K}\left( t,\mathbf{x,\beta }\left( t\right) \right) =\mathsf{a}%
\left( t\right) \exp \left( i\mathbf{\beta }\left( t\right) \cdot \mathbf{x}%
\right) .%
\end{array}
\label{Eq1.2}
\end{equation}%
\cite{Kelvin87} and \cite{Orr07} were the first to use such waves to analyze
the stability of the steady motions of an incompressible fluid.

Affine problems are not artificial constructs. They appear organically in
several situations, for example, when the linear description of the
underlying physical mechanism is either exact or provides an excellent
approximation to reality or when the evolution in the phase space is
studied; see Chapter \ref{Chap3} below.

Subsequently and independently, affine PDEs and the associated wave
functions were used by many researchers in various areas, including the
theory of stochastic processes, physics, biology, and mathematical finance,
to mention a few. The Ornstein-Uhlenbeck (OU) and Feller processes are the
simplest but extremely important examples of affine processes; see \cite%
{Uhlenbeck30, Chandrasekhar43, Feller51, Feller52}. For financial
applications of affine processes see \cite{Duffie96, Duffie00,
Dai00,Lipton01, Duffie03, Sepp07, Lipton08, Filipovic09}, among others.

The book uses Kelvin waves of the form (\ref{Eq1.2}) to study transition
probability density functions (t.p.d.fs) for affine stochastic processes.
These processes can be either degenerate, i.e., have more independent
components than the sources of uncertainty, or non-degenerate, when every
component has its source of uncertainty. Recall that the t.p.d.f. for a
stochastic process describes the likelihood of a system transitioning from
one state to another over a specified period. Knowing the iterated t.p.d.f.
is fundamental for understanding the dynamics and behavior of stochastic
processes over time and is tantamount to knowing the process itself.

In the book, Kelvin waves are also used to solve several essential and
intricate problems occurring in financial applications, such as pricing
options with stochastic volatility, path-dependent options, and Asian
options with geometric averaging, among many others.

The main objective is to link various financial engineering topics with
their counterparts in hydrodynamics and molecular physics and showcase the
interdisciplinary nature of quantitative finance and economic modeling.
Finding such connections allows us to understand better how to model, price,
and risk-manage various financial instruments, derive several new results,
and provide additional intuition regarding their salient features. This work
continues the previous efforts in this direction; see \cite{Lipton08,
Lipton18}, Chapter 12.

There are several approaches one can use to solve affine equations
efficiently. For instance, Lie symmetries are a powerful tool for studying
certain classes of affine equations. Numerous authors describe general
techniques based on Lie symmetries; see, e.g., \cite{Ovsiannikov82,
Ibragimov85, Olver86, Bluman1989}, while their specific applications to
affine equations are covered by \cite{Berest93, Aksenov95, Craddock04,
Craddock12, Kovalenko14}, among many others. However, Lie symmetry
techniques are exceedingly cumbersome and might be challenging to use in
practice, especially when complicated affine equations are considered.

Laplace transform of spatial variables can be used in some cases, for
instance, for Feller processes; see, e.g., \cite{Feller51, Feller52}.
However, they are hard to use for solving generic affine equations.

Reductions of a given equation to a the simpler, solvable form is another
powerful method that can be successfully used in many instances; see, e.g., 
\cite{Chandrasekhar43, Carr02, Lipton14, Lipton18}, Chapter 9. Although the
reduction method is quite powerful, experience suggests it is often hard to
use in practice.

Finally, the affine ansatz based on Kelvin waves provides yet another
approach, which is the focus of the present book; see also \cite{Duffie96,
Dai00, Duffie03, Lipton08, Filipovic09, Lipton18}, Chapter 12. Undoubtedly,
the affine framework, also known as the affine ansatz, is the most potent
among the abovementioned techniques due to its comprehensive nature,
versatility, and (relative) ease of use, even in complex situations. In
practice, applications of Kelvin waves consist of three steps:

\begin{itemize}
\item Effective separation variables for the evolution problems with
pseudo-differential generators linearly dependent on spatial coordinates;

\item Solving ODEs parametrized by time-dependent wave vectors; see Eq. (\ref%
{Eq1.2});

\item Aggregating their solutions together to get the solution to the
original problem.
\end{itemize}

However, despite being a ruthlessly efficient tool, Kelvin waves have
limitations - using them to solve evolution problems supplied with external
boundary conditions is challenging. This exciting topic is actively
researched now; it will be discussed elsewhere in due course.

\subsection{Main Results\label{Sec12}}

This book develops a coherent, unified mathematical framework using Kelvin
waves as a powerful and versatile tool for studying t.p.d.fs in the context
of generic affine processes. It discovers previously hidden connections
among large classes of apparently unrelated problems from hydrodynamics,
molecular physics, and financial engineering. All these problems require
solving affine (pseudo-) differential equations, i.e., equations with
coefficients, which linearly depend on spatial variables. The book discusses
some classical results and derives several original ones related to:

\begin{itemize}
\item small wave-like perturbations of linear flows of ideal and viscous
fluids described by Euler and Navier-Stokes equations, respectively;

\item motions of free and harmonically bound particles under the impact of
random external white-noise forces described by the Klein-Kramers equations
and the hypoelliptic Kolmogorov equation, which play an essential role in
statistical physics;

\item Gaussian and non-Gaussian affine processes, such as the
Ornstein-Uhlenbeck and Feller processes, which are the archetypal
mean-reverting processes, and their generalizations;

\item dynamics of financial markets, particularly derivative products.
\end{itemize}

To solve some of the more complicated problems, one must augment primary
processes by introducing subordinate processes for auxiliary variables, such
as integrals over the original stochastic variable, and develop a uniform
mathematical formalism to construct t.p.d.fs for the abovementioned
processes.

Quite unexpectedly, the analysis identifies and rectifies an error in the
original solution of the Kolmogorov equation. The rectified solution is
dimensionally correct, properly scales when the process parameters change,
and agrees with numerical results.

Furthermore, this book derives many original results and extends and
reinterprets some well-known ones. For instance, it develops a concise and
efficient expression for t.p.d.fs in the case of processes with stochastic
volatility. Moreover, the analysis reveals an unexpected similarity between
the propagation of vorticity in two-dimensional flows of viscous
incompressible fluid and the motion of a harmonically bound particle, which
is used to find a new explicit expression for the vorticity of a
two-dimensional flow in terms of the Gaussian density.

Finally, the book applies the new methodology to various financial
engineering topics, such as pricing options with stochastic volatility,
options with path-dependent volatility, Asian options, volatility and
variance swaps, options on stocks with path-dependent volatility, and bonds
and bond options. In contrast to the classical approach, the book treats
primary fixed-income products, such as bonds and bond options, as
path-dependent, allowing us to gain additional intuition regarding such
products' pricing and risk management. It also highlights the flexibility of
the interdisciplinary framework by incorporating additional complexities
into the picture and with jump-diffusion processes and, more generally,
processes driven by affine pseudo-differential processes frequently used in
financial applications.

\subsection{Book Structure\label{Sec13}}

Chapter \ref{Chap2} introduces Kelvin waves. Section \ref{Sec21} introduces
the Euler equations, which describe the dynamics of a perfect fluid,
alongside the Navier-Stokes equation for viscous incompressible fluids.
Section \ref{Sec22} discusses the exact equilibria of these equations,
focusing on states where velocity varies linearly and pressure quadratically
with spatial coordinates, referred to as linear flows. Section \ref{Sec23}
illustrates that the renowned Kelvin waves provide solutions to the
linearized Euler and Navier-Stokes equations for small perturbations of the
linear flows. This section also explores the use of Kelvin waves in
analyzing the stability of these flows.

The book uses Kelvin waves as a fundamental tool in the analytical arsenal,
demonstrating their applicability across various study areas. For instance,
they allow one to discover profound and surprising links between the viscous
two-dimensional vorticity equations and the Klein-Kramers equation, a
cornerstone of stochastic physics; see Section \ref{Sec66}. This connection
results in a novel formula representing vorticity as a Gaussian density and
the stream function as the solution to the associated Poisson equation.

Chapter \ref{Chap3} investigates the degenerate stochastic process
introduced by Kolmogorov in 1934, alongside the associated Fokker-Planck
equation and its solution proposed by Kolmogorov. Further connections
between the Kolmogorov and Klein-Kramers equations are explored in Chapter %
\ref{Chap4}. To start with, Chapter \ref{Chap3} summarizes Kolmogorov's
original findings. Surprisingly, the Fokker-Planck equation, as used by
Kolmogorov in his seminal paper, is inconsistent with his initial
assumptions regarding the underlying process. Moreover, his proposed
solution has dimensional inconsistencies and, as a result, does not satisfy
the Fokker-Planck equation and initial conditions. However, there is a
silver lining; Kolmogorov's solution can be corrected via several
complementary methods, which the Chapter outlines. It concludes with an
example of a representative corrected solution to the Kolmogorov problem.

Chapter \ref{Chap4} explores a selection of representative affine stochastic
processes in statistical physics. First, it introduces the Langevin
equation, which describes the dynamics of an underdamped Brownian particle
in a potential field. Following this, it derives the Klein-Kramers equation,
capturing the probabilistic aspects of the motion of such a particle. It
turns out that the Kolmogorov equation derived in Chapter \ref{Chap3} is a
particular case of the Klein-Kramers equation. The Chapter presents
Chandrasekhar's solutions to the Klein-Kramers equations describing free and
harmonically bound particles. The Klein-Kramers equation is inherently
degenerate, with white noise impacting the particle's velocity but not its
position. It is shown in Chapter \ref{Chap8} that many path-dependent
problems share this characteristic in mathematical finance. For instance,
financial variables like the geometric price averages, which serve as the
underlying for a particular class of Asian options, can be conceptualized as
path integrals, fitting into the category of degenerate stochastic processes.

Chapter \ref{Chap5} describes backward (Kolmogorov) and forward
(Fokker-Planck) equations for t.p.d.fs of multidimensional stochastic
jump-diffusion processes. The Chapter explains the significance of studying
t.p.d.fs. It sets up the general framework for Kolmogorov and Fokker-Planck
equations and identifies the subset of affine stochastic processes amenable
to analysis using the Kelvin-wave formalism. Subsequently, the Chapter
introduces an augmentation technique, providing a natural approach to tackle
degenerate problems. Finally, it illustrates methods for transforming
specific non-affine processes into affine form through coordinate
transformations, enhancing the scope of problems accessible by the
Kelvin-wave methodology.

Chapter \ref{Chap6} studies Gaussian stochastic processes. It introduces a
general formula for regular Gaussian processes, accommodating both
degenerate scenarios and non-degenerate cases, as in Kolmogorov's example.
It expands this formula to address the practically significant scenario of
killed Gaussian processes, followed by several illustrative examples. Then,
the Chapter presents the derivation of the t.p.d.f. for the Kolmogorov
process with time-varying coefficients and explores the OU process with
time-dependent coefficients and its extension, the augmented OU process,
which models the combined dynamics of the process and its integral. Although
the results are classical, their derivation through Kelvin wave expansions
provides a novel and enriching angle, offering an alternative viewpoint for
understanding and deriving these established results. Next, the Chapter
examines free and harmonically bound particles, contrasting the Kelvin wave
method with Chandrasekhar's classical approach. Finally, it revisits the
basic concepts introduced in Chapter \ref{Chap2}, demonstrating the akin
nature of the temporal-spatial evolution of vorticity in the two-dimensional
flow of a viscous fluid to the dynamics of a harmonically bound particle.
This finding is intriguing and unexpected, forging a connection between
seemingly unrelated physical phenomena.

Chapter \ref{Chap7} considers non-Gaussian processes. It starts a general
formula for non-Gaussian dynamics, accommodating degenerate and
non-degenerate processes. Then, it expands this formula to killed processes.
Several interesting examples are studied. These examples include a
Kolmogorov process driven by anomalous diffusion, Feller processes with
constant and time-dependent coefficients and degenerate and non-degenerate
augmented Feller processes. A novel method for investigating finite-time
explosions of t.p.d.fs for augmented Feller processes is developed as a
helpful byproduct of the analysis. In addition, arithmetic Brownian motions
with path-dependent volatility and degenerate and non-degenerate arithmetic
Brownian motions with stochastic volatility are analyzed in detail.

Chapter \ref{Chap8} illustrates the application of the methodology to
financial engineering. To start with, it lays the foundation of financial
engineering, providing a primer for the uninitiated. Then, the Chapter
introduces the geometric Brownian motion, a staple in financial modeling,
and discusses the modifications necessary to reflect the complexities of
financial markets better. Several traditional models such as Bachelier,
Black-Scholes, Heston, and Stein-Stein models, and a novel path-dependent
volatility model, are explored via the Kelvin wave formalism. In addition,
it is shown how to price Asian options with geometric averaging via the
Kolmogorov's solution described in Chapter \ref{Chap3}. Besides, volatility
and variance swaps and swaptions, bonds and bond options are investigated by
linking financial formulas to those used in physics for underdamped Brownian
motion.

Succinct Chapter \ref{Chap9} outlines potential future expansions of the
work presented in this book and summarizes the conclusions.

A note on notation: Given the wide-ranging scope of this book, from
hydrodynamics to molecular physics, probability theory, and financial
engineering, adopting a unified notation system is impractical. Each field
has its conventions carved in stone, leading to inevitable variations in
notation. Notation is designed for consistency within and, where possible,
across sections. However, readers are encouraged to remain vigilant to
maintain coherence in their understanding.

Finally, this book is a revised and expanded version of \cite{Lipton23a}.

\section{Fluid Flows\label{Chap2}}

\subsection{Euler and Navier-Stockes Equations\label{Sec21}}

Hydrodynamics studies how fluids (liquids and gases) move, primarily relying
on fluid motion's fundamental equations: the Euler and Navier-Stokes
equations, with the Euler equations applicable to inviscid (frictionless)
flow and the Navier-Stokes equations describing viscous fluids.
Hydrodynamics has numerous applications across various fields, including
engineering, astrophysics, oceanography, and climate change, among many
others.

Recall that the Euler system of partial differential equations (PDEs)
describing the motion of an inviscid, incompressible fluid has the form:%
\begin{equation}
\begin{array}{c}
\frac{\partial \mathbf{V}}{\partial t}+(\mathbf{V\cdot }\nabla )\mathbf{V+}%
\nabla \left( \frac{P}{\rho }\right) =0, \\ 
\\ 
\nabla \cdot \mathbf{V=}0\mathbf{,}%
\end{array}
\label{Eq2.1}
\end{equation}%
where $t$ is time, $\mathbf{x}$ is the position, \textbf{$V$}$\left( t,%
\mathbf{x}\right) $ is the velocity vector, $P\left( t,\mathbf{x}\right) $
is the pressure, $\rho $ is the constant density, $\nabla $ is the gradient,
and $\cdot $ denotes the scalar product; see, e.g., \cite{Chandrasekhar61}.
In Cartesian coordinates, Eqs (\ref{Eq2.1}) can be written as follows:%
\begin{equation}
\begin{array}{c}
\frac{\partial V_{i}}{\partial t}+V_{j}\frac{\partial V_{i}}{\partial x_{j}}+%
\frac{\partial }{\partial x_{i}}\left( \frac{P}{\rho }\right) =0, \\ 
\\ 
\frac{\partial V_{i}}{\partial x_{i}}=0.%
\end{array}
\label{Eq2.2}
\end{equation}%
Here and below Einstein's summation convention over repeated indices is used.

The motion of the incompressible viscous fluid is described by the classical
Navier-Stokes equations of the form:%
\begin{equation}
\begin{array}{c}
\frac{\partial \mathbf{V}}{\partial t}+(\mathbf{V\cdot }\nabla )\mathbf{V-}%
\nu \Delta \mathbf{V+}\nabla \left( \frac{P}{\rho }\right) =0, \\ 
\\ 
\nabla \cdot \mathbf{V=}0\mathbf{,}%
\end{array}
\label{Eq2.11a}
\end{equation}%
where $\nu $ is the kinematic viscosity; see, e.g., \cite{Chandrasekhar61}.
Explicitly,%
\begin{equation}
\begin{array}{c}
\frac{\partial V_{i}}{\partial t}+V_{j}\frac{\partial V_{i}}{\partial x_{j}}%
-\nu \frac{\partial ^{2}V_{i}}{\partial x_{j}\partial x_{j}}+\frac{\partial 
}{\partial x_{i}}\left( \frac{P}{\rho }\right) =0, \\ 
\\ 
\frac{\partial V_{i}}{\partial x_{i}}=0.%
\end{array}
\label{Eq2.11b}
\end{equation}%
The diffusive term $\mathbf{-}\nu \Delta \mathbf{V}$ in Eqs (\ref{Eq2.11b})
describes frictions ignored in Eqs (\ref{Eq2.11a}). Due to their greater
generality, the Navier-Stokes equations are fundamental in understanding
important phenomena, such as the transition from laminar to turbulent flow.

\subsection{Linear Flows\label{Sec22}}

This section studies exact solutions of the Euler and Navier-Stockes
equations known as linear flows. These solutions are valuable for several
reasons: (a) exact solutions provide precise, analytical descriptions of
fluid flow patterns under specific conditions; (b) they serve as benchmarks
for understanding fundamental hydrodynamics phenomena like wave propagation;
(c) they provide a bridge which is crucial for more complex studies by
simplifying the inherently complex and nonlinear nature of hydrodynamics,
and making it possible to understand the behavior of more general fluid
flows. Linear solutions of the Euler and Navier-Stokes equations help to
study fluid flow stability. This understanding is crucial in predicting and
controlling flow behavior in various engineering applications, from
aerospace to hydraulic engineering. By starting with linear solutions, one
can incrementally introduce nonlinear effects, allowing for a systematic
study of nonlinear phenomena in hydrodynamics. This approach can uncover the
mechanisms behind complex flows, including turbulence and chaotic flow
behaviors. Exact linear solutions of the Euler equations provide a clear,
analytical framework for exploring the behavior of fluids and validating
more complicated models.

It is easy to show that Eqs (\ref{Eq2.1}) have a family of solutions $\left( 
\mathbf{V}\left( t,\mathbf{x}\right) ,P\left( t,\mathbf{x}\right) \right) $,
linearly depending on spatial coordinates:%
\begin{equation}
\begin{array}{c}
\mathbf{V}\left( t,\mathbf{x}\right) =\mathfrak{L}\left( t\right) \mathbf{%
x,\ \ \ }\frac{P\left( t,\mathbf{x}\right) }{\rho }=\frac{P_{0}}{\rho }+%
\frac{1}{2}\mathfrak{M}\left( t\right) \mathbf{x\cdot x,}%
\end{array}
\label{Eq2.3}
\end{equation}%
where the $3\times 3$ matrices $\mathfrak{L}\left( t\right) $, $\mathfrak{M}%
\left( t\right) $, such that 
\begin{equation}
\begin{array}{c}
\frac{d\mathfrak{L}\left( t\right) }{dt}+\mathfrak{L}^{2}\left( t\right) +%
\mathfrak{M}\left( t\right) =0, \\ 
\\ 
\mathrm{Tr}\left( \mathfrak{L}\left( t\right) \right) =0,\ \ \ \mathfrak{M}%
\left( t\right) =\mathfrak{M}^{\ast }\left( t\right) .%
\end{array}
\label{Eq2.4}
\end{equation}
It is clear that linear flows, given by Eq. (\ref{Eq2.3}), are unaffected by
viscosity, hence they satisfy Eqs (\ref{Eq2.12}).

Flows (\ref{Eq2.3}) have stagnation points at the origin. Typical examples
are planar flows of the form%
\begin{equation}
\begin{array}{c}
V_{1}=\frac{1}{2}\left( sx_{1}-wx_{2}\right) ,\ \ \ V_{2}=\frac{1}{2}\left(
wx_{1}-sx_{2}\right) ,\ \ \ V_{3}=0, \\ 
\\ 
\frac{P}{\rho }=\frac{P_{0}}{\rho }+\frac{1}{4}\left( w^{2}-s^{2}\right)
\left( x_{1}^{2}+x_{2}^{2}\right) ,%
\end{array}
\label{Eq2.5}
\end{equation}%
these flows are elliptic when $s<w$, and hyperbolic otherwise; see, e.g., 
\cite{Friedlander03}.

\subsection{Kelvin Waves in an Incompressible Fluid\label{Sec23}}

The study of small perturbations of exact solutions of the Euler and
Navier-Stokes equations is the core of the stability analysis in fluid
dynamics. Examining their behavior is essential for predicting how fluid
flows evolve under slight disturbances. One can determine whether a
particular flow is stable or unstable by introducing small perturbations to
an exact solution and observing the system's response. If these
perturbations grow over time, the flow is considered unstable; if they decay
or remain bounded, the flow is stable. One of this analysis's most critical
applications is understanding the transition from laminar (smooth and
orderly) to turbulent (chaotic and unpredictable) flows. Small perturbations
can exhibit exponential growth, leading to the onset of turbulence. For more
detailed investigations, direct numerical simulations of the perturbed
Navier-Stokes equations can be used to study the nonlinear evolution of
perturbations. This approach can capture the complete transition from
initial instability to fully developed turbulence, offering insights into
the complex interactions that drive flow dynamics. The study of
perturbations offers theoretical insights into the fundamental nature of
fluid dynamics, including the mechanisms of flow instability, transition,
and turbulence structure. It helps in developing reduced-order models and
theories that explain complex fluid phenomena. Here, Kelvin waves are used
as the primary tool for studying small perturbations of linear flows. In the
rest of the book, Kelvin waves are used for other purposes. This section is
dedicated to their brief description.

\begin{figure}[tbp]
\begin{center}
\subfigure[]{\includegraphics[width = 0.35 \textwidth]
{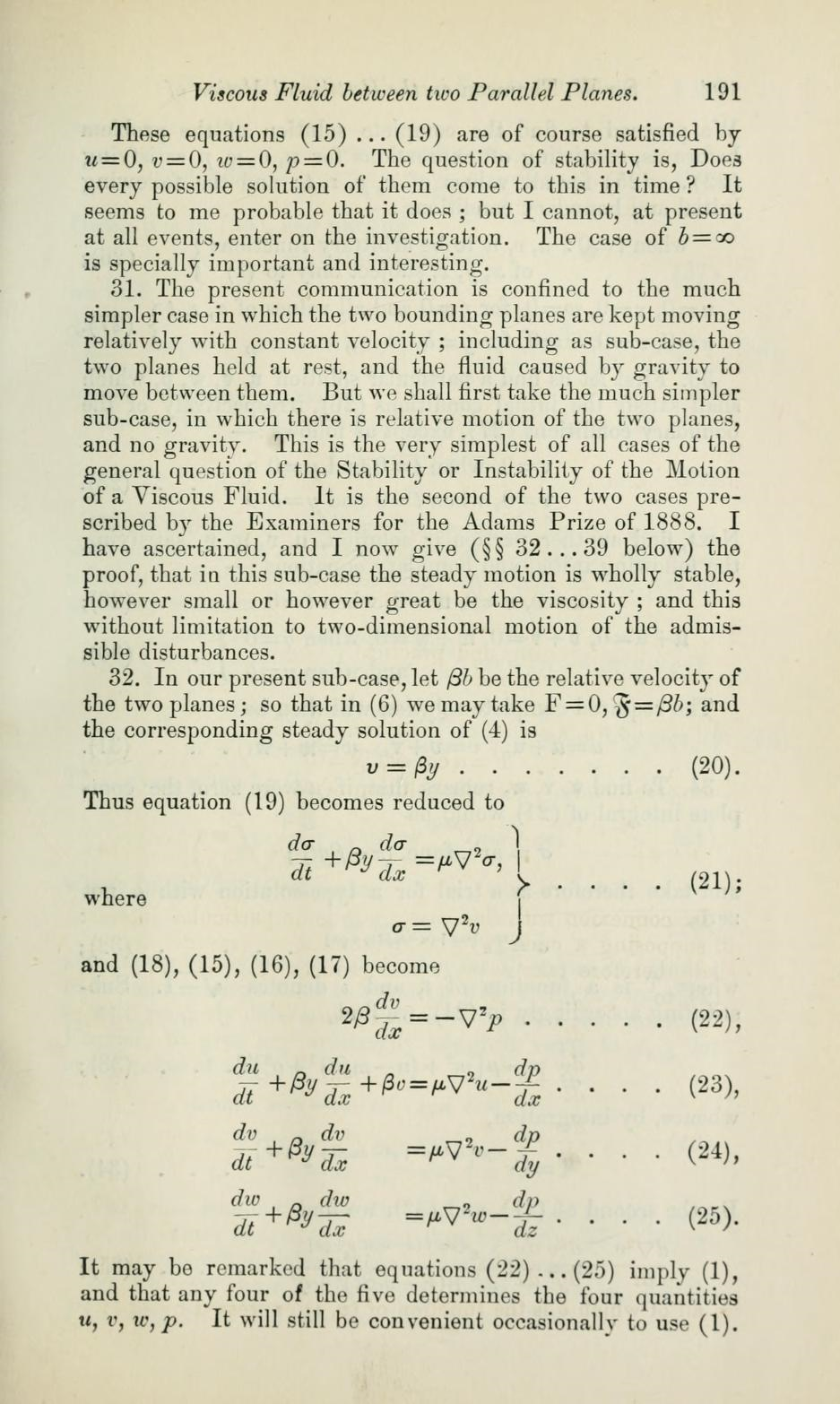}}
\subfigure[]{\includegraphics[width = 0.35 \textwidth]
{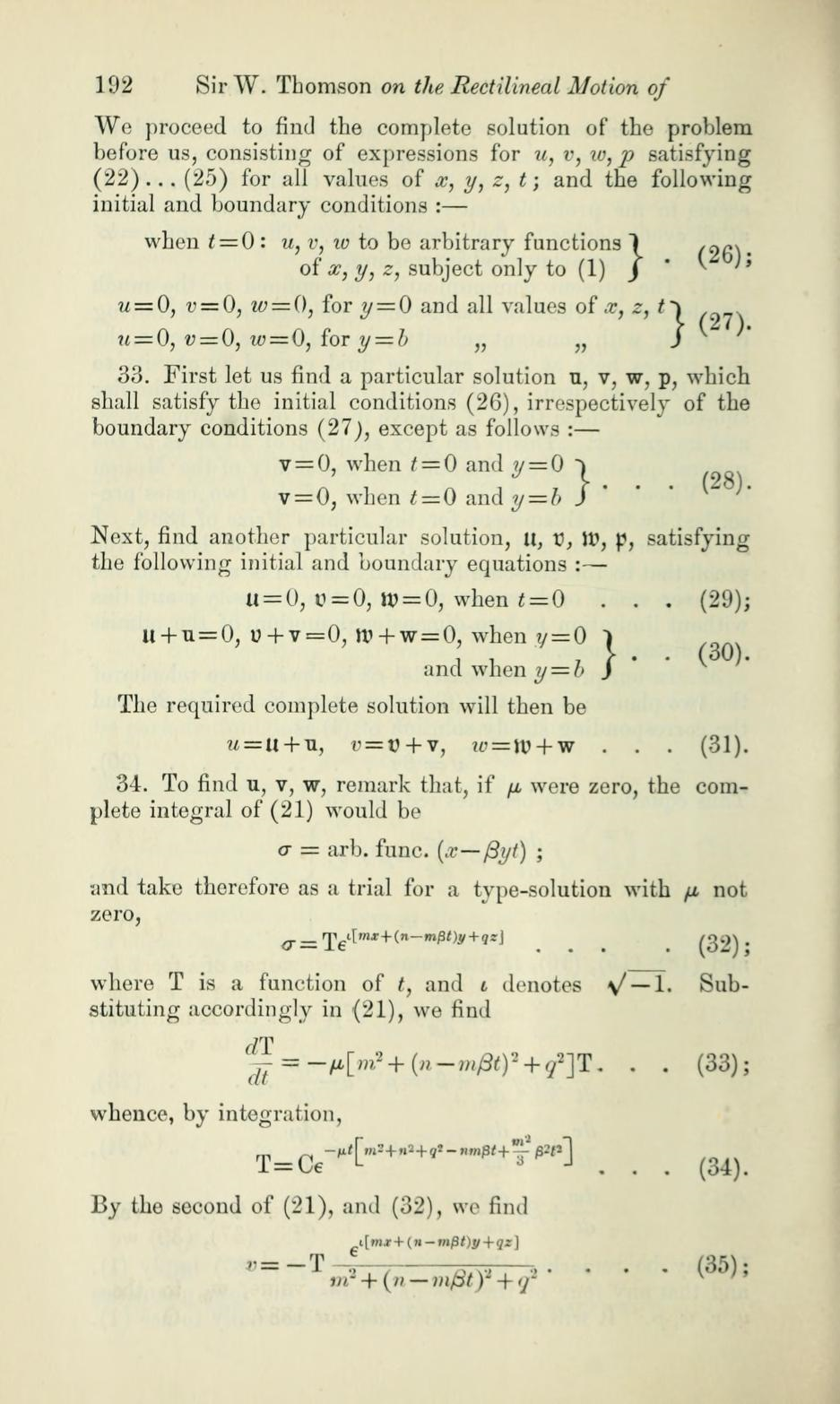}}
\end{center}
\caption{An excerpt from the original Kelvin's paper, where Kelvin waves are
introduced for the first time; see \protect\cite{Kelvin87}. Public domain.}
\label{Fig2.1}
\end{figure}

It is necessary to the behavior of perturbations of solutions given by\ Eq. (%
\ref{Eq2.3}), which are denoted by $\left( \mathbf{v}\left( t,\mathbf{x}%
\right) ,p\left( t,\mathbf{x}\right) \right) $. By neglecting the quadratic
term $($\textbf{$v$}$\mathbf{\cdot }\nabla )$\textbf{$v$}, one can write the
system of PDEs for $\left( \mathbf{v,}p\right) $ as follows:%
\begin{equation}
\begin{array}{c}
\frac{\partial \mathbf{v}}{\partial t}+(\mathfrak{L}\left( t\right) \mathbf{%
x\cdot }\nabla )\mathbf{v+}\mathfrak{L}\left( t\right) \mathbf{v+}\nabla
\left( \frac{p}{\rho }\right) =0, \\ 
\\ 
\nabla \cdot \mathbf{v=}0\mathbf{.}%
\end{array}
\label{Eq2.6}
\end{equation}%
It has been known for a long time that linear PDEs (\ref{Eq2.6}) have
wave-like solutions of the form:%
\begin{equation}
\begin{array}{c}
\left( \mathbf{v}\left( t,\mathbf{x}\right) ,\frac{p\left( t,\mathbf{x}%
\right) }{\rho }\right) =\left( \mathbf{a}\left( t\right) ,a\left( t\right)
\right) \exp \left( i\mathbf{\beta }\left( t\right) \cdot \left( \mathbf{x-r}%
\left( t\right) \right) \right) ,%
\end{array}
\label{Eq2.7}
\end{equation}%
where $\left( \mathbf{a}\left( t\right) ,a\left( t\right) \right) $ are
time-dependent amplitudes, and $\mathbf{\beta }\left( t\right) $ is the
time-dependent wave vector; see \cite{Kelvin87, Orr07, Craik86,
Friedlander03}. In this book, these solutions are called the Kelvin waves.
It should be emphasized that the so-called affine ansatz is a special
instance of Kelvin wave. This observation allows one to discover
similarities among seemingly unrelated topics, which, in turn, facilitates
their holistic and comprehensive study. An excerpt from the original
Kelvin's paper is shown in Figure \ref{Fig2.1}.

As one can see from Figure \ref{Fig2.1}, Kelvin considered the special case
of the so-called shear linear flow of the form 
\begin{equation}
\begin{array}{c}
\mathbf{V}\left( t,\mathbf{x}\right) =\left( V_{1}\left( x_{2}\right)
,0,0\right) =\left( l_{12}x_{2},0,0\right) ,%
\end{array}
\label{Eq2.7b}
\end{equation}%
between two plates, $x_{2}=0$ and $x_{2}=L$, the first one at rest and the
second one moving in parallel.

The triplet $\mathbf{r}\left( t\right) $, $\mathbf{\beta }\left( t\right) $, 
$\mathbf{a}\left( t\right) $ satisfies the following system of ODEs:%
\begin{equation}
\begin{array}{c}
\frac{d\mathbf{r}\left( t\right) }{dt}-\mathfrak{L}\left( t\right) \mathbf{r}%
\left( t\right) =0,\ \ \ \mathbf{r}\left( 0\right) =\mathbf{r}_{0}, \\ 
\\ 
\frac{d\mathbf{\beta }\left( t\right) }{dt}+\mathfrak{L}^{\ast }\left(
t\right) \mathbf{\beta }\left( t\right) =0,\ \ \ \mathbf{\beta }\left(
0\right) =\mathbf{\beta }_{0}, \\ 
\\ 
\frac{d\mathbf{a}\left( t\right) }{dt}+\mathfrak{L}\left( t\right) \mathbf{a}%
\left( t\right) -2\frac{\mathfrak{L}\left( t\right) \mathbf{a}\left(
t\right) \cdot \mathbf{\beta }\left( t\right) }{\mathbf{\beta }\left(
t\right) \cdot \mathbf{\beta }\left( t\right) }\mathbf{\beta }\left(
t\right) =0,\ \mathbf{a}\left( 0\right) =\mathbf{a}_{0}, \\ 
\\ 
\ \mathbf{\beta }_{0}\cdot \mathbf{a}_{0}=0.%
\end{array}
\label{Eq2.8}
\end{equation}%
Here and below, the superscript $\ast $ stands for transpose. The
corresponding $p\left( t\right) $ can be found via the incompressibility
condition. It is easy to show that for $t\geq 0$, 
\begin{equation}
\begin{array}{c}
\mathbf{\beta }\left( t\right) \cdot \mathbf{r}\left( t\right) =\mathbf{%
\beta }_{0}\cdot \mathbf{r}_{0},\ \ \ \mathbf{\beta }\left( t\right) \cdot 
\mathbf{a}\left( t\right) =0.%
\end{array}
\label{Eq2.9}
\end{equation}
Thus, the Kelvin-wave formalism results in ingenious separation of variables
and allows us to solve a system of ODEs (\ref{Eq2.8}), rather than PDEs (\ref%
{Eq2.6}).

Typically, Eqs (\ref{Eq2.8}) are used to study the stability of the linear
flow. Such a flow is unstable whenever $\left\Vert \mathbf{a}\left( t\right)
\right\Vert \rightarrow \infty $ for some choices of $\mathbf{\beta }_{0},%
\mathbf{a}_{0}$; see \cite{Bayly86, Lifschitz95, Bayly96}. Moreover, it can
be shown that the same instabilities occur in general three-dimensional
flows, because locally they are equivalent to linear flows; see \cite%
{Lifschitz91a, Friedlander91, Lifschitz91b, Friedlander03}.

Interestingly, \cite{Chandrasekhar61} pointed out that the superposition of
the linear flow (\ref{Eq2.3}) and the Kelvin wave (\ref{Eq2.7}), i.e.,%
\begin{equation}
\begin{array}{c}
\mathbf{\tilde{V}}\left( t,\mathbf{x}\right) =\mathfrak{L}\left( t\right) 
\mathbf{x+v}\left( t,\mathbf{x}\right) , \\ 
\\ 
\frac{\tilde{P}\left( t,\mathbf{x}\right) }{\rho }=\frac{1}{2}\mathfrak{M}%
\left( t\right) \mathbf{x\cdot x+}\frac{p\left( t,\mathbf{x}\right) }{\rho },%
\end{array}
\label{Eq2.10}
\end{equation}%
satisfies the nonlinear Euler equations (\ref{Eq2.1}) since the nonlinear
term $(\mathbf{v\cdot \nabla })\mathbf{v}$ vanishes identically due to
incompressibility.\footnote{%
Thus, even the greatest minds occasionally can be myopic -- it took eighty
years for fluid dynamists to connect the dots and observe that $\left( 
\mathbf{\tilde{v}},\tilde{p}/\rho \right) $ solve the nonlinear Euler
equations.} Studying secondary instabilities of flows with elliptic
streamlines, i.e., instabilities of Kelvin waves is an important and
intricate topic; see \cite{Fabijonas97}.

Viscosity does affect small perturbations of linear flows. For viscous
incompressible fluids, Kelvin waves are governed by the following equations:%
\begin{equation}
\begin{array}{c}
\frac{\partial \mathbf{v}}{\partial t}+(\mathfrak{L}\left( t\right) \mathbf{%
x\cdot }\nabla )\mathbf{v+}\mathfrak{L}\left( t\right) \mathbf{v\mathbf{-}}%
\nu \Delta \mathbf{v+}\nabla \left( \frac{p}{\rho }\right) =0, \\ 
\\ 
\nabla \cdot \mathbf{v=}0\mathbf{.}%
\end{array}
\label{Eq2.12}
\end{equation}%
The viscous version of Eqs (\ref{Eq2.8}) has the form; see \cite{Lifschitz91}%
:%
\begin{equation}
\begin{array}{c}
\frac{d\mathbf{r}\left( t\right) }{dt}-\mathfrak{L}\left( t\right) \mathbf{r}%
\left( t\right) =0,\ \ \ \mathbf{r}\left( 0\right) =\mathbf{r}_{0}, \\ 
\\ 
\frac{d\mathbf{\beta }\left( t\right) }{dt}+\mathfrak{L}^{\ast }\left(
t\right) \mathbf{\beta }\left( t\right) =0,\ \ \ \mathbf{\beta }\left(
0\right) =\mathbf{\beta }_{0}, \\ 
\\ 
\frac{d\mathbf{a}\left( t\right) }{dt}+\mathfrak{L}\left( t\right) \mathbf{a}%
\left( t\right) -2\frac{\mathfrak{L}\left( t\right) \mathbf{a}\left(
t\right) \cdot \mathbf{\beta }\left( t\right) }{\mathbf{\beta }\left(
t\right) \cdot \mathbf{\beta }\left( t\right) }\mathbf{\beta }\left(
t\right) +\nu \left\vert \mathbf{\beta }\left( t\right) \right\vert ^{2}%
\mathbf{a}\left( t\right) =0,\ \mathbf{a}\left( 0\right) =\mathbf{a}_{0}, \\ 
\\ 
\ \mathbf{\beta }_{0}\cdot \mathbf{a}_{0}=0.%
\end{array}
\label{Eq2.13}
\end{equation}
It is shown in Section \ref{Sec65} that in the two-dimensional case, the
Navier-Stokes equations for small perturbations of linear flows are more or
less identical to the Fokker-Planck equations for harmonically bound
articles, which is surprising.

\begin{figure}[tbp]
\begin{center}
\includegraphics[width=0.7\textwidth]
{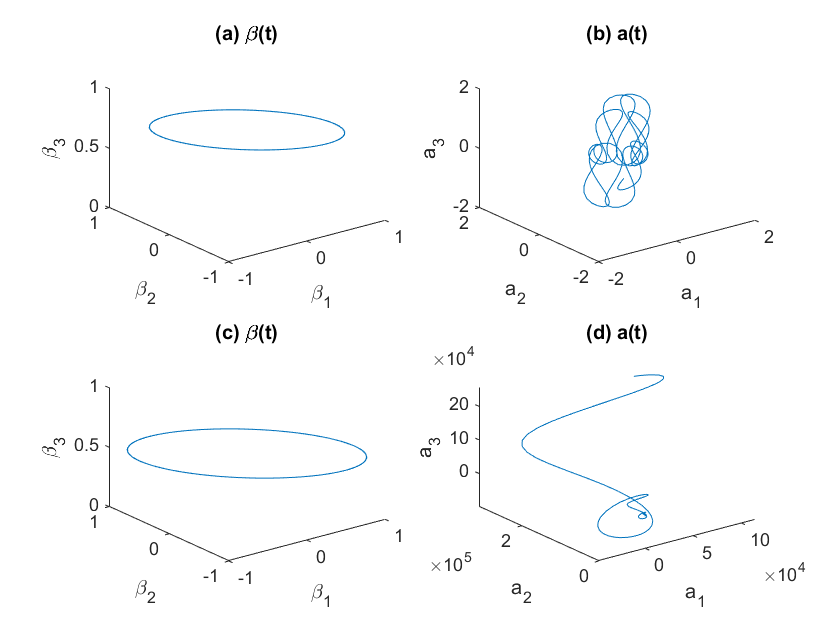}
\end{center}
\par
\vspace{-10pt}
\caption{Kelvin waves corresponding to two different orientations of the
initial wave vector $\mathbf{\protect\beta }\left( 0\right) $ and $\mathbf{a}%
\left( 0\right) $. (a), (b) $\mathbf{\protect\beta }(0)=\left( \sin \left( 
\protect\pi /4\right) ,0,\cos \left( \protect\pi /4\right) \right) $, $%
\mathbf{a}\left( 0\right) =\left( 0,\sin \left( \protect\pi /4\right)
,0\right) $; (c), (d) $\mathbf{\protect\beta }(0)=\left( \sin \left( \protect%
\pi /3\right) ,0,\cos \left( \protect\pi /3\right) \right) $, $\mathbf{a}%
\left( 0\right) =\left( 0,\sin \left( \protect\pi /3\right) ,0\right) $.
Other parameters are as follows: $T=100$, $\protect\omega =1$, $s=0.5$. In
the first case, $\mathbf{a}\left( t\right) $ stays bounded, while $\mathbf{a}%
\left( t\right) $ explodes explodes in the second case. This explosion means
that the underlying elliptic flow is unstable. Own graphics.}
\label{Fig2.2}
\end{figure}

\begin{figure}[tbp]
\begin{center}
\includegraphics[width=0.7\textwidth]
{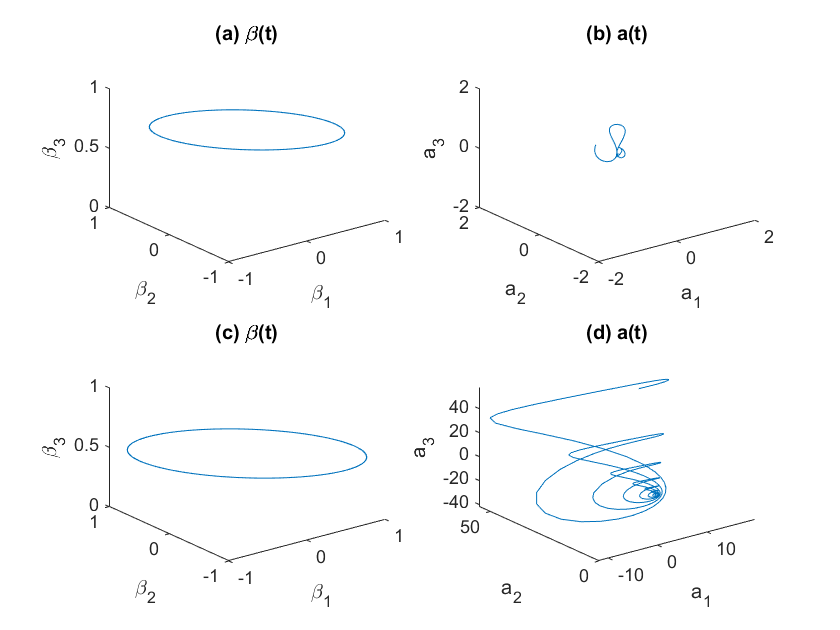}
\end{center}
\par
\vspace{-10pt}
\caption{Kelvin waves in the viscous fluid with viscosity $\protect\nu =0.07$%
. Other parameters and initial conditions are the same as in Figure \protect
\ref{Fig2.2}. Viscosity dampens the instability but, generally, does not
suppress it entirely. Own graphics.}
\label{Fig2.3}
\end{figure}

The evolution of a typical Kelvin wave parameters triplet $\mathbf{r}\left(
t\right) $, $\mathbf{\beta }\left( t\right) $, $\mathbf{a}\left( t\right) $
is illustrated in Figure \ref{Fig2.2}. 
The impact of viscosity is illustrated in Figure \ref{Fig2.3}. 
These Figures show that depending on the initial orientation of the wave
vector $\mathbf{\beta }\left( t\right) $, the amplitude $\mathbf{a}\left(
t\right) $ can be either bounded or unbounded. For elliptic flows, unbounded
amplitudes are always present for specific orientations, so all of them are
unstable; see \cite{Bayly86, Bayly96, Friedlander03} and references therein.

\section{Kolmogorov Stochastic Process\label{Chap3}}

\subsection{Background\label{Sec30}}

The Kolmogorov equation studies the evolution of a particle in the phase
space. The particle's position and velocity evolve in time due to the
interplay between the deterministic drift and stochastic force affecting
only its velocity. Since only the particle's velocity is affected by the
random force, the PDE describing the evolution of the t.p.d.f. in the phase
space is degenerate. The Kolmogorov equation is a particular case of the
Klein-Kramers equation studied in Chapter \ref{Chap4}.

The significance of the Kolmogorov equation lies in its ability to model the
intricate balance between deterministic behavior and stochastic dynamics,
providing a basic framework for studying the evolution of systems in phase
space. It has important applications in various fields, including physics
for understanding particle dynamics, finance for modeling asset prices, and
beyond. It demonstrates the profound interplay between stochastic processes
and differential equations.

The Kolmogorov equation is hypoelliptic; as such, it serves as a prototype
for a broad class of hypoelliptic PDEs. Although it does not meet the exact
criteria for ellipticity (due to the second-order derivatives not being
present in all directions of the phase space), the solutions to the equation
are still smooth, which is particularly important in the context of
stochastic processes, where hypoellipticity ensures that the probability
density function remains smooth and well-behaved, facilitating the analysis
of the system's dynamics over time.

\subsection{Summary of Kolmogorov's paper\label{Sec31}}

In a remarkable (and remarkably concise) note, Kolmogorov considers a system
of particles in $n$-dimensional space with coordinates $q_{1},...,q_{n}$,
and velocities $\dot{q}_{1},...,\dot{q}_{n}$, assumes the probability
density function%
\begin{equation*}
\begin{array}{c}
g\left( t,q_{1},...,q_{n},\dot{q}_{1},...,\dot{q}_{n},t^{\prime
},q_{1}^{\prime },...,q_{n}^{\prime },\dot{q}_{1}^{\prime },...,\dot{q}%
_{n}^{\prime }\right)%
\end{array}%
\end{equation*}%
exist for some time $t^{\prime }>t$, and reveals (without any explanation)
an analytical expression for $g$ in the one-dimensional case; see \cite%
{Kolmogorov34}. This note is the third in a series of papers; the previous
two being \cite{Kolmogorov31, Kolmogorov33}.

Kolmogorov makes the following natural assumptions:%
\begin{equation}
\begin{array}{c}
\mathbf{E}\left\vert \Delta q_{i}-\dot{q}_{i}\Delta t\right\vert =o\left(
\Delta t\right) ,%
\end{array}
\label{Eq3.1}
\end{equation}%
\begin{equation}
\begin{array}{c}
\mathbf{E}\left( \Delta q_{i}\right) ^{2}=o\left( \Delta t\right) ,%
\end{array}
\label{Eq3.2}
\end{equation}%
where $\Delta t=t^{\prime }-t$. Eqs (\ref{Eq3.1}), (\ref{Eq3.2}) imply%
\begin{equation}
\begin{array}{c}
\mathbf{E}\left( \Delta q_{i}\right) =\dot{q}_{i}\Delta t+o\left( \Delta
t\right) ,%
\end{array}
\label{Eq3.3}
\end{equation}%
\begin{equation}
\begin{array}{c}
\mathbf{E}\left( \Delta q_{i}\Delta q_{j}\right) \leq \sqrt{\mathbf{E}\left(
\Delta q_{i}\right) ^{2}\mathbf{E}\left( \Delta q_{j}\right) ^{2}}=o\left(
\Delta t\right) .%
\end{array}
\label{Eq3.4}
\end{equation}%
Furthermore, under very general assumptions, the following relationships hold%
\begin{equation}
\begin{array}{c}
\mathbf{E}\left( \Delta \dot{q}_{i}\right) =f_{i}\left( t,q,\dot{q}\right)
\Delta t+o\left( \Delta t\right) ,%
\end{array}
\label{Eq3.5}
\end{equation}%
\begin{equation}
\begin{array}{c}
\mathbf{E}\left( \Delta \dot{q}_{i}\right) ^{2}=k_{ii}\left( t,q,\dot{q}%
\right) \Delta t+o\left( \Delta t\right) ,%
\end{array}
\label{Eq3.6}
\end{equation}%
\begin{equation}
\begin{array}{c}
\mathbf{E}\left( \Delta \dot{q}_{i}\Delta \dot{q}_{j}\right) =k_{ij}\left(
t,q,\dot{q}\right) \Delta t+o\left( \Delta t\right) ,%
\end{array}
\label{Eq3.7}
\end{equation}%
where $f,k$ are continuous functions. Eqs (\ref{Eq3.2}), (\ref{Eq3.6}) imply%
\begin{equation}
\begin{array}{c}
\mathbf{E}\left( \Delta \dot{q}_{i}\Delta \dot{q}_{j}\right) \leq \sqrt{%
\mathbf{E}\left( \Delta \dot{q}_{i}\right) ^{2}\mathbf{E}\left( \Delta \dot{q%
}_{j}\right) ^{2}}=o\left( \Delta t\right) .%
\end{array}
\label{Eq3.8}
\end{equation}%
Under some natural physical assumptions, it follows that $g$ satisfies the
following differential equation of the Fokker-Planck type:%
\begin{equation}
\begin{array}{c}
\frac{\partial g}{\partial t^{\prime }}=-\sum \dot{q}_{i}^{\prime }\frac{%
\partial g}{\partial q_{i}^{\prime }}-\sum \frac{\partial }{\partial \dot{q}%
_{i}^{\prime }}\left\{ f_{i}\left( t,q,\dot{q}\right) g\right\} +\sum \sum 
\frac{\partial ^{2}}{\partial \dot{q}_{i}^{\prime }\partial \dot{q}%
_{j}^{\prime }}\left\{ k\left( t,q,\dot{q}\right) g\right\} .%
\end{array}
\label{Eq3.9}
\end{equation}%
In the one-dimensional case, one has%
\begin{equation}
\begin{array}{c}
\frac{\partial g}{\partial t^{\prime }}=-\dot{q}^{\prime }\frac{\partial g}{%
\partial q^{\prime }}-\frac{\partial }{\partial \dot{q}^{\prime }}\left\{
f\left( t,q,\dot{q}\right) g\right\} +\frac{\partial ^{2}}{\partial \dot{q}%
^{\prime 2}}\left\{ k\left( t,q,\dot{q}\right) g\right\} .%
\end{array}
\label{Eq3.10}
\end{equation}%
These equations are known as ultra-parabolic Fokker-Plank- Kolmogorov
equations due to their degeneracy.

When $f$ and $k$ are constants, Eq. (\ref{Eq3.10}) becomes%
\begin{equation}
\begin{array}{c}
\frac{\partial g}{\partial t^{\prime }}=-\dot{q}^{\prime }\frac{\partial g}{%
\partial q^{\prime }}-f\frac{\partial g}{\partial \dot{q}^{\prime }}+k\frac{%
\partial ^{2}g}{\partial \dot{q}^{\prime 2}}.%
\end{array}
\label{Eq3.10a}
\end{equation}%
The corresponding fundamental solution of has the form:%
\begin{equation}
\begin{array}{c}
g=\frac{2\sqrt{3}}{\pi k^{2}\left( t^{\prime }-t\right) ^{2}}\exp \left\{ -%
\frac{\left( \dot{q}^{\prime }-\dot{q}-f\left( t^{\prime }-t\right) \right)
^{2}}{4k\left( t^{\prime }-t\right) }-\frac{3\left( q^{\prime }-q-\frac{\dot{%
q}^{\prime }+\dot{q}}{2}\left( t^{\prime }-t\right) \right) ^{2}}{%
k^{3}\left( t^{\prime }-t\right) ^{3}}\right\} .%
\end{array}
\label{Eq3.11}
\end{equation}%
One can see that $\Delta \dot{q}$ is of the order $\left( \Delta t\right)
^{1/2}$. At the same time%
\begin{equation}
\begin{array}{c}
\Delta q=\dot{q}\Delta t+O\left( \Delta t\right) ^{3/2}.%
\end{array}
\label{Eq3.12}
\end{equation}%
One can prove that a similar relation holds for the general Eq. (\ref{Eq3.9}%
).

\begin{figure}[tbp]
\begin{center}
\subfigure[]{\includegraphics[width = 0.4 \textwidth]
{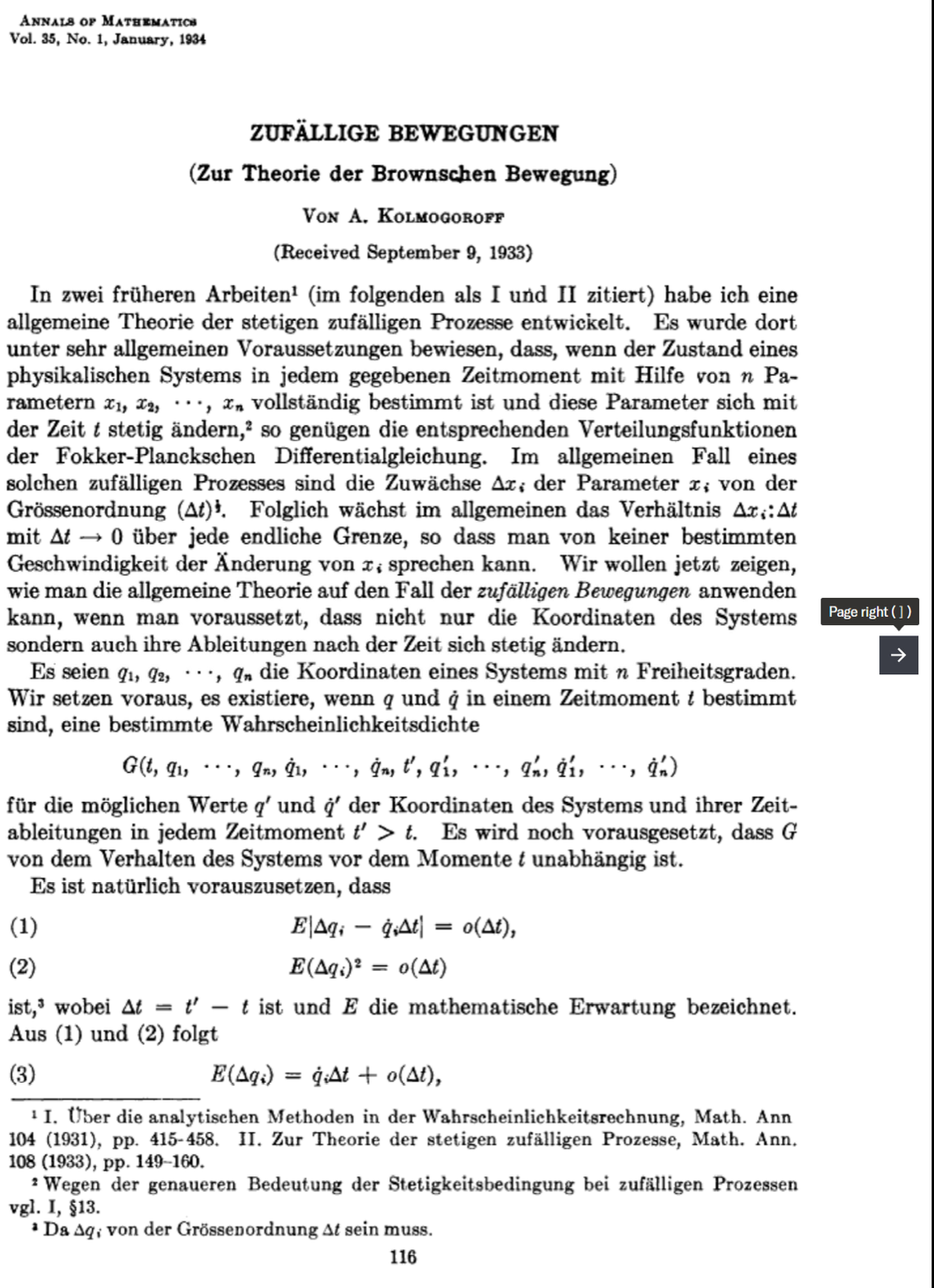}}
\subfigure[]{\includegraphics[width = 0.4 \textwidth]
{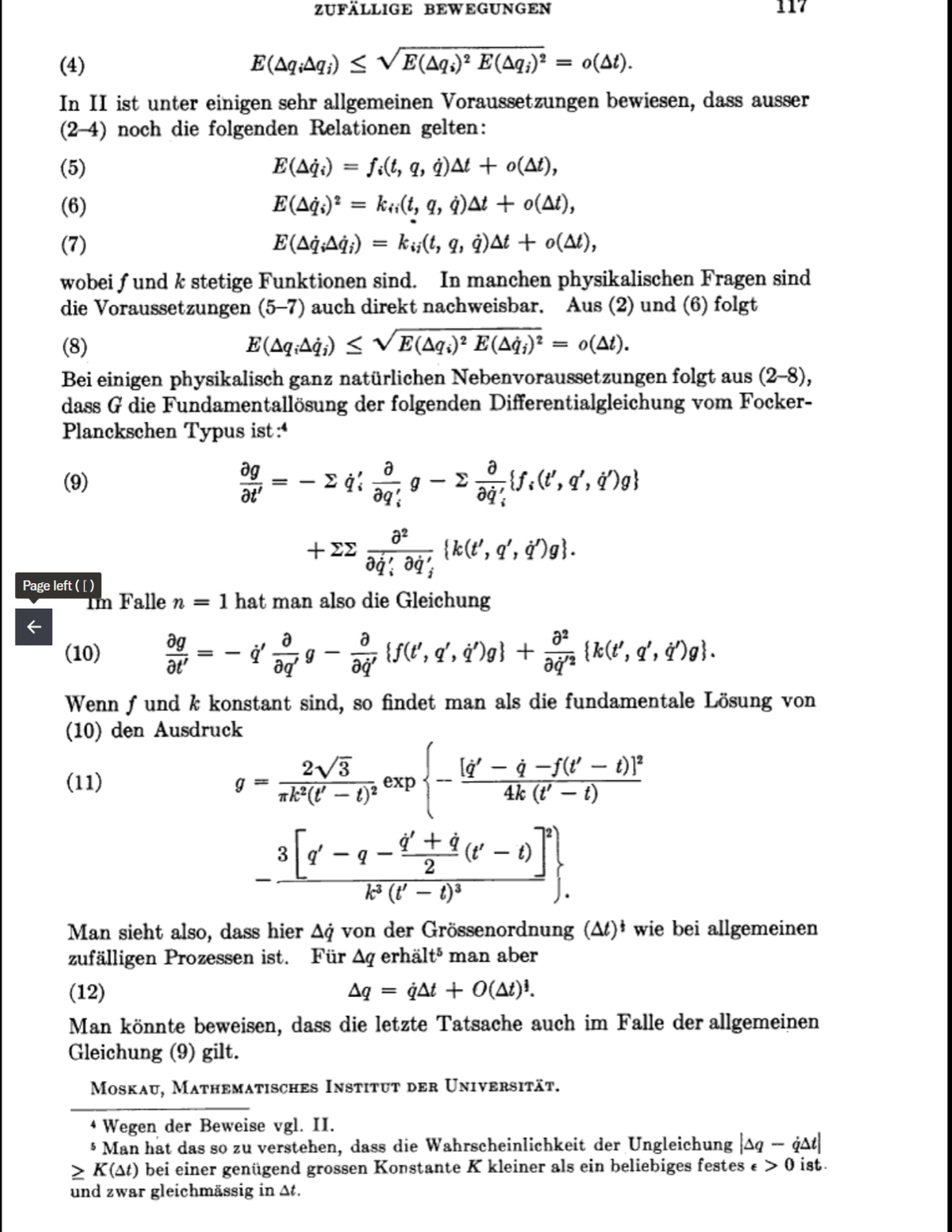}}
\end{center}
\caption{The original Kolmogorov's paper, presented here for the inquisitive
reader to enjoy; see \protect\cite{Kolmogorov34}. Reproduced by kind
permission of the Editors of Annals of Mathematics. \copyright Annals of
Mathematics.}
\label{Fig3.1}
\end{figure}

The original Kolmogorov paper is shown in Figure \ref{Fig3.1}.

Kolmogorov equations fascinated mathematicians for a long time and generated
a great deal of research; see, e.g., \cite{Weber51, Hormander67, Kuptsov72,
Lanconelli02, Pascucci05, Ivasishen10, Duong18}, among others.

It is worth mentioning that physicists derived Eqs (\ref{Eq3.9}), (\ref%
{Eq3.10}) at least a decade earlier than Kolmogorov, see Chapter \ref{Chap4}
below.

\subsection{Challenge and Response\label{Sec32}}

Despite its undoubting brilliance, the original Kolmogorov's paper has
several issues.

First, Eqs (\ref{Eq3.9}), (\ref{Eq3.10}) \emph{are not} the Fokker-Planck
equations associated with the process described by Eqs (\ref{Eq3.5}) - (\ref%
{Eq3.7}), since they miss the prefactor $1/2$ in front of the diffusion
terms. The corrected multivariate equation has the form:%
\begin{equation}
\begin{array}{c}
\frac{\partial g}{\partial t^{\prime }}=-\sum \dot{q}_{i}^{\prime }\frac{%
\partial }{\partial q_{i}^{\prime }}g-\sum \frac{\partial }{\partial \dot{q}%
_{i}^{\prime }}\left\{ f_{i}\left( t,q,\dot{q}\right) g\right\} +\frac{1}{2}%
\sum \sum \frac{\partial ^{2}}{\partial \dot{q}_{i}^{\prime }\partial \dot{q}%
_{j}^{\prime }}\left\{ \left\{ k\left( t,q,\dot{q}\right) g\right\} \right\}
,%
\end{array}
\label{Eq3.13}
\end{equation}%
while the corresponding one-dimensional equation has the form:%
\begin{equation}
\begin{array}{c}
\frac{\partial g}{\partial t^{\prime }}=-\dot{q}^{\prime }\frac{\partial }{%
\partial q^{\prime }}g-\frac{\partial }{\partial \dot{q}^{\prime }}\left\{
f\left( t,q,\dot{q}\right) g\right\} +\frac{1}{2}\frac{\partial ^{2}}{%
\partial \dot{q}^{\prime 2}}\left\{ \left\{ k\left( t,q,\dot{q}\right)
g\right\} \right\} .%
\end{array}
\label{Eq3.14}
\end{equation}%
Alternatively, Eqs (\ref{Eq3.6}), (\ref{Eq3.7}) can be altered as follows:%
\begin{equation}
\begin{array}{c}
\mathbf{E}\left( \Delta \dot{q}_{i}\right) ^{2}=2k_{ii}\left( t,q,\dot{q}%
\right) \Delta t+o\left( \Delta t\right) ,%
\end{array}
\label{Eq3.6a}
\end{equation}%
\begin{equation}
\begin{array}{c}
\mathbf{E}\left( \Delta \dot{q}_{i}\Delta \dot{q}_{j}\right) =2k_{ij}\left(
t,q,\dot{q}\right) \Delta t+o\left( \Delta t\right) .%
\end{array}
\label{Eq3.7a}
\end{equation}%
Below, the Fokker-Planck equation is updated.

Second, $g$ given by Eq. (\ref{Eq3.11}) does not solve Eq. (\ref{Eq3.10}).
It also does not satisfy the (implicit) initial condition%
\begin{equation}
\begin{array}{c}
g\left( t,q,\dot{q},t,q^{\prime },\dot{q}^{\prime }\right) =\delta \left(
q^{\prime }-q\right) \delta \left( \dot{q}^{\prime }-\dot{q}\right) ,%
\end{array}
\label{Eq3.15}
\end{equation}%
where $\delta \left( .\right) $ is the Dirac $\delta $-function. The fact
that expression (\ref{Eq3.11}) does not solve Eq. (\ref{Eq3.10}) can be
verified by substitution. However, it is easier to verify this statement via
dimensional analysis. The dimensions of the corresponding variables and
coefficients are as follows:%
\begin{equation}
\begin{array}{c}
\left[ t\right] =\left[ t^{\prime }\right] =T,\ \ \ \left[ q\right] =\left[
q^{\prime }\right] =L,\ \ \ \left[ \dot{q}\right] =\left[ \dot{q}^{\prime }%
\right] =\frac{L}{T},\ \ \ \left[ g\right] =\frac{T}{L^{2}},\ \ \ \left[ f%
\right] =\frac{L}{T^{2}},\ \ \ \left[ k\right] =\frac{L^{2}}{T^{3}}.%
\end{array}
\label{Eq3.16}
\end{equation}%
It is easy to show that $g$ is scale-invariant, so that%
\begin{equation}
\begin{array}{c}
g\left( \lambda ^{2}t,\lambda ^{3}q,\lambda \dot{q},\lambda ^{2}t^{\prime
},\lambda ^{3}q^{\prime },\lambda \dot{q}^{\prime };\lambda ^{-1}f,k\right)
=\lambda ^{-4}g\left( t,q,\dot{q},t^{\prime },q^{\prime },\dot{q}^{\prime
};f,k\right) .%
\end{array}
\label{Eq3.17}
\end{equation}%
The original Kolmogorov formula contains two typos, making it dimensionally
incorrect since the term 
\begin{equation*}
\begin{array}{c}
\frac{3\left( q^{\prime }-q-\frac{\dot{q}^{\prime }+\dot{q}}{2}\left(
t^{\prime }-t\right) \right) ^{2}}{k^{3}\left( t^{\prime }-t\right) ^{3}}%
\end{array}%
\end{equation*}%
in the exponent is not nondimensional, as it should be, and has dimension $%
T^{6}L^{-1}$, while the prefactor 
\begin{equation*}
\begin{array}{c}
\frac{2\sqrt{3}}{\pi k^{2}\left( t^{\prime }-t\right) ^{2}}%
\end{array}%
\end{equation*}%
has dimension $T^{4}L^{-1}$, instead of the right dimension $TL^{-2}$.

Third, due to yet another typo, the solution given by Eq. (\ref{Eq3.11})
does not converge to the initial condition in the limit $t^{\prime
}\rightarrow t$. Indeed, asymptotically, one has%
\begin{equation}
\begin{array}{c}
g\sim H\left( \frac{k^{3}\left( t^{\prime }-t\right) ^{3}}{6},q^{\prime
}-q\right) H\left( 2k\left( t^{\prime }-t\right) ,\dot{q}^{\prime }-\dot{q}%
\right) \rightarrow 4\delta \left( q^{\prime }-q\right) \delta \left( \dot{q}%
^{\prime }-\dot{q}\right) ,%
\end{array}
\label{Eq3.18}
\end{equation}%
where $H\left( \mu ,\nu \right) $ is the standard heat kernel:%
\begin{equation}
\begin{array}{c}
H\left( \mu ,\nu \right) =\frac{\exp \left( -\frac{\nu ^{2}}{2\mu }\right) }{%
\sqrt{2\pi \mu }}.%
\end{array}
\label{Eq3.19}
\end{equation}%
However, not all is lost. Dimensional analysis shows that the correct
solution $g\left( t,q,\dot{q},t^{\prime },q^{\prime },\dot{q}^{\prime
};f,k\right) $ of Eq. (\ref{Eq3.10}) has the form:%
\begin{equation}
\begin{array}{c}
g=\frac{\sqrt{3}}{2\pi k\left( t^{\prime }-t\right) ^{2}}\exp \left\{ -\frac{%
\left( \dot{q}^{\prime }-\dot{q}-f\left( t^{\prime }-t\right) \right) ^{2}}{%
4k\left( t^{\prime }-t\right) }-\frac{3\left( q^{\prime }-q-\frac{\dot{q}%
^{\prime }+\dot{q}}{2}\left( t^{\prime }-t\right) \right) ^{2}}{k\left(
t^{\prime }-t\right) ^{3}}\right\} ,%
\end{array}
\label{Eq3.20}
\end{equation}%
which is not far from Kolmogorov's formula. Similarly, the correct solution
of Eq. (\ref{Eq3.14}) has the form:%
\begin{equation}
\begin{array}{c}
g=\frac{\sqrt{3}}{\pi k\left( t^{\prime }-t\right) ^{2}}\exp \left\{ -\frac{%
\left( \dot{q}^{\prime }-\dot{q}-f\left( t^{\prime }-t\right) \right) ^{2}}{%
2k\left( t^{\prime }-t\right) }-\frac{6\left( q^{\prime }-q-\frac{\dot{q}%
^{\prime }+\dot{q}}{2}\left( t^{\prime }-t\right) \right) ^{2}}{k\left(
t^{\prime }-t\right) ^{3}}\right\} .%
\end{array}
\label{Eq3.21}
\end{equation}

\subsection{Direct Verification\label{Sec33}}

In order to avoid confusion, from now on, the notation is changed to make
the formulas easier to read. Specifically, it is assumed that $\bar{x}$
represents the position of a particle at time $\bar{t}$ and $x$ its position
at time $t$, while $\bar{y}$ represents its velocity at time $\bar{t}$, and $%
y$ its velocity at time $t$, so that 
\begin{equation}
\begin{array}{c}
\left( t,q,\dot{q}\right) \rightarrow \left( t,x,y\right) ,\ \ \ \left(
t^{\prime },q^{\prime },\dot{q}^{\prime }\right) \rightarrow \left( \bar{t},%
\bar{x},\bar{y}\right) .%
\end{array}
\label{Eq3.22}
\end{equation}%
One of our objectives is deriving the (corrected) Kolmogorov formula from
first principles using Kelvin waves. Subsequently, it is shown how to use it
in the financial mathematics context. The governing SDE can be written as%
\begin{equation}
\begin{array}{c}
d\hat{x}_{t}=\hat{y}_{t}dt,\ \ \ \hat{x}_{t}=x, \\ 
\\ 
d\hat{y}_{t}=bdt+\sigma d\hat{W}_{t},\ \ \ \hat{y}_{t}=y.%
\end{array}
\label{Eq3.23}
\end{equation}%
The corresponding Fokker-Planck- Kolmogorov problem for the t.p.d.f. $\varpi
\left( t,x,y,\bar{t},\bar{x},\bar{y}\right) $ has the form: 
\begin{equation}
\begin{array}{c}
\varpi _{\bar{t}}\left( t,x,y,\bar{t},\bar{x},\bar{y}\right) -\frac{1}{2}%
\sigma ^{2}\varpi _{\bar{y}\bar{y}}\left( t,x,y,\bar{t},\bar{x},\bar{y}%
\right) \\ 
\\ 
+\bar{y}\varpi _{\bar{x}}\left( t,x,y,\bar{t},\bar{x},\bar{y}\right)
+b\varpi _{\bar{y}}\left( t,x,y,\bar{t},\bar{x},\bar{y}\right) =0, \\ 
\\ 
\varpi \left( t,x,y,t,\bar{x},\bar{y}\right) =\delta \left( \bar{x}-x\right)
\delta \left( \bar{y}-y\right) .%
\end{array}
\label{Eq3.24}
\end{equation}%
The solution of Eqs (\ref{Eq3.24}) is as follows:%
\begin{equation}
\begin{array}{c}
\varpi \left( t,x,y,\bar{t},\bar{x},\bar{y}\right) =\frac{\sqrt{3}}{\pi
\sigma ^{2}T^{2}}\exp \left( -\Phi \left( t,x,y,\bar{t},\bar{x},\bar{y}%
\right) \right) ,%
\end{array}
\label{Eq3.25}
\end{equation}%
where%
\begin{equation}
\begin{array}{c}
\Phi \left( t,x,y,\bar{t},\bar{x},\bar{y}\right) =\frac{\left( \bar{y}%
-y-bT\right) ^{2}}{2\sigma ^{2}T}+\frac{6\left( \bar{x}-x-\frac{\left( \bar{y%
}+y\right) T}{2}\right) ^{2}}{\sigma ^{2}T^{3}}=\frac{A^{2}}{2}+6B^{2},%
\end{array}
\label{Eq3.26}
\end{equation}%
and%
\begin{equation}
\begin{array}{c}
A=\frac{\left( \bar{y}-y-bT\right) }{\sqrt{\sigma ^{2}T}},\ \ \left[ A\right]
=1,\ \ \ B=\frac{\left( \bar{x}-x-\frac{\left( \bar{y}+y\right) T}{2}\right) 
}{\sqrt{\sigma ^{2}T^{3}}},\ \ \ \left[ B\right] =1.%
\end{array}
\label{Eq3.27}
\end{equation}%
Here and below, the following shorthand notation is used:%
\begin{equation}
\begin{array}{c}
T=\bar{t}-t.%
\end{array}
\label{Eq4.7a}
\end{equation}%
Let us check that $\varpi $ satisfies the Fokker-Plank equation and the
initial conditions. A simple calculation yields:%
\begin{equation}
\begin{array}{c}
\Phi _{\bar{t}}=-\left( \frac{A^{2}}{2T}+\frac{bA}{\sqrt{\sigma ^{2}T}}+%
\frac{18B^{2}}{T}+\frac{6\left( \bar{y}+y\right) B}{\sqrt{\sigma ^{2}T^{3}}}%
\right) , \\ 
\\ 
\ \Phi _{\bar{x}}=\frac{12B}{\sqrt{\sigma ^{2}T^{3}}},\ \ \ \Phi _{\bar{y}}=%
\frac{A-6B}{\sqrt{\sigma ^{2}T}},\ \ \ \Phi _{\bar{y}\bar{y}}=\frac{4}{%
\sigma ^{2}T},%
\end{array}
\label{Eq3.28}
\end{equation}%
\begin{equation}
\begin{array}{c}
\frac{\varpi _{\bar{t}}}{\varpi }=-\frac{2}{T}-\Phi _{\bar{t}},\ \ \ \frac{%
\varpi _{\bar{x}}}{\varpi }=-\Phi _{\bar{x}},\ \ \ \frac{\varpi _{\bar{y}}}{%
\varpi }=-\Phi _{\bar{y}},\ \ \ \frac{\varpi _{\bar{y}\bar{y}}}{\varpi }%
=-\Phi _{\bar{y}\bar{y}}+\Phi _{\bar{x}}^{2},%
\end{array}
\label{Eq3.29}
\end{equation}%
so that 
\begin{equation}
\begin{array}{c}
\varpi _{\bar{t}}^{K}-\frac{1}{2}\sigma ^{2}\varpi _{\bar{y}\bar{y}}^{K}+%
\bar{y}\varpi _{\bar{x}}^{K}+b\varpi _{\bar{y}}^{K} \\ 
\\ 
=\varpi ^{K}\left( -\frac{2}{T}-\Phi _{\bar{t}}+\frac{1}{2}\sigma ^{2}\left(
\Phi _{\bar{y}\bar{y}}-\Phi _{\bar{y}}^{2}\right) -\bar{y}\Phi _{\bar{x}%
}-b\Phi _{\bar{y}}\right) \\ 
\\ 
=\varpi ^{K}\left( -\frac{2}{T}+\frac{A^{2}}{2T}+\frac{bA}{\sqrt{\sigma ^{2}T%
}}+\frac{18B^{2}}{T}+\frac{6\left( \bar{y}+y\right) B}{\sqrt{\sigma ^{2}T^{3}%
}}\right. \\ 
\\ 
\left. +\frac{2}{T}-\frac{\left( A-6B\right) ^{2}}{2T}-\frac{12\bar{y}B}{%
\sqrt{\sigma ^{2}T^{3}}}-\frac{b\left( A-6B\right) }{\sqrt{\sigma ^{2}T}}%
\right) =0.%
\end{array}
\label{Eq3.30}
\end{equation}%
When $T\rightarrow 0$ one has the following asymptotic expression:%
\begin{equation}
\begin{array}{c}
\varpi ^{K}\left( t,x,y,\bar{t},\bar{x},\bar{y}\right) \sim H\left( \frac{%
\sigma ^{2}T^{3}}{12},\bar{x}-x\right) H\left( \sigma ^{2}T,\bar{y}-y\right)
\rightarrow \delta \left( \bar{x}-x\right) \delta \left( \bar{y}-y\right) .%
\end{array}
\label{Eq3.31}
\end{equation}

\subsection{Solution via Kelvin Waves\label{Sec34}}

Now, Kolmogorov's formula is derived by using Kelvin waves (or an affine
ansatz), which requires solving the problem of the form:%
\begin{equation}
\begin{array}{c}
\mathcal{K}_{\bar{t}}\left( t,x,y,\bar{t},\bar{x},\bar{y},k,l\right) -\frac{1%
}{2}\sigma ^{2}\mathcal{K}_{\bar{y}\bar{y}}\left( t,x,y,\bar{t},\bar{x},\bar{%
y},k,l\right) \\ 
\\ 
+\bar{y}\mathcal{K}_{\bar{x}}\left( t,x,y,\bar{t},\bar{x},\bar{y},k,l\right)
+b\mathcal{K}_{\bar{y}}\left( t,x,y,\bar{t},\bar{x},\bar{y},k,l\right) =0,
\\ 
\\ 
\mathcal{K}\left( t,\bar{x},\bar{y},t,x,y,k,l\right) =\exp \left( ik\left( 
\bar{x}-x\right) +il\left( \bar{y}-y\right) \right) .%
\end{array}
\label{Eq3.32}
\end{equation}%
Here%
\begin{equation}
\begin{array}{c}
\ \left[ k\right] =\frac{1}{L},\ \ \ \left[ l\right] =\frac{T}{L},\ \ \left[ 
\mathcal{K}\right] =1.%
\end{array}
\label{Eq3.33}
\end{equation}%
By using the well-known results concerning the inverse Fourier transform of
the $\delta $-function, one gets the following expression for the t.p.d.f. $%
\varpi \left( t,x,y,\bar{t},\bar{x},\bar{y}\right) $:%
\begin{equation}
\begin{array}{c}
\varpi \left( t,x,y,\bar{t},\bar{x},\bar{y}\right) =\frac{1}{\left( 2\pi
\right) ^{2}}\int_{-\infty }^{\infty }\int_{-\infty }^{\infty }\mathcal{K}%
\left( t,x,y,\bar{t},\bar{x},\bar{y},k,l\right) dkdl.%
\end{array}
\label{Eq3.34}
\end{equation}%
To calculate $\mathcal{K}$, one can use the affine ansatz and represent it
in the form:%
\begin{equation}
\begin{array}{c}
\mathcal{K}\left( t,x,y,\bar{t},\bar{x},\bar{y},k,l\right) =\exp \left( \Psi
\left( t,x,y,\bar{t},\bar{x},\bar{y},k,l\right) \right) ,%
\end{array}
\label{Eq3.35}
\end{equation}%
where%
\begin{equation}
\begin{array}{c}
\Psi \left( t,x,y,\bar{t},\bar{x},\bar{y},k,l\right) =\alpha \left( t,\bar{t}%
\right) +ik\left( \bar{x}-x\right) +i\gamma \left( t,\bar{t}\right) \bar{y}%
-ily.%
\end{array}
\label{Eq3.36}
\end{equation}%
and%
\begin{equation}
\begin{array}{c}
\frac{\mathcal{K}_{\bar{t}}}{\mathcal{K}}=\Psi _{\bar{t}}=\left( \alpha _{%
\bar{t}}\left( t,\bar{t}\right) +i\gamma _{\bar{t}}\left( t,\bar{t}\right) 
\bar{y}\right) ,\ \ \ \frac{\mathcal{K}_{\bar{x}}}{\mathcal{K}}=\Psi _{\bar{x%
}}=ik,\ \ \  \\ 
\\ 
\frac{\mathcal{K}_{\bar{y}}}{\mathcal{K}}=\Psi _{\bar{y}}=i\gamma \left( t,%
\bar{t}\right) ,\ \ \ \frac{\mathcal{K}_{\bar{y}\bar{y}}}{\mathcal{K}}=\Psi
_{\bar{y}}^{2}=-\gamma ^{2}\left( t,\bar{t}\right) .%
\end{array}
\label{Eq3.37}
\end{equation}%
Accordingly,%
\begin{equation}
\begin{array}{c}
\alpha _{\bar{t}}\left( t,\bar{t}\right) +\frac{1}{2}\sigma ^{2}\gamma
^{2}\left( t,\bar{t}\right) +i\gamma _{\bar{t}}\left( t,\bar{t}\right) \bar{y%
}+ik\bar{y}+ib\gamma \left( t,\bar{t}\right) =0, \\ 
\\ 
\alpha \left( t,t\right) =0,\ \ \ \gamma \left( t,t\right) =l,%
\end{array}
\label{Eq3.38}
\end{equation}%
so that%
\begin{equation}
\begin{array}{c}
\alpha _{\bar{t}}\left( t,\bar{t}\right) +\frac{1}{2}\sigma ^{2}\gamma
^{2}\left( t,\bar{t}\right) +ib\gamma \left( t,\bar{t}\right) =0,\ \ \
\alpha \left( t,t\right) =0, \\ 
\\ 
\gamma _{\bar{t}}\left( t,\bar{t}\right) +k=0,\ \ \ \gamma \left( t,t\right)
=l.%
\end{array}
\label{Eq3.39}
\end{equation}%
Straightforward calculation shows:%
\begin{equation}
\begin{array}{c}
\gamma \left( t,\bar{t}\right) =-kT+l, \\ 
\\ 
\alpha \left( t,\bar{t}\right) =-\frac{1}{2}\sigma ^{2}\left( \frac{%
k^{2}T^{3}}{3}-klT^{2}+l^{2}T\right) -ib\left( -\frac{kT^{2}}{2}+lT\right) .%
\end{array}
\label{Eq3.40}
\end{equation}%
Equations (\ref{Eq3.34}), (\ref{Eq3.35}), (\ref{Eq3.36}), (\ref{Eq3.40})
yield:%
\begin{equation}
\begin{array}{c}
\varpi \left( t,x,y,\bar{t},\bar{x},\bar{y}\right) \\ 
\\ 
=\frac{1}{\left( 2\pi \right) ^{2}}\int_{-\infty }^{\infty }\int_{-\infty
}^{\infty }\exp \left( -\frac{1}{2}\sigma ^{2}\left( \frac{k^{2}T^{3}}{3}%
-klT^{2}+l^{2}T\right) \right. \\ 
\\ 
\left. +ik\left( \bar{x}-x-\bar{y}T+\frac{bT^{2}}{2}\right) +il\left( \bar{y}%
-y-bT\right) \right) dkdl.%
\end{array}
\label{Eq3.41}
\end{equation}%
It is clear that $\varpi \left( t,x,y,\bar{t},\bar{x},\bar{y}\right) $ can
be viewed as the characteristic function of the Gaussian density in the $%
\left( k,l\right) $ space, evaluated at the point $\left( \bar{x}-x-\bar{y}%
T\right. $ $\left. +\frac{bT^{2}}{2},\bar{y}-y-bT\right) $:%
\begin{equation}
\begin{array}{c}
\varpi \left( t,x,y,\bar{t},\bar{x},\bar{y}\right) \\ 
\\ 
=\frac{\left( \det \left( \mathfrak{C}\right) \right) ^{1/2}}{2\pi }%
\int_{-\infty }^{\infty }\int_{-\infty }^{\infty }G\left( T,k,l\right) \\ 
\\ 
\times \exp \left( ik\left( \bar{x}-x-\bar{y}T+\frac{bT^{2}}{2}\right)
+il\left( \bar{y}-y-bT\right) \right) dkdl,%
\end{array}
\label{Eq3.42}
\end{equation}%
where%
\begin{equation}
\begin{array}{c}
G\left( T,k,l\right) =\frac{1}{2\pi \left( \det \left( \mathfrak{C}\right)
\right) ^{1/2}}\exp \left( -\frac{1}{2}\left( 
\begin{array}{c}
k \\ 
l%
\end{array}%
\right) \cdot \mathfrak{C}^{-1}\left( T\right) \left( 
\begin{array}{c}
k \\ 
l%
\end{array}%
\right) \right) ,%
\end{array}
\label{Eq3.43}
\end{equation}%
and%
\begin{equation}
\begin{array}{c}
\mathfrak{C}\left( T\right) \mathbb{=}\left( 
\begin{array}{cc}
\frac{12}{\sigma ^{2}T^{3}} & \frac{6}{\sigma ^{2}T^{2}} \\ 
\frac{6}{\sigma ^{2}T^{2}} & \frac{4}{\sigma ^{2}T}%
\end{array}%
\right) , \\ 
\\ 
\det \left( \mathfrak{C}\left( T\right) \right) =\frac{12}{\sigma ^{4}T^{4}}.%
\end{array}
\label{Eq3.44}
\end{equation}%
As before, $\cdot $ denotes the scalar product. Accordingly,%
\begin{equation}
\begin{array}{c}
\varpi \left( t,x,y,\bar{t},\bar{x},\bar{y}\right) =\frac{\sqrt{3}}{\pi
\sigma ^{2}T^{2}}\exp \left( -\Omega \left( t,x,y,\bar{t},\bar{x},\bar{y}%
\right) \right) ,%
\end{array}
\label{Eq3.45}
\end{equation}%
where%
\begin{equation}
\begin{array}{c}
\Omega \left( t,x,y,\bar{t},\bar{x},\bar{y}\right) = \\ 
\\ 
=\frac{1}{2}\left( 
\begin{array}{c}
\bar{x}-x-\bar{y}T+\frac{bT^{2}}{2} \\ 
\bar{y}-y-bT%
\end{array}%
\right) \cdot \mathfrak{C}\left( 
\begin{array}{c}
\bar{x}-x-\bar{y}T+\frac{bT^{2}}{2} \\ 
\bar{y}-y-bT%
\end{array}%
\right) \\ 
\\ 
=\frac{6\left( \bar{x}-x-\bar{y}T+\frac{bT^{2}}{2}\right) ^{2}}{\sigma
^{2}T^{3}}+\frac{6\left( \bar{x}-x-\bar{y}T+\frac{bT^{2}}{2}\right) \left( 
\bar{y}-y-bT\right) }{\sigma ^{2}T^{2}}+\frac{2\left( \bar{y}-y-bT\right)
^{2}}{\sigma ^{2}T} \\ 
\\ 
=\frac{A^{2}}{2}+6B^{2}=\Phi \left( \bar{t},\bar{x},\bar{y}\,,x,y\right) ,%
\end{array}
\label{Eq3.46}
\end{equation}%
as expected. This calculation completes the derivation of the corrected
Kolmogorov formula.

Note that the t.p.d.f. $\varpi $ is a bivariate Gaussian distribution.
Completing the square, one can write:%
\begin{equation}
\begin{array}{c}
\Phi =\frac{\left( \bar{y}-y-bT\right) ^{2}}{2\sigma ^{2}T}+\frac{6\left( 
\bar{x}-x-\frac{\left( \bar{y}+y\right) T}{2}\right) ^{2}}{\sigma ^{2}T^{3}}
\\ 
\\ 
=\frac{6}{\sigma ^{2}T^{3}}\left( \bar{x}-p\right) ^{2}-\frac{6}{\sigma
^{2}T^{2}}\left( \bar{x}-p\right) \left( \bar{y}-q\right) +\frac{2}{\sigma
^{2}T}\left( \bar{y}-q\right) ^{2},%
\end{array}
\label{Eq3.47}
\end{equation}%
and represent $\varpi $ the form:%
\begin{equation}
\begin{array}{c}
\varpi \left( t,x,y,\bar{t},\bar{x},\bar{y}\right) =\frac{\exp \left( -\frac{%
1}{2\left( 1-\rho ^{2}\right) }\left( \frac{\left( \bar{x}-p\right) ^{2}}{%
\sigma _{x}^{2}}-\frac{2\rho \left( \bar{x}-p\right) \left( \bar{y}-q\right) 
}{\sigma _{x}\sigma _{y}}+\frac{\left( \bar{y}-q\right) ^{2}}{\sigma _{y}^{2}%
}\right) \right) }{2\pi \sigma _{x}\sigma _{y}\sqrt{1-\rho ^{2}}},%
\end{array}
\label{Eq3.48}
\end{equation}%
where%
\begin{equation}
\begin{array}{c}
\sigma _{x}=\sqrt{\frac{\sigma ^{2}T^{3}}{3}},\ \ \ \sigma _{y}=\sqrt{\sigma
^{2}T},\ \ \ \rho =\frac{\sqrt{3}}{2}, \\ 
\\ 
p=x+yT+\frac{bT^{2}}{2},\ \ q=y+bT.%
\end{array}
\label{Eq3.49}
\end{equation}%
Eq. (\ref{Eq3.25}) can be derived by using the Hankel transform. Since%
\begin{equation}
\begin{array}{c}
\sigma ^{-2}\mathfrak{C}^{-1}\left( T\right) =\left( 
\begin{array}{cc}
\frac{T^{3}}{3} & -\frac{T^{2}}{2} \\ 
-\frac{T^{2}}{2} & T%
\end{array}%
\right) =\left( 
\begin{array}{cc}
\frac{T^{3/2}}{2} & -\frac{T^{1/2}}{2} \\ 
-\frac{T^{3/2}}{2\sqrt{3}} & \frac{\sqrt{3}T^{1/2}}{2}%
\end{array}%
\right) ^{\ast }\left( 
\begin{array}{cc}
\frac{T^{3/2}}{2} & -\frac{T^{1/2}}{2} \\ 
-\frac{T^{3/2}}{2\sqrt{3}} & \frac{\sqrt{3}T^{1/2}}{2}%
\end{array}%
\right) ,%
\end{array}
\label{Eq3.50}
\end{equation}%
one can introduce%
\begin{equation}
\begin{array}{c}
\left( 
\begin{array}{c}
\bar{k} \\ 
\bar{l}%
\end{array}%
\right) =\left( 
\begin{array}{cc}
\frac{T^{3/2}}{2} & -\frac{T^{1/2}}{2} \\ 
-\frac{T^{3/2}}{2\sqrt{3}} & \frac{\sqrt{3}T^{1/2}}{2}%
\end{array}%
\right) \left( 
\begin{array}{c}
k \\ 
l%
\end{array}%
\right) , \\ 
\\ 
\left( 
\begin{array}{c}
k \\ 
l%
\end{array}%
\right) =\left( 
\begin{array}{cc}
3T^{-3/2} & \sqrt{3}T^{-3/2} \\ 
T^{-1/2} & \sqrt{3}T^{-1/2}%
\end{array}%
\right) \left( 
\begin{array}{c}
\bar{k} \\ 
\bar{l}%
\end{array}%
\right) ,%
\end{array}
\label{Eq3.51}
\end{equation}%
and rewrite Eq. (\ref{Eq3.42}) as follows:%
\begin{equation}
\begin{array}{c}
\varpi \left( t,x,y,\bar{t},\bar{x},\bar{y}\right) =\frac{\sqrt{3}}{2\pi
^{2}T^{2}}\int_{-\infty }^{\infty }\int_{-\infty }^{\infty }\exp \left( -%
\frac{1}{2}\sigma ^{2}\left( \bar{k}^{2}+\bar{l}^{2}\right) \right. \\ 
\\ 
\left. +i\left( 
\begin{array}{c}
\bar{k} \\ 
\bar{l}%
\end{array}%
\right) \cdot \left( 
\begin{array}{cc}
3T^{-3/2} & T^{-1/2} \\ 
\sqrt{3}T^{-3/2} & \sqrt{3}T^{-1/2}%
\end{array}%
\right) \left( 
\begin{array}{c}
\bar{x}-x-\bar{y}T+\frac{bT^{2}}{2} \\ 
\bar{y}-y-bT%
\end{array}%
\right) \right) d\bar{k}d\bar{l} \\ 
\\ 
=\frac{\sqrt{3}}{2\pi ^{2}T^{2}}\int_{-\infty }^{\infty }\int_{-\infty
}^{\infty }\exp \left( -\frac{1}{2}\sigma ^{2}\left( \bar{k}^{2}+\bar{l}%
^{2}\right) \right. \\ 
\\ 
\left. +i\bar{k}\frac{3\left( \bar{x}-x-\bar{y}T+\frac{bT^{2}}{2}\right)
+\left( \bar{y}-y-bT\right) T}{T^{3/2}}+i\bar{l}\frac{\sqrt{3}\left( \bar{x}%
-x-\bar{y}T+\frac{bT^{2}}{2}\right) +\sqrt{3}\left( \bar{y}-y-bT\right) T}{%
T^{3/2}}\right) d\bar{k}d\bar{l}%
\end{array}
\label{Eq3.52a}
\end{equation}%
Thus, $\varpi \left( t,x,y,\bar{t},\bar{x},\bar{y}\right) $ is the Fourier
transform of a\emph{\ radially-symmetric} function of $\left( \bar{k},\bar{l}%
\right) ^{\ast }$. Accordingly, it can be calculated via the Hankel
transform of the function $\exp \left( -\sigma ^{2}\bar{r}^{2}/2\right) :$%
\begin{equation}
\begin{array}{c}
\varpi \left( t,x,y,\bar{t},\bar{x},\bar{y}\right) =\frac{\sqrt{3}}{\pi T^{2}%
}\mathcal{H}_{0}\left[ e^{-\frac{\sigma ^{2}\bar{r}^{2}}{2}}\right] \left( 
\bar{s}\right) =\frac{\sqrt{3}}{\pi \sigma ^{2}T^{2}}e^{-\frac{\bar{s}^{2}}{%
2\sigma ^{2}}},%
\end{array}
\label{Eq3.52b}
\end{equation}%
where%
\begin{equation}
\begin{array}{c}
\bar{r}^{2}=\bar{k}^{2}+\bar{l}^{2}, \\ 
\\ 
\bar{s}^{2}=\frac{4\left( 3\left( \bar{x}-x-\bar{y}T+\frac{bT^{2}}{2}\right)
^{2}+3\left( \bar{x}-x-\bar{y}T+\frac{bT^{2}}{2}\right) \left( \bar{y}%
-y-bT\right) T+\left( \bar{y}-y-bT\right) ^{2}T^{2}\right) }{T^{3}}.%
\end{array}
\label{Eq3.52c}
\end{equation}%
see, e.g., \cite{Piessens00}. As expected, the corresponding expression
coincides with the one given by Eq. (\ref{Eq3.48}).

\subsection{Solution via Coordinate Transform\label{Sec35}}

This section briefly considers the method of coordinate transformations,
reducing the original Fokker-Planck equation for the Kolmogorov problem to a
Fokker-Planck equation with spatially-independent coefficients. To this end,
the following ansatz is used:%
\begin{equation}
\begin{array}{c}
\left( \tilde{x},\tilde{y}\right) =\left( \bar{x}-\left( \bar{t}-t\right) 
\bar{y},\bar{y}\right) .%
\end{array}
\label{Eq3.53}
\end{equation}%
This choice is explained in more detail in Chapter \ref{Chap6}.
Straightforward calculation yields%
\begin{equation}
\begin{array}{c}
\frac{\partial }{\partial \bar{t}}=\frac{\partial }{\partial \bar{t}}-\tilde{%
y}\frac{\partial }{\partial \tilde{x}},\ \ \ \frac{\partial }{\partial \bar{x%
}}=\frac{\partial }{\partial \tilde{x}},\ \ \ \frac{\partial }{\partial y}%
=-\left( \bar{t}-t\right) \frac{\partial }{\partial \tilde{x}}+\frac{%
\partial }{\partial \tilde{y}},%
\end{array}
\label{Eq3.54}
\end{equation}%
so that Eq. (\ref{Eq3.24}) becomes%
\begin{equation}
\begin{array}{c}
\left( \frac{\partial }{\partial \bar{t}}-\tilde{y}\frac{\partial }{\partial 
\tilde{x}}\right) \varpi \left( t,x,y,\bar{t},\tilde{x},\tilde{y}\right) -%
\frac{1}{2}\sigma ^{2}\left( -\left( \bar{t}-t\right) \frac{\partial }{%
\partial \tilde{x}}+\frac{\partial }{\partial \tilde{y}}\right) ^{2}\varpi
\left( t,x,y,\bar{t},\tilde{x},\tilde{y}\right) \\ 
\\ 
+\tilde{y}\varpi _{\tilde{x}}\left( t,x,y,\bar{t},\tilde{x},\tilde{y}\right)
+b\left( -\left( \bar{t}-t\right) \frac{\partial }{\partial \tilde{x}}+\frac{%
\partial }{\partial \tilde{y}}\right) \varpi \left( t,x,y,\bar{t},\tilde{x},%
\tilde{y}\right) =0, \\ 
\\ 
\varpi \left( t,x,y,t,\tilde{x},\tilde{y}\right) =\delta \left( \tilde{x}%
-x\right) \delta \left( \tilde{y}-y\right) .%
\end{array}
\label{Eq3.55}
\end{equation}%
Further calculations show that coefficients of the above equation are
spatially independent:%
\begin{equation}
\begin{array}{c}
\frac{\partial }{\partial \bar{t}}\varpi \left( t,x,y,\bar{t},\tilde{x},%
\tilde{y}\right) -\frac{1}{2}\sigma ^{2}\left( -\left( \bar{t}-t\right) 
\frac{\partial }{\partial \tilde{x}}+\frac{\partial }{\partial \tilde{y}}%
\right) ^{2}\varpi \left( t,x,y,\bar{t},\tilde{x},\tilde{y}\right) \\ 
\\ 
+b\left( -\left( \bar{t}-t\right) \frac{\partial }{\partial \tilde{x}}+\frac{%
\partial }{\partial \tilde{y}}\right) \varpi \left( t,x,y,\bar{t},\tilde{x},%
\tilde{y}\right) =0, \\ 
\\ 
\varpi \left( t,x,y,t,\tilde{x},\tilde{y}\right) =\delta \left( \tilde{x}%
-x\right) \delta \left( \tilde{y}-y\right) .%
\end{array}
\label{Eq3.56}
\end{equation}%
Accordingly, one can use the classical Fourier transform and represent the
solution of Eq. (\ref{Eq3.56}) in the form%
\begin{equation}
\begin{array}{c}
\varpi \left( t,x,y,\bar{t},\tilde{x},\tilde{y}\right) \\ 
\\ 
=\frac{1}{\left( 2\pi \right) ^{2}}\int_{-\infty }^{\infty }\int_{-\infty
}^{\infty }\exp \left( -\frac{1}{2}\sigma ^{2}\left( \frac{k^{2}T^{3}}{3}%
-klT^{2}+l^{2}T\right) \right. \\ 
\\ 
\left. +ik\left( \tilde{x}-x+\frac{bT^{2}}{2}\right) +il\left( \tilde{y}%
-y-bT\right) \right) dkdl,%
\end{array}
\label{Eq3.57}
\end{equation}%
similar to Eq. (\ref{Eq3.41}). Thus, $\varpi $ has the form given by Eq. (%
\ref{Eq3.48}) with $\left( \bar{x},\bar{y}\right) $ replaced by \ $\left( 
\tilde{x},\tilde{y}\right) $. The exact form is recovered once $\left( 
\tilde{x},\tilde{y}\right) $ are expressed in terms of $\left( \bar{x},\bar{y%
}\right) $ by virtue of Eq. (\ref{Eq3.53}).

\subsection{A Representative Example\label{Sec36}}

A typical solution of the Kolmogorov equation is illustrated in Figure \ref%
{Fig3.2}. 
\begin{figure}[tbp]
\begin{center}
\includegraphics[width=0.8\textwidth]
{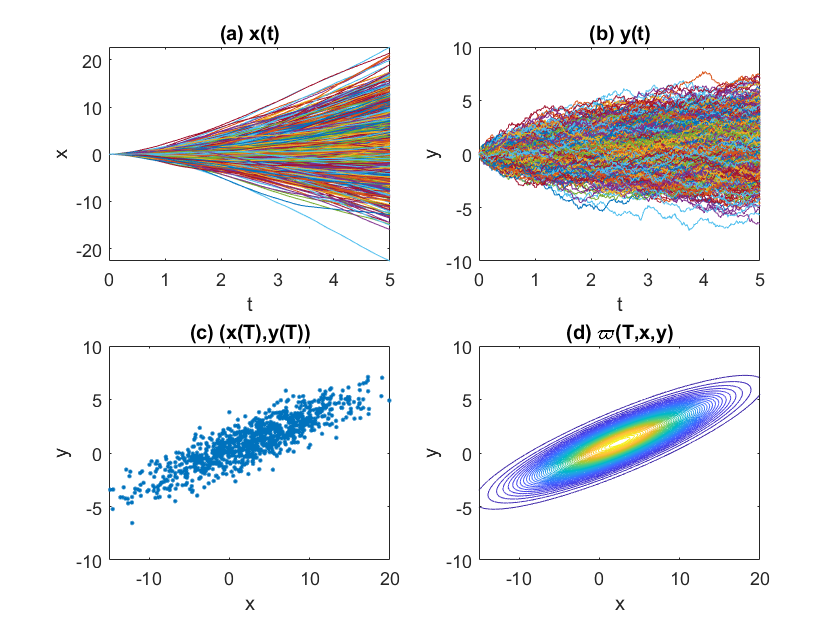}
\end{center}
\par
\vspace{-10pt}
\caption{A thousand trajectories of a typical Kolmogorov process. Parameters
are as follows: $T=5$, $dt=0.01$, $f=0.2$, $\protect\sigma =0.8$. (a) $%
x\left( t\right) $, (b) $y\left( t\right) $, (c) $\left( \bar{x}\left(
T\right) ,\bar{y}\left( T\right) \right) $, (d) contour lines of $\protect%
\varpi \left( 0,0,0,T,\tilde{x},\tilde{y}\right) $. Own graphics.}
\label{Fig3.2}
\end{figure}
This Figure clearly shows that there is a good agreement between a Monte
Carlo simulation of the stochastic process $\left( \hat{x}_{t},\hat{y}%
_{t}\right) $ given by Eqs (\ref{Eq3.23}) and the corrected Kolmogorov
formula (\ref{Eq3.21}).

\section{Klein-Kramers Stochastic Process\label{Chap4}}

\subsection{Background\label{Sec40}}

The Klein-Kramers equation plays a vital role in statistical physics by
offering a detailed mathematical framework for studying the dynamics of
particles in a viscous, random medium. Specifically, it describes the
evolution of the t.p.d.f. of a particle's momentum and position in the phase
plane, accounting for deterministic forces arising from potential and
stochastic thermal forces arising from random collisions with the medium's
molecules. This equation is particularly important for studying
non-equilibrium systems, which cannot be analyzed via traditional
equilibrium statistical mechanics tools. By incorporating frictional forces,
which tend to dampen the motion of particles, potential forces, which push
them deterministically, and random thermal forces, which inject randomness
into the system, the Klein-Kramers equation bridges the gap between
microscopic laws of motion and the macroscopic observable phenomena, such as
diffusion, thermal conductivity, and viscosity. Moreover, the Klein-Kramers
equation serves as a foundation for exploring more complex phenomena in
non-equilibrium statistical mechanics, including the study of transition
state theory in macrokinetics of chemical reactions, the behavior of
particles in external fields, and the exploration of noise-induced
transitions and stochastic resonance in physical and biological systems. It
also arises in financial engineering, for instance, in pricing volatility
and variance swaps.

\subsection{Langevin Equation\label{Sec41}}

Start with the Langevin equation for particles moving in a potential field
and impacted by random forces; see \cite{Langevin08}. This section uses the
standard notation, rather that the original notation used in \cite%
{Chandrasekhar43}. Hopefully, the diligent reader will not be easily
confused. The stochastic Langevin equation describes the evolution of
systems under the influence of deterministic forces and random fluctuations.
Because of its versatility, it is widely used in physics and other
disciplines to model the dynamics of particles subjected to systematic
forces derived from potential energy and random forces representing thermal
fluctuations. This equation describes a particle experiencing frictional
resistance proportional to its velocity (a deterministic component) and
random kicks from the surrounding molecules (a stochastic component
capturing the essence of Brownian motion). The Langevin equation thus
provides a robust framework for studying the behavior of systems subject to
noise, enabling insights into phenomena such as diffusion, thermal
equilibrium, and the statistical properties of microscopic systems.

Consider an underdamped Brownian particle. In contrast to the standard
Brownian motion, which is overdamped, it is assumed that the frictions are
finite, so that one must treat particle's velocity as an independent degree
of freedom. Hence, particle's state is described by a pair $\left( \mathbf{x}%
,\mathbf{y}\right) $, where $\mathbf{x}$ and $\mathbf{y}$ are its position
and velocity, respectively. Consider a $d$-dimensional space, with $d=1$ and 
$d=3$ of particular interest and write the corresponding Langevin equations
in the form:%
\begin{equation}
\begin{array}{c}
\frac{d\mathbf{\hat{x}}_{t}}{dt}=\mathbf{\hat{y}}_{t},\ \ \ \mathbf{\hat{x}}%
_{t}=\mathbf{x,} \\ 
\\ 
\frac{d\mathbf{\hat{y}}_{t}}{dt}=-\kappa \mathbf{\hat{y}}_{t}\mathbf{-}\frac{%
\nabla V\left( \mathbf{\hat{x}}_{t}\right) }{m}+\sqrt{\frac{2\kappa k_{B}T}{m%
}}\frac{d\mathbf{\hat{W}}_{t}}{dt},\ \ \mathbf{\hat{y}}_{t}=\mathbf{y},\ 
\end{array}
\label{Eq4.1}
\end{equation}%
where $\mathbf{\hat{W}}_{t}$ is a standard $d$-dimensional Wiener process.
Here $m$ is the particle mass, $\kappa $ is the friction coefficient, $k_{B}$
is the Boltzmann constant, $T$ is the temperature, $V\left( \mathbf{x}%
\right) $ is the external potential, $d\mathbf{\hat{W}}_{t}/dt$ is a $d$%
-dimensional Gaussian white noise. Below, the ratio $\kappa k_{B}T/m$ is
denoted as $a$.

Of course, one can rewrite Eqs (\ref{Eq4.1}) as a system of stochastic
differential equations (SDEs):%
\begin{equation}
\begin{array}{c}
d\mathbf{\hat{x}}_{t}=\mathbf{\hat{y}}_{t}dt,\ \ \ \mathbf{\hat{x}}_{t}=%
\mathbf{x,} \\ 
\\ 
d\mathbf{\hat{y}}_{t}=-\kappa \mathbf{\hat{y}}_{t}dt\mathbf{-}\frac{\nabla
V\left( \mathbf{\hat{x}}_{t}\right) }{m}dt+\sqrt{2a}d\mathbf{\hat{W}}_{t},\
\ \ \mathbf{\hat{y}}_{t}=\mathbf{y}.%
\end{array}
\label{Eq4.2}
\end{equation}%
For a $1$-dimensional particle, Eq. (\ref{Eq4.2}) becomes:%
\begin{equation}
\begin{array}{c}
d\hat{x}_{t}=\hat{y}_{t}dt,\ \ \ \hat{x}_{t}=x\mathbf{,} \\ 
\\ 
d\hat{y}_{t}=-\kappa \hat{y}_{t}dt\mathbf{-}\frac{V_{x}\left( \hat{x}%
_{t}\right) }{m}dt+\sqrt{2a}d\hat{W}_{t},\ \ \ \hat{y}_{t}=y.%
\end{array}
\label{Eq4.2b}
\end{equation}%
It is clear that the Kolmogorov equation (\ref{Eq3.10a}) is a special case
of Eq. (\ref{Eq4.2b}) with $\kappa =0$, $V\left( x\right) =mfx$, $k=a$.

\subsection{Klein-Kramers Equation\label{Sec42}}

Fokker, Planck, and their numerous followers derived and studied the forward
parabolic equation for the t.p.d.f. $\varpi \left( t,\mathbf{x,y},\bar{t},%
\mathbf{\bar{x},\bar{y}}\right) $ associated with a stochastic process. For
the stochastic process governed by SDEs (\ref{Eq4.2}), the corresponding
equation, called the Klein-Kramers equation, has the form:%
\begin{equation}
\begin{array}{c}
\varpi _{\bar{t}}-a\varpi _{\mathbf{\bar{y}\bar{y}}}+\mathbf{\bar{y}}\varpi
_{\mathbf{\bar{x}}}-\left( \left( \kappa \mathbf{\bar{y}}+\frac{\nabla
V\left( \mathbf{\bar{x}}\right) }{m}\right) \varpi \right) _{\mathbf{\bar{y}}%
}=0, \\ 
\\ 
\varpi \left( t,\mathbf{x,y},t,\mathbf{\bar{x},\bar{y}}\right) =\delta
\left( \mathbf{\bar{x}}-\mathbf{x}\right) \delta \left( \mathbf{\bar{y}}-%
\mathbf{y}\right) .%
\end{array}
\label{Eq4.3}
\end{equation}%
The backward parabolic Kolmogorov equation can be written as follows:%
\begin{equation}
\begin{array}{c}
\varpi _{t}+a\varpi _{\mathbf{yy}}+\mathbf{y}\varpi _{\mathbf{x}}-\left(
\kappa \mathbf{y}+\frac{\nabla V\left( \mathbf{x}\right) }{m}\right) \varpi
_{\mathbf{y}}=0, \\ 
\\ 
\varpi \left( \bar{t},\mathbf{x,y},\bar{t},\mathbf{\bar{x},\bar{y}}\right)
=\delta \left( \mathbf{\bar{x}}-\mathbf{x}\right) \delta \left( \mathbf{\bar{%
y}}-\mathbf{y}\right) .%
\end{array}
\label{Eq4.4}
\end{equation}%
Details are given in \cite{Fokker14, Planck17, Klein21, Chapman28,
Kolmogorov31, Kolmogorov33, Kolmogorov34, Kramers40, Chandrasekhar43,
Risken89, Hanggi90}, as well as a multitude of subsequent sources. For
fascinating historical details see \cite{Ebeling08}. Klein-Kramers Equation
(occasionally called Klein-Kramers-Chandrasekhar equation) describes the
dynamics of a particle's probability distribution in phase space (position
and momentum) for systems subjected to friction and random forces, typically
at the mesoscopic scale. The Klein-Kramers equation provides a comprehensive
framework for modeling and understanding complex systems far from
equilibrium, linking microscopic physics with macroscopic observables.
Accordingly, it is used in various fields, such as materials science,
chemistry, and astrophysics, to predict the evolution of systems over time,
accounting for both deterministic dynamics and the effects of randomness.

\subsection{Chandrasekhar's Solutions\label{Sec43}}

In a well-known survey article, \cite{Chandrasekhar43} described elegant
solutions of Eq. (\ref{Eq4.3}) for a free particle and a harmonically bound
particle, which he derived by using ingenious changes of coordinates. For a
free particle, \cite{Chandrasekhar43} writes the corresponding Klein-Kramers
equation as follows:%
\begin{equation}
\begin{array}{c}
\varpi _{\bar{t}}-a\varpi _{\bar{y}\bar{y}}+\bar{y}\varpi _{\bar{x}}-\kappa 
\bar{y}\varpi _{\bar{y}}-\kappa \varpi =0, \\ 
\\ 
\varpi \left( t,x\mathbf{,}y,t,\bar{x}\mathbf{,}\bar{y}\right) =\delta
\left( \bar{x}-x\right) \delta \left( \bar{y}-y\right) .%
\end{array}
\label{Eq4.5}
\end{equation}%
By using ingenious coordinate transforms, he shows that%
\begin{equation}
\begin{array}{c}
\varpi =\frac{1}{2\pi \left( FG-H^{2}\right) ^{1/2}}\exp \left( -\frac{%
\left( FR^{2}-2HRS+GS^{2}\right) }{2\left( FG-H^{2}\right) }\right) ,%
\end{array}
\label{Eq4.6}
\end{equation}%
where%
\begin{equation}
\begin{array}{c}
R=\bar{y}-e^{-\kappa T}y, \\ 
\\ 
S=\left( \bar{x}-x\right) -\frac{\left( 1-e^{-\kappa T}\right) }{\kappa }y,
\\ 
\\ 
F=\frac{a}{\kappa ^{3}}\left( -3+4e^{-\kappa T}-e^{-2\kappa T}+2\kappa
T\right) , \\ 
\\ 
G=\frac{a}{\kappa }\left( 1-e^{-2\kappa T}\right) , \\ 
\\ 
H=\frac{a}{\kappa ^{2}}\left( 1-e^{-\kappa T}\right) ^{2}.%
\end{array}
\label{Eq4.7}
\end{equation}%
Here the original Chandrasekhar's notation is slightly changed to make the
exposition more internally consistent.

Since it is assumed that stochastic drivers are uncorrelated, the t.p.d.f. $%
\varpi ^{\left( 3\right) }$ can be presented as a product of three $1$%
-dimensional t.p.d.f. $\varpi ^{\left( 1\right) }$:%
\begin{equation}
\begin{array}{c}
\varpi ^{\left( 3\right) }=\varpi _{1}^{\left( 1\right) }\varpi _{2}^{\left(
1\right) }\varpi _{3}^{\left( 1\right) }=\frac{1}{8\pi ^{3}\left(
FG-H^{2}\right) ^{3/2}}\exp \left( -\frac{\left( F\left\vert \mathbf{R}%
\right\vert ^{2}-2H\mathbf{R}\cdot \mathbf{S}+G\left\vert \mathbf{S}%
\right\vert ^{2}\right) }{2\left( FG-H^{2}\right) }\right) ,%
\end{array}
\label{Eq4.8}
\end{equation}%
where $\left\vert \mathbf{R}\right\vert ^{2}=\left( \bar{y}_{1}-e^{-\kappa
T}y_{1}\right) ^{2}+\left( \bar{y}_{2}-e^{-\kappa T}y_{2}\right) ^{2}+\left( 
\bar{y}_{3}-e^{-\kappa T}y_{3}\right) ^{2}$, etc.

Chandrasekhar generalized Eq. (\ref{Eq4.6}) to the case of harmonically
bound particles. We shall revisit Chandrasekhar's formulas for free and
bound particles below. While \cite{Chandrasekhar43} stopped at Eq. (\ref%
{Eq4.6}), for practical applications, it is more useful to represent the
exponent as an explicit quadratic form of $\bar{x}$ and $\bar{y}$, which is
done in Section \ref{Sec65}.

\section{Transition Probability Densities for Stochastic Processes\label%
{Chap5}}

\subsection{Motivation\label{Sec51}}

The problems considered in Chapters \ref{Chap3} and \ref{Chap4} are used
below to develop a general theory. For that, one needs to know some
foundational information about stochastic processes discussed in this
Chapter. Stochastic processes play a crucial role across various scientific
disciplines, which is fundamental for modeling systems influenced by
randomness and uncertainty. These processes are pivotal in fields ranging
from physics and chemistry to biology, economics, and financial engineering.
They help to understand phenomena where outcomes are not deterministic but
probabilistic, capturing the dynamics of complex systems over time. The
analysis of stochastic processes enables scientists and engineers to predict
behavior, assess risks, and make informed decisions based on the likelihood
of future events.

The backward Kolmogorov and forward Fokker-Planck equations offer a
mathematical description of how systems evolve under the influence of
stochastic factors. This capability to model the t.p.d.fs of diverse
processes underlines the equations' fundamental importance in scientific
research and practical applications across disciplines.

The Kolmogorov and Fokker-Planck equations are adjoint partial differential
equations that describe how the probability density of a system's state
evolves in time. The Kolmogorov equation focuses on calculating the expected
value at a given time of random outcomes, which become known sometime in the
future. Conversely, the Fokker-Planck equation is concerned with the
evolution of the conditional probability density function of a process's
state at a future time, given its current state.

The Kolmogorov and Fokker-Planck equations are applied in physics and
chemistry to study the random motion of particles in fluids, the statistical
behavior of thermodynamic systems, and the kinetics of chemical reactions.
In biology, these equations model population dynamics, genetic variation,
and the spread of diseases, among other processes, providing insights into
how randomness affects biological phenomena. In financial engineering, they
are used to model the evolution of asset prices, interest rates, and other
economic indicators, underpinning the valuation of derivatives and the
management of financial risks.

\subsection{Backward and Forward Equations\label{Sec52}}

Start with a jump-diffusion process driven by the SDE of the form%
\begin{equation}
\begin{array}{c}
d\hat{z}_{t}=b\left( t,\hat{z}_{t}\right) dt+\sigma \left( t,\hat{z}%
_{t}\right) d\hat{W}_{t}+\upsilon d\hat{\Pi}_{t}\left( t,\hat{z}_{t}\right)
,\ \ \ \hat{z}_{t}=z,%
\end{array}
\label{Eq5.1}
\end{equation}%
with smooth coefficients $b,\sigma $. This process is driven by the standard
Wiener process $\hat{W}_{t}$ and the Poisson process $\hat{\Pi}_{t}\left(
t,z\right) $ with intensity $\lambda \left( t,z\right) $ such that 
\begin{equation}
\begin{array}{c}
\mathrm{E}\left\{ \left. d\Pi _{t}\left( t,z\right) \right\vert \hat{z}%
_{t}=z\right\} =\lambda \left( t,z\right) dt,%
\end{array}
\label{Eq5.2}
\end{equation}%
while $\upsilon $ is drawn from a distribution with density $\phi \left(
\upsilon ,t,z\right) $, which (in general) is $\left( t,z\right) $-dependent.

More generally, it is possible to consider the so-called general compound or
marked Poisson processes, such that $\upsilon =\upsilon \left( t,z,q\right) $%
, where $\upsilon $ is monotonic in $z$, and $q$ is a random mark variable
drawn from a distribution with density $\phi \left( q,t,z\right) $, which
(in general) is $\left( t,z\right) $-dependent. However, since this book is
interested in a particular class of stochastic processes, solvable via
Kelvin waves ansatz this generalization is not particularly useful.

It is well-known that for suitable test functions $\tilde{u}\left( z\right) $
the expectation%
\begin{equation}
\begin{array}{c}
u\left( t,z\right) =E\left\{ \left. \tilde{u}\left( \hat{z}_{\bar{t}}\right)
\right\vert \hat{z}_{t}=z\right\} ,%
\end{array}
\label{Eq5.3}
\end{equation}%
solves the following integro-differential backward Kolmogorov problem:%
\begin{equation}
\begin{array}{c}
u_{t}\left( t,z\right) +a\left( t,z\right) u_{zz}\left( t,z\right) +b\left(
t,z\right) u_{z}\left( t,z\right) \\ 
\\ 
+\lambda \left( t,z\right) \int\limits_{-\infty }^{\infty }u\left(
t,z+\upsilon \right) \phi \left( \upsilon ,t,z\right) d\upsilon -\lambda
\left( t,z\right) u\left( t,z\right) =0, \\ 
\\ 
u\left( \bar{t},z\right) =\tilde{u}\left( z\right) ,%
\end{array}
\label{Eq5.4}
\end{equation}%
where%
\begin{equation}
\begin{array}{c}
a\left( t,z\right) =\frac{1}{2}\sigma ^{2}\left( t,z\right) .%
\end{array}
\label{Eq5.4a}
\end{equation}
In particular, the t.p.d.f. $\varpi \left( t,z,\bar{t},\bar{z}\right) $ such
that 
\begin{equation}
\begin{array}{c}
\mathrm{\Pr ob}\left\{ \left. \bar{z}<\hat{z}_{\bar{t}}<\bar{z}+d\bar{z}%
\right\vert \hat{z}_{t}=z\right\} =\varpi \left( t,z,\bar{t},\bar{z}\right) d%
\bar{z},%
\end{array}
\label{Eq5.5}
\end{equation}%
solves the following backward Kolmogorov problem:%
\begin{equation}
\begin{array}{c}
\varpi _{t}\left( t,z\right) +a\left( t,z\right) \varpi _{zz}\left(
t,z\right) +b\left( t,z\right) \varpi _{z}\left( t,z\right) - \\ 
\\ 
+\lambda \left( t,z\right) \int\limits_{-\infty }^{\infty }\varpi \left(
t,z+\upsilon \right) \phi \left( \upsilon ,t,z\right) d\upsilon -\lambda
\left( t,z\right) \varpi \left( t,z\right) =0, \\ 
\\ 
\varpi \left( \bar{t},z,\bar{t},\bar{z}\right) =\delta \left( z-\bar{z}%
\right) .%
\end{array}
\label{Eq5.6}
\end{equation}
It is possible to derive a forward problem for $\varpi \left( t,z,\bar{t},%
\bar{z}\right) $, which $\varpi $ satisfies as a function of $\left( \bar{t},%
\bar{z}\right) $, which is called Fokker-Planck or forward Kolmogorov
problem. This problem has the form:%
\begin{equation}
\begin{array}{c}
\varpi _{\bar{t}}\left( \bar{t},\bar{z}\right) -\left( a\left( \bar{t},\bar{z%
}\right) \varpi \left( \bar{t},\bar{z}\right) \right) _{\bar{z}\bar{z}%
}+\left( b\left( \bar{t},\bar{z}\right) \varpi \left( \bar{t},\bar{z}\right)
\right) _{\bar{z}} \\ 
\\ 
-\int\limits_{-\infty }^{\infty }\lambda \left( \bar{t},\bar{z}-\upsilon
\right) \varpi \left( \bar{t},\bar{z}-\upsilon \right) \phi \left( \bar{t},%
\bar{z}-\upsilon ,\upsilon \right) d\upsilon +\lambda \left( \bar{t},\bar{z}%
\right) \varpi \left( \bar{t},\bar{z}\right) =0, \\ 
\\ 
\varpi \left( t,z,t,\bar{z}\right) =\delta \left( \bar{z}-z\right) .%
\end{array}
\label{Eq5.7}
\end{equation}
One can generalize backward Kolmogorov and forward Fokker-Planck equation to
the multi-dimensional case. The underlying $n_{z}$-dimensional process $%
\mathbf{\hat{z}}_{t}=\left[ \hat{z}_{i,t}\right] $, $i=1,...,n_{z}$, has the
form%
\begin{equation}
\begin{array}{c}
d\mathbf{\hat{z}}_{t}=\mathbf{b}\left( t,\mathbf{\hat{z}}_{t}\right) dt+%
\mathbf{\Sigma }\left( t,\mathbf{\hat{z}}_{t}\right) d\mathbf{\hat{W}}_{t}+%
\mathbf{\upsilon }d\mathbf{\hat{\Pi}}_{t}\left( t,\mathbf{z}_{t}\right) ,%
\end{array}
\label{Eq5.8}
\end{equation}%
where \textbf{$\hat{W}$}$_{t}=\left[ \hat{W}_{j,t}\right] $ is an $n_{W}$%
-dimensional Wiener process, $j=1,...,n_{W}$, and $\mathbf{\hat{\Pi}}=\left[ 
\hat{\Pi}_{k,t}\right] $ is an $n_{\Pi }$-dimensional state-dependent
Poisson process, $k=1,...,n_{\Pi }$ with intensity $\mathbf{\lambda }$. The
corresponding state-dependent coefficients are as follows: 
\begin{equation}
\begin{array}{c}
\mathbf{b}\left( t,\mathbf{z}\right) =\left[ b_{i}\left( t,\mathbf{z}\right) %
\right] ,\ \ \ i=1,...,n_{z}, \\ 
\\ 
\mathbf{\Sigma }\left( t,\mathbf{z}\right) =\left[ \Sigma _{ij}\left( t,%
\mathbf{z}\right) \right] ,\ \ \ \ i=1,...,n_{z},\ \ \ \ j=1,...,n_{W}, \\ 
\\ 
\mathbf{\lambda }\left( t,\mathbf{z}\right) =\left[ \lambda _{i}\left( t,%
\mathbf{z}\right) \right] ,\ \ \ i=1,...,n_{\Pi }, \\ 
\\ 
\mathbf{\upsilon }=\left[ \upsilon _{ij}\right] ,\ \ \ \ i=1,...,n_{z},\ \ \
\ k=1,...,n_{\Pi },%
\end{array}
\label{Eq5.9}
\end{equation}%
while $\upsilon _{k}$ are drawn from distributions with densities $\phi
_{k}\left( \upsilon ,t,\mathbf{z}\right) $, which (in general) are $\left( t,%
\mathbf{z}\right) $-dependent. Explicitly, Eqs (\ref{Eq5.8}) can be written
as follows:%
\begin{equation}
\begin{array}{c}
d\hat{z}_{i,t}=b_{i}\left( t,\mathbf{\hat{z}}_{t}\right) dt+\mathbf{\Sigma }%
_{ij}\left( t,\mathbf{\hat{z}}_{t}\right) d\hat{W}_{j,t}+\upsilon _{ik}d\hat{%
\Pi}_{k}\left( t,\mathbf{\hat{z}}_{t}\right) .%
\end{array}
\label{Eq5.8a}
\end{equation}
The backward and forward equations for the t.p.d.f. $\varpi $ can be written
as follows:%
\begin{equation}
\begin{array}{c}
\varpi _{t}\left( t,\mathbf{z}\right) +a_{ij}\left( t,\mathbf{z}\right)
\varpi _{z_{i}z_{j}}\left( t,\mathbf{z}\right) +b_{i}\left( t,\mathbf{z}%
\right) \varpi _{z_{i}}\left( t,\mathbf{z}\right) \\ 
\\ 
+\lambda _{k}\left( t,\mathbf{z}\right) \int\limits_{-\infty }^{\infty
}\varpi \left( t,\mathbf{z}+\mathbf{\upsilon }_{k}\right) \phi _{k}\left( 
\mathbf{\upsilon }_{k},t,z\right) d\mathbf{\upsilon }_{k}-\Lambda \left( t,%
\mathbf{z}\right) \varpi \left( t,\mathbf{z}\right) =0, \\ 
\\ 
\varpi \left( \bar{t},\mathbf{z},\bar{t},\mathbf{\bar{z}}\right) =\delta
\left( \mathbf{z}-\mathbf{\bar{z}}\right) ,%
\end{array}
\label{Eq5.10}
\end{equation}%
\begin{equation}
\begin{array}{c}
\mathfrak{A}\left( t,\mathbf{z}\right) =\left( a_{ii^{\prime }}\left( t,%
\mathbf{z}\right) \right) =\frac{1}{2}\mathbf{\Sigma \left( t,z\right)
\Sigma }^{\ast }\left( t,\mathbf{z}\right) =\frac{1}{2}\sigma _{ij}\left( t,%
\mathbf{z}\right) \sigma _{i^{\prime }j}\left( t,\mathbf{z}\right) , \\ 
\\ 
\Lambda \left( t,\mathbf{z}\right) =\sum\limits_{k=1}^{n_{\Pi }}\lambda
_{k}\left( t,\mathbf{z}\right) ,%
\end{array}
\label{Eq5.11}
\end{equation}%
For the generic terminal condition $\tilde{u}\left( \mathbf{z}\right) $, the
corresponding backward problem has the form:%
\begin{equation}
\begin{array}{c}
u_{t}\left( t,\mathbf{z}\right) +a_{ij}\left( t,\mathbf{z}\right)
u_{z_{i}z_{j}}\left( t,\mathbf{z}\right) +b_{i}\left( t,z\right)
u_{z_{i}}\left( t,\mathbf{z}\right) \\ 
\\ 
+\lambda _{k}\left( t,\mathbf{z}\right) \int\limits_{-\infty }^{\infty
}u\left( t,\mathbf{z}+\upsilon _{k}\right) \phi _{k}\left( \upsilon _{k},t,%
\mathbf{z}\right) d\upsilon _{k}-\Lambda \left( t,\mathbf{z}\right) u\left(
t,\mathbf{z}\right) =0, \\ 
\\ 
u\left( \bar{t},\mathbf{z}\right) =\tilde{u}\left( \mathbf{z}\right) ,%
\end{array}
\label{Eq5.12}
\end{equation}%
The forward equations for the t.p.d.f. $\varpi $ can be written as follows:%
\begin{equation}
\begin{array}{c}
\varpi _{\bar{t}}\left( \bar{t},\mathbf{\bar{z}}\right) -\left( a_{ij}\left( 
\bar{t},\mathbf{\bar{z}}\right) \varpi \left( \bar{t},\mathbf{\bar{z}}%
\right) \right) _{\bar{z}_{i}\bar{z}_{j}}+\left( b_{i}\left( \bar{t},\mathbf{%
\bar{z}}\right) \varpi \left( \bar{t},\mathbf{\bar{z}}\right) \right) _{\bar{%
z}_{i}} \\ 
\\ 
-\int\limits_{-\infty }^{\infty }\lambda _{k}\left( \bar{t},\mathbf{\bar{z}}%
-\upsilon _{k}\right) \varpi \left( \bar{t},\mathbf{\bar{z}}-\mathbf{%
\upsilon }_{k}\right) \phi _{k}\left( \upsilon _{k},\bar{t},\mathbf{\bar{z}}-%
\mathbf{\upsilon }_{k}\right) d\mathbf{\upsilon }_{k}+\Lambda \left( \bar{t},%
\mathbf{\bar{z}}\right) \varpi \left( \bar{t},\mathbf{\bar{z}}\right) =0, \\ 
\\ 
\varpi \left( t,\mathbf{z},t,\mathbf{\bar{z}}\right) =\delta \left( \mathbf{%
\bar{z}}-\mathbf{z}\right) .%
\end{array}
\label{Eq5.13}
\end{equation}
Further details can be found in \cite{Bharucha60, Feller71, Gihman72,
Arnold74, Hanson07}, among others.

Although, depending on the actual problem at hand, it might be preferable to
work with either backward or forward problem, experience suggests that in
the context of mathematical finance the backward problem is easier to deal
with, not least because they are meaningful for the generic terminal value $%
\tilde{u}\left( \bar{z}\right) $.

Since the above definitions are very general, it is necessary to be more
specific in defining the class of problems which can be solved by using
Kelvin waves. Consider processes, such that%
\begin{equation}
\begin{array}{c}
\mathfrak{A}\left( t,\mathbf{z}\right) =\mathfrak{A}^{0}\left( t\right)
+z_{i}\mathfrak{A}^{i}\left( t\right) ,\ \ \ \mathbf{b}\left( t,\mathbf{z}%
\right) =\mathbf{b}^{0}\left( t\right) +z_{i}\mathbf{b}^{i}\left( t\right) ,
\\ 
\\ 
\mathbf{\lambda }\left( t,\mathbf{z}\right) =\mathbf{\lambda }^{0}\left(
t\right) +z_{i}\mathbf{\lambda }^{i}\left( t\right) ,\ \ \ \mathbf{\phi }%
\left( \upsilon ,t,\mathbf{z}\right) =\mathbf{\phi }\left( \upsilon
,t\right) ,%
\end{array}
\label{Eq5.14}
\end{equation}%
so that the corresponding backward Kolmogorov problem has the form:%
\begin{equation}
\begin{array}{c}
u_{t}\left( t,\mathbf{z}\right) +\left( a_{ij}^{0}\left( t\right)
+z_{l}a_{ij}^{l}\left( t\right) \right) u_{z_{i}z_{j}}\left( t,\mathbf{z}%
\right) +\left( b_{i}^{0}\left( t\right) +z_{l}b_{i}^{l}\left( t\right)
\right) u_{z_{i}}\left( t,\mathbf{z}\right) \\ 
\\ 
+\left( \lambda _{k}^{0}\left( t\right) +z_{l}\lambda _{k}^{l}\left(
t\right) \right) \int\limits_{-\infty }^{\infty }u\left( t,\mathbf{z}+%
\mathbf{\upsilon }_{k}\right) \phi _{k}\left( \mathbf{\upsilon }%
_{k},t\right) d\upsilon _{k} \\ 
\\ 
-\left( \Lambda ^{0}\left( t\right) +z_{l}\Lambda ^{l}\left( t\right)
\right) u\left( t,\mathbf{z}\right) =0, \\ 
\\ 
u\left( \bar{t},\mathbf{z}\right) =\tilde{u}\left( \mathbf{z}\right) .%
\end{array}
\label{Eq5.15}
\end{equation}%
Symbolically, Eq (\ref{Eq5.15}) can be written as follows:%
\begin{equation}
\begin{array}{c}
u_{t}\left( t,\mathbf{z}\right) +\mathcal{L}^{\left( 0\right) }\left[ u%
\right] \left( t,\mathbf{z}\right) +\sum\limits_{l=1}^{n_{z}}z_{l}\mathcal{L}%
^{\left( l\right) }\left[ u\right] \left( t,\mathbf{z}\right) =0, \\ 
\\ 
u\left( \bar{t},\mathbf{z}\right) =\tilde{u}\left( \mathbf{z}\right) ,%
\end{array}
\label{Eq5.16}
\end{equation}%
where $\mathcal{L}^{\left( 0\right) }$, $\mathcal{L}^{\left( l\right) }$ are
spatially-homogeneous operators, with coefficients depending only on time
(at most):%
\begin{equation}
\begin{array}{c}
\mathcal{L}^{\left( 0\right) }\left[ u\right] \left( t,\mathbf{z}\right)
=a_{ij}^{0}\left( t\right) u_{z_{i}z_{j}}\left( t,\mathbf{z}\right)
+b_{i}^{0}\left( t\right) u_{z_{i}}\left( t,\mathbf{z}\right) \\ 
\\ 
+\lambda _{k}^{0}\left( t\right) \int\limits_{-\infty }^{\infty }u\left( t,%
\mathbf{z}+\mathbf{\upsilon }_{k}\right) \phi _{k}\left( \mathbf{\upsilon }%
_{k},t\right) d\upsilon _{k}-\Lambda ^{0}\left( t\right) u\left( t,\mathbf{z}%
\right) , \\ 
\\ 
\mathcal{L}^{\left( l\right) }\left[ u\right] \left( t,\mathbf{z}\right)
=a_{ij}^{l}\left( t\right) u_{z_{i}z_{j}}\left( t,\mathbf{z}\right)
+b_{i}^{l}\left( t\right) u_{z_{i}}\left( t,\mathbf{z}\right) \\ 
\\ 
+\lambda _{k}^{l}\left( t\right) \int\limits_{-\infty }^{\infty }u\left( t,%
\mathbf{z}+\mathbf{\upsilon }_{k}\right) \phi _{k}\left( \mathbf{\upsilon }%
_{k},t\right) d\upsilon _{k}-\Lambda ^{l}\left( t\right) u\left( t,\mathbf{z}%
\right) .%
\end{array}
\label{Eq5.16a}
\end{equation}%
For the t.p.d.f. $\varpi $, one has%
\begin{equation}
\begin{array}{c}
\varpi _{t}\left( t,\mathbf{z}\right) +\mathcal{L}^{\left( 0\right) }\varpi
\left( t,\mathbf{z}\right) +\sum\limits_{l=1}^{n_{z}}z_{l}\mathcal{L}%
^{\left( l\right) }\left[ u\right] u\left( t,\mathbf{z}\right) =0, \\ 
\\ 
\varpi \left( \bar{t},\mathbf{z,}\bar{t}\mathbf{,\bar{z}}\right) =\delta
\left( \mathbf{z-\bar{z}}\right) .%
\end{array}
\label{Eq5.17}
\end{equation}%
Moreover, to cover interesting and important cases, such as anomalous
diffusions, and the likes, generalize Eq. (\ref{Eq5.16}) and consider
pseudo-differential operators $\mathcal{L}^{\left( \bar{l}\right) }$, $\bar{l%
}=0,...,n_{z}$. Recall that a translationally-invariant pseudo-differential
operator $\mathcal{L}$ is defined as follows:%
\begin{equation}
\begin{array}{c}
\mathcal{L}\left[ u\right] \left( \mathbf{z}\right) =\frac{1}{\left( 2\pi
\right) ^{n_{z}}}\int\limits_{-\infty }^{\infty }\int\limits_{-\infty
}^{\infty }\mathrm{L}\left( \mathbf{m}\right) u\left( \mathbf{z}^{\prime
}\right) e^{i\mathbf{m}\left( \mathbf{z}-\mathbf{z}^{\prime }\right) }d%
\mathbf{z}^{\prime }d\mathbf{m,}%
\end{array}
\label{Eq5.18}
\end{equation}%
where $\mathrm{L}\left( \mathbf{m}\right) $ is called the symbol of a
pseudo-differential operator; see, e.g., \cite{Cordes95, Wong14}. It is
clear that all diffusion operators belong to this category, and so do
jump-diffusion operators. The symbol of the operator $\mathcal{L}^{\left( 
\bar{l}\right) }\left( t\right) $ 
\begin{equation}
\begin{array}{c}
\mathrm{L}^{\left( \bar{l}\right) }\left( t,\mathbf{m}\right)
=-a_{ij}^{\left( \bar{l}\right) }\left( t\right) m_{i}m_{j}+ib_{i}^{\left( 
\bar{l}\right) }\left( t\right) m_{i}+\lambda _{k}^{\left( \bar{l}\right)
}\left( t\right) \psi _{k}\left( t,\mathbf{m}\right) -\Lambda ^{\left( \bar{l%
}\right) }\left( t\right) ,%
\end{array}
\label{Eq5.19}
\end{equation}%
where $\psi _{k}\left( \mathbf{m}\right) $ is the characteristic function of 
$\phi _{k}\left( \mathbf{\upsilon }\right) $:%
\begin{equation}
\begin{array}{c}
\psi _{k}\left( t,\mathbf{m}\right) =\int\limits_{-\infty }^{\infty }e^{i%
\mathbf{m\upsilon }_{k}}\phi _{k}\left( t,\mathbf{\upsilon }_{k}\right) d%
\mathbf{\upsilon }_{k}.%
\end{array}
\label{Eq5.20}
\end{equation}%
While frequently studied in the pure and applied mathematical context, in
the financial engineering context pseudo-differential operators are seldom
discussed; see, however \cite{Jacob01}.

By definition, Fourier and Kelvin modes are eignefunctions of the operators $%
\mathcal{L}^{\left( 0\right) }$, $\mathcal{L}^{\left( l\right) }$.
Accordingly, when all $\mathcal{L}^{\left( l\right) }=0$, one can solve the
corresponding backward problem vie the standard Fourier modes $\mathcal{F}$
given by\ Eq. (\ref{Eq1.1}):%
\begin{equation}
\begin{array}{c}
u\left( t,\mathbf{z}\right) =\frac{1}{\left( 2\pi \right) ^{n_{z}}}%
\int\limits_{-\infty }^{\infty }\int\limits_{-\infty }^{\infty }\tilde{u}%
\left( \mathbf{z}^{\prime }\right) e^{\alpha \left( t,\bar{t},\mathbf{m}%
\right) +i\mathbf{m}\left( \mathbf{z}-\mathbf{z}^{\prime }\right) }d\mathbf{z%
}^{\prime }d\mathbf{m},%
\end{array}
\label{Eq5.21}
\end{equation}%
where%
\begin{equation}
\begin{array}{c}
\alpha _{t}\left( t,\bar{t},\mathbf{m}\right) +\mathrm{L}^{\left( 0\right)
}\left( t,\mathbf{m}\right) =0,\ \ \ \alpha \left( \bar{t},\bar{t},\mathbf{m}%
\right) =0,%
\end{array}
\label{Eq5.22}
\end{equation}%
so that%
\begin{equation}
\begin{array}{c}
\alpha \left( t,\bar{t},\mathbf{m}\right) =\int\limits_{t}^{\bar{t}}\mathrm{L%
}^{\left( 0\right) }\left( s,\mathbf{m}\right) ds%
\end{array}
\label{Eq5.23}
\end{equation}%
However, in general, one needs to use Kelvin modes $\mathcal{K}$, given by
Eq. (\ref{Eq1.2}):%
\begin{equation}
\begin{array}{c}
u\left( t,\mathbf{z}\right) =\frac{1}{\left( 2\pi \right) ^{n_{z}}}%
\int\limits_{-\infty }^{\infty }\int\limits_{-\infty }^{\infty }\tilde{u}%
\left( \mathbf{z}^{\prime }\right) e^{\alpha \left( t,\bar{t},\mathbf{m}%
\right) +i\mathbf{\delta }\left( t,\bar{t},\mathbf{m}\right) \mathbf{z}-i%
\mathbf{mz}^{\prime }}d\mathbf{z}^{\prime }d\mathbf{m},%
\end{array}
\label{Eq5.24}
\end{equation}%
where%
\begin{equation}
\begin{array}{c}
\alpha _{t}\left( t,\bar{t},\mathbf{m}\right) +\mathrm{L}^{\left( 0\right)
}\left( t,\mathbf{\delta }\left( t,\bar{t},\mathbf{m}\right) \right) =0,\ \
\ \alpha \left( \bar{t},\bar{t},\mathbf{m}\right) =0, \\ 
\\ 
\delta _{l,t}\left( t,\bar{t},\mathbf{m}\right) +\mathrm{L}^{\left( l\right)
}\left( t,\mathbf{\delta }\left( t,\bar{t},\mathbf{m}\right) \right) =0,\ \
\ \mathbf{\delta }\left( \bar{t},\bar{t},\mathbf{m}\right) =\mathbf{m}.%
\end{array}
\label{Eq5.25}
\end{equation}%
Of course, finding explicit solution of ODEs (\ref{Eq5.25}) is possible only
in exceptional cases, some of which are discussed below. However, it is 
\emph{always} possible to solve them numerically, which is much easier than
trying to solve the corresponding PDEs directly.

As mentioned earlier, three archetypal stochastic processes are arithmetic
Wiener processes (or Brownian motions), Ornstein-Uhlenbeck (OU) and Feller
processes are; see \cite{Uhlenbeck30, Chandrasekhar43, Feller51, Feller52}.
These processes are described by the following SDEs:%
\begin{equation}
\begin{array}{c}
d\hat{y}_{t}=\chi dt+\varepsilon d\hat{W}_{t},\ \ \ \hat{y}_{t}=y,%
\end{array}
\label{Eq5.26a}
\end{equation}%
\begin{equation}
\begin{array}{c}
d\hat{y}_{t}=\left( \chi -\kappa \hat{y}_{t}\right) dt+\varepsilon d\hat{W}%
_{t},\ \ \ \hat{y}_{t}=y,%
\end{array}
\label{Eq5.26b}
\end{equation}%
\begin{equation}
\begin{array}{c}
d\hat{y}_{t}=\left( \chi -\kappa \hat{y}_{t}\right) dt+\varepsilon \sqrt{%
\hat{y}_{t}}d\hat{W}_{t},\ \ \ \hat{y}_{t}=y,%
\end{array}
\label{Eq5.26c}
\end{equation}%
respectively. It is clear that the corresponding $\mathcal{L}^{\left(
0\right) },\mathcal{L}^{\left( 1\right) }$ are:%
\begin{equation}
\begin{array}{c}
\mathcal{L}^{\left( 0\right) }\left[ u\right] =\frac{1}{2}\varepsilon
^{2}u_{yy}+\chi u_{y},\ \ \ \mathcal{L}^{\left( 1\right) }\left[ u\right] =0,%
\end{array}
\label{Eq5.27}
\end{equation}%
\begin{equation}
\begin{array}{c}
\mathcal{L}^{\left( 0\right) }\left[ u\right] =\frac{1}{2}\varepsilon
^{2}u_{yy}+\chi u_{y},\ \ \ \mathcal{L}^{\left( 1\right) }\left[ u\right]
=-\kappa u_{y},%
\end{array}
\label{Eq5.28}
\end{equation}%
\begin{equation}
\begin{array}{c}
\mathcal{L}^{\left( 0\right) }\left[ u\right] =\chi u_{y},\ \ \ \mathcal{L}%
^{\left( 1\right) }\left[ u\right] =\frac{1}{2}\varepsilon ^{2}u_{yy}-\kappa
u_{y}.%
\end{array}
\label{Eq5.29}
\end{equation}%
There are important differences among these processes. For an arithmetic
Brownian motion, the operator $\mathcal{L}^{\left( 0\right) }$ is a
second-order differential operator, while $\mathcal{L}^{\left( 1\right) }$
is zero, and the process is defined on the whole axis. For an OU process the
operator $\mathcal{L}^{\left( 0\right) }$ is a second-order differential
operator, while $\mathcal{L}^{\left( 1\right) }$ is a first-order operator;
accordingly, this process is defined on the entire axis. In contrast, for a
Feller process $\mathcal{L}^{\left( 0\right) }$ is a first-order
differential operator, while $\mathcal{L}^{\left( 1\right) }$ is a
second-order operator; hence, the process is only defined on a positive
semi-axis.\footnote{%
A Feller process might or might not be able to reach zero, which depends on
the magnitude of the ratio $2\chi /\varepsilon ^{2}$.}

\subsection{Augmentation Procedure\label{Sec53}}

While covering a lot of useful applications, OU and Feller processes are not
sufficient to study all the practically important problems. Hence, one needs
to enrich them via the so-called augmentation procedure; see \cite{Lipton01}%
. The underlying idea is straightforward. Given a stochastic process, say,
an arithmetic Brownian motion, or an OU process, one can expand it by
introducing additional stochastic variables depending on the original
process. For example, an augmented Brownian motion (\ref{Eq5.26a}) becomes a
one-dimensional Kolmogorov process:%
\begin{equation}
\begin{array}{c}
d\hat{x}_{t}=\hat{y}_{t}dt,\ \ \ \hat{x}_{t}=x, \\ 
\\ 
d\hat{y}_{t}=\chi dt+\varepsilon d\hat{W}_{t},\ \ \ \hat{y}_{t}=y.%
\end{array}
\label{Eq5.30}
\end{equation}%
Similarly, one can augment OU and Feller processes as follows: 
\begin{equation}
\begin{array}{c}
d\hat{x}_{t}=\hat{y}_{t}dt,\ \ \ x_{t}=x, \\ 
\\ 
d\hat{y}_{t}=\left( \chi -\kappa \hat{y}_{t}\right) dt+\varepsilon d\hat{W}%
_{t},\ \ \ y_{t}=y,%
\end{array}
\label{Eq5.31}
\end{equation}%
\begin{equation}
\begin{array}{c}
d\hat{x}_{t}=\hat{y}_{t}dt,\ \ \ x_{t}=x, \\ 
\\ 
d\hat{y}_{t}=\left( \chi -\kappa \hat{y}_{t}\right) dt+\varepsilon \sqrt{%
\hat{y}_{t}}d\hat{W}_{t},\ \ \ y_{t}=y.%
\end{array}
\label{Eq5.32}
\end{equation}%
respectively. Of course, many other possibilities are practically important.
Below, the book analyzes several practically relevant and mathematically
interesting augmented stochastic processes.\qquad

\subsection{Reduction Procedure\label{Sec54}}

Stochastic processes, which are not inherently affine, can often be
transformed into an affine form through appropriate modifications. While
some transformations are readily apparent, others demand significant effort
and inspiration to identify, as highlighted by \cite{Carr02} and referenced
works.

Consider the geometric Brownian motion, the cornerstone of mathematical
finance and other disciplines. The associated stochastic process is not
affine and is described by:%
\begin{equation}
\begin{array}{c}
d\hat{X}_{t}=\mu \left( t\right) \hat{X}_{t}dt+\nu \left( t\right) \hat{X}%
_{t}d\hat{W}_{t},\ \ \ \ \hat{X}_{t}=X.%
\end{array}
\label{Eq5.33}
\end{equation}%
Applying a logarithmic transformation,%
\begin{equation}
\begin{array}{c}
\hat{X}_{t}\rightarrow \hat{x}_{t}=\ln \left( \hat{X}_{t}\right) ,%
\end{array}
\label{Eq5.34}
\end{equation}%
converts it into an arithmetic Brownian motion, which is affine:%
\begin{equation}
\begin{array}{c}
d\hat{x}_{t}=\left( \mu \left( t\right) -\frac{1}{2}\nu ^{2}\left( t\right)
\right) dt+\nu \left( t\right) d\hat{W}_{t},\ \ \ \ \hat{x}_{t}=x=\ln \left(
X\right) .%
\end{array}
\label{Eq5.35}
\end{equation}%
This example illustrates that, with some ingenuity, even non-affine
processes like the geometric Brownian motion can be adapted for use with the
existing analytical frameworks.

Another helpful example is transforming the Rayleigh process into the Feller
process. Recall that the Rayleigh process describes a stochastic process on
the positive semi-axis. We write this process as follows:%
\begin{equation}
\begin{array}{c}
d\hat{\sigma}_{t}=\left( \frac{A}{\hat{\sigma}_{t}}-B\hat{\sigma}_{t}\right)
dt+Cd\hat{W}_{t},\ \ \ \hat{\sigma}_{t}=\sigma ,%
\end{array}
\label{Eq5.36}
\end{equation}%
where $A,B,C>0$. Define $\hat{v}_{t}=\hat{\sigma}_{t}^{2}$, then, according
to Ito's lemma, the dynamics of the process $\hat{v}_{t}$ has the form:%
\begin{equation}
\begin{array}{c}
d\hat{v}_{t}=\left( 2A+C^{2}-2B\hat{v}_{t}\right) dt+2C\sqrt{\hat{v}_{t}}d%
\hat{W}_{t},\ \ \ \hat{v}_{t}=v=\sigma ^{2}.%
\end{array}
\label{Eq5.37}
\end{equation}%
In financial applications considered in Chapter \ref{Chap8}, the pair $%
\sigma $, $v$ represents the volatility and variance of a price process.

\section{Gaussian Stochastic Processes\label{Chap6}}

\subsection{Regular Gaussian Processes\label{Sec61}}

Consider the governing system of SDEs which might or might not be
degenerate, and write the governing system of SDEs as follows:%
\begin{equation}
\begin{array}{c}
d\mathbf{\hat{z}}_{t}=\left( \mathbf{b}+\mathfrak{B}\mathbf{\hat{z}}%
_{t}\right) dt+\mathbf{\Sigma }d\mathbf{\hat{W}}_{t},\ \ \ \mathbf{\hat{z}}%
_{t}=\mathbf{z},%
\end{array}
\label{Eq6.1}
\end{equation}%
where $\mathbf{\hat{z}}_{t}\mathbf{,b}$ are $\left( M\times 1\right) $
vectors, $\mathfrak{B}$ and $\mathbf{\Sigma }$ are $\left( M\times M\right) $
matrices. Below, it is assumed that the corresponding coefficients are
time-dependent.

The Fokker-Plank equation has the form:%
\begin{equation}
\begin{array}{c}
\varpi _{\bar{t}}\left( t,\mathbf{z,}\bar{t},\mathbf{\bar{z}}\right) -\sum
\sum \mathfrak{A}\varpi _{\mathbf{\bar{z}\bar{z}}}\left( t,\mathbf{z,}\bar{t}%
,\mathbf{\bar{z}}\right) \\ 
\\ 
+\left( \mathbf{b}+\mathfrak{B}\mathbf{\bar{z}}\right) \cdot \varpi _{%
\mathbf{\bar{z}}}\left( t,\mathbf{z,}\bar{t},\mathbf{\bar{z}}\right)
+b\varpi \left( t,\mathbf{z,}\bar{t},\mathbf{\bar{z}}\right) =0, \\ 
\\ 
\varpi \left( t,\mathbf{z},t,\mathbf{\bar{z}}\right) =\delta \left( \mathbf{%
\bar{z}}-\mathbf{z}\right) ,%
\end{array}
\label{Eq6.2}
\end{equation}%
where, in agreement with the general Eq (\ref{Eq5.11}), $\mathfrak{A}$ is
proportional to the covariance matrix, 
\begin{equation}
\begin{array}{c}
\mathfrak{A}=\left( a_{mm^{\prime }}\right) =\frac{1}{2}\mathbf{\Sigma
\Sigma }^{\ast }=\frac{1}{2}\sigma _{mk}\sigma _{m^{\prime }k},\ \ \ b=%
\mathrm{Tr}\left( \mathfrak{B}\right) =b_{mm}.%
\end{array}
\label{Eq6.3}
\end{equation}%
Recall that Eistein's summation rule is used throughout the book. Explicitly,%
\begin{equation}
\begin{array}{c}
\partial _{\bar{t}}\varpi -a_{mm^{\prime }}\partial _{\bar{z}_{m}}\partial _{%
\bar{z}_{m^{\prime }}}\varpi +\left( b_{m}+b_{mm^{\prime }}\bar{z}%
_{m^{\prime }}\right) \partial _{\bar{z}_{m}}\varpi +b\varpi =0, \\ 
\\ 
\varpi \left( t,\mathbf{z},t,\mathbf{\bar{z}}\right) =\delta \left( \mathbf{%
\bar{z}}-\mathbf{z}\right) .%
\end{array}
\label{Eq6.4}
\end{equation}
The general Kolmogorov-type SDE solvable via the Kelvin (or affine) ansatz,
can be written in the form:%
\begin{equation}
\begin{array}{c}
d\mathbf{\hat{x}}_{t}=\left( \mathbf{b}^{\left( x\right) }+\mathfrak{B}%
^{\left( xx\right) }\mathbf{\hat{x}}_{t}+\mathfrak{B}^{\left( xy\right) }%
\mathbf{\hat{y}}_{t}\right) dt,\ \ \ \mathbf{\hat{x}}_{t}=\mathbf{x,} \\ 
\\ 
d\mathbf{\hat{y}}_{t}=\left( \mathbf{b}^{\left( y\right) }+\mathfrak{B}%
^{\left( yx\right) }\mathbf{\hat{x}}_{t}+\mathfrak{B}^{\left( yy\right) }%
\mathbf{\hat{y}}_{t}\right) dt+\mathfrak{\mathbf{\Sigma }}^{\left( yy\right)
}d\mathbf{\hat{W}}_{t}^{\left( y\right) },\ \ \ \mathbf{y}_{t}=\mathbf{y,}%
\end{array}
\label{Eq6.23}
\end{equation}%
where $\mathbf{\hat{x}}_{_{t}}$ and $\mathbf{b}^{\left( x\right) }$ are $%
\left( K\times 1\right) $ column vectors, $\mathbf{\hat{y}}_{t}$ and $%
\mathbf{b}^{\left( y\right) }$ are $\left( L\times 1\right) $ column
vectors, $\mathfrak{B}^{\left( xx\right) }$, $\mathfrak{B}^{\left( xy\right)
}$, $\mathfrak{B}^{\left( yx\right) }$, $\mathfrak{B}^{\left( yy\right) }$,
and $\mathbf{\Sigma }^{\left( yy\right) }$ are $\left( K\times K\right) $, $%
\left( K\times L\right) $, $\left( L\times K\right) $, $\left( L\times
L\right) $, and $\left( L\times L\right) $ matrices, respectively. Below, it
is assumed that the corresponding coefficients are time-dependent. As usual, 
\textbf{$\hat{W}$}$_{t}$ is a standard $L$-dimensional Brownian motion.

More compactly, one can write the system of SDEs as follows:%
\begin{equation}
\begin{array}{cc}
d\mathbf{\hat{z}}_{t}=\left( \mathbf{b}^{\left( z\right) }+\mathfrak{B}%
^{\left( zz\right) }\mathbf{\hat{z}}_{t}\right) dt+\left( 
\begin{array}{c}
0 \\ 
\mathbf{\Sigma }^{\left( yy\right) }d\mathbf{\hat{W}}_{t}^{\left( y\right) }%
\end{array}%
\right) , & \mathbf{\hat{z}}_{t}=\left( 
\begin{array}{c}
\mathbf{x} \\ 
\mathbf{y}%
\end{array}%
\right) ,%
\end{array}
\label{Eq6.24}
\end{equation}%
where%
\begin{equation}
\begin{array}{c}
\mathbf{\hat{z}}_{t}=\left( 
\begin{array}{c}
\mathbf{\hat{x}}_{t} \\ 
\mathbf{\hat{y}}_{t}%
\end{array}%
\right) ,\ \ \ \mathbf{b}^{\left( z\right) }=\left( 
\begin{array}{c}
\mathbf{b}^{\left( x\right) } \\ 
\mathbf{b}^{\left( y\right) }%
\end{array}%
\right) ,\ \ \ \mathfrak{B}^{\left( zz\right) }=\left( 
\begin{array}{cc}
\mathfrak{B}^{\left( xx\right) } & \mathfrak{B}^{\left( xy\right) } \\ 
\mathfrak{B}^{\left( yx\right) } & \mathfrak{B}^{\left( yy\right) }%
\end{array}%
\right) ,%
\end{array}
\label{Eq6.25}
\end{equation}%
so that $\mathbf{\hat{z}}_{t}$ and $\mathbf{b}^{\left( z\right) }$ are $%
\left( M\times 1\right) $ column vectors, and $\mathfrak{B}^{\left(
zz\right) }$ is a $\left( M\times M\right) $ matrix, with $M=K+L$. In
addition, define a scalar $b^{\left( z\right) }=\mathrm{Tr}\left( \mathfrak{B%
}^{\left( zz\right) }\right) =\mathrm{Tr}\left( \mathfrak{B}^{\left(
xx\right) }\right) +\mathrm{Tr}\left( \mathfrak{B}^{\left( yx\right)
}\right) $.

The corresponding Fokker-Plank problem has the form:%
\begin{equation}
\begin{array}{c}
\varpi _{\bar{t}}\left( t,\mathbf{z,}\bar{t},\mathbf{\bar{z}}\right) -\dsum
\dsum \mathfrak{A}\varpi _{\mathbf{yy}}\left( t,\mathbf{z,}\bar{t},\mathbf{%
\bar{z}}\right) \\ 
\\ 
+\left( \mathbf{b}^{\left( z\right) }+\mathfrak{B}^{\left( z\right) }\mathbf{%
\bar{z}}\right) \cdot \varpi _{\mathbf{\bar{z}}}\left( t,\mathbf{z,}\bar{t},%
\mathbf{\bar{z}}\right) +b^{\left( z\right) }\varpi \left( t,\mathbf{z,}\bar{%
t},\mathbf{\bar{z}}\right) =0, \\ 
\\ 
\varpi \left( t,\mathbf{z},t,\mathbf{\bar{z}}\right) =\delta \left( \mathbf{x%
}-\mathbf{x}\right) \delta \left( \mathbf{y}-\mathbf{y}\right) ,%
\end{array}
\label{Eq6.26}
\end{equation}%
where $\mathfrak{A}$ has the form: 
\begin{equation}
\begin{array}{c}
\mathfrak{A}=\left( a_{ll^{\prime }}\right) =\frac{1}{2}\sigma _{l\bar{l}%
}\sigma _{l^{\prime }\bar{l}}=\frac{1}{2}\mathbf{\Sigma }^{\left( yy\right) }%
\mathbf{\Sigma }^{\left( yy\right) \ast }.%
\end{array}
\label{Eq6.27}
\end{equation}%
Explicitly,%
\begin{equation}
\begin{array}{c}
\partial _{\bar{t}}\varpi -a_{ll^{\prime }}\partial _{\bar{z}_{K+l}}\partial
_{\bar{z}_{K+l^{\prime }}}\varpi +\left( b_{m}^{\left( z\right)
}+b_{mm^{\prime }}^{\left( zz\right) }\bar{z}_{m^{\prime }}\right) \partial
_{\bar{z}_{m}}\varpi +b^{\left( z\right) }\varpi =0,%
\end{array}
\label{Eq6.28}
\end{equation}

\subsubsection{Solution via Kelvin Waves}

By using the Kelvin-inspired ansatz, one can represent $\varpi $ in the form:%
\begin{equation}
\begin{array}{c}
\varpi \left( t,\mathbf{z,}\bar{t},\mathbf{\bar{z}}\right) =\frac{1}{\left(
2\pi \right) ^{M}}\int_{-\infty }^{\infty }...\int_{-\infty }^{\infty }%
\mathcal{K}\left( t,\mathbf{z,}\bar{t},\mathbf{\bar{z},m}\right) d\mathbf{m},
\\ 
\\ 
\mathcal{K}\left( t,\mathbf{z,}\bar{t},\mathbf{\bar{z},m}\right) =\exp
\left( \Psi \left( t,\mathbf{z,}\bar{t},\mathbf{\bar{z},m}\right) \right) ,
\\ 
\\ 
\Psi \left( t,\mathbf{z,}\bar{t},\mathbf{\bar{z},m}\right) =\alpha \left( t,%
\bar{t}\right) +i\mathbf{\delta }\left( t,\bar{t}\right) \cdot \mathbf{\bar{z%
}}-i\mathbf{m}\cdot \mathbf{z},%
\end{array}
\label{Eq6.6}
\end{equation}%
where $\mathbf{m}$ is $\left( M\times 1\right) $ column vector, $\mathbf{%
\delta }~$is $\left( M\times 1\right) $ column vector, and%
\begin{equation}
\begin{array}{c}
\alpha \left( t,t\right) =0,\ \ \ \mathbf{\delta }\left( t,t\right) =\mathbf{%
m}.%
\end{array}
\label{Eq6.7}
\end{equation}%
Accordingly:%
\begin{equation}
\begin{array}{c}
\frac{\mathcal{K}_{\bar{t}}}{\mathcal{K}}=\Psi _{\bar{t}}=\left( \alpha _{%
\bar{t}}\left( t,\bar{t}\right) +i\mathbf{\delta }_{\bar{t}}\left( t,\bar{t}%
\right) \cdot \mathbf{\bar{z}}\right) , \\ 
\\ 
\frac{\mathcal{K}_{\mathbf{\bar{z}}}}{\mathcal{K}}=\Psi _{\mathbf{\bar{z}}}=i%
\mathbf{\delta }\left( t,\bar{t}\right) ,\ \ \ \frac{\mathcal{K}_{\mathbf{%
\bar{z}\bar{z}}}}{\mathcal{K}}=\Psi _{\mathbf{\bar{z}}}^{2}=-\mathbf{\delta }%
\left( t,\bar{t}\right) \mathbf{\delta }^{\ast }\left( t,\bar{t}\right) .%
\end{array}
\label{Eq6.8}
\end{equation}%
The coupled equations for $\alpha ,\mathbf{\delta }$ have the form:%
\begin{equation}
\begin{array}{c}
\alpha _{\bar{t}}\left( t,\bar{t}\right) +i\mathbf{\delta }_{\bar{t}}\left(
t,\bar{t}\right) \cdot \mathbf{\bar{z}}+\mathbf{\delta }\left( t,\bar{t}%
\right) \cdot \mathfrak{A}\mathbf{\delta }\left( t,\bar{t}\right) +i\mathbf{%
\delta }\left( t,\bar{t}\right) \cdot \left( \mathbf{b}+\mathfrak{B}\mathbf{%
\bar{z}}\right) +b=0,%
\end{array}
\label{Eq6.9}
\end{equation}%
so that,%
\begin{equation}
\begin{array}{c}
\alpha _{\bar{t}}\left( t,\bar{t}\right) +\mathbf{\delta }\left( t,\bar{t}%
\right) \cdot \mathfrak{A}\mathbf{\delta }\left( t,\bar{t}\right) +i\mathbf{%
\delta }\left( t,\bar{t}\right) \cdot \mathbf{b}+b=0,\ \ \ \alpha \left(
t,t\right) =0,%
\end{array}
\label{Eq6.10}
\end{equation}%
\begin{equation}
\begin{array}{c}
\mathbf{\delta }_{\bar{t}}\left( t,\bar{t}\right) +\mathfrak{B}^{\ast }%
\mathbf{\delta }\left( t,\bar{t}\right) =0,\ \ \ \mathbf{\delta }\left(
t,t\right) =\mathbf{m}.%
\end{array}
\label{Eq6.11}
\end{equation}%
Let $\mathfrak{L}\left( t,\bar{t}\right) $ is the fundamental solution of
the homogeneous system of ODEs (\ref{Eq6.11}), i.e., the matrix such that%
\begin{equation}
\begin{array}{c}
\partial _{\bar{t}}\mathbf{\mathfrak{L}}\left( t,\bar{t}\right) +\mathfrak{B}%
^{\ast }\left( \bar{t}\right) \mathbf{\mathfrak{L}}\left( t,\bar{t}\right)
=0,\ \ \ \mathbf{\mathfrak{L}}\left( t,t\right) =\mathfrak{I},%
\end{array}
\label{Eq6.12}
\end{equation}%
The solution of Eq. (\ref{Eq6.11}) has the form:%
\begin{equation}
\begin{array}{c}
\mathbf{\delta }\left( t,\bar{t}\right) =\mathbf{\mathfrak{L}}\left( t,\bar{t%
}\right) \mathbf{m}.%
\end{array}
\label{Eq6.13}
\end{equation}%
Thus,%
\begin{equation}
\begin{array}{c}
\alpha \left( t,\bar{t}\right) =-\frac{1}{2}\mathbf{m}\cdot \mathfrak{C}%
^{-1}\left( t,\bar{t}\right) \mathbf{m}-i\mathbf{m\cdot d}\left( t,\bar{t}%
\right) -\varsigma \left( t,\bar{t}\right) ,%
\end{array}
\label{Eq6.14}
\end{equation}%
where $\mathfrak{C}^{-1}$ is an $M\times M$ positive-definite matrix of the
form:%
\begin{equation}
\begin{array}{c}
\mathfrak{C}^{-1}\left( t,\bar{t}\right) =2\int_{t}^{\bar{t}}\mathfrak{L}%
^{\ast }\left( t,s\right) \mathfrak{A}\left( s\right) \mathfrak{L}\left(
t,s\right) ds,%
\end{array}
\label{Eq6.15}
\end{equation}%
while $\mathbf{d}$ is an $\left( M\times 1\right) $ column vector,%
\begin{equation}
\begin{array}{c}
\mathbf{d}\left( t,\bar{t}\right) =\int_{t}^{\bar{t}}\mathfrak{L}^{\ast
}\left( t,s\right) \mathbf{b}\left( s\right) ds,%
\end{array}
\label{Eq6.16}
\end{equation}%
and $\varsigma $ is a scalar,%
\begin{equation}
\begin{array}{c}
\varsigma \left( t,\bar{t}\right) =\int_{t}^{\bar{t}}b\left( s\right) ds.%
\end{array}
\label{Eq6.17}
\end{equation}%
Accordingly,%
\begin{equation}
\begin{array}{c}
\Psi \left( t,\bar{t},\mathbf{\bar{z}},\mathbf{m}\right) =-\frac{1}{2}%
\mathbf{m}\cdot \mathfrak{C}^{-1}\left( t,\bar{t}\right) \mathbf{m}+i\mathbf{%
m}\cdot \left( \mathfrak{L}^{\ast }\left( t,\bar{t}\right) \mathbf{\bar{z}}-%
\mathbf{d}\left( t,\bar{t}\right) -\mathbf{z}\right) -\varsigma \left( t,%
\bar{t}\right) .%
\end{array}
\label{Eq6.18}
\end{equation}%
Thus, 
\begin{equation}
\begin{array}{c}
\varpi \left( t,\mathbf{z},\bar{t},\mathbf{\bar{z}}\right) =\frac{\det
\left( \mathfrak{C}\left( t,\bar{t}\right) \right) ^{1/2}\exp \left(
-\varsigma \left( t,\bar{t}\right) \right) }{\left( 2\pi \right) ^{M/2}} \\ 
\\ 
\times \int_{-\infty }^{\infty }...\int_{-\infty }^{\infty }G\left( t,\bar{t}%
,\mathbf{m}\right) \exp \left( i\mathbf{m}\cdot \left( \mathfrak{L}^{\ast
}\left( t,\bar{t}\right) \mathbf{\bar{z}}-\mathbf{d}\left( t,\bar{t}\right) -%
\mathbf{z}\right) \right) d\mathbf{m},%
\end{array}
\label{Eq6.19}
\end{equation}%
where $G\left( t,\bar{t},\mathbf{m}\right) $ is the density of a
multivariate Gaussian distribution in the $\mathbf{m}$-space. It is clear
that $\varpi \left( t,\mathbf{z,}\bar{t},\mathbf{\bar{z}}\right) $ is
proportional to the characteristic function of $G$ evaluated at the point $%
\left( \mathfrak{L}^{\ast }\left( t,\bar{t}\right) \mathbf{\bar{z}}-\mathbf{d%
}\left( t,\bar{t}\right) -\mathbf{z}\right) $,so that%
\begin{equation}
\begin{array}{c}
\varpi \left( t,\mathbf{z,}\bar{t},\mathbf{\bar{z}}\right) =\frac{\det
\left( \mathfrak{C}\left( t,\bar{t}\right) \right) ^{1/2}\exp \left(
-\varsigma \left( t,\bar{t}\right) \right) }{\left( 2\pi \right) ^{M/2}} \\ 
\\ 
\times \exp \left( -\frac{1}{2}\left( \mathfrak{L}^{\ast }\left( t,\bar{t}%
\right) \mathbf{\bar{z}}-\mathbf{d}\left( t,\bar{t}\right) -\mathbf{z}%
\right) \mathcal{\cdot }\mathfrak{C}\left( t,\bar{t}\right) \left( \mathfrak{%
L}^{\ast }\left( t,\bar{t}\right) \mathbf{\bar{z}}-\mathbf{d}\left( t,\bar{t}%
\right) -\mathbf{z}\right) \right) .%
\end{array}
\label{Eq6.20}
\end{equation}%
Thus, $\varpi $ can be represented it the form:%
\begin{equation}
\begin{array}{c}
\varpi \left( t,\mathbf{z,}\bar{t},\mathbf{\bar{z}}\right) =\mathrm{N}\left( 
\mathbf{r}\left( t,\bar{t}\right) \mathbf{,}\mathfrak{H}\left( t,\bar{t}%
\right) \right) ,%
\end{array}
\label{Eq6.21}
\end{equation}%
where%
\begin{equation}
\begin{array}{c}
\mathfrak{H}\left( t,\bar{t}\right) =\left( \mathfrak{L}^{\ast }\left( t,%
\bar{t}\right) \right) ^{-1}\mathfrak{C}^{-1}\left( t,\bar{t}\right) 
\mathfrak{L}^{-1}\left( t,\bar{t}\right) , \\ 
\\ 
\mathbf{r}\left( t,\bar{t}\right) =\left( \mathfrak{L}^{\ast }\left( t,\bar{t%
}\right) \right) ^{-1}\left( \mathbf{d}\left( t,\bar{t}\right) +\mathbf{z}%
\right) .%
\end{array}
\label{Eq6.22}
\end{equation}%
These results are applicable to the general Kolmogorov-type SDE solvable via
the Kelvin (or affine) ansatz, which have the form (\ref{Eq6.23}). By using
the same Kelvin ansatz as before, one can represent $\varpi $ in the form (%
\ref{Eq6.6}):%
\begin{equation}
\begin{array}{c}
\varpi \left( t,\mathbf{z,}\bar{t},\mathbf{\bar{z}}\right) =\frac{1}{\left(
2\pi \right) ^{M}}\int_{-\infty }^{\infty }...\int_{-\infty }^{\infty }%
\mathcal{K}\left( t,\mathbf{z,}\bar{t},\mathbf{\bar{z},m}\right) d\mathbf{m},
\\ 
\\ 
\mathcal{K}\left( t,\mathbf{z,}\bar{t},\mathbf{\bar{z},m}\right) =\exp
\left( \Psi \left( t,\mathbf{z,}\bar{t},\mathbf{\bar{z},m}\right) \right) ,
\\ 
\\ 
\Psi \left( t,\mathbf{z,}\bar{t},\mathbf{\bar{z},m}\right) =\alpha \left( t,%
\bar{t}\right) +i\mathbf{\delta }\left( t,\bar{t}\right) \cdot \mathbf{\bar{z%
}}-i\mathbf{m}\cdot \mathbf{z},%
\end{array}
\label{Eq6.29}
\end{equation}%
where $\mathbf{m}$ is an $\left( M\times 1\right) $ column vector, $\mathbf{m%
}=\left( \mathbf{k,l}\right) ^{\ast }$, $\mathbf{k}$ is an $\left( K\times
1\right) $ column vector, $\mathbf{l}~$is an $\left( L\times 1\right) $
column vector, $\mathbf{\delta }$ is an $\left( M\times 1\right) $ column
vector, $\ \mathbf{\delta }=\left( \mathbf{\beta },\mathbf{\gamma }\right)
^{\ast }$, $\mathbf{\beta }~$is an $\left( K\times 1\right) $ column vector$%
, $ $\mathbf{\gamma }~$is an $\left( L\times 1\right) $ column vector, and%
\begin{equation}
\begin{array}{c}
\alpha \left( t,t\right) =0,\ \ \ \mathbf{\delta }\left( t,t\right) =\left( 
\mathbf{\beta }\left( t,t\right) ,\mathbf{\gamma }\left( t,t\right) \right)
^{\ast }=\mathbf{m}=\left( \mathbf{k,l}\right) ^{\ast }.%
\end{array}
\label{Eq6.30}
\end{equation}%
As before:%
\begin{equation}
\begin{array}{c}
\frac{\mathcal{K}_{t}}{\mathcal{K}}=\Psi _{t}=\left( \alpha _{\bar{t}}\left(
t,\bar{t}\right) +i\mathbf{\delta }_{\bar{t}}\left( t,\bar{t}\right) \cdot 
\mathbf{\bar{z}}\right) ,\ \ \ \frac{\mathcal{K}_{\mathbf{x}}}{\mathcal{K}}%
=\Psi _{\mathbf{x}}=i\mathbf{\beta }\left( t,\bar{t}\right) , \\ 
\\ 
\frac{\mathcal{K}_{\mathbf{y}}}{\mathcal{K}}=\Psi _{\mathbf{y}}=i\mathbf{%
\gamma }\left( t,\bar{t}\right) ,\ \ \ \frac{\mathcal{K}_{\mathbf{yy}}}{%
\mathcal{K}}=\Psi _{\mathbf{y}}^{2}=-\mathbf{\gamma }\left( t,\bar{t}\right) 
\mathbf{\gamma }^{\ast }\left( t,\bar{t}\right) .%
\end{array}
\label{Eq6.31}
\end{equation}%
The equations for $\alpha ,\mathbf{\delta }$ have the form:%
\begin{equation}
\begin{array}{c}
\alpha _{\bar{t}}\left( t,\bar{t}\right) +i\mathbf{\delta }_{\bar{t}}\left(
t,\bar{t}\right) \cdot \mathbf{\bar{z}}+\mathbf{\gamma }\left( t,\bar{t}%
\right) \cdot \mathfrak{A}\mathbf{\gamma }\left( t,\bar{t}\right) +i\mathbf{%
\delta }\left( t,\bar{t}\right) \cdot \left( \mathbf{b}^{\left( z\right) }+%
\mathfrak{B}^{\left( zz\right) }\mathbf{\bar{z}}\right) +b^{\left( z\right)
}=0.%
\end{array}
\label{Eq6.32}
\end{equation}%
Accordingly,%
\begin{equation}
\begin{array}{c}
\alpha _{\bar{t}}\left( t,\bar{t}\right) +\mathbf{\gamma }\left( t,\bar{t}%
\right) \cdot \mathfrak{A}\mathbf{\gamma }\left( t,\bar{t}\right) +i\mathbf{%
\delta }\left( t,\bar{t}\right) \cdot \mathbf{b}^{\left( z\right)
}+b^{\left( z\right) }=0,\ \ \ \alpha \left( t,t\right) =0,%
\end{array}
\label{Eq6.33}
\end{equation}%
\begin{equation}
\begin{array}{c}
\mathbf{\delta }_{\bar{t}}\left( t,\bar{t}\right) +\mathfrak{B}^{\left(
zz\right) \ast }\mathbf{\delta }\left( t,\bar{t}\right) =0,\ \ \ \mathbf{%
\delta }\left( t,t\right) =\mathbf{m}=\left( \mathbf{k,l}\right) ^{\ast }.%
\end{array}
\label{Eq6.34}
\end{equation}%
Let $\mathfrak{L}\left( t,\bar{t}\right) $ is the fundamental solution of
the homogeneous system of ODEs (\ref{Eq6.34}), i.e., the matrix such that%
\begin{equation}
\begin{array}{c}
\partial _{\bar{t}}\mathbf{\mathfrak{L}}\left( t,\bar{t}\right) +\mathfrak{B}%
^{\left( zz\right) \ast }\left( \bar{t}\right) \mathbf{\mathfrak{L}}\left( t,%
\bar{t}\right) =0,\ \ \ \mathbf{\mathfrak{L}}\left( t,t\right) =\mathfrak{I},%
\end{array}
\label{Eq6.35}
\end{equation}%
where $\mathfrak{I}$ is the identity matrix. The well-known Liouville's
formula yields%
\begin{equation}
\begin{array}{c}
\det \left( \mathfrak{L}\left( t,\bar{t}\right) \right) =\exp \left(
-\int_{t}^{\bar{t}}b^{\left( z\right) }\left( s\right) ds\right) .%
\end{array}
\label{Eq6.36}
\end{equation}%
The solution of Eq. (\ref{Eq6.33}) is%
\begin{equation}
\begin{array}{c}
\mathbf{\delta }\left( t,\bar{t}\right) =\mathbf{\mathfrak{L}}\left( t,\bar{t%
}\right) \mathbf{m}.%
\end{array}
\label{Eq6.37}
\end{equation}%
It is convenient to write $\mathfrak{L}\left( t,\bar{t}\right) $ in the
block form:%
\begin{equation}
\begin{array}{c}
\mathfrak{L}\left( t,\bar{t}\right) =\left( 
\begin{array}{cc}
\mathfrak{L}^{\left( xx\right) }\left( t,\bar{t}\right) & \mathfrak{L}%
^{\left( xy\right) }\left( t,\bar{t}\right) \\ 
\mathfrak{L}^{\left( yx\right) }\left( t,\bar{t}\right) & \mathfrak{L}%
^{\left( yy\right) }\left( t,\bar{t}\right)%
\end{array}%
\right) .%
\end{array}
\label{Eq6.38}
\end{equation}%
It follows from Eq. (\ref{Eq6.34}) that%
\begin{equation}
\begin{array}{c}
\alpha \left( t,\bar{t}\right) =-\frac{1}{2}\mathbf{m}\cdot \mathfrak{C}%
^{-1}\left( t,\bar{t}\right) \mathbf{m}-i\mathbf{m\cdot d}^{\left( z\right)
}\left( t,\bar{t}\right) -\varsigma \left( t,\bar{t}\right) ,%
\end{array}
\label{Eq6.39}
\end{equation}%
where $\mathfrak{C}^{-1}$ is an $M\times M$ positive-definite matrix split
into four blocks of the form:%
\begin{equation}
\begin{array}{c}
\mathfrak{C}^{-1}\left( t,\bar{t}\right) \\ 
\\ 
=2\left( 
\begin{array}{cc}
\int_{t}^{\bar{t}}\mathfrak{L}^{\left( yx\right) \ast }\left( t,s\right) 
\mathfrak{A}\left( s\right) \mathfrak{L}^{\left( yx\right) }\left(
t,s\right) ds & \int_{t}^{\bar{t}}\mathfrak{L}^{\left( yx\right) \ast
}\left( t,s\right) \mathfrak{A}\left( s\right) \mathfrak{L}^{\left(
yy\right) }\left( t,s\right) ds \\ 
\int_{t}^{\bar{t}}\mathfrak{L}^{\left( yy\right) \ast }\left( t,s\right) 
\mathfrak{A}\left( s\right) \mathfrak{L}^{\left( yx\right) }\left(
t,s\right) ds & \int_{t}^{\bar{t}}\mathfrak{L}^{\left( yy\right) \ast
}\left( t,s\right) \mathfrak{A}\left( s\right) \mathfrak{L}^{\left(
yy\right) }\left( t,s\right) ds%
\end{array}%
\right) ,%
\end{array}
\label{Eq6.40}
\end{equation}%
while $\mathbf{d}^{\left( z\right) }=\left( \mathbf{d}^{\left( x\right) },%
\mathbf{d}^{\left( y\right) }\right) ^{\ast }$, $\mathbf{d}^{\left( x\right)
}$ and $\mathbf{d}^{\left( y\right) }$ are $\left( M\times 1\right) $ and $%
\left( N\times 1\right) $ column vectors, and $\varsigma $ is a scalar:%
\begin{equation}
\begin{array}{c}
\mathbf{d}^{\left( z\right) }\left( t,\bar{t}\right) =\int_{t}^{\bar{t}}%
\mathfrak{L}^{\ast }\left( t,s\right) \mathbf{b}^{\left( z\right) }\left(
s\right) ds,%
\end{array}
\label{Eq6.41}
\end{equation}%
\begin{equation}
\begin{array}{c}
\varsigma \left( t,\bar{t}\right) =\int_{t}^{\bar{t}}b^{\left( z\right)
}\left( s\right) ds.%
\end{array}
\label{Eq6.42}
\end{equation}%
Accordingly,%
\begin{equation}
\begin{array}{c}
\Psi \left( t,\mathbf{z,}\bar{t},\mathbf{\bar{z}},\mathbf{m}\right) \\ 
\\ 
=-\frac{1}{2}\mathbf{m}\cdot \mathfrak{C}^{-1}\left( t,\bar{t}\right) 
\mathbf{m}+i\mathbf{\mathfrak{L}}\left( t,\bar{t}\right) \mathbf{m}\cdot 
\mathbf{\bar{z}}-i\mathbf{m}\cdot \left( \mathbf{d}^{\left( z\right) }\left(
t,\bar{t}\right) +\mathbf{z}\right) -\varsigma \left( t,\bar{t}\right) \\ 
\\ 
=-\frac{1}{2}\mathbf{m}\cdot \mathfrak{C}^{-1}\left( t,\bar{t}\right) 
\mathbf{m}+i\mathbf{m}\cdot \left( \mathfrak{L}^{\ast }\left( t,\bar{t}%
\right) \mathbf{\bar{z}-d}^{\left( z\right) }\left( t,\bar{t}\right) \mathbf{%
-z}\right) -\varsigma \left( t,\bar{t}\right) .%
\end{array}
\label{Eq6.43}
\end{equation}%
Thus, 
\begin{equation}
\begin{array}{c}
\varpi \left( t,\mathbf{z,}\bar{t},\mathbf{\bar{z}}\right) =\frac{\det
\left( \mathfrak{C}\right) ^{1/2}\exp \left( -\varsigma \left( t,\bar{t}%
\right) \right) }{\left( 2\pi \right) ^{M/2}}\int_{-\infty }^{\infty
}...\int_{-\infty }^{\infty }G\left( t,\bar{t},\mathbf{m}\right) \\ 
\\ 
\times \exp \left( i\mathbf{m}\cdot \left( \mathfrak{L}^{\ast }\left( t,\bar{%
t}\right) \mathbf{\bar{z}-d}^{\left( z\right) }\left( t,\bar{t}\right) 
\mathbf{-z}\right) \right) d\mathbf{m},%
\end{array}
\label{Eq6.44}
\end{equation}%
where $G\left( t,\bar{t},\mathbf{m}\right) $ is the density of a
multivariate Gaussian distribution in the $\mathbf{m}$-space. It is clear
that $\varpi \left( t,\mathbf{z,}\bar{t},\mathbf{\bar{z}}\right) $ is
proportional to the characteristic function of $G$ evaluated at the point $%
\left( \mathfrak{L}^{\ast }\left( t,\bar{t}\right) \mathbf{\bar{z}-d}%
^{\left( z\right) }\left( t,\bar{t}\right) \mathbf{-z}\right) $, so that%
\begin{equation}
\begin{array}{c}
\varpi \left( t,\mathbf{z,}\bar{t},\mathbf{\bar{z}}\right) =\frac{\det
\left( \mathfrak{C}\right) ^{1/2}\exp \left( -\varsigma \left( t,\bar{t}%
\right) \right) }{\left( 2\pi \right) ^{M/2}} \\ 
\\ 
\times \exp \left( -\frac{1}{2}\left( \mathfrak{L}^{\ast }\left( t,\bar{t}%
\right) \mathbf{\bar{z}-d}^{\left( z\right) }\left( t,\bar{t}\right) \mathbf{%
-z}\right) \cdot \mathfrak{C}\left( \mathfrak{L}^{\ast }\left( t,\bar{t}%
\right) \mathbf{\bar{z}-d}^{\left( z\right) }\left( t,\bar{t}\right) \mathbf{%
-z}\right) \right) .%
\end{array}
\label{Eq6.45}
\end{equation}%
By using Eq. (\ref{Eq6.36}) one can rewrite (\ref{Eq6.45}) it the standard
Gaussian form:%
\begin{equation}
\begin{array}{c}
\varpi \left( t,\mathbf{z,}\bar{t},\mathbf{\bar{z}}\right) =\mathrm{N}\left( 
\mathbf{r}\left( t,\bar{t}\right) ,\mathfrak{H}\left( t,\bar{t}\right)
\right) ,%
\end{array}
\label{Eq6.46}
\end{equation}%
where the covariance matrix $\mathfrak{H}$ and the mean $\mathbf{r}$ are as
follows: 
\begin{equation}
\begin{array}{c}
\mathfrak{H}\left( t,\bar{t}\right) =\left( \mathfrak{L}^{\ast }\left( t,%
\bar{t}\right) \right) ^{-1}\mathfrak{C}^{-1}\left( t,\bar{t}\right) 
\mathfrak{L}^{-1}\left( t,\bar{t}\right) , \\ 
\\ 
\mathbf{r}\left( t,\bar{t}\right) \mathbf{=}\left( \mathfrak{L}^{\ast
}\left( t,\bar{t}\right) \right) ^{-1}\left( \mathbf{d}^{\left( z\right)
}\left( t,\bar{t}\right) +\mathbf{z}\right) .%
\end{array}
\label{Eq6.47}
\end{equation}

\subsubsection{Solution via Coordinate Transform}

Consider the Fokker-Planck problem (\ref{Eq6.4}). Introduce new variables:%
\begin{equation}
\begin{array}{c}
\left( \bar{t},\mathbf{\bar{z}}\right) \rightarrow \left( \bar{t},\mathbf{%
\tilde{z}}\right) =\left( \bar{t},\mathfrak{R}\left( \bar{t}\right) \mathbf{%
\bar{z}}\right) ,\ \ \ \tilde{z}_{m}=r_{mm^{\prime }}\left( \bar{t}\right) 
\bar{z}_{m^{\prime }},\ \ \ r_{mm^{\prime }}\left( 0\right) =\delta
_{mm^{\prime }}.%
\end{array}
\label{Eq6.48}
\end{equation}%
Then%
\begin{equation}
\begin{array}{c}
\partial _{\bar{t}}=\partial _{\bar{t}}+\partial _{\bar{t}}r_{mm^{\prime }}%
\bar{z}_{m^{\prime }}\partial _{\tilde{z}_{m}},\ \ \ \partial _{\bar{z}%
_{m}}=r_{m^{\prime }m}\partial _{\tilde{z}_{m^{\prime }}}.%
\end{array}
\label{Eq6.49}
\end{equation}%
The transformed Fokker-Planck problem becomes%
\begin{equation}
\begin{array}{c}
\partial _{\bar{t}}\tilde{\varpi}-a_{mm^{\prime }}r_{nm}r_{n^{\prime
}m^{\prime }}\partial _{\tilde{z}_{n}}\partial _{\tilde{z}_{n^{\prime }}}%
\tilde{\varpi}+\left( \left( b_{mm^{\prime }}\bar{z}_{m^{\prime
}}+b_{m}\right) r_{nm}+\partial _{\bar{t}}r_{nm^{\prime }}\bar{z}_{m^{\prime
}}\right) \partial _{\tilde{z}_{n}}\tilde{\varpi}+b\tilde{\varpi}=0, \\ 
\\ 
\tilde{\varpi}\left( t,\mathbf{z},t,\mathbf{\tilde{z}}\right) =\delta \left( 
\mathbf{\tilde{z}}-\mathbf{z}\right) .%
\end{array}
\label{Eq6.50}
\end{equation}%
To simplify the drift term, it is required that%
\begin{equation}
\begin{array}{c}
\partial _{\bar{t}}r_{mm^{\prime }}\left( t,\bar{t}\right) +b_{nm^{\prime
}}\left( t,\bar{t}\right) r_{mn}\left( t,\bar{t}\right) =0,\ \ \
r_{mm^{\prime }}\left( t,t\right) =\delta _{nm^{\prime }}.%
\end{array}
\label{Eq6.51}
\end{equation}%
In matrix notation:%
\begin{equation}
\begin{array}{c}
\partial _{\bar{t}}\mathfrak{R}\left( t,\bar{t}\right) +\mathfrak{R}\left( t,%
\bar{t}\right) \mathfrak{B}\left( t\right) =0,\ \ \ \ \mathfrak{R}\left(
t,t\right) =\mathfrak{I}.%
\end{array}
\label{Eq6.52}
\end{equation}%
Thus, $\mathfrak{R}=\mathfrak{L}^{\ast }$, $r_{mm^{\prime }}=l_{m^{\prime
}m} $, where $\mathfrak{L}$ is given by Eq. (\ref{Eq6.35}). It is easy to
see that $\tilde{\varpi}$ satisfies the Fokker-Planck problem of the
following form:%
\begin{equation}
\begin{array}{c}
\partial _{\bar{t}}\tilde{\varpi}-\tilde{a}_{nn^{\prime }}\left( t,\bar{t}%
\right) \partial _{\tilde{z}_{n}}\partial _{\tilde{z}_{n^{\prime }}}\tilde{%
\varpi}+\tilde{b}_{n}\left( t,\bar{t}\right) \partial _{\tilde{z}_{n}}\tilde{%
\varpi}+b\left( t,\bar{t}\right) \tilde{\varpi}=0, \\ 
\\ 
\tilde{\varpi}\left( t,\mathbf{z},t,\mathbf{\tilde{z}}\right) =\delta \left( 
\mathbf{\tilde{z}}-\mathbf{z}\right) ,%
\end{array}
\label{Eq6.55}
\end{equation}%
with%
\begin{equation}
\begin{array}{c}
\tilde{a}_{nn^{\prime }}\left( t,\bar{t}\right) =l_{mn}\left( t,\bar{t}%
\right) a_{mm^{\prime }}\left( t,\bar{t}\right) l_{m^{\prime }n^{\prime
}}\left( t,\bar{t}\right) , \\ 
\\ 
\tilde{b}_{n}\left( \bar{t}\right) =l_{m^{\prime }n}\left( t,\bar{t}\right)
b_{m^{\prime }}\left( \bar{t}\right) .%
\end{array}
\label{Eq6.56}
\end{equation}%
In matrix notation:%
\begin{equation}
\begin{array}{c}
\widetilde{\mathfrak{A}}=\mathfrak{L}^{\ast }\left( t,\bar{t}\right) 
\mathfrak{A}\left( t,\bar{t}\right) \mathfrak{L}\left( t,\bar{t}\right) ,\ \
\ \widetilde{\mathbf{b}}=\mathfrak{L}^{\ast }\left( t,\bar{t}\right) \mathbf{%
b}.%
\end{array}
\label{Eq6.56b}
\end{equation}%
Accordingly, 
\begin{equation}
\begin{array}{c}
\tilde{\varpi}\left( t,\mathbf{z},\bar{t},\mathbf{\tilde{z}}\right) =\exp
\left( -\int_{t}^{\bar{t}}b\left( s\right) ds\right) \mathrm{N}\left( \left. 
\tilde{z}\right\vert \mathbf{z}+\int_{t}^{\bar{t}}\tilde{b}_{n}\left(
s\right) ds,\int_{t}^{\bar{t}}\widetilde{\mathfrak{C}}\left( s\right)
ds\right) .%
\end{array}
\label{Eq6.60}
\end{equation}%
Reverting back to the original variables, $\left( \bar{t},\mathbf{\tilde{z}}%
\right) \rightarrow \left( \bar{t},\mathbf{\bar{z}}\right) $, one recovers
Eq. (\ref{Eq6.46}), as expected.

\subsection{Killed Gaussian Processes\label{Sec62}}

Consider a process governed by a system of SDEs (\ref{Eq6.1}), which is
killed with intensity $\bar{c}$ linearly depending of $\mathbf{\bar{z}}$,
namely, 
\begin{equation}
\begin{array}{c}
\bar{c}=c+\mathbf{c}\cdot \mathbf{\bar{z}},%
\end{array}
\label{Eq6.61}
\end{equation}%
where $c$ is a scalar, and $\mathbf{c}^{\left( z\right) }$ is a $\left(
M\times 1\right) $ column vector. Thus, $\bar{c}$ is the intensity at which
the process goes into a "killed" state at some random time. The Fokker-Plank
equation for a killed process has the form:%
\begin{equation}
\begin{array}{c}
\varpi _{\bar{t}}\left( t,\mathbf{z,}\bar{t},\mathbf{\bar{z}}\right) -\dsum
\dsum \mathfrak{A}\varpi _{\mathbf{\bar{z}\bar{z}}}\left( t,\mathbf{z,}\bar{t%
},\mathbf{\bar{z}}\right) \\ 
\\ 
+\left( \mathbf{b}+\mathfrak{B}\mathbf{\bar{z}}\right) \cdot \varpi _{%
\mathbf{\bar{z}}}\left( t,\mathbf{z,}\bar{t},\mathbf{\bar{z}}\right) +\left(
b+c+\mathbf{c\cdot \bar{z}}\right) \varpi \left( t,\mathbf{z,}\bar{t},%
\mathbf{\bar{z}}\right) =0, \\ 
\\ 
\varpi \left( t,\mathbf{\bar{z}},t,\mathbf{z}\right) =\delta \left( \mathbf{%
\bar{z}}-\mathbf{z}\right) .%
\end{array}
\label{Eq6.62}
\end{equation}%
Explicitly,%
\begin{equation}
\begin{array}{c}
\varpi _{\bar{t}}-a_{mm^{\prime }}\varpi _{\bar{z}_{m}\bar{z}_{m^{\prime
}}}+\left( b_{m}+b_{mm^{\prime }}\bar{z}_{m^{\prime }}\right) \varpi _{\bar{z%
}_{m}}+\left( b+c+c_{m}\bar{z}_{m}\right) \varpi =0, \\ 
\\ 
\varpi \left( t,\mathbf{\bar{z}},t,\mathbf{z}\right) =\delta \left( \mathbf{%
\bar{z}}-\mathbf{z}\right) .%
\end{array}
\label{Eq6.63}
\end{equation}%
This problem can be solved by the same technique as before.

\subsubsection{Solution via Kelvin Waves}

The familiar Kelvin ansatz yields:%
\begin{equation}
\begin{array}{c}
\alpha _{\bar{t}}\left( t,\bar{t}\right) +\mathbf{\delta }\left( t,\bar{t}%
\right) \cdot \mathfrak{A}\mathbf{\delta }\left( t,\bar{t}\right) +i\mathbf{%
\delta }\left( t,\bar{t}\right) \cdot \mathbf{b}+b+c=0,\ \ \ \alpha \left(
t,t\right) =0,%
\end{array}
\label{Eq6.64}
\end{equation}%
\begin{equation}
\begin{array}{c}
\mathbf{\delta }_{\bar{t}}\left( t,\bar{t}\right) +\mathfrak{B}^{\ast }%
\mathbf{\delta }\left( t,\bar{t}\right) -i\mathbf{c}=0,\ \ \ \mathbf{\delta }%
\left( t,t\right) =\mathbf{m.}%
\end{array}
\label{Eq6.65}
\end{equation}%
Let $\mathfrak{L}\left( t,\bar{t}\right) $ is the fundamental solution of
the homogeneous system of ODEs (\ref{Eq6.65}), i.e., the matrix such that%
\begin{equation}
\begin{array}{c}
\partial _{\bar{t}}\mathbf{\mathfrak{L}}\left( t,\bar{t}\right) +\mathfrak{B}%
^{\ast }\left( \bar{t}\right) \mathbf{\mathfrak{L}}\left( t,\bar{t}\right)
=0,\ \ \ \mathbf{\mathfrak{L}}\left( t,t\right) =\mathfrak{I},%
\end{array}
\label{Eq6.66}
\end{equation}%
The solution of Eq. (\ref{Eq6.65}) has the form:%
\begin{equation}
\begin{array}{c}
\mathbf{\delta }\left( t,\bar{t}\right) =\mathbf{\mathfrak{L}}\left( t,\bar{t%
}\right) \mathbf{m}+i\mathfrak{L}\left( t,\bar{t}\right) \int_{t}^{\bar{t}}%
\mathbf{\mathfrak{L}}^{-1}\left( t,s\right) \mathbf{c}\left( s\right)
ds\equiv \mathbf{\mathfrak{L}}\left( t,\bar{t}\right) \left( \mathbf{m}+i%
\mathbf{e}\left( t,\bar{t}\right) \right) , \\ 
\\ 
\mathbf{e}\left( t,\bar{t}\right) =\int_{t}^{\bar{t}}\mathbf{\mathfrak{L}}%
^{-1}\left( t,s\right) \mathbf{c}\left( s\right) ds.%
\end{array}
\label{Eq6.67}
\end{equation}%
Thus,%
\begin{equation}
\begin{array}{c}
\alpha =-\frac{1}{2}\mathbf{m}\cdot \mathfrak{C}^{-1}\mathbf{m}-i\mathbf{%
m\cdot d}-\varsigma ,%
\end{array}
\label{Eq6.68}
\end{equation}%
where $\mathfrak{C}^{-1}$ is an $M\times M$ positive-definite matrix of the
form:%
\begin{equation}
\begin{array}{c}
\mathfrak{C}^{-1}\left( t,\bar{t}\right) =2\int_{t}^{\bar{t}}\mathfrak{L}%
^{\ast }\left( t,s\right) \mathfrak{A}\left( s\right) \mathfrak{L}\left(
t,s\right) ds,%
\end{array}
\label{Eq6.69}
\end{equation}%
while $\mathbf{d}$ is an $\left( M\times 1\right) $ column vector,%
\begin{equation}
\begin{array}{c}
\mathbf{d}\left( t,\bar{t}\right) =\int_{t}^{\bar{t}}\mathfrak{L}^{\ast
}\left( t,s\right) \left( \mathbf{b}\left( s\right) +\mathfrak{A}\left(
s\right) \mathfrak{L}\left( t,s\right) \mathbf{e}\left( s\right) \right) ds,%
\end{array}
\label{Eq6.70}
\end{equation}%
and $\varsigma =\varsigma _{0}+\varsigma _{1}$ is a scalar,%
\begin{equation}
\begin{array}{c}
\varsigma _{0}\left( t,\bar{t}\right) =\int_{t}^{\bar{t}}b\left( s\right) ds,
\\ 
\\ 
\varsigma _{1}\left( t,\bar{t}\right) =\int_{t}^{\bar{t}}\left( c\left(
s\right) -\frac{1}{2}\mathbf{e}\left( t,s\right) \cdot \mathfrak{L}^{\ast
}\left( t,s\right) \mathfrak{A}\left( s\right) \mathfrak{L}\left( t,s\right) 
\mathbf{e}\left( s\right) -\mathbf{e}\left( t,s\right) \cdot \mathfrak{L}%
^{\ast }\left( t,s\right) \mathbf{b}\left( s\right) \right) ds.%
\end{array}
\label{Eq6.71}
\end{equation}%
Accordingly,%
\begin{equation}
\begin{array}{c}
\Psi \left( t,\bar{t},\mathbf{\bar{z}},\mathbf{m}\right) =-\frac{1}{2}%
\mathbf{m}\cdot \mathfrak{C}^{-1}\left( t,\bar{t}\right) \mathbf{m}+i\mathbf{%
m}\cdot \left( \mathfrak{L}^{\ast }\left( t,\bar{t}\right) \mathbf{\bar{z}}-%
\mathbf{d}\left( t,\bar{t}\right) -\mathbf{z}\right) \\ 
\\ 
-\mathfrak{L}\left( t,\bar{t}\right) \mathbf{e}\left( t,\bar{t}\right) \cdot 
\mathbf{\bar{z}}-\varsigma \left( t,\bar{t}\right) .%
\end{array}
\label{Eq6.72}
\end{equation}%
Thus, 
\begin{equation}
\begin{array}{c}
\varpi \left( t,\mathbf{z},\bar{t},\mathbf{\bar{z}}\right) =\frac{\det
\left( \mathfrak{C}\left( t,\bar{t}\right) \right) ^{1/2}\exp \left( -%
\mathfrak{L}\left( t,\bar{t}\right) \mathbf{e}\left( t,\bar{t}\right) \cdot 
\mathbf{\bar{z}}-\varsigma _{0}\left( t,\bar{t}\right) -\varsigma _{1}\left(
t,\bar{t}\right) \right) }{\left( 2\pi \right) ^{M/2}} \\ 
\\ 
\times \int_{-\infty }^{\infty }...\int_{-\infty }^{\infty }G\left( t,\bar{t}%
,\mathbf{m}\right) \exp \left( i\mathbf{m}\cdot \left( \mathfrak{L}^{\ast
}\left( t,\bar{t}\right) \mathbf{\bar{z}}-\mathbf{d}\left( t,\bar{t}\right) -%
\mathbf{z}\right) \right) d\mathbf{m},%
\end{array}
\label{Eq6.73}
\end{equation}%
where $G\left( t,\bar{t},\mathbf{m}\right) $ is the density of a
multivariate Gaussian distribution in the $\mathbf{m}$-space. It is clear
that $\varpi \left( t,\mathbf{z,}\bar{t},\mathbf{\bar{z}}\right) $ is
proportional to the characteristic function of $G$ evaluated at the point $%
\left( \mathfrak{L}^{\ast }\left( t,\bar{t}\right) \mathbf{\bar{z}}-\mathbf{d%
}\left( t,\bar{t}\right) -\mathbf{z}\right) $,so that%
\begin{equation}
\begin{array}{c}
\varpi \left( t,\mathbf{z,}\bar{t},\mathbf{\bar{z}}\right) =\frac{\det
\left( \mathfrak{C}\left( t,\bar{t}\right) \right) ^{1/2}\exp \left( -%
\mathfrak{L}\left( t,\bar{t}\right) \mathbf{e}\left( t,\bar{t}\right) \cdot 
\mathbf{\bar{z}}-\varsigma _{0}\left( t,\bar{t}\right) -\varsigma _{1}\left(
t,\bar{t}\right) \right) }{\left( 2\pi \right) ^{M/2}} \\ 
\\ 
\times \exp \left( -\frac{1}{2}\left( \mathfrak{L}^{\ast }\left( t,\bar{t}%
\right) \mathbf{\bar{z}}-\mathbf{d}\left( t,\bar{t}\right) -\mathbf{z}%
\right) \mathcal{\cdot }\mathfrak{C}\left( t,\bar{t}\right) \left( \mathfrak{%
L}^{\ast }\left( t,\bar{t}\right) \mathbf{\bar{z}}-\mathbf{d}\left( t,\bar{t}%
\right) -\mathbf{z}\right) \right) .%
\end{array}
\label{Eq6.74}
\end{equation}%
It is often convenient to rewrite Eq. (\ref{Eq6.74}) as follows:%
\begin{equation}
\begin{array}{c}
\varpi \left( t,\mathbf{z,}\bar{t},\mathbf{\bar{z}}\right) =Q\left( t,\bar{t}%
,\mathbf{\bar{z}}\right) \mathrm{N}\left( \mathbf{q}\left( t,\bar{t}\right) 
\mathbf{,}\mathfrak{H}\left( t,\bar{t}\right) \right) ,%
\end{array}
\label{Eq6.75}
\end{equation}%
where%
\begin{equation}
\begin{array}{c}
\mathfrak{H}\left( t,\bar{t}\right) =\left( \mathfrak{L}^{\ast }\left( t,%
\bar{t}\right) \right) ^{-1}\mathfrak{C}^{-1}\left( t,\bar{t}\right) 
\mathfrak{L}^{-1}\left( t,\bar{t}\right) , \\ 
\\ 
\mathbf{q}\left( t,\bar{t}\right) =\left( \mathfrak{L}^{\ast }\left( t,\bar{t%
}\right) \right) ^{-1}\left( \mathbf{d}\left( t,\bar{t}\right) +\mathbf{z}%
\right) , \\ 
\\ 
Q\left( t,\bar{t},\mathbf{\bar{z}}\right) =\exp \left( -\mathfrak{L}\left( t,%
\bar{t}\right) \mathbf{e}\left( t,\bar{t}\right) \cdot \mathbf{\bar{z}}%
-\varsigma _{1}\left( t,\bar{t}\right) \right) .%
\end{array}
\label{Eq6.76}
\end{equation}%
As could be expected, the probability $\varpi $ is no longer conserved due
to a prefactor $Q$, reflecting the fact that the process is killed with
intensity $\bar{c}$.

It is worth noting that $Q$ depends on $\mathbf{\bar{z}}$ but does not
depend on $\mathbf{z}$. Completing the square, one can represent $\varpi $
is the form:

\begin{equation}
\begin{array}{c}
\varpi \left( t,\mathbf{z,}\bar{t},\mathbf{\bar{z}}\right) =R\left( t,%
\mathbf{z,}\bar{t}\right) \mathrm{N}\left( \mathbf{r}\left( t,\bar{t}\right) 
\mathbf{,}\mathfrak{H}\left( t,\bar{t}\right) \right) ,%
\end{array}
\label{Eq6.77}
\end{equation}%
where 
\begin{equation}
\begin{array}{c}
\mathbf{r}\left( t,\bar{t}\right) =\left( \mathfrak{L}^{\ast }\left( t,\bar{t%
}\right) \right) ^{-1}\left( \mathbf{d}\left( t,\bar{t}\right) +\mathbf{z-}%
\mathfrak{C}^{-1}\left( t,\bar{t}\right) \mathbf{e}\left( t,\bar{t}\right)
\right) \\ 
\\ 
=\mathbf{q}\left( t,\bar{t}\right) -\mathfrak{H}\left( t,\bar{t}\right) 
\mathfrak{L}\left( t,\bar{t}\right) \mathbf{e}\left( t,\bar{t}\right) , \\ 
\\ 
R\left( t,\mathbf{z,}\bar{t}\right) =\exp \left( -\mathbf{e}\left( t,\bar{t}%
\right) \cdot \left( \mathbf{d}\left( t,\bar{t}\right) +\mathbf{z}\right) +%
\frac{1}{2}\mathbf{e}\left( t,\bar{t}\right) \cdot \mathfrak{C}^{-1}\left( t,%
\bar{t}\right) \mathbf{e}\left( t,\bar{t}\right) -\varsigma _{1}\left( t,%
\bar{t}\right) \right) .%
\end{array}
\label{Eq6.78}
\end{equation}%
It is clear that $R$ depends on $\mathbf{z}$ but does not depend on $\mathbf{%
\bar{z}}$. Accordingly, Eq. (\ref{Eq6.77}) is easier to use than Eq. (\ref%
{Eq6.75}) when future expectations are calculated.

The same formulas can be derived via the method of coordinate transforms.
Details are left to the interested reader.

\subsection{Example: Kolmogorov Process\label{Sec63}}

Extend the Kolmogorov formula to the case when $b$ and $\sigma $ are
functions of time, $b\left( t\right) $ and $\sigma \left( t\right) $. The
corresponding SDE has the form:%
\begin{equation}
\begin{array}{c}
d\hat{x}_{t}=\hat{y}_{t}dt,\ \ \ \hat{x}_{t}=x, \\ 
\\ 
d\hat{y}_{t}=b\left( t\right) dt+\sigma \left( t\right) d\hat{W}_{t},\ \ \ 
\hat{y}_{t}=y.%
\end{array}
\label{Eq6.79}
\end{equation}%
Accordingly, Eq. (\ref{Eq6.35}) can be written as follows:%
\begin{equation}
\begin{array}{c}
\mathfrak{L}^{\prime }\left( t,\bar{t}\right) +\left( 
\begin{array}{cc}
0 & 0 \\ 
1 & 0%
\end{array}%
\right) \mathfrak{L}\left( t,\bar{t}\right) =0,\ \ \mathfrak{L}\left(
t,t\right) =\left( 
\begin{array}{cc}
1 & 0 \\ 
0 & 1%
\end{array}%
\right) ,%
\end{array}
\label{Eq6.80}
\end{equation}%
so that%
\begin{equation}
\begin{array}{c}
\mathfrak{L}\left( t,\bar{t}\right) =\left( 
\begin{array}{cc}
1 & 0 \\ 
-T & 1%
\end{array}%
\right) ,\ \ \ \mathfrak{L}^{-1}\left( t,\bar{t}\right) =\left( 
\begin{array}{cc}
1 & 0 \\ 
T & 1%
\end{array}%
\right) .%
\end{array}
\label{Eq6.81}
\end{equation}%
Once $\mathfrak{L}\left( t,\bar{t}\right) $ is known, one can compute $%
\mathfrak{C}^{-1}\left( t,\bar{t}\right) $, $\mathbf{d}^{\left( z\right)
}\left( t,\bar{t}\right) $, $\varsigma \left( t,\bar{t}\right) $: 
\begin{equation}
\begin{array}{c}
\mathfrak{C}^{-1}\left( t,\bar{t}\right) =\left( 
\begin{array}{cc}
\psi _{2}\left( t,\bar{t}\right) & -\psi _{1}\left( t,\bar{t}\right) \\ 
-\psi _{1}\left( t,\bar{t}\right) & \psi _{0}\left( t,\bar{t}\right)%
\end{array}%
\right) , \\ 
\\ 
\mathbf{d}^{\left( z\right) }\left( t,\bar{t}\right) =\left( 
\begin{array}{c}
d^{\left( x\right) }\left( t,\bar{t}\right) \\ 
d^{\left( y\right) }\left( t,\bar{t}\right)%
\end{array}%
\right) =\left( 
\begin{array}{c}
-\phi _{1}\left( t,\bar{t}\right) \\ 
\phi _{0}\left( t,\bar{t}\right)%
\end{array}%
\right) , \\ 
\\ 
\varsigma \left( t,\bar{t}\right) =0,%
\end{array}
\label{Eq6.82}
\end{equation}%
where%
\begin{equation}
\begin{array}{c}
\phi _{i}\left( t,\bar{t}\right) =\int_{t}^{\bar{t}}\left( s-t\right)
^{i}b\left( s\right) ds,\ \ \ \ \ \psi _{i}\left( t,\bar{t}\right)
=\int_{t}^{\bar{t}}\left( s-t\right) ^{i}\sigma ^{2}\left( s\right) ds.%
\end{array}
\label{Eq6.83}
\end{equation}%
Next, the covariance matrix $\mathfrak{H}\left( t,\bar{t}\right) $, and the
mean $\mathbf{r}\left( t,\bar{t}\right) $ are calculated as follows:%
\begin{equation}
\begin{array}{c}
\mathfrak{H}\left( t,\bar{t}\right) =\left( \mathfrak{L}^{\ast }\left( t,%
\bar{t}\right) \right) ^{-1}\mathfrak{C}^{-1}\left( t,\bar{t}\right) 
\mathfrak{L}^{-1}\left( t,\bar{t}\right) \\ 
\\ 
=\left( 
\begin{array}{cc}
1 & T \\ 
0 & 1%
\end{array}%
\right) \left( 
\begin{array}{cc}
\psi _{2}\left( t,\bar{t}\right) & -\psi _{1}\left( t,\bar{t}\right) \\ 
-\psi _{1}\left( t,\bar{t}\right) & \psi _{0}\left( t,\bar{t}\right)%
\end{array}%
\right) \left( 
\begin{array}{cc}
1 & 0 \\ 
T & 1%
\end{array}%
\right) \\ 
\\ 
=\left( 
\begin{array}{cc}
\psi _{0}\left( t,\bar{t}\right) T^{2}-2\psi _{1}\left( t,\bar{t}\right)
T+\psi _{2}\left( t,\bar{t}\right) & \psi _{0}\left( t,\bar{t}\right) T-\psi
_{1}\left( t,\bar{t}\right) \\ 
\psi _{0}\left( t,\bar{t}\right) T-\psi _{1}\left( t,\bar{t}\right) & \psi
_{0}\left( t,\bar{t}\right)%
\end{array}%
\right) ,%
\end{array}
\label{Eq6.84}
\end{equation}%
\begin{equation}
\begin{array}{c}
\mathbf{r}\left( t,\bar{t}\right) =\left( 
\begin{array}{c}
-\phi _{1}\left( t,\bar{t}\right) +x+\left( \phi _{0}\left( t,\bar{t}\right)
+y\right) T \\ 
\phi _{0}\left( t,\bar{t}\right) +y%
\end{array}%
\right) .%
\end{array}
\label{Eq6.85}
\end{equation}%
Accordingly, $\varpi \left( t,x,y,\bar{t},\bar{x},\bar{y}\right) $ is a
bivariate Gaussian distribution of the form (\ref{Eq6.21}), with 
\begin{equation}
\begin{array}{c}
\sigma _{x}^{2}\left( t,\bar{t}\right) =\psi _{0}\left( t,\bar{t}\right)
T^{2}-2\psi _{1}\left( t,\bar{t}\right) T+\psi _{2}\left( t,\bar{t}\right)
,\ \ \ \sigma _{y}^{2}=\psi _{0}\left( t,\bar{t}\right) , \\ 
\\ 
\rho \left( t,\bar{t}\right) =\frac{\left( \psi _{0}\left( t,\bar{t}\right)
T-\psi _{1}\left( t,\bar{t}\right) \right) }{\sqrt{\psi _{0}\left( t,\bar{t}%
\right) \left( \psi _{0}\left( t,\bar{t}\right) T^{2}-2\psi _{1}\left( t,%
\bar{t}\right) T+\psi _{2}\left( t,\bar{t}\right) \right) }}, \\ 
\\ 
p\left( t,\bar{t}\right) =-\phi _{1}\left( t,\bar{t}\right) +x+\left( \phi
_{0}\left( t,\bar{t}\right) +y\right) T,\ \ \ q\left( t,\bar{t}\right) =\phi
_{0}\left( t,\bar{t}\right) +y.%
\end{array}
\label{Eq6.86}
\end{equation}%
It is left to the interested reader to verify that Eq. (\ref{Eq6.86})
coincides with Eq. (\ref{Eq3.48}) when $\sigma $ and $b$ are constant.
Therefore, the classical Kolmogorov solution can be extended to the case of
time-dependent parameters.

\subsection{Example: OU Process\label{Sec64}}

\subsubsection{OU Process}

It is worth deriving the well-known t.p.d.f. for the OU process using Kelvin
waves for benchmarking purposes. The following SDE governs the OU process:%
\begin{equation}
\begin{array}{c}
d\hat{y}_{t}=\left( \chi \left( t\right) -\kappa \left( t\right) \hat{y}%
_{t}\right) dt+\varepsilon \left( t\right) d\hat{W}_{t},\ \ \ \bar{y}_{t}=y.%
\end{array}
\label{Eq6.87}
\end{equation}%
Equivalently,%
\begin{equation}
\begin{array}{c}
d\hat{y}_{t}=\kappa \left( t\right) \left( \theta \left( t\right) -\hat{y}%
_{t}\right) dt+\varepsilon \left( t\right) d\hat{W}_{t},\ \ \ \bar{y}_{t}=y,%
\end{array}
\label{Eq6.87a}
\end{equation}%
where $\theta \left( t\right) =\chi \left( t\right) /\kappa \left( t\right) $%
.

The corresponding Fokker-Planck problem has the form:

\begin{equation}
\begin{array}{c}
\varpi _{\bar{t}}\left( t,y,\bar{t},\bar{y}\right) -\frac{1}{2}\varepsilon
^{2}\varpi _{\bar{y}\bar{y}}\left( t,y,\bar{t},\bar{y}\right) +\left( \chi
\left( \bar{t}\right) -\kappa \left( \bar{t}\right) \bar{y}\right) \varpi _{%
\bar{y}}\left( t,y,\bar{t},\bar{y}\right) -\kappa \left( \bar{t}\right)
\varpi \left( t,y,\bar{t},\bar{y}\right) =0, \\ 
\\ 
\varpi \left( t,y,\bar{t},\bar{y}\right) =\delta \left( \bar{y}-y\right) .%
\end{array}
\label{Eq6.88}
\end{equation}%
The associated function $\mathcal{K}\left( t,y,\bar{t},\bar{y},l\right) $
has the form:%
\begin{equation}
\begin{array}{c}
\mathcal{K}=\exp \left( \alpha \left( t,\bar{t}\right) +i\gamma \left( t,%
\bar{t}\right) \bar{y}-ily\right) ,%
\end{array}
\label{Eq6.89}
\end{equation}%
so that%
\begin{equation}
\begin{array}{c}
\alpha _{\bar{t}}\left( t,\bar{t}\right) +\frac{1}{2}\varepsilon ^{2}\left( 
\bar{t}\right) \gamma ^{2}\left( t,\bar{t}\right) +i\chi \left( \bar{t}%
\right) \gamma \left( t,\bar{t}\right) -\kappa \left( \bar{t}\right) =0,\ \
\ \alpha \left( t,t\right) =0, \\ 
\\ 
\gamma _{\bar{t}}\left( t,\bar{t}\right) -\kappa \left( \bar{t}\right)
\gamma \left( t,\bar{t}\right) =0,\ \ \ \gamma \left( t,t\right) =l.%
\end{array}
\label{Eq6.90}
\end{equation}%
Thus, 
\begin{equation}
\begin{array}{c}
\gamma \left( t,\bar{t}\right) =e^{\eta \left( t,\bar{t}\right) }l, \\ 
\\ 
\alpha \left( t,\bar{t}\right) =-\frac{1}{2}\psi _{0}\left( t,\bar{t}\right)
l^{2}-\left( \int_{t}^{\bar{t}}e^{\eta \left( t,s\right) }\chi \left(
s\right) ds\right) il+\eta \left( t,\bar{t}\right) .%
\end{array}
\label{Eq6.91}
\end{equation}%
where%
\begin{equation}
\begin{array}{c}
\eta \left( t,\bar{t}\right) =\dint\limits_{t}^{\bar{t}}\kappa \left(
s\right) ds,%
\end{array}
\label{Eq6.92}
\end{equation}%
\begin{equation}
\begin{array}{c}
\psi _{0}\left( t,\bar{t}\right) =\dint\limits_{t}^{\bar{t}}e^{2\eta \left(
t,s\right) }\varepsilon ^{2}\left( s\right) ds.%
\end{array}
\label{Eq6.93}
\end{equation}%
Since the same quantities will appear regularly throughout the book, it is
convenient to introduce the following notation%
\begin{equation}
\begin{array}{c}
\mathsf{A}_{\kappa }\left( t,\bar{t}\right) =e^{-\eta \left( t,\bar{t}%
\right) },\ \ \ \mathsf{B}_{\kappa }\left( t,\bar{t}\right) =\int_{t}^{\bar{t%
}}e^{-\eta \left( t,s\right) }ds,\ \ \ \mathsf{\bar{B}}_{\kappa }\left( t,%
\bar{t}\right) =\int_{t}^{\bar{t}}e^{-\eta \left( s,\bar{t}\right) }ds, \\ 
\\ 
\mathsf{A}_{-\kappa }\left( t,\bar{t}\right) =e^{\eta \left( t,\bar{t}%
\right) },\ \ \ \mathsf{B}_{-\kappa }\left( t,\bar{t}\right) =\int_{t}^{\bar{%
t}}e^{\eta \left( t,s\right) }ds,\ \ \ \mathsf{\bar{B}}_{-\kappa }\left( t,%
\bar{t}\right) =\int_{t}^{\bar{t}}e^{\eta \left( s,\bar{t}\right) }ds,%
\end{array}
\label{Eq6.100b}
\end{equation}%
In particular, for constant $\kappa $, one has%
\begin{equation}
\begin{array}{c}
\mathsf{A}_{\kappa }\left( t,\bar{t}\right) =e^{-\kappa T}=\mathsf{A}%
_{\kappa }\left( T\right) ,\ \ \ \mathsf{A}_{-\kappa }\left( t,\bar{t}%
\right) =e^{\kappa T}=\mathsf{A}_{-\kappa }\left( T\right) , \\ 
\\ 
\mathsf{B}_{\kappa }\left( t,\bar{t}\right) =\mathsf{\bar{B}}_{\kappa
}\left( t,\bar{t}\right) =\frac{1-e^{-\kappa T}}{\kappa }=\mathsf{B}_{\kappa
}\left( T\right) =\mathsf{\bar{B}}_{\kappa }\left( T\right) , \\ 
\\ 
\mathsf{B}_{-\kappa }\left( t,\bar{t}\right) =\mathsf{\bar{B}}_{-\kappa
}\left( t,\bar{t}\right) =\frac{e^{\kappa T}-1}{\kappa }=\mathsf{B}_{-\kappa
}\left( T\right) =\mathsf{\bar{B}}_{-\kappa }\left( T\right) ,%
\end{array}
\label{Eq6.101c}
\end{equation}%
and%
\begin{equation}
\begin{array}{c}
\mathsf{A}_{0}\left( t,\bar{t}\right) =1,\ \ \ \mathsf{B}_{0}\left( t,\bar{t}%
\right) =\mathsf{\bar{B}}_{0}\left( t,\bar{t}\right) =T.%
\end{array}
\label{Eq6.101e}
\end{equation}%
In this notation, $\psi _{0}$ can be written as follows: 
\begin{equation}
\begin{array}{c}
\psi _{0}\left( t,\bar{t}\right) =\dint\limits_{t}^{\bar{t}}\mathsf{A}%
_{-2\kappa }\left( t,s\right) \varepsilon ^{2}\left( s\right) ds.%
\end{array}
\label{Eq6.101f}
\end{equation}%
Thus, the following well-known expression is obtained:%
\begin{equation}
\begin{array}{c}
\varpi \left( t,y,\bar{t},\bar{y}\right) \\ 
\\ 
=\frac{1}{2\pi }\int_{-\infty }^{\infty }\exp \left( -\frac{\psi _{0}\left(
t,\bar{t}\right) l^{2}}{2}+\left( e^{\eta \left( t,\bar{t}\right) }\bar{y}%
-\int_{t}^{\bar{t}}e^{\eta \left( t,s\right) }\chi \left( s\right)
ds-y\right) il+\eta \left( t,\bar{t}\right) \right) dl \\ 
\\ 
=\frac{\mathsf{A}_{-\kappa }\left( t,\bar{t}\right) }{\sqrt{2\pi \psi
_{0}\left( t,\bar{t}\right) }}\exp \left( -\frac{\left( \mathsf{A}_{-\kappa
}\left( t,\bar{t}\right) \bar{y}-\int_{t}^{\bar{t}}\mathsf{A}_{-\kappa
}\left( t,s\right) \chi \left( s\right) ds-y\right) ^{2}}{2\psi _{0}\left( t,%
\bar{t}\right) }\right) \\ 
\\ 
=\frac{1}{\sqrt{2\pi \hat{\psi}_{0}\left( t,\bar{t}\right) }}\exp \left( -%
\frac{\left( \bar{y}-\int_{t}^{\bar{t}}\mathsf{A}_{\kappa }\left( t,s\right)
\chi \left( s\right) ds-\mathsf{A}_{\kappa }\left( t,\bar{t}\right) y\right)
^{2}}{2\hat{\psi}_{0}\left( t,\bar{t}\right) }\right) ,%
\end{array}
\label{Eq6.94}
\end{equation}%
where%
\begin{equation}
\begin{array}{c}
\hat{\psi}_{0}\left( t,\bar{t}\right) =\mathsf{A}_{2\kappa }\left( t,\bar{t}%
\right) \psi _{0}\left( t,\bar{t}\right) =\dint\limits_{t}^{\bar{t}}\mathsf{A%
}_{2\kappa }\left( s,\bar{t}\right) \varepsilon ^{2}\left( s\right) ds.%
\end{array}
\label{Eq6.95}
\end{equation}%
For further discussion, see the original paper by \cite{Uhlenbeck30}, as
well as \cite{Chandrasekhar43, Risken89},\ and references therein.

For time-independent parameters, Eq. (\ref{Eq6.94}) has the form:%
\begin{equation}
\begin{array}{c}
\varpi \left( t,y,\bar{t},\bar{y}\right) =\frac{1}{\sqrt{2\pi \Sigma
^{2}\left( t,\bar{t}\right) }}\exp \left( -\frac{\left( \bar{y}-\theta -%
\mathsf{A}_{\kappa }\left( T\right) \left( y-\theta \right) \right) ^{2}}{%
2\Sigma ^{2}\left( t,\bar{t}\right) }\right) ,%
\end{array}
\label{Eq6.96}
\end{equation}%
with%
\begin{equation}
\begin{array}{c}
\Sigma ^{2}\left( t,\bar{t}\right) =\frac{\varepsilon ^{2}\left(
1-e^{-2\kappa T}\right) }{2\kappa }=\varepsilon ^{2}\mathsf{B}_{2\kappa
}\left( T\right) .%
\end{array}
\label{Eq6.97}
\end{equation}

\subsubsection{Gaussian Augmented OU Process}

This subsection considers an augmented one-dimensional OU process of the
form:

\begin{equation}
\begin{array}{c}
d\hat{x}_{t}=\hat{y}_{t}dt,\ \ \ \hat{x}_{t}=x, \\ 
\\ 
d\hat{y}_{t}=\left( \chi \left( t\right) -\kappa \left( t\right) \hat{y}%
_{t}\right) dt+\varepsilon \left( t\right) d\hat{W}_{t},\ \ \ \hat{y}_{t}=y.%
\end{array}
\label{Eq6.98}
\end{equation}%
To align the analysis with the existing body of work, switch from the
general notation, used above, to a specific one customary used for the OU
process. Here and below, the word \textquotedblleft
augmentation\textquotedblright\ means that one expands the original process
by incorporating its integral or other path-dependent characteristics, such
as running maximum or minimum as part of the process; see Chapter \ref{Chap5}
above. The augmentation is a very useful tool. In particular, in financial
engineering it is used for handling large classes of path-dependent options;
details can be found in \cite{Lipton01}, Chapter 13.

For an OU\ process, Eq. (\ref{Eq6.35}) can be written as follows:%
\begin{equation}
\begin{array}{c}
\mathfrak{L}_{\bar{t}}\left( t,\bar{t}\right) +\left( 
\begin{array}{cc}
0 & 0 \\ 
1 & -\kappa \left( \bar{t}\right)%
\end{array}%
\right) \mathfrak{L}\left( t,\bar{t}\right) =0,\ \ \mathfrak{L}\left(
t,t\right) =\left( 
\begin{array}{cc}
1 & 0 \\ 
0 & 1%
\end{array}%
\right) ,%
\end{array}
\label{Eq6.99}
\end{equation}%
so that%
\begin{equation}
\begin{array}{c}
\mathfrak{L}\left( t,\bar{t}\right) =\left( 
\begin{array}{cc}
1 & 0 \\ 
-\mathsf{\bar{B}}_{-\kappa }\left( t,\bar{t}\right) & \mathsf{A}_{-\kappa
}\left( t,\bar{t}\right)%
\end{array}%
\right) ,\ \ \ \mathfrak{L}^{-1}\left( t,\bar{t}\right) =\left( 
\begin{array}{cc}
1 & 0 \\ 
\mathsf{B}_{\kappa }\left( t,\bar{t}\right) & \mathsf{A}_{\kappa }\left( t,%
\bar{t}\right)%
\end{array}%
\right) .%
\end{array}
\label{Eq6.101}
\end{equation}%
Now, one can compute $\mathfrak{C}^{-1}\left( t,\bar{t}\right) $, $\mathbf{d}%
^{\left( z\right) }\left( t,\bar{t}\right) $, and $\varsigma \left( t,\bar{t}%
\right) $:%
\begin{equation}
\begin{array}{c}
\mathfrak{C}^{-1}\left( t,\bar{t}\right) =\left( 
\begin{array}{cc}
\psi _{2}\left( t,\bar{t}\right) & -\psi _{1}\left( t,\bar{t}\right) \\ 
-\psi _{1}\left( t,\bar{t}\right) & \psi _{0}\left( t,\bar{t}\right)%
\end{array}%
\right) ,%
\end{array}
\label{Eq6.101a}
\end{equation}%
where%
\begin{equation}
\begin{array}{c}
\psi _{0}\left( t,\bar{t}\right) =\dint\limits_{t}^{\bar{t}}\mathsf{A}%
_{-\kappa }^{2}\left( t,s\right) \varepsilon ^{2}\left( s\right) ds, \\ 
\\ 
\psi _{1}\left( t,\bar{t}\right) =-\dint\limits_{t}^{\bar{t}}\mathsf{\bar{B}}%
_{-\kappa }\left( t,s\right) \mathsf{A}_{-\kappa }\left( t,s\right)
\varepsilon ^{2}\left( s\right) ds, \\ 
\\ 
\psi _{2}\left( t,\bar{t}\right) =\dint\limits_{t}^{\bar{t}}\mathsf{\bar{B}}%
_{-\kappa }^{2}\left( t,s\right) \varepsilon ^{2}\left( s\right) ds.%
\end{array}
\label{Eq6.102}
\end{equation}%
\begin{equation}
\begin{array}{c}
\mathbf{d}^{\left( z\right) }\left( t,\bar{t}\right) =\left( 
\begin{array}{c}
d^{\left( x\right) }\left( t,\bar{t}\right) \\ 
d^{\left( y\right) }\left( t,\bar{t}\right)%
\end{array}%
\right) =\left( 
\begin{array}{c}
-\int_{t}^{\bar{t}}\mathsf{\bar{B}}_{-\kappa }\left( t,s\right) \chi \left(
s\right) ds \\ 
\int_{t}^{\bar{t}}\mathsf{A}_{-\kappa }\left( t,s\right) \chi \left(
s\right) ds,%
\end{array}%
\right) ,%
\end{array}
\label{Eq6.103}
\end{equation}%
\begin{equation}
\begin{array}{c}
\varsigma \left( t,\bar{t}\right) =-\eta \left( t,\bar{t}\right) .%
\end{array}
\label{Eq6.104}
\end{equation}%
Next, one can calculate the covariance matrix $\mathfrak{H}\left( t,\bar{t}%
\right) $, and mean vector $\mathbf{r}\left( t,\bar{t}\right) $ as follows:%
\begin{equation}
\begin{array}{c}
\mathfrak{H}\left( t,\bar{t}\right) =\left( \mathfrak{L}^{\ast }\left( t,%
\bar{t}\right) \right) ^{-1}\mathfrak{C}^{-1}\left( t,\bar{t}\right) 
\mathfrak{L}^{-1}\left( t,\bar{t}\right) \\ 
\\ 
=\left( 
\begin{array}{cc}
1 & \ \mathsf{B}_{\kappa }\left( t,\bar{t}\right) \\ 
0 & \ \mathsf{A}_{\kappa }\left( t,\bar{t}\right)%
\end{array}%
\right) \left( 
\begin{array}{cc}
\psi _{2}\left( t,\bar{t}\right) & -\psi _{1}\left( t,\bar{t}\right) \\ 
-\psi _{1}\left( t,\bar{t}\right) & \psi _{0}\left( t,\bar{t}\right)%
\end{array}%
\right) \left( 
\begin{array}{cc}
1 & 0 \\ 
\ \mathsf{B}_{\kappa }\left( t,\bar{t}\right) & \ \mathsf{A}_{\kappa }\left(
t,\bar{t}\right)%
\end{array}%
\right) \\ 
\\ 
=\left( 
\begin{array}{cc}
h_{0}\left( t,\bar{t}\right) & h_{1}\left( t,\bar{t}\right) \\ 
h_{1}\left( t,\bar{t}\right) & h_{2}\left( t,\bar{t}\right)%
\end{array}%
\right) ,%
\end{array}
\label{Eq6.105}
\end{equation}%
\begin{equation}
\begin{array}{c}
\mathbf{r}\left( t,\bar{t}\right) =\left( \mathfrak{L}^{\ast }\left( t,\bar{t%
}\right) \right) ^{-1}\left( \mathbf{d}^{\left( z\right) }\left( t,\bar{t}%
\right) +\left( 
\begin{array}{c}
x \\ 
y%
\end{array}%
\right) \right) =\left( 
\begin{array}{c}
p\left( t,\bar{t}\right) \\ 
q\left( t,\bar{t}\right)%
\end{array}%
\right) .%
\end{array}
\label{Eq6.107}
\end{equation}%
Here%
\begin{equation}
\begin{array}{c}
h_{0}\left( t,\bar{t}\right) =\psi _{0}\ \mathsf{B}_{\kappa }^{2}\left( t,%
\bar{t}\right) -2\psi _{1}\ \mathsf{B}_{\kappa }\left( t,\bar{t}\right)
+\psi _{2}, \\ 
\\ 
h_{1}\left( t,\bar{t}\right) =\left( \psi _{0}\ \mathsf{B}_{\kappa }\left( t,%
\bar{t}\right) -\psi _{1}\right) \ \mathsf{A}_{\kappa }\left( t,\bar{t}%
\right) , \\ 
\\ 
h_{2}\left( t,\bar{t}\right) =\psi _{0}\mathsf{A}_{\kappa }^{2}\left( t,\bar{%
t}\right) ,%
\end{array}
\label{Eq6.106}
\end{equation}%
\begin{equation}
\begin{array}{c}
p\left( t,\bar{t}\right) =-\int_{t}^{\bar{t}}\mathsf{\bar{B}}_{-\kappa
}\left( t,s\right) \chi \left( s\right) ds+x+\mathsf{B}_{\kappa }\left( t,%
\bar{t}\right) \left( \int_{t}^{\bar{t}}\mathsf{A}_{-\kappa }\left(
t,s\right) \chi \left( s\right) ds+y\right) , \\ 
\\ 
q\left( t,\bar{t}\right) =\mathsf{A}_{\kappa }\left( t,\bar{t}\right) \left(
\int_{t}^{\bar{t}}\mathsf{A}_{-\kappa }\left( t,s\right) \chi \left(
s\right) ds+y\right) .%
\end{array}
\label{Eq6.108}
\end{equation}%
Thus, $\varpi \left( t,x,y,\bar{t},\bar{x},\bar{y}\right) $ is a bivariate
Gaussian distribution of the form (\ref{Eq6.21}) with the covariance matrix $%
\mathfrak{H}$, given by Eq. (\ref{Eq6.105}) centered at the point $\mathbf{r=%
}\left( p,q\right) ^{\ast }$ given by Eq. (\ref{Eq6.107}). Explicitly, one
has%
\begin{equation}
\begin{array}{c}
\sigma _{x}^{2}\left( t,\bar{t}\right) =h_{0}\left( t,\bar{t}\right) ,\ \ \
\sigma _{y}^{2}\left( t,\bar{t}\right) =h_{2}\left( t,\bar{t}\right) ,\ \ \
\rho \left( t,\bar{t}\right) =\frac{h_{1}\left( t,\bar{t}\right) }{\sqrt{%
h_{0}\left( t,\bar{t}\right) h_{2}\left( t,\bar{t}\right) }}.%
\end{array}
\label{Eq6.109}
\end{equation}%
When $\chi ,\kappa ,\theta ,\varepsilon $ are constant, the above formulas
become significantly simpler. Namely,%
\begin{equation}
\begin{array}{c}
\mathfrak{L}\left( T\right) =\left( 
\begin{array}{cc}
1 & 0 \\ 
-\mathsf{B}_{-\kappa }\left( T\right) & \mathsf{A}_{-\kappa }\left( T\right)%
\end{array}%
\right) ,\ \ \ \mathfrak{L}^{-1}\left( T\right) =\left( 
\begin{array}{cc}
1 & 0 \\ 
\mathsf{B}_{\kappa }\left( T\right) & \mathsf{A}_{\kappa }\left( T\right)%
\end{array}%
\right) ,%
\end{array}
\label{Eq6.110}
\end{equation}%
\begin{equation}
\begin{array}{c}
\mathfrak{C}^{-1}\left( T\right) =\left( 
\begin{array}{cc}
\frac{\varepsilon ^{2}}{\kappa ^{2}}\left( \mathsf{B}_{0}\left( T\right) -2%
\mathsf{B}_{-\kappa }\left( T\right) +\mathsf{B}_{-2\kappa }\left( T\right)
\right) & -\frac{\varepsilon ^{2}}{2}\mathsf{B}_{-\kappa }\left( T\right) \\ 
-\frac{\varepsilon ^{2}}{2}\mathsf{B}_{-\kappa }\left( T\right) & 
\varepsilon ^{2}\mathsf{B}_{-2\kappa }\left( T\right)%
\end{array}%
\right) ,%
\end{array}
\label{Eq6.111}
\end{equation}%
\begin{equation}
\begin{array}{c}
\mathbf{d}^{\left( z\right) }\left( T\right) =\left( 
\begin{array}{c}
\left( T-\mathsf{B}_{-\kappa }\left( T\right) \right) \theta \\ 
\mathsf{B}_{-\kappa }\left( T\right) \chi%
\end{array}%
\right) ,%
\end{array}
\label{Eq6.112}
\end{equation}%
\begin{equation}
\begin{array}{c}
\varsigma \left( T\right) =-\kappa T,%
\end{array}
\label{Eq6.113}
\end{equation}%
\begin{equation}
\begin{array}{c}
\mathfrak{H}\left( T\right) =\left( 
\begin{array}{cc}
\frac{\varepsilon ^{2}}{\kappa ^{2}}\left( \mathsf{B}_{0}\left( T\right) -2%
\mathsf{B}_{\kappa }\left( T\right) +\mathsf{B}_{2\kappa }\left( T\right)
\right) & \frac{\varepsilon ^{2}}{2}\mathsf{B}_{\kappa }\left( T\right) \\ 
\frac{\varepsilon ^{2}}{2}\mathsf{B}_{\kappa }\left( T\right) & \varepsilon
^{2}\mathsf{B}_{2\kappa }\left( T\right)%
\end{array}%
\right) ,%
\end{array}
\label{Eq6.114}
\end{equation}%
\begin{equation}
\begin{array}{c}
\mathbf{r}\left( T\right) =\left( 
\begin{array}{c}
x+\theta T-\mathsf{B}_{\kappa }\left( T\right) \left( \theta -y\right) \\ 
\\ 
\theta -\mathsf{A}_{\kappa }\left( T\right) \left( \theta -y\right)%
\end{array}%
\right) .%
\end{array}
\label{Eq6.115}
\end{equation}%
Thus, when coefficients are constant, $\varpi \left( t,x,y,\bar{t},\bar{x},%
\bar{y}\right) $ is a bivariate Gaussian distribution of the form (\ref%
{Eq6.21}) with the covariance matrix $\mathfrak{H}$, given by Eq. (\ref%
{Eq6.114}) and the mean vector $\mathbf{r=}\left( p,q\right) ^{\ast }$ given
by Eq. (\ref{Eq6.115}).

Calculate the marginal distribution of $\bar{x}$, denoted by $\varpi
^{\left( x\right) }\left( t,y,\bar{t},\bar{x}\right) $, which is used on
several occasions below. It is well known that marginal distributions of a
multivariate Gaussian distribution are also Gaussian, so that:%
\begin{equation}
\begin{array}{c}
\varpi ^{\left( x\right) }\left( t,y,\bar{t},\bar{x}\right) =\frac{1}{\sqrt{%
2\pi h_{0}\left( t,\bar{t}\right) }}\exp \left( \frac{\left( \bar{x}-p\left(
t,\bar{t}\right) \right) ^{2}}{2h_{0}\left( t,\bar{t}\right) }\right) ,%
\end{array}
\label{Eq6.116}
\end{equation}%
where $h_{0}$ is given by Eqs (\ref{Eq6.114}). At the same time, the density
of marginal distribution for $\bar{y}$ has the form:%
\begin{equation}
\begin{array}{c}
\varpi ^{\left( y\right) }\left( t,y,\bar{t},\bar{y}\right) =\frac{1}{\sqrt{%
2\pi h_{2}\left( t,\bar{t}\right) }}\exp \left( \frac{\left( \bar{y}-q\left(
t,\bar{t}\right) \right) ^{2}}{2h_{2}\left( t,\bar{t}\right) }\right) ,%
\end{array}
\label{Eq6.117}
\end{equation}%
where $h_{2}$ is given by Eqs (\ref{Eq6.114}), which is the familiar density
of the OU process derived in the previous section.

\subsection{Example: Diffusion of Free and Harmonically Bound Particles\label%
{Sec65}}

The above results can be used to revisit the motion of free and harmonically
bound particles considered in Chapter \ref{Chap3}.

\begin{figure}[tbp]
\begin{center}
\includegraphics[width=0.8\textwidth]
{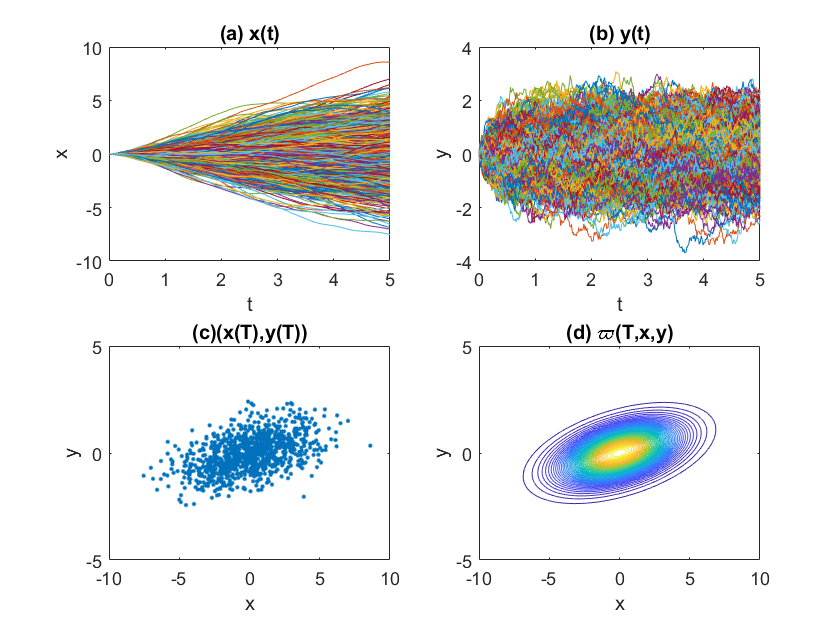}
\end{center}
\par
\vspace{-10pt}
\caption{A thousand trajectories of a typical free particle. Parameters are
as follows: $T=5$, $dt=0.01$, $\protect\kappa =0.8$, $\protect\sigma =1.0$.
(a) $x\left( t\right) $, (b) $y\left( t\right) $, (c) $\left( \bar{x}\left(
T\right) ,\bar{y}\left( T\right) \right) $, (d) contour lines of $\protect%
\varpi \left( 0,0,0,T,\tilde{x},\tilde{y}\right) $. Own graphics.}
\label{Fig6.1}
\end{figure}

To describe a free particle, it is assumed that $\chi =0$. Eq. (\ref{Eq6.114}%
) does not change, while Eq. (\ref{Eq6.115}) can be simplified as follows:%
\begin{equation}
\begin{array}{c}
\left( 
\begin{array}{c}
p\left( T\right) \\ 
q\left( T\right)%
\end{array}%
\right) =\left( 
\begin{array}{c}
x+\mathsf{B}_{\kappa }\left( T\right) y \\ 
\mathsf{A}_{\kappa }\left( T\right) y%
\end{array}%
\right) .%
\end{array}
\label{Eq6.118}
\end{equation}%
It is clear that Eqs (\ref{Eq4.6}), (\ref{Eq4.7}) and (\ref{Eq6.114}), (\ref%
{Eq6.118}) are in agreement. A typical free particle behavior is illustrated
in Figure \ref{Fig6.1}. 

Analysis of a harmonically bound particle requires additional efforts. In
the case in question, Eq. (\ref{Eq6.35}) can be written as follows:%
\begin{equation}
\begin{array}{c}
\mathfrak{L}^{\prime }\left( t,\bar{t}\right) +\left( 
\begin{array}{cc}
0 & -\omega ^{2} \\ 
1 & -\kappa%
\end{array}%
\right) \mathfrak{L}\left( t,\bar{t}\right) =0,\ \ \mathfrak{L}\left(
t,t\right) =\left( 
\begin{array}{cc}
1 & 0 \\ 
0 & 1%
\end{array}%
\right) .%
\end{array}
\label{Eq6.119}
\end{equation}%
The corresponding characteristic equation and its solutions are as follows:%
\begin{equation}
\begin{array}{c}
\lambda ^{2}-\kappa \lambda +\omega ^{2}=0,%
\end{array}
\label{Eq6.120}
\end{equation}%
\begin{equation}
\begin{array}{c}
\lambda _{\pm }=\mu \pm \zeta , \\ 
\\ 
\mu =\frac{\kappa }{2},\ \ \ \zeta =\frac{\sqrt{\kappa ^{2}-4\omega ^{2}}}{2}%
.%
\end{array}
\label{Eq6.120a}
\end{equation}%
Introduce%
\begin{equation}
\begin{array}{c}
\mathsf{E}_{0}\left( T\right) =e^{\mu T}=e^{\kappa T/2},\ \ \ \mathsf{E}%
_{\pm }\left( T\right) =e^{\pm \zeta T}.%
\end{array}
\label{Eq6.125}
\end{equation}%
It is left to the reader to check that%
\begin{equation}
\begin{array}{c}
\mathfrak{L}=\frac{\mathsf{E}_{0}}{\sqrt{\kappa ^{2}-4\omega ^{2}}}\left( 
\begin{array}{cc}
-\left( \lambda _{-}\mathsf{E}_{+}-\lambda _{+}\mathsf{E}_{-}\right) & 
\omega ^{2}\left( \mathsf{E}_{+}-\mathsf{E}_{-}\right) \\ 
-\left( \mathsf{E}_{+}-\mathsf{E}_{-}\right) & \left( \lambda _{+}\mathsf{E}%
_{+}-\lambda _{-}\mathsf{E}_{-}\right)%
\end{array}%
\right) ,%
\end{array}
\label{Eq6.122}
\end{equation}%
\begin{equation}
\begin{array}{c}
\mathfrak{L}^{-1}=\frac{\mathsf{E}_{0}^{-1}}{\sqrt{\kappa ^{2}-4\omega ^{2}}}%
\left( 
\begin{array}{cc}
\left( \lambda _{+}\mathsf{E}_{+}-\lambda _{-}\mathsf{E}_{-}\right) & 
-\omega ^{2}\left( \mathsf{E}_{+}-\mathsf{E}_{-}\right) \\ 
\left( \mathsf{E}_{+}-\mathsf{E}_{-}\right) & -\left( \lambda _{-}\mathsf{E}%
_{+}-\lambda _{+}\mathsf{E}_{-}\right)%
\end{array}%
\right) ,%
\end{array}
\label{Eq6.123}
\end{equation}%
\begin{equation}
\begin{array}{c}
\det \mathfrak{L=}\left( \det \mathfrak{L}^{-1}\right) ^{-1}\mathfrak{=}%
\mathsf{E}_{0}^{2}=e^{\kappa T},%
\end{array}
\label{Eq6.124}
\end{equation}%
Accordingly,%
\begin{equation}
\begin{array}{c}
\varpi \left( t,\mathbf{z,}\bar{t},\mathbf{\bar{z}}\right) =\mathrm{N}\left( 
\mathbf{r}\left( t,\bar{t}\right) ,\mathfrak{H}\left( t,\bar{t}\right)
\right) ,%
\end{array}
\label{Eq6.126}
\end{equation}%
with%
\begin{equation}
\begin{array}{c}
\mathfrak{H}=\left( \mathfrak{L}^{\ast }\right) ^{-1}\mathfrak{C}^{-1}%
\mathfrak{L}^{-1}, \\ 
\\ 
\mathbf{r=}\left( \mathfrak{L}^{\ast }\right) ^{-1}\mathbf{z}.%
\end{array}
\label{Eq6.127}
\end{equation}%
Here, 
\begin{equation}
\begin{array}{c}
\mathfrak{C}^{-1}=\left( 
\begin{array}{cc}
\psi _{2} & -\psi _{1} \\ 
-\psi _{1} & \psi _{0}%
\end{array}%
\right) ,%
\end{array}
\label{Eq6.130}
\end{equation}%
where%
\begin{equation}
\begin{array}{c}
\psi _{0}=\frac{\varepsilon ^{2}}{\left( \kappa ^{2}-4\omega ^{2}\right) }%
\int_{t}^{\bar{t}}\left( \lambda _{+}e^{\lambda _{+}\left( s-t\right)
}-\lambda _{-}e^{\lambda _{-}\left( s-t\right) }\right) ^{2}ds \\ 
\\ 
=\frac{\varepsilon ^{2}}{2\kappa \left( \kappa ^{2}-4\omega ^{2}\right) }%
\left( \mathsf{E}_{0}^{2}\left( \kappa \lambda _{+}\mathsf{E}%
_{+}^{2}-4\omega ^{2}+\kappa \lambda _{-}\mathsf{E}_{-}^{2}\right) -\left(
\kappa ^{2}-4\omega ^{2}\right) \right) , \\ 
\\ 
\psi _{1}=\frac{\varepsilon ^{2}}{\left( \kappa ^{2}-4\omega ^{2}\right) }%
\int_{t}^{\bar{t}}\left( e^{\lambda _{+}\left( s-t\right) }-e^{\lambda
_{-}\left( s-t\right) }\right) \left( \lambda _{+}e^{\lambda _{+}\left(
s-t\right) }-\lambda _{-}e^{\lambda _{-}\left( s-t\right) }\right) ds \\ 
\\ 
=\frac{\varepsilon ^{2}}{2\left( \kappa ^{2}-4\omega ^{2}\right) }\mathsf{E}%
_{0}^{2}\left( \mathsf{E}_{+}-\mathsf{E}_{-}\right) ^{2}, \\ 
\\ 
\psi _{2}=\frac{\varepsilon ^{2}}{\left( \kappa ^{2}-4\omega ^{2}\right) }%
\int_{t}^{\bar{t}}\left( e^{\lambda _{+}\left( s-t\right) }-e^{\lambda
_{-}\left( s-t\right) }\right) ^{2}ds \\ 
\\ 
=\frac{\varepsilon ^{2}}{2\kappa \omega ^{2}\left( \kappa ^{2}-4\omega
^{2}\right) }\left( \mathsf{E}_{0}^{2}\left( \kappa \lambda _{-}\mathsf{E}%
_{+}^{2}-4\omega ^{2}+\kappa \lambda _{+}\mathsf{E}_{-}^{2}\right) -\left(
\kappa ^{2}-4\omega ^{2}\right) \right) .%
\end{array}
\label{Eq6.131}
\end{equation}%
Further,%
\begin{equation}
\begin{array}{c}
\mathfrak{H}=\frac{\mathsf{E}_{0}^{-2}}{\left( \kappa ^{2}-4\omega
^{2}\right) }\left( 
\begin{array}{cc}
\left( \lambda _{+}\mathsf{E}_{+}-\lambda _{-}\mathsf{E}_{-}\right) & \left( 
\mathsf{E}_{+}-\mathsf{E}_{-}\right) \\ 
-\omega ^{2}\left( \mathsf{E}_{+}-\mathsf{E}_{-}\right) & -\left( \lambda
_{-}\mathsf{E}_{+}-\lambda _{+}\mathsf{E}_{-}\right)%
\end{array}%
\right) \\ 
\\ 
\times \left( 
\begin{array}{cc}
\psi _{2} & -\psi _{1} \\ 
-\psi _{1} & \psi _{0}%
\end{array}%
\right) \left( 
\begin{array}{cc}
\left( \lambda _{+}\mathsf{E}_{+}-\lambda _{-}\mathsf{E}_{-}\right) & 
-\omega ^{2}\left( \mathsf{E}_{+}-\mathsf{E}_{-}\right) \\ 
\left( \mathsf{E}_{+}-\mathsf{E}_{-}\right) & -\left( \lambda _{-}\mathsf{E}%
_{+}-\lambda _{+}\mathsf{E}_{-}\right)%
\end{array}%
\right) .%
\end{array}
\label{Eq6.132}
\end{equation}%
Straightforward but tedious calculation yields%
\begin{equation}
\begin{array}{c}
h_{0}=\frac{\varepsilon ^{2}}{2\kappa \omega ^{2}}\left( 1-\frac{\mathsf{E}%
_{0}^{-2}\left( \omega ^{2}\left( \mathsf{E}_{+}-\mathsf{E}_{-}\right)
^{2}+\left( \lambda _{+}\mathsf{E}_{+}-\lambda _{-}\mathsf{E}_{-}\right)
^{2}\right) }{\left( \kappa ^{2}-4\omega ^{2}\right) }\right) , \\ 
\\ 
h_{1}=\frac{\varepsilon ^{2}}{2}\frac{\mathsf{E}_{0}^{-2}\left( \mathsf{E}%
_{+}-\mathsf{E}_{-}\right) ^{2}}{\left( \kappa ^{2}-4\omega ^{2}\right) },
\\ 
\\ 
h_{2}=\frac{\varepsilon ^{2}}{2\kappa }\left( 1-\frac{\mathsf{E}%
_{0}^{-2}\left( \omega ^{2}\left( \mathsf{E}_{+}-\mathsf{E}_{-}\right)
^{2}+\left( \lambda _{-}\mathsf{E}_{+}-\lambda _{+}\mathsf{E}_{-}\right)
^{2}\right) }{\left( \kappa ^{2}-4\omega ^{2}\right) }\right) .%
\end{array}
\label{Eq6.134}
\end{equation}%
In the limit $\omega ^{2}\rightarrow 0$,%
\begin{equation}
\begin{array}{c}
h_{0}=\frac{\varepsilon ^{2}}{\kappa ^{2}}\left( \mathsf{B}_{0}-2\mathsf{B}%
_{\kappa }+\mathsf{B}_{2\kappa }\right) ,\ \ \ h_{1}=\frac{\varepsilon ^{2}}{%
2}\mathsf{B}_{\kappa }^{2},\ \ \ h_{2}=\varepsilon ^{2}\mathsf{B}_{2\kappa },%
\end{array}
\label{Eq6.135}
\end{equation}%
so that Eqs. (\ref{Eq6.114}) and (\ref{Eq6.135}) are in agreement.

\begin{figure}[tbp]
\begin{center}
\includegraphics[width=0.8\textwidth]
{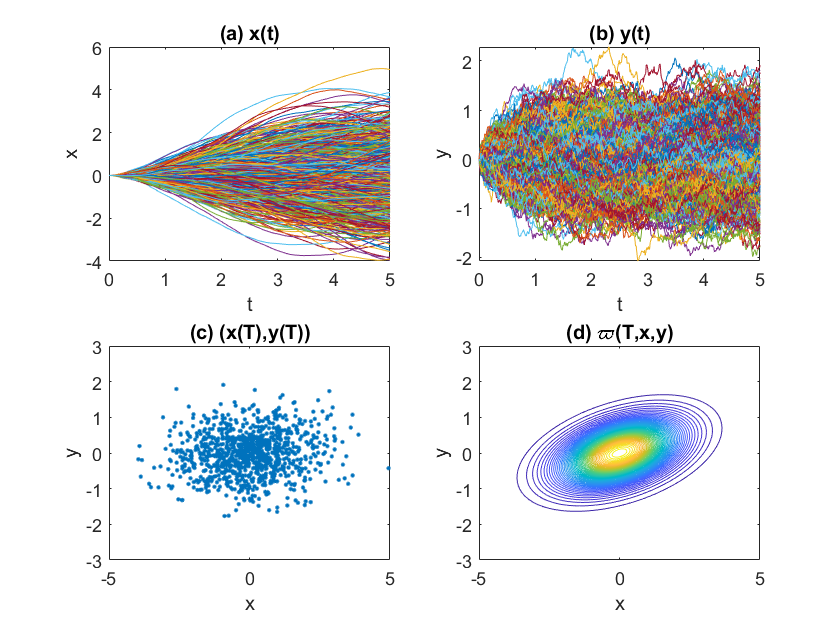}
\end{center}
\par
\vspace{-10pt}
\caption{A thousand trajectories of a harmonically bounded particle.
Parameters are as follows: $T=5$, $dt=0.01$, $\protect\kappa =0.2$, $\protect%
\omega =0.5 $, $\protect\sigma =0.5$. (a) $x\left( t\right) $, (b) $y\left(
t\right) $, (c) $\left( \bar{x}\left( T\right) ,\bar{y}\left( T\right)
\right) $, (d) contour lines of $\protect\varpi \left( 0,0,0,T,\tilde{x},%
\tilde{y}\right) $. Own graphics.}
\label{Fig6.2}
\end{figure}

Here%
\begin{equation}
\begin{array}{c}
\mathbf{r=}\left( 
\begin{array}{c}
p \\ 
q%
\end{array}%
\right) =\left( 
\begin{array}{c}
\frac{\mathsf{E}_{0}^{-1}\left( \left( \lambda _{+}\mathsf{E}_{+}-\lambda
_{-}\mathsf{E}_{-}\right) x+\left( \mathsf{E}_{+}-\mathsf{E}_{-}\right)
y\right) }{\sqrt{\kappa ^{2}-4\omega ^{2}}} \\ 
-\frac{\mathsf{E}_{0}^{-1}\left( \omega ^{2}\left( \mathsf{E}_{+}-\mathsf{E}%
_{-}\right) x+\left( \lambda _{-}\mathsf{E}_{+}-\lambda _{+}\mathsf{E}%
_{-}\right) y\right) }{\sqrt{\kappa ^{2}-4\omega ^{2}}}%
\end{array}%
\right) .%
\end{array}
\label{Eq6.128}
\end{equation}%
In the limit $\omega ^{2}\rightarrow 0$,%
\begin{equation}
\begin{array}{c}
\mathbf{r=}\left( 
\begin{array}{c}
p \\ 
q%
\end{array}%
\right) =\left( 
\begin{array}{c}
x+\mathsf{B}_{\kappa }\left( T\right) y \\ 
\mathsf{A}_{\kappa }\left( T\right) y%
\end{array}%
\right) .%
\end{array}
\label{Eq6.129}
\end{equation}%
Moreover, while it is easy to show that Chandrasekhar's solution given in 
\cite{Chandrasekhar43} is in agreement with the solution given by Eq. (\ref%
{Eq6.126}), the solution is more convenient from a practical standpoint,
since it is \emph{explicitly} written as a Gaussian density in the $\left( 
\bar{x},\bar{y}\right) $ space. A typical bounded particle behavior is shown
in Figure \ref{Fig6.2}. 

\subsection{Example: Vorticity of Two-Dimensional Flows\label{Sec66}}

Briefly return to the starting point and consider strictly two-dimensional
flows; see \cite{Friedlander03}. Velocity fields of such flows have the form:%
\begin{equation}
\begin{array}{c}
\mathbf{V}\left( \bar{t},\bar{x}_{1},\bar{x}_{2}\right) =\left( V_{1}\left( 
\bar{t},\bar{x}_{1},\bar{x}_{2}\right) ,V_{2}\left( \bar{t},\bar{x}_{1},\bar{%
x}_{2}\right) \right) , \\ 
\\ 
\mathbf{v}\left( \bar{t},\bar{x}_{1},\bar{x}_{2}\right) =\left( v_{1}\left( 
\bar{t},\bar{x}_{1},\bar{x}_{2}\right) ,v_{2}\left( \bar{t},\bar{x}_{1},\bar{%
x}_{2}\right) \right) .%
\end{array}
\label{Eq6.136}
\end{equation}%
By virtue of incompressibility, one can introduce the so-called stream
functions such that%
\begin{equation}
\begin{array}{c}
V_{1}=-\frac{\partial \Psi }{\partial \bar{x}_{2}},\ \ \ V_{2}=\frac{%
\partial \Psi }{\partial \bar{x}_{1}},\ \ \ v_{1}=-\frac{\partial \psi }{%
\partial \bar{x}_{2}},\ \ \ v_{2}=\frac{\partial \psi }{\partial \bar{x}_{1}}%
,%
\end{array}
\label{Eq6.137}
\end{equation}%
and define the scalar vorticity as follows:%
\begin{equation}
\begin{array}{c}
\Omega =\Delta \Psi ,\ \ \ \omega =\Delta \psi .%
\end{array}
\label{Eq6.138}
\end{equation}%
Contour lines of $\Psi $ are called streamlines of the flow.

By using the above definitions, the two-dimensional the Navier-Stokes
equations can be written as equations for the stream and vorticity: 
\begin{equation}
\begin{array}{c}
\frac{\partial \Omega }{\partial \bar{t}}-\frac{\partial \Psi }{\partial 
\bar{x}_{2}}\frac{\partial \Omega }{\partial \bar{x}_{1}}+\frac{\partial
\Psi }{\partial \bar{x}_{1}}\frac{\partial \Omega }{\partial \bar{x}_{2}}%
-\nu \Delta \Omega =0, \\ 
\\ 
\Delta \Psi -\Omega =0.%
\end{array}
\label{Eq6.139}
\end{equation}%
Time-independent quadratic stream functions $\Psi \left( \bar{x}_{1},\bar{x}%
_{2}\right) $ generate exact equilibrium solutions of Eqs (\ref{Eq6.139}).
Consider fields consisting of pure strain and pure rotation. The
corresponding $\Psi $ have the form:%
\begin{equation}
\begin{array}{c}
\Psi \left( \bar{x}_{1},\bar{x}_{2}\right) =\frac{1}{4}\left( w\left( \bar{x}%
_{1}^{2}+\bar{x}_{2}^{2}\right) -2s\bar{x}_{1}\bar{x}_{2}\right) ,%
\end{array}
\label{Eq6.140}
\end{equation}%
where $\omega >s$, to ensure that streamlines are elliptic rather than
hyperbolic, so that%
\begin{equation}
\begin{array}{c}
V_{1}=-\frac{\partial \Psi }{\partial \bar{x}_{2}}=\frac{1}{2}\left( s\bar{x}%
_{1}-w\bar{x}_{2}\right) ,\ \ \ V_{2}=\frac{\partial \Psi }{\partial \bar{x}%
_{1}}=\frac{1}{2}\left( w\bar{x}_{1}-s\bar{x}_{2}\right) .%
\end{array}
\label{Eq6.141}
\end{equation}%
Recall that these flows were introduced in Chapter \ref{Chap2}, Eq. (\ref%
{Eq2.5}).

Small perturbations $\psi $ of the time-independent quadratic stream
function $\Psi $ satisfy the following equations:%
\begin{equation}
\begin{array}{c}
\frac{\partial \omega }{\partial \bar{t}}-\frac{\partial \Psi }{\partial 
\bar{x}_{2}}\frac{\partial \omega }{\partial \bar{x}_{1}}+\frac{\partial
\Psi }{\partial \bar{x}_{1}}\frac{\partial \omega }{\partial \bar{x}_{2}}%
-\nu \Delta \omega =0, \\ 
\\ 
\Delta \psi -\omega =0.%
\end{array}
\label{Eq6.142}
\end{equation}%
It is helpful to study the first equation (\ref{Eq6.142}) in isolation, by
writing it explicitly as follows:%
\begin{equation}
\begin{array}{c}
\frac{\partial \omega }{\partial \bar{t}}+\frac{1}{2}\left( s\bar{x}_{1}-w%
\bar{x}_{2}\right) \frac{\partial \omega }{\partial \bar{x}_{1}}+\frac{1}{2}%
\left( w\bar{x}_{1}-s\bar{x}_{2}\right) \frac{\partial \omega }{\partial 
\bar{x}_{2}}-\nu \Delta \omega =0,%
\end{array}
\label{Eq6.143}
\end{equation}%
and supplying it with the initial condition at time $t$:%
\begin{equation}
\begin{array}{c}
\omega \left( t,\bar{x}_{1},\bar{x}_{2}\right) =\delta \left( \bar{x}%
_{1}-x_{1}\right) \delta \left( \bar{x}_{2}-x_{2}\right) .%
\end{array}
\label{Eq6.144}
\end{equation}%
Once the solution of Eqs(\ref{Eq6.143}), (\ref{Eq6.144}) is found, one can
find $\psi $ by solving the corresponding Laplace equation.

Surprisingly, this equation is identical to the Fokker-Planck equation
associated with the following SDEs for $\mathbf{\hat{z}}_{t}=\left( \hat{x}%
_{1t},\hat{x}_{2t}\right) $:%
\begin{equation}
\begin{array}{c}
d\mathbf{\hat{z}}_{t}=\mathfrak{B}\mathbf{\hat{z}}_{t}dt+\mathbf{\Sigma }d%
\mathbf{\hat{W}}_{t},\ \ \ \mathbf{\hat{z}}_{t}=\left( 
\begin{array}{c}
x_{1} \\ 
x_{2}%
\end{array}%
\right) ,%
\end{array}
\label{Eq6.145}
\end{equation}%
where%
\begin{equation}
\begin{array}{c}
\mathfrak{B=}\frac{1}{2}\left( 
\begin{array}{cc}
s & -w \\ 
w & -s%
\end{array}%
\right) ,\ \ \ \mathbf{\Sigma =}\sqrt{2\nu }\left( 
\begin{array}{cc}
1 & 0 \\ 
0 & 1%
\end{array}%
\right) .%
\end{array}
\label{Eq6.146}
\end{equation}%
Thus, one can use Section \ref{Sec61} results. Eq. (\ref{Eq6.35}) becomes%
\begin{equation}
\begin{array}{c}
\mathfrak{L}^{\prime }\left( t,\bar{t}\right) +\frac{1}{2}\left( 
\begin{array}{cc}
s & w \\ 
-w & -s%
\end{array}%
\right) \mathfrak{L}\left( t,\bar{t}\right) =0,\ \ \mathfrak{L}\left(
t,t\right) =\left( 
\begin{array}{cc}
1 & 0 \\ 
0 & 1%
\end{array}%
\right) .%
\end{array}
\label{Eq6.147a}
\end{equation}%
The corresponding characteristic equation has the form:%
\begin{equation}
\begin{array}{c}
\lambda ^{2}+\frac{1}{4}\left( w^{2}-s^{2}\right) =0.%
\end{array}
\label{Eq6.147b}
\end{equation}%
Its solutions are%
\begin{equation}
\begin{array}{c}
\lambda _{\pm }=\pm \zeta ,\ \ \ \zeta =\frac{i\sqrt{w^{2}-s^{2}}}{2}.%
\end{array}
\label{Eq6.147c}
\end{equation}%
Simple but tedious calculations omitted for the sake of brevity show that%
\begin{equation}
\begin{array}{c}
\mathfrak{L}=\left( 
\begin{array}{cc}
\mathsf{c}_{1}-\frac{s}{2\left\vert \zeta \right\vert }\mathsf{s}_{1} & -%
\frac{w}{2\left\vert \zeta \right\vert }\mathsf{s}_{1} \\ 
\frac{w}{2\left\vert \zeta \right\vert }\mathsf{s}_{1} & \mathsf{c}_{1}+%
\frac{s}{2\left\vert \zeta \right\vert }\mathsf{s}_{1}%
\end{array}%
\right) ,\ \ \ \det \left( \mathfrak{L}\right) =1 \\ 
\\ 
\mathfrak{L}^{-1}=\left( 
\begin{array}{cc}
\mathsf{c}_{1}+\frac{s}{2\left\vert \zeta \right\vert }\mathsf{s}_{1} & 
\frac{w}{2\left\vert \zeta \right\vert }\mathsf{s}_{1} \\ 
-\frac{w}{2\left\vert \zeta \right\vert }\mathsf{s}_{1} & \mathsf{c}_{1}-%
\frac{s}{2\left\vert \zeta \right\vert }\mathsf{s}_{1}%
\end{array}%
\right) ,\ \ \ \det \left( \mathfrak{L}^{-1}\right) =1%
\end{array}
\label{Eq6.148}
\end{equation}%
where%
\begin{equation}
\begin{array}{c}
\mathsf{c}_{1}\left( t,\bar{t}\right) =\cos \left( \left\vert \zeta
\right\vert T\right) ,\ \ \ \mathsf{s}_{1}\left( t,\bar{t}\right) =\sin
\left( \left\vert \zeta \right\vert T\right) .%
\end{array}
\label{Eq6.149}
\end{equation}%
Next, Eq. (\ref{Eq6.40}) yields%
\begin{equation}
\begin{array}{c}
\mathfrak{C}^{-1}=2\nu \int_{t}^{\bar{t}}\mathfrak{L}^{\ast }\left(
t,s\right) \mathfrak{L}\left( t,s\right) ds=\left( 
\begin{array}{cc}
\psi _{2} & -\psi _{1} \\ 
-\psi _{1} & \psi _{0}%
\end{array}%
\right) ,%
\end{array}
\label{Eq6.150}
\end{equation}%
where%
\begin{equation}
\begin{array}{c}
\psi _{0}=2\nu \int_{t}^{\bar{t}}\left( 1+\frac{s}{2\left\vert \zeta
\right\vert }\mathsf{s}_{2}\left( t,s\right) +\frac{s^{2}}{4\left\vert \zeta
\right\vert ^{2}}\left( 1-\mathsf{c}_{2}\left( t,s\right) \right) \right) ds
\\ 
\\ 
=2\nu \left( \left( 1+\frac{s^{2}}{4\left\vert \zeta \right\vert ^{2}}%
\right) T-\frac{s}{4\left\vert \zeta \right\vert ^{2}}\mathsf{c}_{2}-\frac{%
s^{2}}{8\left\vert \zeta \right\vert ^{3}}\mathsf{s}_{2}\right) , \\ 
\\ 
\psi _{1}=-\frac{\nu sw}{2\left\vert \zeta \right\vert ^{2}}\int_{t}^{\bar{t}%
}\left( 1-\mathsf{c}_{2}\left( t,s\right) \right) ds=-\frac{\nu sw}{%
2\left\vert \zeta \right\vert ^{2}}\left( T-\frac{1}{2\left\vert \zeta
\right\vert }\mathsf{s}_{2}\right) , \\ 
\\ 
\psi _{2}=2\nu \int_{t}^{\bar{t}}\left( 1-\frac{s}{2\left\vert \zeta
\right\vert }\mathsf{s}_{2}\left( t,s\right) +\frac{s^{2}}{4\left\vert \zeta
\right\vert ^{2}}\left( 1-\mathsf{c}_{2}\left( t,s\right) \right) \right) ds
\\ 
\\ 
=2\nu \left( \left( 1+\frac{s^{2}}{4\left\vert \zeta \right\vert ^{2}}%
\right) T+\frac{s}{4\left\vert \zeta \right\vert ^{2}}\mathsf{c}_{2}-\frac{%
s^{2}}{8\left\vert \zeta \right\vert ^{3}}\mathsf{s}_{2}\right) ,%
\end{array}
\label{Eq6.151}
\end{equation}%
and%
\begin{equation}
\begin{array}{c}
\mathsf{c}_{2}\left( t,\bar{t}\right) =\cos \left( 2\left\vert \zeta
\right\vert T\right) ,\ \ \ \mathsf{s}_{2}\left( t,\bar{t}\right) =\sin
\left( 2\left\vert \zeta \right\vert T\right) .%
\end{array}
\label{Eq6.152}
\end{equation}%
Finally, Eqs (\ref{Eq6.21}), (\ref{Eq6.22}) yield:%
\begin{equation}
\begin{array}{c}
\omega \left( t,\mathbf{z,}\bar{t},\mathbf{\bar{z}}\right) =\mathrm{N}\left( 
\mathbf{r}\left( t,\bar{t}\right) ,\mathfrak{H}\left( t,\bar{t}\right)
\right) .%
\end{array}
\label{Eq6.153}
\end{equation}%
The corresponding covariance matrix $\mathfrak{H}$ and mean $\mathbf{r}$ are
as follows: 
\begin{equation}
\begin{array}{c}
\mathfrak{H}=\left( 
\begin{array}{cc}
h_{0} & h_{1} \\ 
h_{1} & h_{2}%
\end{array}%
\right) ,%
\end{array}
\label{Eq6.154}
\end{equation}%
where%
\begin{equation}
\begin{array}{c}
h_{0}=\left( \frac{w^{2}}{8\left\vert \zeta \right\vert ^{2}}+\frac{\left(
4\left\vert \zeta \right\vert ^{2}-s^{2}\right) }{8\left\vert \zeta
\right\vert ^{2}}\mathsf{c}_{2}+\frac{s}{2\left\vert \zeta \right\vert }%
\mathsf{s}_{2}\right) \psi _{2} \\ 
\\ 
+\frac{w}{2\left\vert \zeta \right\vert }\left( \frac{s}{2\left\vert \zeta
\right\vert }\left( 1-\mathsf{c}_{2}\right) +\mathsf{s}_{2}\right) \psi _{1}+%
\frac{w^{2}}{8\left\vert \zeta \right\vert ^{2}}\left( 1-\mathsf{c}%
_{2}\right) \psi _{0}, \\ 
\\ 
h_{1}=\frac{w}{4\left\vert \zeta \right\vert }\left( \frac{s}{2\left\vert
\zeta \right\vert }\left( 1-\mathsf{c}_{2}\right) +\mathsf{s}_{2}\right)
\psi _{2} \\ 
\\ 
-\left( 1+\frac{w^{2}}{4\left\vert \zeta \right\vert ^{2}}\left( 1-\mathsf{c}%
_{2}\right) \right) \psi _{1}+\frac{w}{4\left\vert \zeta \right\vert }\left( 
\frac{s}{2\left\vert \zeta \right\vert }\left( 1-\mathsf{c}_{2}\right) -%
\mathsf{s}_{2}\right) \psi _{0}, \\ 
\\ 
h_{2}=\frac{w^{2}}{8\left\vert \zeta \right\vert ^{2}}\left( 1-\mathsf{c}%
_{2}\right) \psi _{2}+\frac{w}{2\left\vert \zeta \right\vert }\left( \frac{s%
}{2\left\vert \zeta \right\vert }\left( 1-\mathsf{c}_{2}\right) -\mathsf{s}%
_{2}\right) \psi _{1} \\ 
\\ 
+\left( \frac{w^{2}}{8\left\vert \zeta \right\vert ^{2}}+\frac{\left(
4\left\vert \zeta \right\vert ^{2}-s^{2}\right) }{8\left\vert \zeta
\right\vert ^{2}}\mathsf{c}_{2}-\frac{s}{2\left\vert \zeta \right\vert }%
\mathsf{s}_{2}\right) \psi _{0},%
\end{array}
\label{Eq6.155}
\end{equation}%
and%
\begin{equation}
\begin{array}{c}
\mathbf{r=}\left( 
\begin{array}{c}
r_{1} \\ 
r_{2}%
\end{array}%
\right) \mathbf{=}\left( 
\begin{array}{c}
\left( \mathsf{c}_{1}+\frac{s}{2\left\vert \zeta \right\vert }\mathsf{s}%
_{1}\right) x_{1}-\frac{w}{2\left\vert \zeta \right\vert }\mathsf{s}_{1}x_{2}
\\ 
\frac{w}{2\left\vert \zeta \right\vert }\mathsf{s}_{1}x_{1}+\left( \mathsf{c}%
_{1}-\frac{s}{2\left\vert \zeta \right\vert }\mathsf{s}_{1}\right) x_{2}%
\end{array}%
\right) .%
\end{array}
\label{Eq6.157}
\end{equation}%
Eqs (\ref{Eq6.155}) are symmetric, namely $h_{0}\rightarrow h_{2}$ when $%
\left( a,b\right) \rightarrow \left( -a,-b\right) $ and $\left( \psi
_{0},\psi _{2}\right) \rightarrow \left( \psi _{2},\psi _{0}\right) $. The
second Eq. (\ref{Eq6.139}), which is a static Poisson equation, allows us to
find $\psi $, since $\omega $ is known. Its analytical solution is not easy
to derive and is not presented here due to the lack of space. However, the
special case of purely rotational flow, $s=0$, can be done easily; see Eq. (%
\ref{Eq6.165}) below.

It is interesting to note that 
\begin{equation}
\begin{array}{c}
\Psi \left( r_{1},r_{2}\right) =\Psi \left( x_{1},x_{2}\right) ,%
\end{array}
\label{Eq6.158}
\end{equation}%
so that the location of the Gaussian distribution $\omega $ moves along
streamlines of the flow defined by the stream function $\Psi $.

When the flow is purely rotational, so that $s=0$, the above formulas
considerably simplify.\ Specifically, one has:%
\begin{equation}
\begin{array}{c}
\psi _{0}=2\nu T,\ \ \psi _{1}=0,\ \ \ \ \ \psi _{2}=2\nu T, \\ 
\\ 
h_{0}=2\nu T,\ \ \psi _{1}=0,\ \ \ \ \ h_{2}=2\nu T, \\ 
\\ 
r_{1}=\mathsf{c}_{1}x_{1}-\mathsf{s}_{1}x_{2},\ \ \ r_{2}=\mathsf{s}%
_{1}x_{1}+\mathsf{c}_{1}x_{2},%
\end{array}
\label{Eq6.159}
\end{equation}%
so that%
\begin{equation}
\begin{array}{c}
\omega \left( t,x_{1},x_{2},\bar{t},\bar{x}_{1},\bar{x}_{2}\right) \\ 
\\ 
=\frac{1}{4\pi \nu T}\exp \left( -\frac{\left( \bar{x}_{1}-\mathsf{c}%
_{1}x_{1}+\mathsf{s}_{1}x_{2}\right) ^{2}+\left( \bar{x}_{2}-\mathsf{s}%
_{1}x_{1}-\mathsf{c}_{1}x_{2}\right) ^{2}}{4\nu T}\right) .%
\end{array}
\label{Eq6.160}
\end{equation}%
The stream function $\psi $ can be calculated directly by solving the
corresponding Poisson equation.\footnote{%
We are grateful to Andrey Itkin for pointing this out.} To start, notice
that \emph{both} $\omega $ and $\psi $ are rotational symmetric around the
point $\left( x_{1},x_{2}\right) $. Thus, $\omega $ and $\psi $ have the
form:%
\begin{equation}
\begin{array}{c}
\omega =\omega \left( R\right) =\frac{1}{4\pi \nu T}\exp \left( -\frac{R^{2}%
}{2}\right) ,\ \ \ \psi =\psi \left( R\right) ,%
\end{array}
\label{Eq6.162}
\end{equation}%
where 
\begin{equation}
\begin{array}{c}
R^{2}=\frac{\left( \bar{x}_{1}-\mathsf{c}_{1}x_{1}+\mathsf{s}%
_{1}x_{2}\right) ^{2}+\left( \bar{x}_{2}-\mathsf{s}_{1}x_{1}-\mathsf{c}%
_{1}x_{2}\right) ^{2}}{2\nu T}.%
\end{array}
\label{Eq6.161}
\end{equation}%
Then $\psi \left( R\right) $ solves a radially symmetric Poisson equation of
the form:%
\begin{equation}
\begin{array}{c}
\frac{1}{R}\left( R\psi _{R}\left( R\right) \right) _{R}=\frac{1}{2\pi }\exp
\left( -\frac{R^{2}}{2}\right) .%
\end{array}
\label{Eq6.163}
\end{equation}%
Thus,%
\begin{equation}
\begin{array}{c}
R\psi _{R}\left( R\right) =-\frac{1}{2\pi }\exp \left( -\frac{R^{2}}{2}%
\right) +C,%
\end{array}
\label{Eq6.164}
\end{equation}%
where $C$ is an arbitrary constant. Next,%
\begin{equation}
\begin{array}{c}
\psi \left( R\right) =\frac{1}{2\pi }\left( \ln \left( R\right) +\frac{1}{2}%
E_{1}\left( \frac{R^{2}}{2}\right) \right) ,%
\end{array}
\label{Eq6.165}
\end{equation}%
where the choice of $C$ guarantees that $\psi $ has the right behavior when $%
R\rightarrow 0$ and $R\rightarrow \infty $. Here $E_{1}\left( \eta \right) $
is the exponential integral of the form:%
\begin{equation}
\begin{array}{c}
E_{1}\left( \eta \right) =\int\limits_{\eta }^{\infty }\frac{e^{-\eta
^{\prime }}}{\eta ^{\prime }}d\eta ^{\prime }.%
\end{array}
\label{Eq6.166}
\end{equation}

\section{Non-Gaussian Stochastic Processes\label{Chap7}}

\subsection{Regular Non-Gaussian Processes\label{Sec71}}

In many situations, it is useful to consider processes governed by more
general SDEs of the following form:%
\begin{equation}
\begin{array}{c}
d\mathbf{\hat{z}}_{t}=\left( \mathbf{b}\left( t\right) +\mathfrak{B}\left(
t\right) \mathbf{\hat{z}}_{t}\right) dt \\ 
\\ 
+\mathfrak{\mathbf{\Sigma }}\left( t\right) \left( \mathrm{diag}\left( 
\mathbf{d}^{\left( 0\right) }\left( t\right) +\mathfrak{D}\left( t\right) 
\mathbf{\hat{z}}_{t}\right) \right) ^{1/2}d\mathbf{\hat{W}}_{t}^{\left(
z\right) }, \\ 
\\ 
\mathbf{\hat{z}}_{t}=\mathbf{z.}%
\end{array}
\label{Eq7.0}
\end{equation}%
Here, in addition to the functions $\mathbf{b}\left( t\right) $, $\mathfrak{B%
}\left( t\right) $ introduced in the previous section, define an $\left(
M\times 1\right) $ column vector $\mathbf{d}^{\left( 0\right) }$, and an $%
\left( M\times M\right) $ matrix $\mathfrak{D}$. It is convenient to
introduce auxiliary vectors $\mathbf{d}^{\left( i\right) }$ equal to the $i$%
-th column of $\mathfrak{D}$.

Since the corresponding $\left( M\times M\right) $ covariance matrix $%
\mathfrak{A}$ has the form:%
\begin{equation}
\begin{array}{c}
\mathfrak{A=}\frac{1}{2}\mathbf{\Sigma }\left( \mathrm{diag}\left( \mathbf{d}%
^{\left( 0\right) }\left( t\right) +\mathfrak{D}\left( t\right) \mathbf{z}%
_{t}\right) \right) ^{1/2}\left( \mathbf{\Sigma }\left( \mathrm{diag}\left( 
\mathbf{d}^{\left( 0\right) }\left( t\right) +\mathfrak{D}\left( t\right) 
\mathbf{z}_{t}\right) \right) ^{1/2}\right) ^{\ast },%
\end{array}
\label{Eq7.3}
\end{equation}%
it linearly depends on $\mathbf{z}$:%
\begin{equation}
\begin{array}{c}
\mathfrak{A=}\frac{1}{2}\mathbf{\Sigma }\left( \mathrm{diag}\left( \mathbf{d}%
^{\left( 0\right) }+\mathfrak{D}\mathbf{z}\right) \right) \mathbf{\Sigma }%
^{\ast }=\frac{1}{2}\mathfrak{A}^{\left( 0\right) }+\frac{1}{2}\mathfrak{A}%
^{\left( m\right) }z_{m},%
\end{array}
\label{Eq7.4}
\end{equation}%
where%
\begin{equation}
\begin{array}{c}
\mathfrak{A}^{\left( 0\right) }=\frac{1}{2}\mathbf{\Sigma }\mathrm{diag}%
\left( \mathbf{d}^{\left( 0\right) }\right) \Sigma ^{\ast },\ \ \ \mathfrak{A%
}^{\left( i\right) }=\frac{1}{2}\mathbf{\Sigma }\mathrm{diag}\left( \mathbf{d%
}^{\left( i\right) }\right) \Sigma ^{\ast }.%
\end{array}
\label{Eq7.5}
\end{equation}%
In contrast to the Gaussian case, Eqs (\ref{Eq7.3}) have to be defined in
the domain $\boldsymbol{D}$ such that 
\begin{equation}
\boldsymbol{D=}\left\{ \left. \mathbf{z}\right\vert \mathbf{d}^{\left(
0\right) }+\mathfrak{D}\mathbf{z\geq 0}\right\} ,  \label{Eq7.6}
\end{equation}%
rather than in the whole space. In financial engineering, covariance
matrices of the form (\ref{Eq7.3}) were introduced by \cite{Dai00}, and
discussed by \cite{Duffie03, Filipovic09} and many others.

The corresponding Fokker-Plank problem has the form: 
\begin{equation}
\begin{array}{c}
\varpi _{\bar{t}}\left( t,\mathbf{z,}\bar{t},\mathbf{\bar{z}}\right) -\dsum
\dsum \left( \mathfrak{A}^{0}+\bar{z}_{m}\mathfrak{A}^{\left( m\right)
}\right) \varpi _{\mathbf{\bar{z}\bar{z}}}\left( t,\mathbf{z,}\bar{t},%
\mathbf{\bar{z}}\right) \\ 
\\ 
+\left( \mathbf{\hat{b}}+\mathfrak{B}\mathbf{\bar{z}}\right) \cdot \varpi _{%
\mathbf{\bar{z}}}\left( t,\mathbf{z,}\bar{t},\mathbf{\bar{z}}\right)
+b\varpi \left( t,\mathbf{z,}\bar{t},\mathbf{\bar{z}}\right) =0, \\ 
\\ 
\varpi \left( t,\mathbf{z},t,\mathbf{\bar{z}}\right) =\delta \left( \mathbf{%
\bar{z}}-\mathbf{z}\right) ,%
\end{array}
\label{Eq7.7a}
\end{equation}%
where%
\begin{equation}
\begin{array}{c}
\hat{b}_{m}\mathbf{=}b_{m}\mathbf{-}\left( 2a_{mm}^{\left( m\right)
}+a_{mm^{\prime }}^{\left( m^{\prime }\right) }+a_{m^{\prime }m}^{\left(
m^{\prime }\right) }\right) \mathbf{,\ \ }\text{(no summation over }m\text{),%
} \\ 
\\ 
b=\mathrm{Tr}\left( \mathfrak{B}\right) .%
\end{array}
\label{Eq7.7b}
\end{equation}%
Eq (\ref{Eq6.6}) expressing $\varpi $ in terms of $\mathcal{K}$ holds. The
equations for $\alpha ,\mathbf{\delta }$ have the form:%
\begin{equation}
\begin{array}{c}
\alpha _{\bar{t}}\left( t,\bar{t}\right) +i\mathbf{\delta }_{\bar{t}}\left(
t,\bar{t}\right) \cdot \mathbf{\bar{z}}+\mathbf{\delta }\left( t,\bar{t}%
\right) \cdot \mathfrak{A}\mathbf{\delta }\left( t,\bar{t}\right) +i\mathbf{%
\delta }\left( t,\bar{t}\right) \cdot \left( \mathbf{\hat{b}}+\mathfrak{B}%
\mathbf{\bar{z}}\right) +b=0,%
\end{array}
\label{Eq7.8a}
\end{equation}%
or, more explicitly,%
\begin{equation}
\begin{array}{c}
\alpha _{\bar{t}}\left( t,\bar{t}\right) +i\mathbf{\delta }_{\bar{t}}\left(
t,\bar{t}\right) \cdot \mathbf{\bar{z}}+\mathbf{\delta }\left( t,\bar{t}%
\right) \cdot \mathfrak{A}^{\left( 0\right) }\mathbf{\delta }\left( t,\bar{t}%
\right) +\mathbf{\delta }\left( t,\bar{t}\right) \cdot \mathfrak{A}^{\left(
k\right) }\mathbf{\delta }\left( t,\bar{t}\right) \bar{z}_{k} \\ 
\\ 
+i\mathbf{\delta }\left( t,\bar{t}\right) \cdot \left( \mathbf{\hat{b}}+%
\mathfrak{B}\mathbf{\bar{z}}\right) +b=0.%
\end{array}
\label{Eq7.9a}
\end{equation}%
Thus, the system of ODEs for $\alpha ,\mathbf{\delta =}\left( \mathbf{\beta }%
,\mathbf{\gamma }\right) $ can be written as follows:%
\begin{equation}
\begin{array}{c}
\alpha _{\bar{t}}\left( t,\bar{t}\right) +\mathbf{\delta }\left( t,\bar{t}%
\right) \cdot \mathfrak{A}^{\left( 0\right) }\left( t,\bar{t}\right) \mathbf{%
\delta }\left( t,\bar{t}\right) +i\mathbf{\delta }\left( t,\bar{t}\right)
\cdot \mathbf{\hat{b}}\left( t,\bar{t}\right) +b\left( t,\bar{t}\right) =0,\
\ \ \alpha \left( t,t\right) =0, \\ 
\\ 
i\delta _{i}^{\prime }\left( t,\bar{t}\right) +\mathbf{\delta }\left( t,\bar{%
t}\right) \cdot \mathfrak{A}^{\left( i\right) }\mathbf{\delta }\left( t,\bar{%
t}\right) +i\mathfrak{B}_{ij}\delta _{j}\left( t,\bar{t}\right) =0,\ \ \
\delta _{i}\left( t,t\right) =m_{i}.%
\end{array}
\label{Eq7.10a}
\end{equation}%
In the case in question, the equation for $\mathbf{\delta }$ is no longer
linear. Instead, $\mathbf{\delta }$ satisfies the so-called matrix Riccati
equation. Such equations are important for several applications, such as
optimal control. Solving a matrix Riccati equation is quite hard, so it is
more an art than a science; some of the results in this direction are
reported below. However, in the one-dimensional case, the corresponding
Riccati equation can be converted into the second-order ODE, and then solved
explicitly when the coefficients $\mathfrak{A}$, $\mathbf{b}$, $b$ are
time-independent.

In case of an augmented process, one must consider an SDE of the form:%
\begin{equation}
\begin{array}{c}
d\mathbf{\hat{x}}_{t}=\left( \mathbf{b}^{\left( x\right) }\left( t\right) +%
\mathfrak{B}^{\left( xx\right) }\left( t\right) \mathbf{\hat{x}}_{t}+%
\mathfrak{B}^{\left( xy\right) }\left( t\right) \mathbf{\hat{y}}_{t}\right)
dt, \\ 
\\ 
d\mathbf{\hat{y}}_{t}=\left( \mathbf{b}^{\left( y\right) }\left( t\right) +%
\mathfrak{B}^{\left( yx\right) }\left( t\right) \mathbf{\hat{x}}_{t}+%
\mathfrak{B}^{\left( yy\right) }\left( t\right) \mathbf{\hat{y}}_{t}\right)
dt \\ 
\\ 
+\mathfrak{\mathbf{\Sigma }}^{\left( yy\right) }\left( t\right) \left( 
\mathrm{diag}\left( \mathbf{d}^{\left( 0\right) }\left( t\right) +\mathfrak{D%
}\left( t\right) \mathbf{\hat{z}}_{t}\right) \right) ^{1/2}d\mathbf{\hat{W}}%
_{t}^{\left( y\right) }, \\ 
\\ 
\mathbf{\hat{x}}_{t}=\mathbf{x,\ \ \ \hat{y}}_{t}=\mathbf{y,}%
\end{array}
\label{Eq7.1}
\end{equation}%
or, more compactly,%
\begin{equation}
\begin{array}{c}
d\mathbf{\hat{z}}_{t}=\left( \mathbf{b}\left( t\right) +\mathfrak{B}\left(
t\right) \mathbf{\hat{z}}_{t}\right) dt+\left( 
\begin{array}{c}
0 \\ 
\mathbf{\Sigma }^{\left( yy\right) }\left( t\right) \left( \mathrm{diag}%
\left( \mathbf{d}^{\left( 0\right) }\left( t\right) +\mathfrak{D}\left(
t\right) \mathbf{\hat{z}}_{t}\right) \right) ^{1/2}d\mathbf{\hat{W}}%
_{t}^{\left( y\right) }%
\end{array}%
\right) , \\ 
\\ 
\mathbf{\hat{z}}_{t}=\mathbf{z=}\left( 
\begin{array}{c}
\mathbf{x} \\ 
\mathbf{y}%
\end{array}%
\right) .%
\end{array}
\label{Eq7.2}
\end{equation}%
Here $\mathbf{b}^{\left( x\right) }$, $\mathbf{b}^{\left( y\right) }$, $%
\mathbf{b}=\left( \mathbf{b}^{\left( x\right) },\mathbf{b}^{\left( y\right)
}\right) ^{\ast }$, $\mathbf{d}^{\left( 0\right) }$ are column vectors, and $%
\mathfrak{B}^{\left( xx\right) }$, $\mathfrak{B}^{\left( xy\right) }$, $%
\mathfrak{B}^{\left( yx\right) }$, $\mathfrak{B}^{\left( yy\right) }$, $%
\mathfrak{B}$, and $\mathfrak{D}$ are matrices of appropriate dimensions.

The equations for $\alpha ,\mathbf{\delta =}\left( \mathbf{\beta },\mathbf{%
\gamma }\right) $ have the form:%
\begin{equation}
\begin{array}{c}
\alpha _{\bar{t}}\left( t,\bar{t}\right) +i\mathbf{\delta }_{\bar{t}}\left(
t,\bar{t}\right) \cdot \mathbf{\bar{z}}+\mathbf{\gamma }\left( t,\bar{t}%
\right) \cdot \mathfrak{A}\mathbf{\gamma }\left( t,\bar{t}\right) +i\mathbf{%
\delta }\left( t,\bar{t}\right) \cdot \left( \mathbf{\hat{b}}+\mathfrak{B}%
\mathbf{\bar{z}}\right) +b=0,%
\end{array}
\label{Eq7.8b}
\end{equation}%
or, more explicitly,%
\begin{equation}
\begin{array}{c}
\alpha _{\bar{t}}\left( t,\bar{t}\right) +i\mathbf{\delta }_{\bar{t}}\left(
t,\bar{t}\right) \cdot \mathbf{\bar{z}}+\mathbf{\gamma }\left( t,\bar{t}%
\right) \cdot \mathfrak{A}^{\left( 0\right) }\mathbf{\gamma }\left( t,\bar{t}%
\right) +\mathbf{\gamma }\left( t,\bar{t}\right) \cdot \mathfrak{A}^{\left(
k\right) }\mathbf{\gamma }\left( t,\bar{t}\right) \bar{z}_{k} \\ 
\\ 
+i\mathbf{\delta }\left( t,\bar{t}\right) \cdot \left( \mathbf{b}^{\left(
z\right) }+\mathfrak{B}^{\left( zz\right) }\mathbf{\bar{z}}\right) +b=0.%
\end{array}
\label{Eq7.9b}
\end{equation}%
Thus, the system of ODEs for $\alpha ,\mathbf{\delta }$ can be written as
follows:%
\begin{equation}
\begin{array}{c}
\alpha _{\bar{t}}\left( t,\bar{t}\right) +\mathbf{\gamma }\left( t,\bar{t}%
\right) \cdot \mathfrak{A}^{\left( 0\right) }\left( t,\bar{t}\right) \mathbf{%
\gamma }\left( t,\bar{t}\right) +i\mathbf{\delta }\left( t,\bar{t}\right)
\cdot \mathbf{\hat{b}}^{\left( z\right) }\left( t,\bar{t}\right) +b\left( t,%
\bar{t}\right) =0,\ \ \ \alpha \left( t,t\right) =0, \\ 
\\ 
i\mathbf{\delta }_{i,\bar{t}}\left( t,\bar{t}\right) +\mathbf{\gamma }\left(
t,\bar{t}\right) \cdot \mathfrak{A}^{\left( i\right) }\mathbf{\gamma }\left(
t,\bar{t}\right) +i\mathfrak{B}_{ij}^{\left( zz\right) }\delta _{j}\left( t,%
\bar{t}\right) =0,\ \ \ \delta _{i}\left( t,t\right) =m_{i}.%
\end{array}
\label{Eq7.10b}
\end{equation}

\subsection{Killed Non-Gaussian Processes\label{Sec72}}

The non-Gaussian governing SDE has the form:%
\begin{equation}
\begin{array}{c}
d\mathbf{\hat{z}}_{t}=\left( \mathbf{b}+\mathfrak{B}\mathbf{\hat{z}}%
_{t}\right) dt+\mathbf{\Sigma }\left( \mathrm{diag}\left( \mathbf{d}^{\left(
0\right) }+\mathfrak{D}\mathbf{\hat{z}}_{t}\right) \right) ^{1/2}d\mathbf{%
\hat{W}}_{t},%
\end{array}
\label{Eq7.11}
\end{equation}%
where $\mathbf{\hat{z}}_{t}$, $\mathbf{b}$, $\mathbf{d}^{\left( 0\right) }$
are $\left( M\times 1\right) $ vectors, and $\mathbf{\Sigma }$, $\mathfrak{B}
$, $\mathfrak{D}$ are $\left( M\times M\right) $ matrices defined above. As
before, the correlation matrix \textbf{$\Sigma $} can be a full-rank
(non-degenerate) matrix. Once again, it is assumed that the process is
killed with intensity $\bar{c}$ linearly depending of $\mathbf{z}$, namely, 
\begin{equation}
\begin{array}{c}
\bar{c}=c+\mathbf{c}\cdot \mathbf{z},%
\end{array}
\label{Eq7.12}
\end{equation}%
where $c$ is a scalar, and $\mathbf{c}^{\left( z\right) }$ is a $\left(
M\times 1\right) $ column vector.

The corresponding Fokker-Plank problem has the form: 
\begin{equation}
\begin{array}{c}
\varpi _{\bar{t}}\left( t,\mathbf{z,}\bar{t},\mathbf{\bar{z}}\right) -\dsum
\dsum \left( \mathfrak{A}^{0}+\bar{z}_{i}\mathfrak{A}^{i}\right) \varpi _{%
\mathbf{\bar{z}\bar{z}}}\left( t,\mathbf{z,}\bar{t},\mathbf{\bar{z}}\right)
\\ 
\\ 
+\left( \mathbf{\hat{b}}+\mathfrak{B}\mathbf{\bar{z}}\right) \cdot \varpi _{%
\mathbf{\bar{z}}}\left( t,\mathbf{z,}\bar{t},\mathbf{\bar{z}}\right) +\left(
b+c+\mathbf{c}\cdot \mathbf{\bar{z}}\right) \varpi \left( t,\mathbf{z,}\bar{t%
},\mathbf{\bar{z}}\right) =0, \\ 
\\ 
\varpi \left( t,\mathbf{z},t,\mathbf{\bar{z}}\right) =\delta \left( \mathbf{%
\bar{z}}-\mathbf{z}\right) ,%
\end{array}
\label{Eq7.13}
\end{equation}%
The equations for $\alpha ,\mathbf{\delta }$ generalize Eqs (\ref{Eq7.10a}).
They can be written in the form:%
\begin{equation}
\begin{array}{c}
\alpha _{\bar{t}}\left( t,\bar{t}\right) +\mathbf{\delta }\left( t,\bar{t}%
\right) \cdot \mathfrak{A}^{\left( 0\right) }\left( t,\bar{t}\right) \mathbf{%
\delta }\left( t,\bar{t}\right) +i\mathbf{\delta }\left( t,\bar{t}\right)
\cdot \mathbf{b}\left( t,\bar{t}\right) +b\left( t,\bar{t}\right) +c\left( t,%
\bar{t}\right) =0,\ \ \ \alpha \left( t,t\right) =0, \\ 
\\ 
i\delta _{i}^{\prime }\left( t,\bar{t}\right) +\mathbf{\delta }\left( t,\bar{%
t}\right) \cdot \mathfrak{A}^{\left( i\right) }\mathbf{\delta }\left( t,\bar{%
t}\right) +i\mathfrak{B}_{ij}\delta _{j}\left( t,\bar{t}\right) +c_{i}=0,\ \
\ \delta _{i}\left( t,t\right) =m_{i}.%
\end{array}
\label{Eq7.14}
\end{equation}%
As in the case without killing, finding an analytical solution to a
multi-dimensional Riccati equation is generally impossible. However, in the
time-independent one-dimensional case, it can be done. Solution becomes
particularly simple in the special case when $\mathfrak{A}^{\left( 0\right)
}=0$. The most important case is the killed one-dimensional Feller process,
used, for example, to price bonds in the Cox-IngersolI-Ross (CIR) model; see
Chapter \ref{Chap8}.

\subsection{Example: Anomalous Kolmogorov Process\label{Sec73}}

Anomalous diffusion is a phenomenon in which the random motion of particles
or molecules deviates from classical Brownian motion and, as a result,
exhibits non-Gaussian probability distributions, such as power-law or
exponential tails. One can distinguish between subdiffusions (slower
spreading) and superdiffusions (faster spreading). Anomalous diffusion often
involves long-range correlations in particle motion, meaning that the
movement of a particle at a one-time step depends on its previous positions
over longer time scales. Anomalous diffusion frequently displays
scale-invariant properties, meaning that the statistical properties of
motion remain the same across different time or spatial scales. Anomalous
diffusion has applications in physics, chemistry, financial engineering,
biology, and geophysics.

Fractional Brownian motion (fBm) is used to model anomalous diffusion
because it possesses several relevant characteristics. In particular, it
exhibits long memory, which means that the process's future values are
influenced by its past values over long time scales. Additionally, fBm can
produce non-Gaussian behavior while preserving scale-invariance. By
adjusting the Hurst exponent and other parameters, fBm can be tailored to
model different anomalous diffusions, including both subdiffusions and
superdiffusions.

This section studies a fractional Kolmogorov equation of the form:%
\begin{equation}
\begin{array}{c}
\varpi _{\bar{t}}\left( t,x,y,\bar{t},\bar{x},\bar{y}\right) +a\left( -\frac{%
\partial ^{2}}{\partial \bar{y}^{2}}\right) ^{\nu }\varpi \left( t,x,y,\bar{t%
},\bar{x},\bar{y}\right) \\ 
\\ 
+\bar{y}\varpi _{\bar{x}}\left( t,x,y,\bar{t},\bar{x},\bar{y}\right)
+b\varpi _{\bar{y}}\left( t,x,y,\bar{t},\bar{x},\bar{y}\right) =0, \\ 
\\ 
\varpi \left( t,\bar{x},\bar{y},t,x,y\right) =\delta \left( \bar{x}-x\right)
\delta \left( \bar{y}-y\right) .%
\end{array}
\label{Eq7.15}
\end{equation}%
where $0<\nu <1$. The operator $\left( -\left. \partial ^{2}\right/ \partial 
\bar{y}^{2}\right) ^{\nu }$ is a pseudo-differential defined as follows:%
\begin{equation}
\begin{array}{c}
\left( -\frac{\partial ^{2}}{\partial \bar{y}^{2}}\right) ^{\nu }\varpi =%
\mathcal{F}^{-1}\left( \left\vert l\right\vert ^{2\nu }\mathcal{F}\left(
\varpi \right) \right) .%
\end{array}
\label{Eq7.16}
\end{equation}%
Here $\mathcal{F}$ and $\mathcal{F}^{-1}$ denote the direct and inverse
Fourier transforms, respectively. Despite its complexity, problem (\ref%
{Eq7.15}) can be solved by using Kelvin waves. For particular solutions of
the form (\ref{Eq3.34}), (\ref{Eq3.35}), (\ref{Eq3.36}), the corresponding
characteristic equations are%
\begin{equation}
\begin{array}{c}
\alpha _{\bar{t}}\left( t,\bar{t}\right) +a\left\vert \gamma \right\vert
^{2\nu }\left( t,\bar{t}\right) +i\gamma _{\bar{t}}\left( t,\bar{t}\right) 
\bar{y}+ik\bar{y}+ib\gamma \left( t,\bar{t}\right) =0, \\ 
\\ 
\alpha \left( t,t\right) =0,\ \ \ \gamma \left( t,t\right) =l,%
\end{array}
\label{Eq7.17}
\end{equation}%
so that%
\begin{equation}
\begin{array}{c}
\alpha _{\bar{t}}\left( t,\bar{t}\right) +a\left\vert \gamma \right\vert
^{2\nu }\left( t,\bar{t}\right) +ib\gamma \left( t,\bar{t}\right) =0,\ \ \
\alpha \left( t,t\right) =0, \\ 
\\ 
\gamma _{\bar{t}}\left( t,\bar{t}\right) +k=0,\ \ \ \gamma \left( t,t\right)
=l,%
\end{array}
\label{Eq7.18}
\end{equation}%
\begin{equation}
\begin{array}{c}
\gamma \left( t,\bar{t}\right) =-kT+l, \\ 
\\ 
\alpha \left( t,\bar{t}\right) =-a\int\limits_{t}^{\bar{t}}\left\vert
-k\left( s-t\right) +l\right\vert ^{2\nu }ds-ib\left( -\frac{kT^{2}}{2}%
+lT\right) .%
\end{array}
\label{Eq7.19}
\end{equation}%
Thus,%
\begin{equation}
\begin{array}{c}
\Psi =\alpha +ik\left( \bar{x}-x\right) +i\gamma \bar{y}-ily \\ 
\\ 
=-a\int\limits_{t}^{\bar{t}}\left\vert -k\left( s-t\right) +l\right\vert
^{2\nu }ds \\ 
\\ 
+ik\left( \bar{x}-x-\bar{y}T+\frac{bT^{2}}{2}\right) +il\left( \bar{y}%
-y-bT\right) .%
\end{array}
\label{Eq7.20}
\end{equation}%
Now, assume that $\nu =1/2$. The key is to calculate the integral 
\begin{equation}
\begin{array}{c}
\mathcal{I}=\int\limits_{t}^{\bar{t}}\left\vert -k\left( s-t\right)
+l\right\vert ^{2\nu }ds=\int\limits_{0}^{T}\left\vert -ks+l\right\vert ds,%
\end{array}
\label{Eq7.21}
\end{equation}%
for different values of $\left( k,l\right) $. Depending on $\left(
k,l\right) $, this integral can be calculated as follows:%
\begin{equation}
\begin{array}{c}
\mathcal{I}_{1}=\int\limits_{0}^{l/k}\left( -ks+l\right)
ds+\int\limits_{l/k}^{T}\left( ks-l\right) ds=\frac{kT^{2}}{2}-lT+\frac{l^{2}%
}{k},\ \ \ 0\leq k<\infty ,\ \ \ 0\leq l\leq kT, \\ 
\\ 
\mathcal{I}_{2}=-\frac{kT^{2}}{2}+lT,\ \ \ 0\leq k<\infty ,\ \ \ kT\leq
l<\infty , \\ 
\\ 
\mathcal{I}_{3}=-\frac{kT^{2}}{2}+lT,\ \ \ -\infty <k\leq 0,\ \ \ 0\leq
l<\infty , \\ 
\\ 
\mathcal{I}_{4}=\int\limits_{0}^{l/k}\left( ks-l\right)
ds+\int\limits_{l/k}^{T}\left( -ks+l\right) ds=-\frac{kT^{2}}{2}+lT-\frac{%
l^{2}}{k},\ \ \ -\infty <k\leq 0,\ \ \ kT\leq l\leq 0, \\ 
\\ 
\mathcal{I}_{5}=\frac{kT^{2}}{2}-lT,\ \ -\infty <k\leq 0,\ \ \ -\infty
<l\leq kT, \\ 
\\ 
\mathcal{I}_{6}=\frac{kT^{2}}{2}-lT,\ \ 0\leq k<\infty ,\ \ \ -\infty <l\leq
0.%
\end{array}
\label{Eq7.22}
\end{equation}%
Thus,%
\begin{equation}
\begin{array}{c}
\left( 2\pi \right) ^{2}\mathcal{J}_{1}=\int\limits_{0}^{\infty
}\int\limits_{0}^{kT}\exp \left( -a\left( \frac{kT^{2}}{2}-lT+\frac{l^{2}}{k}%
\right) +ikaT^{2}\zeta +ilaT\eta \right) dkdl \\ 
\\ 
=T\int\limits_{0}^{1}\int\limits_{0}^{\infty }\exp \left( \left(
-p+iq\right) k\right) kd\chi dk=-T\int\limits_{0}^{1}\frac{\partial }{%
\partial p}\left( \int\limits_{0}^{\infty }\exp \left( \left( -p+iq\right)
k\right) dk\right) d\chi \\ 
\\ 
=T\int\limits_{0}^{1}\frac{d\chi }{\left( p-iq\right) ^{2}}=\frac{1}{%
a^{2}T^{3}}\int\limits_{0}^{T}\frac{d\chi }{\left( \left( \chi -f_{+}\right)
\left( \chi -f_{-}\right) \right) ^{2}},%
\end{array}
\label{Eq7.23}
\end{equation}%
where $\left( \zeta ,\eta \right) $ are non-dimensional variables: 
\begin{equation}
\begin{array}{c}
\zeta =\frac{\bar{x}-x-\bar{y}T+\frac{bT^{2}}{2}}{aT^{2}},\ \ \ \eta =\frac{%
\bar{y}-y-bT}{aT},%
\end{array}
\label{Eq7.24}
\end{equation}%
\begin{equation}
\begin{array}{c}
l=T\chi k,\ \ \ p\left( \chi \right) =aT^{2}\left( \frac{1}{2}-\chi +\chi
^{2}\right) >0,\ \ \ q\left( \chi \right) =aT^{2}\left( \zeta +\chi \eta
\right) ,%
\end{array}
\label{Eq7.25}
\end{equation}%
and $f_{\pm }$ are roots of the quadratic equation%
\begin{equation}
\begin{array}{c}
\chi ^{2}-\left( 1+i\eta \right) \chi +\left( \frac{1}{2}-i\zeta \right) =0.%
\end{array}
\label{Eq7.26}
\end{equation}%
One can check that%
\begin{equation}
\begin{array}{c}
f_{\pm }=\frac{\left( 1+i\eta \right) \pm \sqrt{\left( 1+i\eta \right)
^{2}-2+4i\zeta }}{2}=\frac{\left( 1+i\eta \right) \pm i\sqrt{D}}{2}, \\ 
\\ 
f_{+}f_{-}=\frac{1}{2}-i\zeta ,\ \ \ f_{+}+f_{-}=1+i\eta ,\ \ \ f_{+}-f_{-}=i%
\sqrt{D},%
\end{array}
\label{Eq7.27}
\end{equation}%
with%
\begin{equation}
\begin{array}{c}
D=1+\eta ^{2}-4i\zeta -2i\eta ,%
\end{array}
\label{Eq7.28}
\end{equation}%
The roots $f_{\pm }$ are never equal since $D$ does not vanish when $\zeta
,\eta $ are real.

Thus, one has:%
\begin{equation}
\begin{array}{c}
\left( 2\pi \right) ^{2}\mathcal{J}_{1}=\frac{1}{a^{2}T^{3}}%
\int\limits_{0}^{1}\frac{d\chi }{\left( \left( \chi -f_{+}\right) \left(
\chi -f_{-}\right) \right) ^{2}}=\frac{1}{a^{2}T^{3}\left(
f_{+}-f_{-}\right) ^{2}}\int\limits_{0}^{1}\left( \frac{1}{\chi -f_{+}}-%
\frac{1}{\chi -f_{-}}\right) ^{2}d\chi \\ 
\\ 
=-\frac{1}{a^{2}T^{3}\left( f_{+}-f_{-}\right) ^{2}}\left( \left( \left( 
\frac{1}{1-f_{+}}+\frac{1}{f_{+}}\right) +\left( \frac{1}{1-f_{-}}+\frac{1}{%
f_{-}}\right) \right) +\frac{2}{\left( f_{+}-f_{-}\right) }\ln \left( \frac{%
f_{-}\left( 1-f_{+}\right) }{f_{+}\left( 1-f_{-}\right) }\right) \right) \\ 
\\ 
=\frac{1}{a^{2}T^{3}D}\left( \frac{4\left( D+2i\zeta +i\eta \right) }{\left(
1-2i\zeta \right) \left( 1-2i\zeta -2i\eta \right) }-\frac{2i}{\sqrt{D}}\ln
\left( \frac{2\zeta +\eta -\sqrt{D}}{2\zeta +\eta +\sqrt{D}}\right) \right) .%
\end{array}
\label{Eq7.29}
\end{equation}%
By symmetry,%
\begin{equation}
\begin{array}{c}
\mathcal{J}_{4}\left( \zeta ,\eta \right) =\mathcal{J}_{1}\left( -\zeta
,-\eta \right) =\overline{\mathcal{J}_{1}\left( \zeta ,\eta \right) }.%
\end{array}
\label{Eq7.30}
\end{equation}%
Next,%
\begin{equation}
\begin{array}{c}
\left( 2\pi \right) ^{2}\mathcal{J}_{2}=\int\limits_{0}^{\infty
}\int\limits_{kT}^{\infty }\exp \left( -a\left( -\frac{kT^{2}}{2}+lT\right)
+ikaT^{2}\zeta +ilaT\eta \right) dkdl \\ 
\\ 
=\frac{1}{aT\left( 1-i\eta \right) }\int\limits_{0}^{\infty }\exp \left( -%
\frac{kaT^{2}}{2}+ikaT^{2}\left( \zeta +\eta \right) \right) dk \\ 
\\ 
=\frac{1}{a^{2}T^{3}\left( \left( 1-i\eta \right) \right) \left( \frac{1}{2}%
-i\left( \zeta +\eta \right) \right) }.%
\end{array}
\label{Eq7.31}
\end{equation}%
Similarly, it is easy to show that%
\begin{equation}
\begin{array}{c}
\left( 2\pi \right) ^{2}\mathcal{J}_{3}=\int\limits_{-\infty
}^{0}\int\limits_{0}^{\infty }\exp \left( -a\left( -\frac{kT^{2}}{2}%
+lT\right) +ikaT^{2}\zeta +ilaT\eta \right) dkdl \\ 
\\ 
=\frac{1}{a^{2}T^{3}\left( 1-i\eta \right) \left( \frac{1}{2}+i\zeta \right) 
},%
\end{array}
\label{Eq7.32}
\end{equation}%
while, by symmetry one gets:%
\begin{equation}
\begin{array}{c}
\mathcal{J}_{5}\left( \zeta ,\eta \right) =\frac{1}{\left( 2\pi \right)
^{2}a^{2}T^{3}\left( 1+i\eta \right) \left( \frac{1}{2}+i\left( \zeta +\eta
\right) \right) }=\overline{\mathcal{J}_{2}\left( \zeta ,\eta \right) }, \\ 
\\ 
\mathcal{J}_{6}\left( \zeta ,\eta \right) =\frac{1}{\left( 2\pi \right)
^{2}a^{2}T^{3}\left( 1+i\eta \right) \left( \frac{1}{2}-i\zeta \right) }=%
\overline{\mathcal{J}_{3}\left( \zeta ,\eta \right) },%
\end{array}
\label{Eq7.35}
\end{equation}%
so that%
\begin{equation}
\begin{array}{c}
\varpi =\frac{1}{\pi ^{2}a^{2}T^{3}}\left( \frac{1}{\left( 1+\eta
^{2}\right) }\left( \frac{\left( 1-2\eta \left( \zeta +\eta \right) \right) 
}{\left( 1+4\left( \zeta +\eta \right) ^{2}\right) }+\frac{\left( 1+2\eta
\zeta \right) }{\left( 1+4\zeta ^{2}\right) }\right) \right. \\ 
\\ 
\left. +\func{Re}\left\{ \frac{1}{D}\left( \frac{2\left( D+2i\zeta +i\eta
\right) }{\left( D-\left( 2\zeta +\eta \right) ^{2}\right) }-\frac{i}{\sqrt{D%
}}\ln \left( \frac{2\zeta +\eta -\sqrt{D}}{2\zeta +\eta +\sqrt{D}}\right)
\right) \right\} \right) ,%
\end{array}
\label{Eq7.37}
\end{equation}%
and%
\begin{equation}
\begin{array}{c}
\varpi \left( \bar{x},\bar{y}\right) d\bar{x}d\bar{y}=\frac{1}{\pi ^{2}a^{2}}%
\left( \frac{1}{\left( 1+\eta ^{2}\right) }\left( \frac{\left( 1-2\eta
\left( \zeta +\eta \right) \right) }{\left( 1+4\left( \zeta +\eta \right)
^{2}\right) }+\frac{\left( 1+2\eta \zeta \right) }{\left( 1+4\zeta
^{2}\right) }\right) \right. \\ 
\\ 
\left. +\func{Re}\left\{ \frac{1}{D}\left( \frac{2\left( D+2i\zeta +i\eta
\right) }{\left( D-\left( 2\zeta +\eta \right) ^{2}\right) }-\frac{i}{\sqrt{D%
}}\ln \left( \frac{2\zeta +\eta -\sqrt{D}}{2\zeta +\eta +\sqrt{D}}\right)
\right) \right\} \right) d\zeta d\eta \\ 
\\ 
\equiv \varpi \left( \zeta ,\eta \right) d\zeta d\eta ,%
\end{array}
\label{Eq7.38}
\end{equation}%
which shows that, as expected, in the non-dimensional variables there is no
explicit dependence on $T$.\footnote{%
In a special case $a=1$, $b=0$, $\xi =0$, $\theta =0$, \cite{He21} attempted
to solve the problem considered above. However, the authors made a severe
error in transitioning from Eq. (2.2) to Eq. (2.3). In contrast to
Kolmogorov's minor error, their error cannot be repaired. Dimensional
analysis shows that the proposed solution is completely incorrect. In our
notation, it has the form:%
\begin{equation*}
\begin{array}{c}
\varpi \left( \bar{x},\bar{y}\right) d\bar{x}d\bar{y}=\frac{72\sqrt{3}}{\pi
^{3}T^{6}\left( 1+4\left( \frac{\bar{y}^{2}}{4T}+\frac{3\left( \bar{x}+\bar{y%
}T/2\right) ^{2}}{T^{3}}\right) \right) ^{7/2}}d\bar{x}d\bar{y}.%
\end{array}%
\end{equation*}%
Introducing rescaled variables, $\zeta =\bar{x}T^{-3/2}$, $\eta =\bar{y}%
T^{-1/2}$, we get%
\begin{equation*}
\begin{array}{c}
\varpi \left( \zeta ,\eta \right) d\zeta d\eta =\frac{72\sqrt{3}}{\pi
^{3}T^{4}\left( 1+4\left( \frac{\eta ^{2}}{4}+3\left( \zeta +\frac{\eta }{2}%
\right) ^{2}\right) \right) ^{7/2}}d\zeta d\eta .%
\end{array}%
\end{equation*}%
This expression explicitly depends on $T$, which is impossible, since its
intergal must be equal to unity.}

\begin{figure}[tbp]
\begin{center}
\includegraphics[width=0.8\textwidth]
{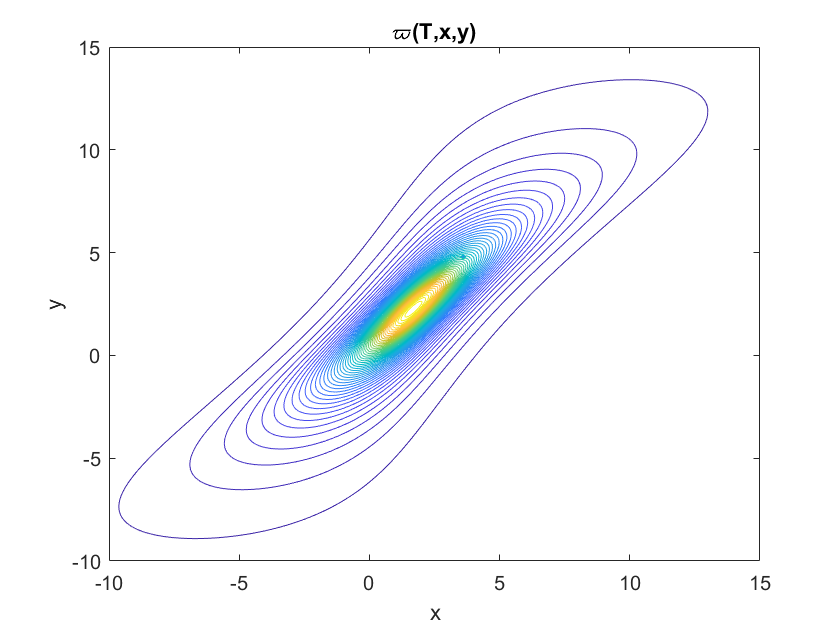}
\end{center}
\par
\vspace{-10pt}
\caption{Contour lines of $\protect\varpi \left( 0,0,0,T,\tilde{x},\tilde{y}%
\right) $ for an anomalous Kolmogorov process with $T=1.5$, $a=2.5$, $b=1.5$%
. Own graphics.}
\label{Fig7.1}
\end{figure}

A typical anomalous Kolmogorov process is depicted in Figure \ref{Fig7.1}. 
The difference between the anomalous diffusion shown in Figure \ref{Fig7.1}
and pure diffusion shown in Figure \ref{Fig3.2} is clear.

It is worth comparing Eqs (\ref{Eq7.37}) and (\ref{Eq3.25}). To this end,
rewrite $\Phi $ given by Eq. (\ref{Eq3.26}) in the form:%
\begin{equation}
\begin{array}{c}
\Phi =\frac{\left( \bar{y}-y-bT\right) ^{2}}{2aT}+\frac{6\left( \bar{x}-x-%
\frac{\left( \bar{y}+y\right) T}{2}\right) ^{2}}{aT^{3}}=\frac{\eta ^{2}}{2}+%
\frac{3\left( 2\zeta +\eta \right) ^{2}}{2},%
\end{array}
\label{Eq7.39}
\end{equation}%
where $\zeta ,\eta $ are non-dimensional variables of the form:%
\begin{equation}
\begin{array}{c}
\zeta =\frac{\bar{x}-x-\bar{y}T+\frac{bT^{2}}{2}}{\sqrt{aT^{3}}},\ \ \ \eta =%
\frac{\bar{y}-y-bT}{\sqrt{aT}},\ 
\end{array}
\label{Eq7.40}
\end{equation}%
and $a$ is the diffusion coefficient; its dimension is $\left[ a\right]
=L^{2}/T^{3}$. Thus,%
\begin{equation}
\begin{array}{c}
\varpi =\frac{\sqrt{3}}{\pi aT^{2}}\exp \left( -\frac{\eta ^{2}+3\left(
2\zeta +\eta \right) ^{2}}{2}\right) ,%
\end{array}
\label{Eq7.41}
\end{equation}%
and%
\begin{equation}
\begin{array}{c}
\varpi \left( \bar{x},\bar{y}\right) d\bar{x}d\bar{y}=\frac{\sqrt{3}}{\pi a}%
\exp \left( -\frac{\eta ^{2}+3\left( 2\zeta +\eta \right) ^{2}}{2}\right)
d\zeta d\eta \equiv \varpi \left( \zeta ,\eta \right) d\zeta d\eta .%
\end{array}
\label{Eq7.42}
\end{equation}%
Comparing Eqs (\ref{Eq7.37}) and (\ref{Eq7.41}), one can see that the
scaling of $\varpi $ and its asymptotic behavior at infinity is completely
different.

\subsection{Example: Feller Process\label{Sec74}}

\subsubsection{Feller Process}

\paragraph{Feller Process with Constant Parameters}

For benchmarking purposes, it is useful to start with deriving the
well-known t.p.d.f. for the Feller process with constant coefficients; see 
\cite{Feller51, Feller52}:%
\begin{equation}
\begin{array}{c}
d\hat{y}_{t}=\left( \chi -\kappa \hat{y}_{t}\right) dt+\varepsilon \sqrt{%
\hat{y}_{t}}d\hat{Z}_{t},\ \ \ \hat{y}_{t}=y.%
\end{array}
\label{Eq7.43}
\end{equation}%
Initially, the process with time-independent parameters is considered; the
time-dependent case is analyzed later in this Section.

To start with, it is assumed that%
\begin{equation}
\begin{array}{c}
\frac{2\chi }{\varepsilon ^{2}}-1\equiv \vartheta >0.%
\end{array}
\label{Eq7.44}
\end{equation}%
This condition guarantees that the process $\hat{y}_{t}$ does not hit zero,
which is one of the main reasons to use the Feller process in practice; it
is relaxed shortly.

The corresponding Fokker-Planck problem has the form:

\begin{equation}
\begin{array}{c}
\varpi _{\bar{t}}-\frac{1}{2}\varepsilon ^{2}\left( \bar{y}\varpi \right) _{%
\bar{y}\bar{y}}+\left( \left( \chi -\kappa \bar{y}\right) \varpi \right) _{%
\bar{y}}=0, \\ 
\\ 
\varpi \left( t,y,t,\bar{y}\right) =\delta \left( \bar{y}-y\right) .%
\end{array}
\label{Eq7.45}
\end{equation}%
This equation can be written as a conservation law:%
\begin{equation}
\begin{array}{c}
\varpi _{\bar{t}}+\mathcal{F}_{\bar{y}}=0, \\ 
\\ 
\varpi \left( t,y,t,\bar{y}\right) =\delta \left( \bar{y}-y\right) ,%
\end{array}
\label{Eq7.45b}
\end{equation}%
where the probability flux$\mathcal{F}$ is given by 
\begin{equation}
\begin{array}{c}
\mathcal{F}=-\frac{1}{2}\varepsilon ^{2}\left( \bar{y}\varpi \right) _{\bar{y%
}}+\left( \chi -\kappa \bar{y}\right) \varpi .%
\end{array}
\label{Eq7.45c}
\end{equation}

However, experience suggests that solving the backward Kolmogorov problem is
more expedient. It can be formulated as follows:%
\begin{equation}
\begin{array}{c}
\varpi _{t}+\frac{1}{2}\varepsilon ^{2}y\varpi _{yy}+\left( \chi -\kappa
y\right) \varpi _{y}=0, \\ 
\\ 
\varpi \left( \bar{t},y,\bar{t},\bar{y}\right) =\delta \left( y-\bar{y}%
\right) .%
\end{array}
\label{Eq7.46}
\end{equation}%
The associated Kelvin wave function $\mathcal{K}\left( t,y,\bar{t},\bar{y}%
,l\right) $ has the form:%
\begin{equation}
\begin{array}{c}
\mathcal{K}=\exp \left( \alpha \left( t,\bar{t}\right) +i\gamma \left( t,%
\bar{t}\right) y-il\bar{y}\right) ,%
\end{array}
\label{Eq7.47}
\end{equation}%
where $\alpha ,\gamma $ solve the following system of backward ODEs:%
\begin{equation}
\begin{array}{c}
\alpha _{t}\left( t,\bar{t}\right) +\chi i\gamma \left( t,\bar{t}\right)
=0,\ \ \ \alpha \left( \bar{t},\bar{t}\right) =0, \\ 
\\ 
i\gamma _{\bar{t}}\left( t,\bar{t}\right) -\frac{1}{2}\varepsilon ^{2}\gamma
^{2}\left( t,\bar{t}\right) -\kappa i\gamma \left( t,\bar{t}\right) =0,\ \ \
\gamma \left( \bar{t},\bar{t}\right) =l.%
\end{array}
\label{Eq7.48}
\end{equation}%
Thus, $\gamma $ solves a nonlinear Riccati equation, which can be linearized
via the standard substitution%
\begin{equation}
\begin{array}{c}
\gamma \left( t,\bar{t}\right) =-\frac{2i\Omega ^{\prime }\left( t,\bar{t}%
\right) }{\varepsilon ^{2}\Omega \left( t,\bar{t}\right) }.%
\end{array}
\label{Eq7.49}
\end{equation}%
As a result, one gets the following equations:%
\begin{equation}
\begin{array}{c}
\Omega _{tt}\left( t,\bar{t}\right) -\kappa \Omega _{t}\left( t,\bar{t}%
\right) =0,\ \ \ \Omega \left( \bar{t},\bar{t}\right) =1,\ \ \ \Omega
^{\prime }\left( \bar{t},\bar{t}\right) =\frac{i\varepsilon ^{2}l}{2},\ \ 
\end{array}
\label{Eq7.50}
\end{equation}%
\begin{equation}
\begin{array}{c}
\alpha _{t}\left( t,\bar{t}\right) +\frac{2\chi }{\varepsilon ^{2}}\left(
\ln \left( \Omega \left( t,\bar{t}\right) \right) \right) _{t}=0,\ \ \
\alpha \left( \bar{t},\bar{t}\right) =0.%
\end{array}
\label{Eq7.51}
\end{equation}%
Accordingly,%
\begin{equation}
\begin{array}{c}
\Omega \left( t,\bar{t}\right) =1-\frac{\varepsilon ^{2}}{2}\mathsf{B}%
_{\kappa }\left( T\right) il,%
\end{array}
\label{Eq7.52}
\end{equation}%
\begin{equation}
\begin{array}{c}
\Omega ^{\prime }\left( t,\bar{t}\right) =\frac{\varepsilon ^{2}}{2}\mathsf{A%
}_{\kappa }\left( T\right) il,%
\end{array}
\label{Eq7.53}
\end{equation}%
\begin{equation}
\begin{array}{c}
\gamma \left( t,\bar{t}\right) =\frac{\mathsf{A}_{\kappa }\left( T\right) l}{%
\left( 1-\frac{\varepsilon ^{2}}{2}\mathsf{B}_{\kappa }\left( T\right)
il\right) },%
\end{array}
\label{Eq7.54}
\end{equation}%
\begin{equation}
\begin{array}{c}
\alpha \left( t,\bar{t}\right) =-\left( \vartheta +1\right) \ln \left( 1-%
\frac{\varepsilon ^{2}}{2}\mathsf{B}_{\kappa }\left( T\right) il\right) ,%
\end{array}
\label{Eq7.55}
\end{equation}%
and%
\begin{equation}
\begin{array}{c}
\mathcal{K}=\exp \left( -\left( \vartheta +1\right) \ln \left( 1-\frac{%
\varepsilon ^{2}}{2}\mathsf{B}_{\kappa }\left( T\right) il\right) +\left( 
\frac{\frac{\varepsilon ^{2}}{2}\mathsf{A}_{\kappa }\left( T\right) }{\left(
1-\frac{\varepsilon ^{2}}{2}\mathsf{B}_{\kappa }\left( T\right) il\right) }y-%
\bar{y}\right) il\right) .%
\end{array}
\label{Eq7.56}
\end{equation}%
To analyze the problem further, it is helpful to define 
\begin{equation}
\begin{array}{c}
M=\frac{2}{\varepsilon ^{2}\mathsf{B}_{\kappa }\left( T\right) },%
\end{array}
\label{Eq7.57}
\end{equation}%
introduce a new variable, $l\rightarrow \hat{l}$:%
\begin{equation}
\begin{array}{c}
\hat{l}=\frac{l}{2M},\ \ \ l=2M\hat{l},%
\end{array}
\label{Eq7.58}
\end{equation}%
and rescale $\mathcal{K}$, $\mathcal{K}dl\rightarrow \mathcal{\hat{K}}d\hat{l%
}$:%
\begin{equation}
\begin{array}{c}
\mathcal{\hat{K}}=2M\exp \left( -M\left( Y+\bar{y}\right) \right) \\ 
\\ 
\times \exp \left( -\left( \vartheta +1\right) \ln \left( 1-2i\hat{l}\right)
+M\left( \frac{Y}{1-2i\hat{l}}+\bar{y}\left( 1-2i\hat{l}\right) \right)
\right) ,%
\end{array}
\label{Eq7.59}
\end{equation}%
where $M$ appears due to the change of variables, and 
\begin{equation}
\begin{array}{c}
Y=e^{-\kappa T}y.%
\end{array}
\label{Eq7.60}
\end{equation}%
Finally,%
\begin{equation}
\begin{array}{c}
\varpi \left( t,y,\bar{t},\bar{y}\right) =\frac{M}{\pi }e^{-M\left( \bar{y}%
+Y\right) }\int_{-\infty }^{\infty }e^{-\left( \vartheta +1\right) \ln
\left( 1-2i\hat{l}\right) +M\left( \frac{Y}{1-2i\hat{l}}+\bar{y}\left( 1-2i%
\hat{l}\right) \right) }d\hat{l}.%
\end{array}
\label{Eq7.61}
\end{equation}%
Eq. (\ref{Eq7.61}) allows us to understand the true meaning of condition (%
\ref{Eq7.44}). When this condition is satisfied, the corresponding integral
converges absolutely when $\hat{l}\rightarrow \pm \infty $. A well-known
formula yields%
\begin{equation}
\begin{array}{c}
\varpi ^{\left( \vartheta \right) }\left( t,y,\bar{t},\bar{y}\right)
=Me^{-M\left( \bar{y}+Y\right) }\left( \frac{\bar{y}}{Y}\right) ^{\vartheta
/2}I_{\vartheta }\left( 2M\sqrt{\bar{y}Y}\right) ,%
\end{array}
\label{Eq7.62}
\end{equation}%
see, e.g., \cite{Lipton01} and references therein. The probability flux $%
\mathcal{F}$ has the form%
\begin{equation}
\begin{array}{c}
\mathcal{F}^{\left( \vartheta \right) }\left( t,y,\bar{t},\bar{y}\right) =-%
\frac{1}{2}\varepsilon ^{2}\left( \bar{y}\varpi ^{\left( \vartheta \right)
}\left( t,y,\bar{t},\bar{y}\right) \right) _{\bar{y}}+\left( \chi -\kappa 
\bar{y}\right) \varpi ^{\left( \vartheta \right) }\left( t,y,\bar{t},\bar{y}%
\right) \\ 
\\ 
=-\frac{1}{2}\varepsilon ^{2}MY\varpi ^{\left( \vartheta +1\right) }\left(
t,y,\bar{t},\bar{y}\right) +\left( \frac{1}{2}\varepsilon ^{2}-\frac{\kappa 
}{M}\right) M\bar{y}\varpi ^{\left( \vartheta \right) }\left( t,y,\bar{t},%
\bar{y}\right) .%
\end{array}
\label{Eq7.62a}
\end{equation}%
It is important to note that the density $\varpi \left( \bar{y}\right) $
integrates to one:%
\begin{equation}
\begin{array}{c}
\int_{0}^{\infty }\varpi \left( t,y,\bar{t},\bar{y}\right) d\bar{y}%
=\int_{0}^{\infty }e^{-u-v}\left( \frac{v}{u}\right) ^{\vartheta
/2}I_{\vartheta }\left( 2\sqrt{uv}\right) dv=1,%
\end{array}
\label{Eq7.63}
\end{equation}%
where $u=MY,\ \ \ v=M\bar{y}$. This fact below is used below.

Using the asymptotic expansion of the modified Bessel function, one can show
that $\varpi ^{\left( \vartheta \right) }$ and $\mathcal{F}$ vanish on the
boundary, since 
\begin{equation}
\begin{array}{c}
\varpi ^{\left( \vartheta \right) }\left( t,y,\bar{t},\bar{y}\right) =\frac{%
Me^{-MY}}{\Gamma \left( \vartheta +1\right) }\left( M\bar{y}\right)
^{\vartheta }\left( 1+O\left( \bar{y}\right) \right) , \\ 
\\ 
\mathcal{F}^{\left( \vartheta \right) }\left( t,y,\bar{t},\bar{y}\right)
=\left( \frac{\varepsilon ^{2}}{2}\left( 1-\frac{MY}{\left( \vartheta
+1\right) }\right) -\frac{\kappa }{M}\right) Me^{-MY}\frac{\left( M\bar{y}%
\right) ^{\left( \vartheta +1\right) }}{\Gamma \left( \vartheta +1\right) }%
\left( 1+O\left( \bar{y}\right) \right) .%
\end{array}
\label{Eq7.62b}
\end{equation}%
Now assume that condition (\ref{Eq7.44}) is violated, so that $-1<\vartheta
<0$. In this case, the integral in Eq. (\ref{Eq7.61}) is no longer
absolutely convergent, so one needs to regularize it. There are two ways of
regularizing the corresponding integral: (I) integration by parts, (II)
change of variables. Not surprisingly, they produce different results.

Start with integration by parts and write%
\begin{equation}
\begin{array}{c}
Int_{\vartheta }\equiv \frac{1}{\pi }\int_{-\infty }^{\infty }e^{-\left(
\vartheta +1\right) \ln \left( 1-2i\hat{l}\right) +\frac{MY}{1-2i\hat{l}}%
}d\left( \frac{e^{M\bar{y}\left( 1-2i\hat{l}\right) }}{-2iM\bar{y}}\right)
\\ 
\\ 
=\frac{1}{\pi }\frac{\left( \vartheta +1\right) }{M\bar{y}}\int_{-\infty
}^{\infty }e^{-\left( \vartheta +2\right) \ln \left( 1-2i\hat{l}\right)
+M\left( \frac{Y}{1-2i\hat{l}}+\bar{y}\left( 1-2i\hat{l}\right) \right) }d%
\hat{l} \\ 
\\ 
+\frac{1}{\pi }\frac{Y}{\bar{y}}\int_{-\infty }^{\infty }e^{-\left(
\vartheta +3\right) \ln \left( 1-2i\hat{l}\right) +M\left( \frac{Y}{1-2i\hat{%
l}}+\bar{y}\left( 1-2i\hat{l}\right) \right) }d\hat{l},%
\end{array}
\label{Eq7.63b}
\end{equation}%
where the integrals are absolutely convergent. Thus, Eq. (\ref{Eq7.62})
yields%
\begin{equation}
\begin{array}{c}
Int_{\vartheta }=\left( \frac{\bar{y}}{Y}\right) ^{\frac{\vartheta }{2}%
}\left( \frac{2\left( \vartheta +1\right) }{Z}I_{\vartheta +1}\left(
Z\right) +I_{\vartheta +2}\left( Z\right) \right) =\left( \frac{\bar{y}}{Y}%
\right) ^{\frac{\vartheta }{2}}I_{\vartheta }\left( Z\right) ,%
\end{array}
\label{Eq7.63c}
\end{equation}%
where $Z=2M\sqrt{\bar{y}Y}$, and a well-known recurrent relation for the
modified Bessel functions is used; \cite{Abramowitz64}, Eq. 9.6.26. Thus,
Eqs (\ref{Eq7.62}), (\ref{Eq7.62a}) hold for $-1<\vartheta <0$:%
\begin{equation}
\begin{array}{c}
\varpi ^{\left( \vartheta ,I\right) }=\varpi ^{\left( \vartheta \right) },\
\ \ \mathcal{F}^{\left( \vartheta ,I\right) }=\mathcal{F}^{\left( \vartheta
\right) }.%
\end{array}
\label{Eq7.63d2}
\end{equation}%
It is important to note that $\varpi ^{\left( \vartheta ,I\right) }\left( 
\bar{y}\rightarrow 0\right) \rightarrow \infty $ when $\vartheta <0$ (the
corresponding singularity is integrable), while $\varpi ^{\left( \vartheta
\right) }$ is bounded at $\bar{y}=0$, when $\vartheta >0$. While the
t.p.d.f. itself blows up at the natural boundary $\bar{y}=0$, the
probability flux $\mathcal{F}^{\left( \vartheta ,I\right) }$ vanishes on the
boundary, so that the total probability of staying on the positive semi-axis
is conserved.

Now, use change of variables to regularize $Int_{\vartheta }$. Specifically,
introduce $\tilde{l}$, such that%
\begin{equation}
\begin{array}{c}
1-2i\hat{l}=\frac{1}{1-2i\tilde{l}},%
\end{array}
\label{Eq7.63e}
\end{equation}%
and formally write $Int_{\vartheta }$ as follows:%
\begin{equation}
\begin{array}{c}
Int_{\vartheta }=\frac{1}{\pi }\int_{-\infty }^{\infty }e^{-\left(
-\vartheta +1\right) \ln \left( 1-2i\tilde{l}\right) +MY\left( 1-2i\tilde{l}%
\right) +\frac{M\bar{y}}{\left( 1-2i\tilde{l}\right) }}d\tilde{l} \\ 
\\ 
=\left( \frac{Y}{\bar{y}}\right) ^{-\frac{\vartheta }{2}}I_{-\vartheta
}\left( 2M\sqrt{\bar{y}Y}\right) =\left( \frac{\bar{y}}{Y}\right) ^{\frac{%
\vartheta }{2}}I_{-\vartheta }\left( 2M\sqrt{\bar{y}Y}\right) .%
\end{array}
\label{Eq7.63f}
\end{equation}%
Accordingly,%
\begin{equation}
\begin{array}{c}
\varpi ^{\left( \vartheta ,II\right) }\left( t,y,\bar{t},\bar{y}\right)
=Me^{-M\left( \bar{y}+Y\right) }\left( \frac{\bar{y}}{Y}\right) ^{\frac{%
\vartheta }{2}}I_{-\vartheta }\left( 2M\sqrt{\bar{y}Y}\right) .%
\end{array}
\label{Eq7.63g}
\end{equation}%
A straightforward calculation yields:%
\begin{equation}
\begin{array}{c}
\mathcal{F}^{\left( \vartheta ,II\right) }\left( t,y,\bar{t},\bar{y}\right)
=Me^{-M\left( \bar{y}+Y\right) }\left( \frac{\bar{y}}{Y}\right) ^{\vartheta
/2}\left( -\frac{1}{2}\varepsilon ^{2}M\sqrt{\bar{y}Y}I_{-\vartheta
+1}\left( 2M\sqrt{\bar{y}Y}\right) \right. \\ 
\\ 
\left. +\left( \frac{1}{2}\varepsilon ^{2}\vartheta +\left( \frac{1}{2}%
\varepsilon ^{2}-\frac{\kappa }{M}\right) M\bar{y}\right) I_{-\vartheta
}\left( 2M\sqrt{\bar{y}Y}\right) \right) .%
\end{array}
\label{Eq7.63h}
\end{equation}%
It is easy to see that both $\varpi ^{\left( \vartheta ,II\right) }$ and $%
\mathcal{F}^{\left( \vartheta ,II\right) }$ are bounded at $\bar{y}=0$:%
\begin{equation}
\begin{array}{c}
\varpi ^{\left( \vartheta ,II\right) }\left( t,y,\bar{t},\bar{y}\right) =%
\frac{Me^{-MY}}{\Gamma \left( -\vartheta +1\right) \left( MY\right)
^{\vartheta }}\left( 1+O\left( \bar{y}\right) \right) , \\ 
\\ 
\mathcal{F}^{\left( \vartheta ,II\right) }\left( t,y,\bar{t},\bar{y}\right) =%
\frac{\varepsilon ^{2}\vartheta Me^{-MY}}{2\Gamma \left( -\vartheta
+1\right) \left( MY\right) ^{\vartheta }}\left( 1+O\left( \bar{y}\right)
\right) .%
\end{array}
\label{Eq7.63i}
\end{equation}%
Since there is a probability flux across the natural boundary $\bar{y}=0$,
the total probability on the positive semi-axis $\left[ 0,\infty \right) $
is less than one.

\begin{figure}[tbp]
\begin{center}
\includegraphics[width=0.8\textwidth]
{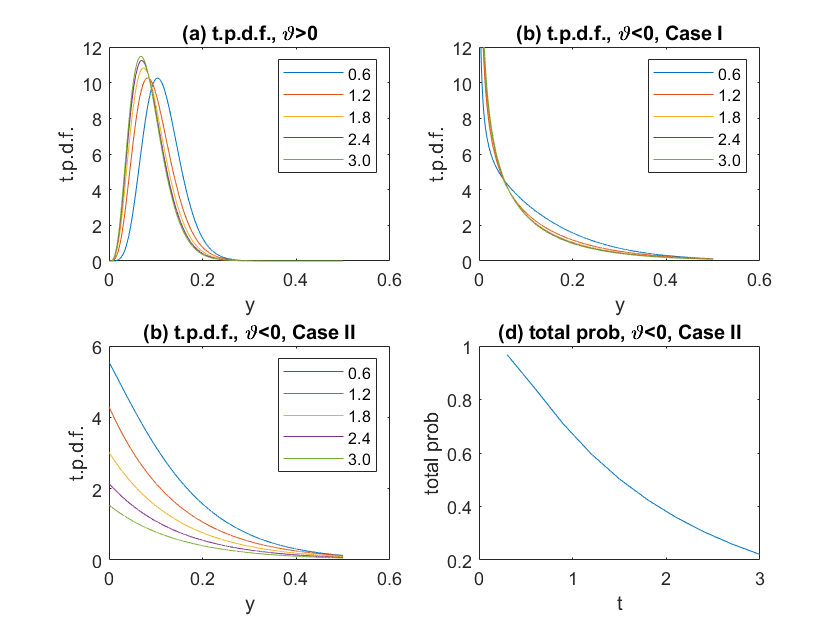}
\end{center}
\par
\vspace{-10pt}
\caption{T.p.d.fs for three Feller processes with different parameters and
regularity conditions. (a) $\protect\chi =0.1$, $\protect\kappa =1.2$, $%
\protect\varepsilon =0.2$, $y_{0}=0.15$, $\bar{t}_{max}=3$; (b), (c) $%
\protect\chi =0.1$, $\protect\kappa =1.2$, $\protect\varepsilon =0.6$, $%
y_{0}=0.15$, $\bar{t}_{max}=3$. For the first and second processes, the
probability of $\bar{y}\geq 0$ is equal to one. For the third process, this
probability, shown as a function of time in (d), is less than one. Own
graphics.}
\label{Fig7.2}
\end{figure}

Representative t.p.d.fs for Feller processes with different values of $%
\vartheta $ are illustrated in Figure \ref{Fig7.2}. 

\paragraph{Feller Process with Time-Dependent Parameters}

Surprisingly, studying the Feller process with time-dependent coefficients
is viewed as a difficult problem, which remains an active area of research;
see, e.g., \cite{Masoliver16, Giorno21} and references therein. However,
using Kelvin wave formalism allows one to find an expression for the
t.p.d.f. in a very natural way.

The process are time-dependent parameters has the form:%
\begin{equation}
\begin{array}{c}
\varpi _{t}+\frac{1}{2}\varepsilon ^{2}\left( t\right) y\varpi _{yy}+\left(
\chi \left( t\right) -\kappa \left( t\right) y\right) \varpi _{y}=0, \\ 
\\ 
\varpi \left( \bar{t},y\right) =\delta \left( y-\bar{y}\right) .%
\end{array}
\label{Eq7.64}
\end{equation}%
Here it is assumed that the following regularity condition is satisfied:%
\begin{equation}
\begin{array}{c}
\vartheta \left( t\right) =\frac{2\chi \left( t\right) }{\varepsilon
^{2}\left( t\right) }-1>0.%
\end{array}
\label{Eq7.65}
\end{equation}%
This condition guarantees that the corresponding integrals converge at
infinity.

As usual, $\varpi $ can be written as a superposition of Kelvin waves of the
form%
\begin{equation}
\begin{array}{c}
\mathcal{K}=\exp \left( \alpha \left( t,\bar{t}\right) +i\gamma \left( t,%
\bar{t}\right) y-il\bar{y}\right) ,%
\end{array}
\label{Eq7.66}
\end{equation}%
where $\alpha ,\gamma $ solve the following system of backward ODEs:%
\begin{equation}
\begin{array}{c}
\alpha _{t}\left( t,\bar{t}\right) +\chi \left( t,\bar{t}\right) i\gamma
\left( t,\bar{t}\right) =0,\ \ \ \alpha \left( \bar{t},\bar{t}\right) =0, \\ 
\\ 
i\gamma _{\bar{t}}\left( t,\bar{t}\right) -\frac{1}{2}\varepsilon ^{2}\left(
t\right) \gamma ^{2}\left( t,\bar{t}\right) -\kappa \left( t\right) i\gamma
\left( t,\bar{t}\right) =0,\ \ \ \gamma \left( \bar{t},\bar{t}\right) =l,%
\end{array}
\label{Eq7.67}
\end{equation}%
Introducing $\Omega \left( t,\bar{t}\right) $, such that 
\begin{equation}
\begin{array}{c}
\gamma \left( t,\bar{t}\right) =-\frac{2i\Omega ^{\prime }\left( t,\bar{t}%
\right) }{\varepsilon ^{2}\left( t\right) \Omega \left( t,\bar{t}\right) },%
\end{array}
\label{Eq7.68}
\end{equation}%
one gets the following second-order equation for $\Omega \left( t,\bar{t}%
\right) $:%
\begin{equation}
\begin{array}{c}
\Omega _{tt}\left( t,\bar{t}\right) -\left( \kappa +2\ln \left( \varepsilon
\right) ^{\prime }\right) \Omega _{t}\left( t,\bar{t}\right) =0,\ \ \ \Omega
\left( \bar{t},\bar{t}\right) =1,\ \ \ \Omega ^{\prime }\left( \bar{t},\bar{t%
}\right) =\frac{\varepsilon ^{2}\left( \bar{t}\right) }{2}il.%
\end{array}
\label{Eq7.69}
\end{equation}%
Solving this equation, one gets:%
\begin{equation}
\begin{array}{c}
\Omega \left( t,\bar{t}\right) =1-\frac{\varepsilon ^{2}\left( t\right) }{2}%
\mathsf{\bar{B}}_{\kappa }\left( t,\bar{t}\right) il, \\ 
\\ 
\Omega _{t}\left( t,\bar{t}\right) =\frac{\varepsilon ^{2}\left( t\right) }{2%
}\left( \mathsf{A}_{\kappa }\left( t,\bar{t}\right) -\frac{2\varepsilon
^{\prime }\left( t\right) }{\varepsilon \left( t\right) }\mathsf{\bar{B}}%
_{\kappa }\left( t,\bar{t}\right) \right) il.%
\end{array}
\label{Eq7.70}
\end{equation}%
Accordingly,%
\begin{equation}
\begin{array}{c}
\gamma \left( t,\bar{t}\right) =\frac{\left( \mathsf{A}_{\kappa }\left( t,%
\bar{t}\right) -\frac{2\varepsilon ^{\prime }\left( t\right) }{\varepsilon
\left( t\right) }\mathsf{\bar{B}}_{\kappa }\left( t,\bar{t}\right) \right) l%
}{\left( 1-\frac{\varepsilon ^{2}\left( t\right) }{2}\mathsf{\bar{B}}%
_{\kappa }\left( t,\bar{t}\right) il\right) },%
\end{array}
\label{Eq7.72}
\end{equation}%
\begin{equation}
\begin{array}{c}
\alpha \left( t,\bar{t}\right) =-\frac{2\chi \left( t\right) }{\varepsilon
^{2}\left( t\right) }\ln \left( 1-\frac{\varepsilon ^{2}\left( t\right) }{2}%
\mathsf{\bar{B}}_{\kappa }\left( t,\bar{t}\right) il\right) \\ 
\\ 
-\int\limits_{t}^{\bar{t}}\left( \frac{2\chi \left( s\right) }{\varepsilon
^{2}\left( s\right) }\right) ^{\prime }\ln \left( 1-\frac{\varepsilon
^{2}\left( s\right) }{2}\mathsf{\bar{B}}_{\kappa }\left( s,\bar{t}\right)
il\right) ds.%
\end{array}
\label{Eq7.73}
\end{equation}%
Thus, Kelving wave becomes%
\begin{equation}
\begin{array}{c}
\mathcal{K}=\exp \left( -\frac{2\chi \left( t\right) }{\varepsilon
^{2}\left( t\right) }\ln \left( 1-\frac{\varepsilon ^{2}\left( t\right) }{2}%
\mathsf{\bar{B}}_{\kappa }\left( t,\bar{t}\right) il\right) \right. \\ 
\\ 
\left. -\int\limits_{t}^{\bar{t}}\left( \frac{2\chi \left( s\right) }{%
\varepsilon ^{2}\left( s\right) }\right) ^{\prime }\ln \left( 1-\frac{%
\varepsilon ^{2}\left( s\right) }{2}\mathsf{\bar{B}}_{\kappa }\left( s,\bar{t%
}\right) il\right) ds+\left( \frac{\left( \mathsf{A}_{\kappa }\left( t,\bar{t%
}\right) -\frac{2\varepsilon ^{\prime }\left( t\right) }{\varepsilon \left(
t\right) }\mathsf{\bar{B}}_{\kappa }\left( t,\bar{t}\right) \right) }{\left(
1-\frac{\varepsilon ^{2}\left( t\right) }{2}\mathsf{\bar{B}}_{\kappa }\left(
t,\bar{t}\right) il\right) }y-\bar{y}\right) il\right) .%
\end{array}
\label{Eq7.74}
\end{equation}%
By analogy with (\ref{Eq7.57}), (\ref{Eq7.58}), (\ref{Eq7.59}), define%
\begin{equation}
\begin{array}{c}
M\left( t,\bar{t}\right) =\frac{2}{\varepsilon ^{2}\left( t\right) \mathsf{%
\bar{B}}_{\kappa }\left( t,\bar{t}\right) },\ \ \ \hat{l}=\frac{l}{2M\left(
t,\bar{t}\right) },\ \ \ l=2M\left( t,\bar{t}\right) \hat{l}%
\end{array}
\label{Eq7.75}
\end{equation}%
and represent $\mathcal{\hat{K}}$ as follows:%
\begin{equation}
\begin{array}{c}
\mathcal{\hat{K}}\left( t,y,\bar{t},\bar{y},\hat{l}\right) =2M\exp \left(
-M\left( t,\bar{t}\right) \left( Y+\bar{y}\right) \right) \exp \left( -\frac{%
2\chi \left( t\right) }{\varepsilon ^{2}\left( t\right) }\ln \left( 1-2i\hat{%
l}\right) \right. \\ 
\\ 
\left. -\int\limits_{t}^{\bar{t}}\left( \frac{2\chi \left( s\right) }{%
\varepsilon ^{2}\left( s\right) }\right) ^{\prime }\ln \left( 1-\frac{%
M\left( t,\bar{t}\right) }{M\left( s,\bar{t}\right) }2i\hat{l}\right)
ds+M\left( t,\bar{t}\right) \left( \frac{Y}{\left( 1-2i\hat{l}\right) }+\bar{%
y}\left( 1-2i\hat{l}\right) \right) \right) ,%
\end{array}
\label{Eq7.76}
\end{equation}%
where%
\begin{equation}
\begin{array}{c}
Y=\left( \mathsf{A}_{\kappa }\left( t,\bar{t}\right) -\frac{4\varepsilon
^{\prime }\left( t\right) }{\varepsilon ^{3}\left( t\right) M\left( t,\bar{t}%
\right) }\right) y.%
\end{array}
\label{Eq7.77}
\end{equation}%
Finally,%
\begin{equation}
\begin{array}{c}
\varpi \left( t,y,\bar{t},\bar{y}\right) =\frac{1}{2\pi }\int\limits_{-%
\infty }^{\infty }\mathcal{\hat{K}}\left( t,y,\bar{t},\bar{y},\hat{l}\right)
d\hat{l}.%
\end{array}
\label{Eq7.78}
\end{equation}%
Therefore, finding $\varpi \left( t,y,\bar{t},\bar{y}\right) $ is reduced to
solving some very simple ODEs and calculating a one-dimensional integral,
which is theoretically appealing and numerically efficient.

\paragraph{Feller Process with Jumps}

Consider a jump-diffusion process $\hat{y}_{t}$ with constant coefficients
governed by the equation: 
\begin{equation}
\begin{array}{c}
d\hat{y}_{t}=\left( \chi -\kappa \hat{y}_{t}\right) dt+\varepsilon \sqrt{%
\hat{y}_{t}}d\hat{Z}_{t}+Jd\hat{\Pi}_{t},\ \ \ \hat{y}_{t}=y,%
\end{array}
\label{Eq7.79}
\end{equation}%
where $\hat{Z}_{t}$ is a standard Wiener process, and $\hat{\Pi}_{t}$ is a
Poisson process with intensity $\lambda $. To preserve tractability, it is
assumed that jumps are positive and exponentially distributed with parameter 
$\phi $; for additional insights, see \cite{Lipton12}.

The backward Kolmogorov problem can be written as:%
\begin{equation}
\begin{array}{c}
\varpi _{t}+\frac{1}{2}\varepsilon ^{2}y\varpi _{yy}+\left( \chi -\kappa
y\right) \varpi _{y}+\lambda \left( \phi \int_{0}^{\infty }\varpi
(t,y+J)e^{-\phi J}dJ-\varpi (t,y)\right) =0, \\ 
\\ 
\varpi \left( \bar{t},y,\bar{t},\bar{y}\right) =\delta \left( y-\bar{y}%
\right) .%
\end{array}
\label{Eq7.80}
\end{equation}%
The corresponding Kelvin wave has the familiar form: 
\begin{equation}
\begin{array}{c}
\mathcal{K}\left( t,y,\bar{t},\bar{y},l\right) =\exp \left\{ \alpha \left( t,%
\bar{t}\right) +i\gamma \left( t,\bar{t}\right) y-il\bar{y}\right\} ,%
\end{array}
\label{Eq7.81}
\end{equation}%
where $\alpha ,\gamma $ satisfy the following system of ODEs: 
\begin{equation}
\begin{array}{c}
\alpha _{t}\left( t,\bar{t}\right) +\chi i\gamma \left( t,\bar{t}\right) +%
\frac{\lambda i\gamma \left( t,\bar{t}\right) }{\phi -i\gamma \left( t,\bar{t%
}\right) },\ \ \ \alpha \left( \bar{t},\bar{t}\right) =0, \\ 
\\ 
i\gamma _{\bar{t}}\left( t,\bar{t}\right) -\frac{1}{2}\varepsilon ^{2}\gamma
^{2}\left( t,\bar{t}\right) -\kappa i\gamma \left( t,\bar{t}\right) =0,\ \ \
\gamma (\bar{t},\bar{t})=l.%
\end{array}
\label{Eq7.82}
\end{equation}%
The expression for $\gamma $ is given by Eq. (\ref{Eq7.54}), while $\alpha $
can be split as follows:%
\begin{equation}
\begin{array}{c}
\alpha \left( t,\bar{t}\right) =\alpha _{0}\left( t,\bar{t}\right) +\lambda
\alpha _{1}\left( t,\bar{t}\right) .%
\end{array}
\label{Eq7.83}
\end{equation}%
In this setting, $\alpha _{0}$ has the familiar form%
\begin{equation}
\begin{array}{c}
\alpha _{0}\left( t,\bar{t}\right) =-(\vartheta +1)\ln \left( 1-\frac{%
\varepsilon ^{2}}{2}\mathsf{\bar{B}}_{\kappa }\left( T\right) il\right) ,%
\end{array}
\label{Eq7.84}
\end{equation}%
while $\alpha _{1}$ can be represented as follows: 
\begin{equation}
\begin{array}{c}
\alpha _{1}\left( t,\bar{t}\right) =\int\limits_{t}^{\bar{t}}\frac{\mathsf{A}%
_{\kappa }\left( \bar{t}-s\right) il}{\phi -\left( \frac{\phi \varepsilon
^{2}}{2}\mathsf{\bar{B}}_{\kappa }\left( \bar{t}-s\right) +\mathsf{A}%
_{\kappa }\left( \bar{t}-s\right) \right) il}ds \\ 
\\ 
=\frac{1}{\left( \kappa -\frac{\phi \varepsilon ^{2}}{2}\right) }\ln \left( 
\frac{\phi -\left( \frac{\phi \varepsilon ^{2}}{2}\mathsf{\bar{B}}_{\kappa
}\left( T\right) +\mathsf{A}_{\kappa }\left( T\right) \right) il}{\phi -il}%
\right) .%
\end{array}
\label{Eq7.85}
\end{equation}%
Thus, jumps do profoundly affect the dynamics of the underlying stochastic
process.

\subsubsection{Augmented Feller Process, I}

This section studies the joint dynamics of a Feller process $\hat{y}_{t}$
and its integral $\hat{x}_{t}$. The corresponding combined process is
described by the following equations:%
\begin{equation}
\begin{array}{c}
d\hat{x}_{t}=\hat{y}_{t}dt,\ \ \ \hat{x}_{t}=x, \\ 
\\ 
d\hat{y}_{t}=\left( \chi -\kappa \hat{y}_{t}\right) dt+\varepsilon \sqrt{%
\hat{y}_{t}}d\hat{Z}_{t},\ \ \ \hat{y}_{t}=y.%
\end{array}
\label{Eq7.87}
\end{equation}%
Depending on the interpretation, these equations can describe the joint
evolution of a particle's position and its velocity, the integral of
variance and variance, among other possibilities.

\begin{figure}[tbp]
\begin{center}
\includegraphics[width=0.8\textwidth]
{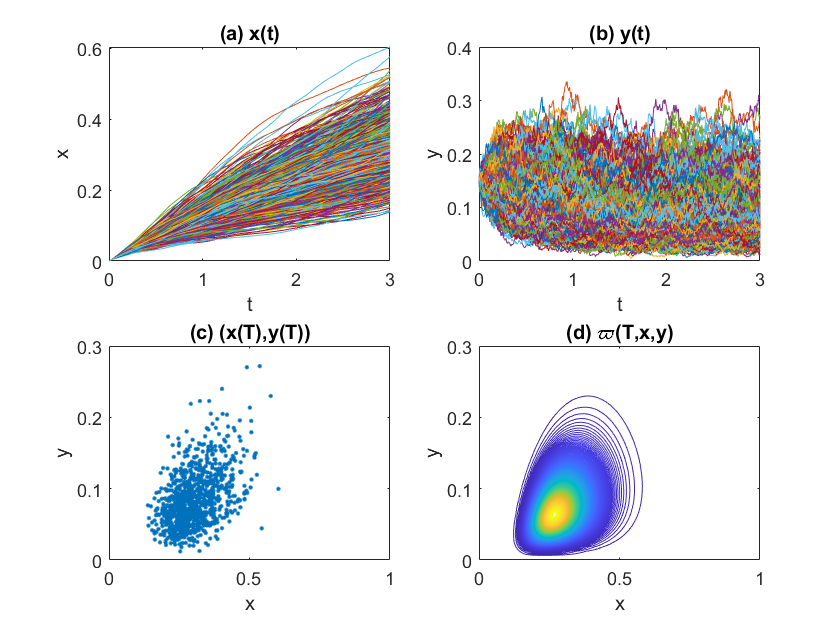}
\end{center}
\par
\vspace{-10pt}
\caption{A thousand trajectories of a representative t.p.d.f. for the
degenerate augmented Feller process. Parameters are $T=3$, $dt=0.01$, $%
\protect\chi =0.1$, $\protect\kappa =1.2$, $\protect\varepsilon =0.2$. $x=0$%
; $y_{0}=0.15$. (a) $x\left( t\right) $, (b) $y\left( t\right) $, (c) $%
\left( \bar{x}\left( T\right) ,\bar{y}\left( T\right) \right) $, (d) contour
lines of $\protect\varpi \left( 0,0.15,0,T,\tilde{x},\tilde{y}\right) $. Own
graphics.}
\label{Fig7.3}
\end{figure}

The forward Fokker-Planck has the form:%
\begin{equation}
\begin{array}{c}
\varpi _{\bar{t}}-\frac{1}{2}\varepsilon ^{2}\left( \bar{y}\varpi \right) _{%
\bar{y}\bar{y}}+\bar{y}\varpi _{\bar{x}}+\left( \left( \chi -\kappa \bar{y}%
\right) \varpi \right) _{\bar{y}}=0, \\ 
\\ 
\varpi \left( t,x,y,t,\bar{x},\bar{y}\right) =\delta \left( \bar{x}-x\right)
\delta \left( \bar{y}-y\right) ,%
\end{array}
\label{Eq7.88}
\end{equation}%
while the backward Kolmogorov problem can be written as follows:%
\begin{equation}
\begin{array}{c}
\varpi _{t}+\frac{1}{2}\varepsilon ^{2}y\varpi _{yy}+y\varpi _{x}+\left(
\chi -\kappa y\right) \varpi _{y}=0 \\ 
\\ 
\varpi \left( \bar{t},x,y,\bar{t},\bar{x},\bar{y}\right) =\delta \left( x-%
\bar{x}\right) \delta \left( y-\bar{y}\right) .%
\end{array}
\label{Eq7.89}
\end{equation}%
Below the backward problem is considered, which allows one to derive the
desired formula more efficiently. The corresponding function $\mathcal{K}$
has the form:%
\begin{equation}
\begin{array}{c}
\mathcal{K}=\exp \left( \alpha \left( t,\bar{t}\right) +ik\left( x-\bar{x}%
\right) +i\gamma \left( t,\bar{t}\right) y-il\bar{y}\right) ,%
\end{array}
\label{Eq7.90}
\end{equation}%
where%
\begin{equation}
\begin{array}{c}
\alpha _{t}\left( t,\bar{t}\right) +i\chi \gamma \left( t,\bar{t}\right)
=0,\ \ \ \alpha \left( \bar{t},\bar{t}\right) =0, \\ 
\\ 
i\gamma _{t}\left( t,\bar{t}\right) -\frac{1}{2}\varepsilon ^{2}\gamma
^{2}\left( t,\bar{t}\right) -i\kappa \gamma \left( t,\bar{t}\right) +ik=0,\
\ \ \gamma \left( \bar{t},\bar{t}\right) =l,%
\end{array}
\label{Eq7.91}
\end{equation}%
As before, one can linearize the Riccati equation for $\gamma $ by using
substitution given by Eq. (\ref{Eq7.49}), with $\Omega \left( t,\bar{t}%
\right) $ solving the second-order equation of the form: 
\begin{equation}
\begin{array}{c}
\Omega _{tt}\left( t,\bar{t}\right) -\kappa \Omega _{t}\left( t,\bar{t}%
\right) +\frac{i\varepsilon ^{2}k}{2}\Omega \left( t,\bar{t}\right) =0,\ \ \
\Omega \left( \bar{t},\bar{t}\right) =1,\ \ \ \Omega ^{\prime }\left( \bar{t}%
,\bar{t}\right) =\frac{i\varepsilon ^{2}l}{2}.%
\end{array}
\label{Eq7.92}
\end{equation}%
One can represent $\Omega \left( t,\bar{t}\right) $ in the form:%
\begin{equation}
\begin{array}{c}
\Omega \left( t,\bar{t}\right) =\omega _{+}e^{\lambda _{+}\left( \bar{t}%
-t\right) }+\omega _{-}e^{\lambda _{+}\left( \bar{t}-t\right) },%
\end{array}
\label{Eq7.93}
\end{equation}%
where $\lambda _{\pm }$ are solutions of the characteristic equation:%
\begin{equation}
\begin{array}{c}
\lambda ^{2}+\kappa \lambda +\frac{i\varepsilon ^{2}k}{2}=0,%
\end{array}
\label{Eq7.94}
\end{equation}%
and $\omega _{\pm }$ satisfy the following system of linear equations:%
\begin{equation}
\begin{array}{c}
\omega _{+}+\omega _{-}=1, \\ 
\\ 
\lambda _{+}\omega _{+}+\lambda _{-}\omega _{-}=-\frac{i\varepsilon ^{2}l}{2}%
.%
\end{array}
\label{Eq7.95}
\end{equation}%
Thus,%
\begin{equation}
\begin{array}{c}
\lambda _{\pm }=\mu \pm \zeta , \\ 
\\ 
\mu =-\frac{\kappa }{2},\ \ \ \zeta =\frac{\sqrt{\kappa ^{2}-2i\varepsilon
^{2}k}}{2},%
\end{array}
\label{Eq7.94b}
\end{equation}%
\begin{equation}
\begin{array}{c}
\omega _{+}=-\frac{\left( 2\lambda _{-}+i\varepsilon ^{2}l\right) }{4\zeta }%
,\ \ \ \omega _{-}=\frac{\left( 2\lambda _{+}+i\varepsilon ^{2}l\right) }{%
4\zeta }.%
\end{array}
\label{Eq7.96}
\end{equation}%
It is useful to note that%
\begin{equation}
\begin{array}{c}
\lambda _{+}\lambda _{-}=\mu ^{2}-\zeta ^{2}=\frac{i\varepsilon ^{2}k}{2}.%
\end{array}
\label{Eq7.97}
\end{equation}%
For the sake of brevity, notation (\ref{Eq6.125}) is used:%
\begin{equation}
\begin{array}{c}
\Omega \left( t,\bar{t}\right) =\frac{\mathsf{E}_{0}\left( -\left( 2\lambda
_{-}+i\varepsilon ^{2}l\right) \mathsf{E}_{+}+\left( 2\lambda
_{+}+i\varepsilon ^{2}l\right) \mathsf{E}_{-}\right) }{4\zeta },%
\end{array}
\label{Eq7.99}
\end{equation}%
\begin{equation}
\begin{array}{c}
\Omega _{t}\left( t,\bar{t}\right) =\frac{\mathsf{E}_{0}\left( \lambda
_{+}\left( 2\lambda _{-}+i\varepsilon ^{2}l\right) \mathsf{E}_{+}-\lambda
_{-}\left( 2\lambda _{+}+i\varepsilon ^{2}l\right) \mathsf{E}_{-}\right) }{%
4\zeta },%
\end{array}
\label{Eq7.100}
\end{equation}%
\begin{equation}
\begin{array}{c}
\gamma =\frac{2i\left( \lambda _{+}\left( 2\lambda _{-}+i\varepsilon
^{2}l\right) \mathsf{E}_{+}-\lambda _{-}\left( 2\lambda _{+}+i\varepsilon
^{2}l\right) \mathsf{E}_{-}\right) }{\varepsilon ^{2}\left( \left( 2\lambda
_{-}+i\varepsilon ^{2}l\right) \mathsf{E}_{+}-\left( 2\lambda
_{+}+i\varepsilon ^{2}l\right) \mathsf{E}_{-}\right) },%
\end{array}
\label{Eq7.101}
\end{equation}%
\begin{equation}
\begin{array}{c}
\alpha =\frac{\chi \kappa T}{\varepsilon ^{2}}-\left( \vartheta +1\right)
\ln \left( \frac{-\left( 2\lambda _{-}+i\varepsilon ^{2}l\right) \mathsf{E}%
_{+}+\left( 2\lambda _{+}+i\varepsilon ^{2}l\right) \mathsf{E}_{-}}{4\zeta }%
\right) .%
\end{array}
\label{Eq7.102}
\end{equation}%
Accordingly, $\mathcal{K}$ can be written in the form:%
\begin{equation}
\begin{array}{c}
\mathcal{K}=\exp \left( \frac{\chi \kappa T}{\varepsilon ^{2}}+ik\left( x-%
\bar{x}\right) \right. \\ 
\\ 
-\left( \vartheta +1\right) \ln \left( \frac{2\left( -\lambda _{-}\mathsf{E}%
_{+}+\lambda _{+}\mathsf{E}_{-}\right) -i\varepsilon ^{2}l\left( \mathsf{E}%
_{+}-\mathsf{E}_{-}\right) }{4\zeta }\right) \\ 
\\ 
\left. +\frac{2\left( 2\lambda _{+}\lambda _{-}\left( \mathsf{E}_{+}-\mathsf{%
E}_{-}\right) +i\varepsilon ^{2}l\left( \lambda _{+}\mathsf{E}_{+}-\lambda
_{-}\mathsf{E}_{-}\right) \right) }{\varepsilon ^{2}\left( 2\left( -\lambda
_{-}\mathsf{E}_{+}+\lambda _{+}\mathsf{E}_{-}\right) -i\varepsilon
^{2}l\left( \mathsf{E}_{+}-\mathsf{E}_{-}\right) \right) }y-il\bar{y}\right)
.%
\end{array}
\label{Eq7.103}
\end{equation}%
Define a new variable $\hat{l}$, such that:%
\begin{equation}
\begin{array}{c}
\hat{l}\mathcal{=}\frac{l}{2M},\ \ \ l=2M\hat{l},%
\end{array}
\label{Eq7.104}
\end{equation}%
where%
\begin{equation}
\begin{array}{c}
M=\frac{2\left( -\lambda _{-}\mathsf{E}_{+}+\lambda _{+}\mathsf{E}%
_{-}\right) }{\varepsilon ^{2}\left( \mathsf{E}_{+}-\mathsf{E}_{-}\right) }.%
\end{array}
\label{Eq7.105}
\end{equation}%
Rescaled $\mathcal{\hat{K}}$ can be factorized as follows:%
\begin{equation}
\begin{array}{c}
\mathcal{\hat{K}}=\mathcal{\hat{K}}_{1}\mathcal{\hat{K}}_{2},%
\end{array}
\label{Eq7.106}
\end{equation}%
where%
\begin{equation}
\begin{array}{c}
\mathcal{\hat{K}}_{1}=\exp \left( \frac{\chi \kappa T}{\varepsilon ^{2}}%
+ik\left( x-\bar{x}\right) -\left( \vartheta +1\right) \ln \left( \frac{%
-\lambda _{-}\mathsf{E}_{+}+\lambda _{+}\mathsf{E}_{-}}{2\zeta }\right)
\right. \\ 
\\ 
\left. +\frac{2\lambda _{+}\lambda _{-}\left( \mathsf{E}_{+}-\mathsf{E}%
_{-}\right) }{\varepsilon ^{2}\left( -\lambda _{-}\mathsf{E}_{+}+\lambda _{+}%
\mathsf{E}_{-}\right) }y\right) , \\ 
\\ 
\mathcal{\hat{K}}_{2}=2M\exp \left( -M\left( Y+\bar{y}\right) \right) \\ 
\\ 
\times \exp \left( -\left( \vartheta +1\right) \ln \left( 1-2i\hat{l}\right)
+M\left( \frac{Y}{1-2i\hat{l}}+\bar{y}\left( 1-2i\hat{l}\right) \right)
\right) ,%
\end{array}
\label{Eq7.107}
\end{equation}%
with%
\begin{equation}
\begin{array}{c}
Y=\frac{4\zeta ^{2}}{\left( -\lambda _{-}\mathsf{E}_{+}+\lambda _{+}\mathsf{E%
}_{-}\right) ^{2}}y.%
\end{array}
\label{Eq7.108}
\end{equation}%
Integration with respect to $\hat{l}$ can be done analytically%
\begin{equation}
\begin{array}{c}
\frac{M}{\pi }\int_{-\infty }^{\infty }\mathcal{\hat{K}}_{2}d\hat{l}%
=Me^{-M\left( \bar{y}+Y\right) }\left( \frac{\bar{y}}{Y}\right) ^{\frac{%
\vartheta }{2}}I_{\vartheta }\left( 2M\sqrt{\bar{y}Y}\right) ,%
\end{array}
\label{Eq7.109}
\end{equation}%
which allows one to calculate $\varpi $ via a single inverse Fourier
transform: 
\begin{equation}
\begin{array}{c}
\varpi =\frac{1}{2\pi }\int_{-\infty }^{\infty }\exp \left( \frac{\chi
\kappa T}{\varepsilon ^{2}}+ik\left( x-\bar{x}\right) -\left( \vartheta
+1\right) \ln \left( \frac{-\lambda _{-}\mathsf{E}_{+}+\lambda _{+}\mathsf{E}%
_{-}}{2\zeta }\right) \right. \\ 
\\ 
\left. +\frac{2\lambda _{+}\lambda _{-}\left( \mathsf{E}_{+}-\mathsf{E}%
_{-}\right) }{\varepsilon ^{2}\left( -\lambda _{-}\mathsf{E}_{+}+\lambda _{+}%
\mathsf{E}_{-}\right) }y\right) Me^{-M\left( \bar{y}+Y\right) }\left( \frac{%
\bar{y}}{Y}\right) ^{\frac{\vartheta }{2}}I_{\vartheta }\left( 2M\sqrt{\bar{y%
}Y}\right) dk.%
\end{array}
\label{Eq7.110}
\end{equation}%
A typical t.p.d.f. for a degenerate augmented Feller process is illustrated
in Figure \ref{Fig7.3}. 

Since the integral over $\bar{y}$ is equal to one, one can represent the
marginal distribution of $\bar{x}$ in the form:%
\begin{equation}
\begin{array}{c}
\varpi ^{\left( x\right) }\left( t,x,y,\bar{t},\bar{x}\right) =\frac{1}{2\pi 
}\int_{-\infty }^{\infty }\digamma \left( t,y,\bar{t},k\right) e^{ik\left( x-%
\bar{x}\right) }dk,%
\end{array}
\label{Eq7.111}
\end{equation}%
where 
\begin{equation}
\begin{array}{c}
\digamma \left( t,y,\bar{t},k\right) =\exp \left( \frac{\chi \kappa T}{%
\varepsilon ^{2}}-\left( \vartheta +1\right) \ln \left( \frac{-\lambda _{-}%
\mathsf{E}_{+}+\lambda _{+}\mathsf{E}_{-}}{2\zeta }\right) +\frac{2\lambda
_{+}\lambda _{-}\left( \mathsf{E}_{+}-\mathsf{E}_{-}\right) }{\varepsilon
^{2}\left( -\lambda _{-}\mathsf{E}_{+}+\lambda _{+}\mathsf{E}_{-}\right) }%
y\right) ,%
\end{array}
\label{Eq7.112}
\end{equation}%
with $\mu ,\zeta $ given by Eqs. (\ref{Eq7.94}). It is easy to check that $%
\varpi ^{\left( x\right) }\left( t,x,y,\bar{t},\bar{x}\right) $ integrates
to one:%
\begin{equation}
\begin{array}{c}
\int_{-\infty }^{\infty }\varpi ^{\left( x\right) }\left( t,x,y,\bar{t},\bar{%
x}\right) d\bar{x}=\frac{1}{2\pi }\int_{-\infty }^{\infty }\int_{-\infty
}^{\infty }\digamma \left( t,y,\bar{t},k\right) \exp \left( ik\left( x-\bar{x%
}\right) \right) dkd\bar{x} \\ 
\\ 
=\int_{-\infty }^{\infty }\digamma \left( t,y,\bar{t},k\right) \delta \left(
k\right) dk=\digamma \left( t,y,\bar{t},0\right) =\exp \left( \frac{\chi
\kappa T}{\varepsilon ^{2}}-\frac{2\chi }{\varepsilon ^{2}}\frac{\kappa T}{2}%
\right) =1.%
\end{array}
\label{Eq7.113}
\end{equation}%
The expected value of $\bar{x}$ has the form:%
\begin{equation}
\begin{array}{c}
X=\int_{-\infty }^{\infty }\varpi ^{\left( x\right) }\left( t,x,y,\bar{t},%
\bar{x}\right) \bar{x}d\bar{x}=x+\int_{-\infty }^{\infty }\varpi ^{\left(
x\right) }\left( t,x,y,\bar{t},\bar{x}\right) \left( \bar{x}-x\right) d\bar{x%
} \\ 
\\ 
=x+\frac{1}{2\pi }\int_{-\infty }^{\infty }\int_{-\infty }^{\infty }\digamma
\left( t,y,\bar{t},k\right) \exp \left( ik\left( \bar{x}-x\right) \right)
\left( \bar{x}-x\right) dkd\bar{x} \\ 
\\ 
=x+\underset{\epsilon \rightarrow 0}{\lim }\left( \frac{d}{d\epsilon }\left( 
\frac{1}{2\pi }\int_{-\infty }^{\infty }\int_{-\infty }^{\infty }\digamma
\left( t,y,\bar{t},k\right) \exp \left( \left( ik+\epsilon \right) \left( 
\bar{x}-x\right) \right) dkd\bar{x}\right) \right) \\ 
\\ 
=x+\underset{\epsilon \rightarrow 0}{\lim }\left( \frac{d}{d\epsilon }\left(
\int_{-\infty }^{\infty }\digamma \left( t,y,\bar{t},k\right) \delta \left(
k-i\epsilon \right) dk\right) \right) \\ 
\\ 
=x+\left. \frac{d}{d\epsilon }\digamma \left( t,y,\bar{t},i\epsilon \right)
\right\vert _{\epsilon =0}.%
\end{array}
\label{Eq7.114}
\end{equation}%
A calculation left to the reader yields%
\begin{equation}
\begin{array}{c}
X=x+\frac{\chi }{\kappa }T-\mathsf{\bar{B}}_{\kappa }\left( T\right) \left( 
\frac{\chi }{\kappa }-y\right) ,%
\end{array}
\label{Eq7.115}
\end{equation}%
which agrees with Eq. (\ref{Eq6.115}).

It is worth noting that $\varpi ^{\left( x\right) }\left( \bar{t},\bar{x}%
\right) $ has fat tails, since some of the exponential moments of $\bar{x}$
have finite-time explosions; see \cite{Andersen07, Friz10} and references
therein.\footnote{%
Of course, it is not surprising that such explosions exist since Riccati
equations are well-known to have solutions exploding in finite time.
Consider the following Riccati initial-value problem%
\begin{equation*}
\begin{array}{c}
f^{\prime }+af^{2}+bf+c=0,\ \ \ f\left( \tau \right) =d,%
\end{array}%
\end{equation*}%
and assume that $b^{2}-4ac<0$. The solution of this initial-value problem
has the form%
\begin{equation*}
\begin{array}{c}
f\left( \tau ,t\right) =-\frac{\left\vert \zeta \right\vert }{a}\tan \left(
\left\vert \zeta \right\vert \left( t-\tau \right) -\arctan \left( \frac{%
2ad+b}{2\left\vert \zeta \right\vert }\right) \right) -\frac{b}{2a},%
\end{array}%
\end{equation*}%
where $\left\vert \zeta \right\vert =\left. \sqrt{4ac-b^{2}}\right/ 2$. The
corresponding blow-up time $t^{\ast }$ has the form%
\begin{equation*}
\begin{array}{c}
t^{\ast }=\tau +\frac{\pi }{2\left\vert \zeta \right\vert }+\frac{1}{%
\left\vert \zeta \right\vert }\arctan \left( \frac{2ad+b}{2\left\vert \zeta
\right\vert }\right) .%
\end{array}%
\end{equation*}%
} Specifically, one needs to analyze if $\mathcal{I}_{p}\left( t,\bar{t}%
\right) $ of the form: \ 
\begin{equation}
\begin{array}{c}
\mathcal{I}_{p}\left( t,\bar{t}\right) =\int_{-\infty }^{\infty }\varpi
^{\left( x\right) }\left( t,x,y,\bar{t},\bar{x}\right) e^{p\bar{x}}d\bar{x}
\\ 
\\ 
=\frac{e^{px}}{2\pi }\int_{-\infty }^{\infty }\int_{-\infty }^{\infty
}\digamma \left( t,y,\bar{t},k\right) \exp \left( ik\left( x-\bar{x}\right)
-p\left( x-\bar{x}\right) \right) dkd\bar{x} \\ 
\\ 
=e^{px}\int_{-\infty }^{\infty }\digamma \left( t,y,\bar{t},k\right) \delta
\left( ik-p\right) dk=e^{px}\digamma \left( t,y,\bar{t},-ip\right) ,%
\end{array}
\label{Eq7.116}
\end{equation}%
blows up for some finite $\bar{t}>t$. Indeed,%
\begin{equation}
\begin{array}{c}
\digamma \left( t,y,\bar{t},-ip\right) \\ 
\\ 
=\exp \left( \frac{\chi \kappa T}{\varepsilon ^{2}}-\left( \vartheta
+1\right) \ln \left( \frac{-\lambda _{-}\mathsf{E}_{+}+\lambda _{+}\mathsf{E}%
_{-}}{2\zeta }\right) +\frac{2\lambda _{+}\lambda _{-}\left( \mathsf{E}_{+}-%
\mathsf{E}_{-}\right) }{\varepsilon ^{2}\left( -\lambda _{-}\mathsf{E}%
_{+}+\lambda _{+}\mathsf{E}_{-}\right) }y\right) ,%
\end{array}
\label{Eq7.117}
\end{equation}%
where%
\begin{equation}
\begin{array}{c}
\lambda _{\pm }=\mu \pm \zeta , \\ 
\\ 
\mu =-\frac{\kappa }{2},\ \ \ \zeta =\frac{\sqrt{\kappa ^{2}-2\varepsilon
^{2}p}}{2}, \\ 
\\ 
\lambda _{+}\lambda _{-}=\frac{\varepsilon ^{2}p}{2}.%
\end{array}
\label{Eq7.118}
\end{equation}%
Thus, when $\zeta >0$ is real:%
\begin{equation}
\begin{array}{c}
\mathcal{I}_{p}=\left( \frac{2\zeta }{-\mu \sinh \left( \left\vert \zeta
\right\vert T\right) +\zeta \cosh \left( \left\vert \zeta \right\vert
T\right) }\right) ^{\left( \vartheta +1\right) }\exp \left( \frac{\chi
\kappa T}{\varepsilon ^{2}}+px+\frac{p\sinh \left( \zeta T\right) }{\left(
-\mu \sinh \left( \left\vert \zeta \right\vert T\right) +\zeta \cosh \left(
\left\vert \zeta \right\vert T\right) \right) }y\right) ,%
\end{array}
\label{Eq7.119}
\end{equation}%
and, when $\zeta =i\left\vert \zeta \right\vert $ is imaginary: 
\begin{equation}
\begin{array}{c}
\mathcal{I}_{p}=\left( \frac{\left\vert \zeta \right\vert }{-\mu \sin \left(
\left\vert \zeta \right\vert T\right) +\left\vert \zeta \right\vert \cos
\left( \left\vert \zeta \right\vert T\right) }\right) ^{\left( \vartheta
+1\right) }\exp \left( \frac{\chi \kappa T}{\varepsilon ^{2}}+px+\frac{p\sin
\left( \left\vert \zeta \right\vert T\right) }{\left( -\mu \sin \left(
\left\vert \zeta \right\vert T\right) +\left\vert \zeta \right\vert \cos
\left( \left\vert \zeta \right\vert T\right) \right) }y\right) .%
\end{array}
\label{Eq7.120}
\end{equation}%
For $p\in \left[ -\infty ,\hat{p}\right] $, $\zeta $ is real, for $p\in %
\left[ \hat{p},\infty \right] $, it is imaginary. Here%
\begin{equation}
\begin{array}{c}
\hat{p}=\frac{\kappa ^{2}}{2\varepsilon ^{2}}>0.%
\end{array}
\label{Eq7.121}
\end{equation}%
There is no blowup when $\zeta $ is real. When $\zeta $ is imaginary, the
blowup time $t^{\ast }$ is the smallest positive root of the equation%
\begin{equation}
\begin{array}{c}
\kappa \sin \left( \sqrt{2\varepsilon ^{2}p-\kappa ^{2}}\left( t^{\ast
}-t\right) \right) +\sqrt{2\varepsilon ^{2}p-\kappa ^{2}}\cos \left( \sqrt{%
2\varepsilon ^{2}p-\kappa ^{2}}\left( t^{\ast }-t\right) \right) =0,%
\end{array}
\label{Eq7.122}
\end{equation}%
\begin{equation}
\begin{array}{c}
t^{\ast }=t+\frac{\pi -\arctan \left( \frac{\sqrt{2\varepsilon ^{2}p-\kappa
^{2}}}{\kappa }\right) }{\sqrt{2\varepsilon ^{2}p-\kappa ^{2}}}.%
\end{array}
\label{Eq7.123}
\end{equation}%
It is clear that $\mathcal{I}_{-1}$ does not blowup. This fact in used in
the next section.

The marginal distribution of $\bar{y}$, $\varpi ^{\left( y\right) }\left( 
\bar{t},\bar{y}\right) $ is the standard Feller distribution given by Eq. (%
\ref{Eq7.62}).

\subsubsection{Augmented Feller Process, II}

This Section studies the joint dynamics of an arithmetic Brownian $\hat{x}%
_{t}$ whose stochastic variance is driven by a Feller process $\hat{y}_{t}$,
and considers the following system of affine SDEs:%
\begin{equation}
\begin{array}{c}
d\hat{x}_{t}=\sqrt{\hat{y}_{t}}d\hat{W}_{t},\ \ \ \hat{x}_{t}=x, \\ 
\\ 
d\hat{y}_{t}=\left( \chi -\kappa \hat{y}_{t}\right) dt+\varepsilon \sqrt{%
\hat{y}_{t}}d\hat{Z}_{t},\ \ \ \hat{y}_{t}=y.%
\end{array}
\label{Eq7.124}
\end{equation}%
Studying such a process is very helpful for finding option prices and
solving other important problems in the financial engineering context.

The associated forward Fokker-Planck problem can be written as follows:%
\begin{equation}
\begin{array}{c}
\varpi _{\bar{t}}-\frac{1}{2}\bar{y}\varpi _{\bar{x}\bar{x}}-\rho
\varepsilon \left( \bar{y}\varpi \right) _{\bar{x}\bar{y}}-\frac{1}{2}%
\varepsilon ^{2}\left( \bar{y}\varpi \right) _{\bar{y}\bar{y}}+\left( \left(
\chi -\kappa \bar{y}\right) \varpi \right) _{\bar{y}}=0, \\ 
\\ 
\varpi \left( t,x,y,t,\bar{x},\bar{y}\right) =\delta \left( \bar{x}-x\right)
\delta \left( \bar{y}-y\right) ,%
\end{array}
\label{Eq7.125}
\end{equation}%
while the backward Kolmogorov problem has the form:%
\begin{equation}
\begin{array}{c}
\varpi _{t}+\frac{1}{2}y\varpi _{xx}+\rho \varepsilon y\varpi _{xy}+\frac{1}{%
2}\varepsilon ^{2}y\varpi _{yy}+\left( \chi -\kappa y\right) \varpi _{y}=0,
\\ 
\\ 
\varpi \left( \bar{t},x,y,\bar{t},\bar{x},\bar{y}\right) =\delta \left( x-%
\bar{x}\right) \delta \left( y-\bar{y}\right) .%
\end{array}
\label{Eq7.126}
\end{equation}%
As before, concentrate on problem (\ref{Eq7.126}).

The Kelvin function $\mathcal{K}$ has the form (\ref{Eq7.90}). The governing
ODEs for $\alpha ,\gamma $ are as follows:%
\begin{equation}
\begin{array}{c}
\alpha _{t}\left( t,\bar{t}\right) +i\chi \gamma \left( t,\bar{t}\right)
=0,\ \ \ \alpha \left( \bar{t},\bar{t}\right) =0, \\ 
\\ 
i\gamma _{t}\left( t,\bar{t}\right) -\frac{1}{2}\varepsilon ^{2}\gamma
^{2}\left( t,\bar{t}\right) -\left( \kappa -i\rho \varepsilon k\right)
i\gamma \left( t,\bar{t}\right) -\frac{1}{2}k^{2}=0,\ \ \ \gamma \left( \bar{%
t},\bar{t}\right) =l.%
\end{array}
\label{Eq7.127}
\end{equation}%
Formulas (\ref{Eq7.99})-(\ref{Eq7.102}) hold; however, the corresponding
characteristic equation is 
\begin{equation}
\begin{array}{c}
\lambda ^{2}+\left( \kappa -i\rho \varepsilon k\right) \lambda -\frac{%
\varepsilon ^{2}}{4}k^{2}=0,%
\end{array}
\label{Eq7.129}
\end{equation}%
so that 
\begin{equation}
\begin{array}{c}
\lambda _{\pm }=\mu \pm \zeta , \\ 
\\ 
\mu =-\frac{1}{2}\left( \kappa -i\rho \varepsilon k\right) ,\ \ \ \zeta =%
\frac{1}{2}\sqrt{\varepsilon ^{2}\bar{\rho}^{2}k^{2}-2i\rho \varepsilon
\kappa k+\kappa ^{2}}, \\ 
\\ 
\lambda _{+}\lambda _{-}=\mu ^{2}-\zeta ^{2}=-\frac{\varepsilon ^{2}}{4}%
k^{2}.%
\end{array}
\label{Eq7.130}
\end{equation}%
where $\bar{\rho}^{2}=1-\rho ^{2}$. Subsequent calculations are very similar
to the ones performed in the previous subsection, so they are omitted for
brevity. The final expressions for $\varpi $ and $\varpi ^{\left( x\right) }$
are given by Eqs (\ref{Eq7.110}), (\ref{Eq7.111}), (\ref{Eq7.112}), with $%
\mu ,\zeta $ given by Eqs (\ref{Eq7.130}). These expressions are similar to
the formulas originally are derived by Lipton as part of his analysis of the
Heston stochastic volatility model; see \cite{Lipton01}.\footnote{%
Despite the fact that these formulas were originally derived by \cite%
{Lipton01}, they are frequently mistakenly attributed to \cite{Dragulescu02}.%
}

\begin{figure}[tbp]
\begin{center}
\includegraphics[width=0.8\textwidth]
{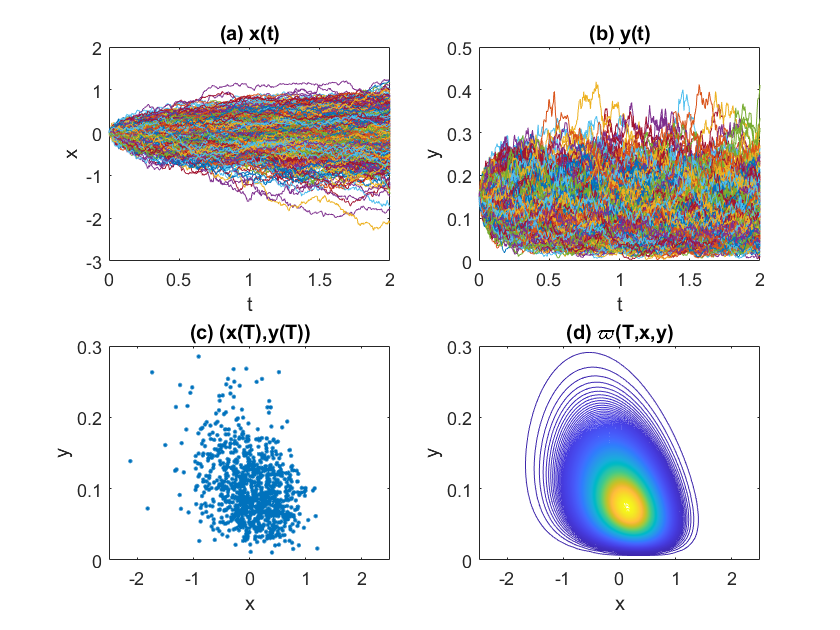}
\end{center}
\par
\vspace{-10pt}
\caption{A thousand trajectories of a representative nondegenerate augmented
Feller process. Parameters are $T=3$, $dt=0.01$, $\protect\chi =0.2$, $%
\protect\kappa =2.0$, $\protect\varepsilon =0.2$, $\protect\rho =-0.5$, $x=0$%
; $y_{0}=0.15$. (a) $x\left( t\right) $, (b) $y\left( t\right) $, (c) $%
\left( \bar{x}\left( T\right) ,\bar{y}\left( T\right) \right) $, (d) contour
lines of $\protect\varpi \left( 0,0.15,0,T,\tilde{x},\tilde{y}\right) $. Own
graphics.}
\label{Fig7.4}
\end{figure}

A typical t.p.d.f. for a non-degenerate augmented Feller process is shown in
Figure \ref{Fig7.4}. 

As before, $\varpi ^{\left( x\right) }\left( t,x,y,\bar{t},\bar{x}\right) $
has fat tails. Consider $\mathcal{I}_{p}\left( t,\bar{t}\right) $ given by
Eq. (\ref{Eq7.116}). The corresponding $\lambda _{\pm }$ have the form:

\begin{equation}
\begin{array}{c}
\lambda _{\pm }=\mu \pm \zeta , \\ 
\\ 
\mu =-\frac{1}{2}\left( \kappa -\rho \varepsilon p\right) ,\ \ \ \zeta =%
\frac{1}{2}\sqrt{-\varepsilon ^{2}\bar{\rho}^{2}p^{2}-2\rho \varepsilon
\kappa p+\kappa ^{2}}, \\ 
\\ 
\lambda _{+}\lambda _{-}=\mu ^{2}-\zeta ^{2}=\frac{\varepsilon ^{2}}{4}p^{2}.%
\end{array}
\label{Eq7.131}
\end{equation}%
Thus, when $\zeta >0$ is real, 
\begin{equation}
\begin{array}{c}
\mathcal{I}_{p}=\left( \frac{\zeta }{-\mu \sinh \left( \zeta T\right) +\zeta
\cosh \left( \zeta T\right) }\right) ^{\left( \vartheta +1\right) }\exp
\left( \frac{2\chi \mu T}{\varepsilon ^{2}}+px+\frac{p\left( p-1\right)
\sinh \left( \zeta T\right) }{2\left( -\mu \sinh \left( \zeta T\right)
+\zeta \cosh \left( \zeta T\right) \right) }y\right) ,%
\end{array}
\label{Eq7.132}
\end{equation}%
and, when $\zeta =i\left\vert \zeta \right\vert $ is imaginary,%
\begin{equation}
\begin{array}{c}
\mathcal{I}_{p}=\left( \frac{\left\vert \zeta \right\vert }{-\mu \sin \left(
\left\vert \zeta \right\vert T\right) +\left\vert \zeta \right\vert \cos
\left( \left\vert \zeta \right\vert T\right) }\right) ^{\left( \vartheta
+1\right) }\exp \left( \frac{2\chi \mu T}{\varepsilon ^{2}}+px+\frac{p\left(
p-1\right) \sin \left( \left\vert \zeta \right\vert T\right) }{2\left( -\mu
\sin \left( \left\vert \zeta \right\vert T\right) +\left\vert \zeta
\right\vert \cos \left( \left\vert \zeta \right\vert T\right) \right) }%
y\right) .%
\end{array}
\label{Eq7.133}
\end{equation}%
One needs to determine when $\zeta $ becomes imaginary. The corresponding
quadratic equation has the form:%
\begin{equation}
\begin{array}{c}
\bar{\rho}^{2}\varepsilon ^{2}p^{2}+2\rho \varepsilon \kappa p-\kappa ^{2}=0,%
\end{array}
\label{Eq7.134}
\end{equation}%
its roots are as follows:%
\begin{equation}
\begin{array}{c}
p_{\pm }=\frac{-\rho \varepsilon \kappa \pm \sqrt{\rho ^{2}\varepsilon
^{2}\kappa ^{2}+\bar{\rho}^{2}\varepsilon ^{2}\kappa ^{2}}}{\bar{\rho}%
^{2}\varepsilon ^{2}}=\frac{\left( -\rho \pm 1\right) \kappa }{\bar{\rho}%
^{2}\varepsilon },%
\end{array}
\label{Eq7.135}
\end{equation}%
so that%
\begin{equation}
\begin{array}{c}
p_{+}>1,\ \ \ p_{-}<1.%
\end{array}
\label{Eq7.137}
\end{equation}%
For $p\in \left[ p_{-},p_{+}\right] $, $\zeta $ is real, for $p\notin \left[
p_{-},p_{+}\right] $, it is imaginary. There is no blowup when $\zeta $ is
real. When $\zeta $ is imaginary, the blowup time $t^{\ast }$ is the
smallest positive root of the equation%
\begin{equation}
\begin{array}{c}
-\mu \sin \left( \left\vert \zeta \right\vert \left( t^{\ast }-t\right)
\right) +\left\vert \zeta \right\vert \cos \left( \left\vert \zeta
\right\vert \left( t^{\ast }-t\right) \right) =0,%
\end{array}
\label{Eq7.138}
\end{equation}%
\begin{equation}
\begin{array}{c}
t^{\ast }=\left\{ 
\begin{array}{cc}
t+\frac{\arctan \left( \frac{\left\vert \zeta \right\vert }{\left\vert \mu
\right\vert }\right) }{\left\vert \zeta \right\vert }, & \mu >0, \\ 
t+\frac{\pi -\arctan \left( \frac{\left\vert \zeta \right\vert }{\left\vert
\mu \right\vert }\right) }{\left\vert \zeta \right\vert }, & \mu <0.%
\end{array}%
\right.%
\end{array}
\label{Eq7.139}
\end{equation}

\subsection{Example: Path-dependent Process\label{Sec75}}

Let $\hat{y}_{t}$ be a stochastic process and $\hat{x}_{t}$ be its moving
average. Then%
\begin{equation}
\begin{array}{c}
\hat{x}_{t}=\kappa \int_{-\infty }^{t}e^{-\kappa \left( t-s\right) }\hat{y}%
_{s}ds.%
\end{array}
\label{Eq7.140}
\end{equation}%
A simple calculation yields:%
\begin{equation}
\begin{array}{c}
d\hat{x}_{t}=\kappa \left( \hat{y}_{t}-\hat{x}_{t}\right) dt.%
\end{array}
\label{Eq7.141}
\end{equation}%
The process $\hat{y}_{t}$ is path-dependent, because its volatility $\hat{%
\sigma}_{t}$ depends on its moving average $\hat{x}_{t}$:%
\begin{equation}
\begin{array}{c}
\hat{\sigma}_{t}=\sqrt{a_{0}+a_{1}\left( \hat{y}_{t}-\hat{x}_{t}\right) },%
\end{array}
\label{Eq7.142}
\end{equation}%
where $a_{0}>0$, $a_{1}<0$, in order to capture the effect of leverage.
Thus, one can write the governing degenerate system of SDEs as follows: 
\begin{equation}
\begin{array}{c}
d\hat{x}_{t}=\kappa \left( \hat{y}_{t}-\hat{x}_{t}\right) dt,\ \ \ \hat{x}%
_{t}=x, \\ 
\\ 
d\hat{y}_{t}=\sqrt{a_{0}+a_{1}\left( \hat{y}_{t}-\hat{x}_{t}\right) }d\hat{W}%
_{t},\ \ \ \hat{y}_{t}=y.%
\end{array}
\label{Eq7.143}
\end{equation}
The Fokker-Planck and Kolmogorov problems are:%
\begin{equation}
\begin{array}{c}
\varpi _{\bar{t}}-\frac{1}{2}\left( \left( a_{0}+a_{1}\left( \bar{y}-\bar{x}%
\right) \right) \varpi \right) _{\bar{y}\bar{y}}+\left( \kappa \left( \bar{y}%
-\bar{x}\right) \varpi \right) _{\bar{x}}=0, \\ 
\\ 
\varpi \left( t,x,y,t,\bar{x},\bar{y}\right) =\delta \left( \bar{x}-x\right)
\delta \left( \bar{y}-y\right) ,%
\end{array}
\label{Eq7.144}
\end{equation}%
\begin{equation}
\begin{array}{c}
\varpi _{t}+\frac{1}{2}\left( a_{0}+a_{1}\left( y-x\right) \right) \varpi
_{yy}+\kappa \left( y-x\right) \varpi _{x}=0, \\ 
\\ 
\varpi \left( \bar{t},x,y,\bar{t},\bar{x},\bar{y}\right) =\delta \left( x-%
\bar{x}\right) \delta \left( y-\bar{y}\right) ,%
\end{array}
\label{Eq7.145}
\end{equation}%
respectively.

A representative Kelvin mode has the form:%
\begin{equation}
\begin{array}{c}
\mathcal{K}=\exp \left( \alpha \left( t,\bar{t}\right) +i\beta \left( t,\bar{%
t}\right) x-ik\bar{x}+i\gamma \left( t,\bar{t}\right) y-il\bar{y}\right) .%
\end{array}
\label{Eq7.146}
\end{equation}%
The system of backward ODEs for $\alpha ,\gamma ,\beta $ is as follows:%
\begin{equation}
\begin{array}{c}
\alpha _{t}\left( t,\bar{t}\right) -\frac{a_{0}}{2}\gamma ^{2}\left( t,\bar{t%
}\right) =0,\ \ \ \alpha \left( \bar{t},\bar{t}\right) =0, \\ 
\\ 
i\beta _{t}\left( t,\bar{t}\right) +\frac{a_{1}}{2}\gamma ^{2}\left( t,\bar{t%
}\right) -i\kappa \beta \left( t,\bar{t}\right) =0,\ \ \ \beta \left( \bar{t}%
,\bar{t}\right) =k, \\ 
\\ 
i\gamma _{t}\left( t,\bar{t}\right) -\frac{a_{1}}{2}\gamma ^{2}\left( t,\bar{%
t}\right) +i\kappa \beta \left( t,\bar{t}\right) =0,\ \ \ \gamma \left( \bar{%
t},\bar{t}\right) =l.%
\end{array}
\label{Eq7.147}
\end{equation}%
Eqs (\ref{Eq7.147}) are matrix Riccati equations, as opposite to scalar
Riccati equations considered above. In general, such equations are very
difficult to solve. However, the case under consideration is one of
relatively rare instances when a matrix Riccati equation can be solved
explicitly. Start with an observation:%
\begin{equation}
\begin{array}{c}
\gamma _{t}\left( t,\bar{t}\right) +\beta _{t}\left( t,\bar{t}\right) =0,%
\end{array}
\label{Eq7.148}
\end{equation}%
so that%
\begin{equation}
\begin{array}{c}
\beta \left( t,\bar{t}\right) =-\gamma \left( t,\bar{t}\right) +k+l.%
\end{array}
\label{Eq7.149}
\end{equation}%
Accordingly,%
\begin{equation}
\begin{array}{c}
i\gamma _{t}\left( t,\bar{t}\right) -\frac{a_{1}}{2}\gamma ^{2}\left( t,\bar{%
t}\right) -i\kappa \gamma \left( t,\bar{t}\right) +i\kappa \left( k+l\right)
=0,\ \ \ \gamma \left( \bar{t},\bar{t}\right) =l.%
\end{array}
\label{Eq7.150}
\end{equation}%
One can use Eqs (\ref{Eq7.99})-(\ref{Eq7.101}) with $\left( \beta ,k\right) $
replaced by $\left( \gamma ,l\right) $, and 
\begin{equation}
\begin{array}{c}
\lambda ^{2}+\kappa \lambda +\frac{ia_{1}\kappa \left( k+l\right) }{2}=0,%
\end{array}
\label{Eq7.151}
\end{equation}%
so that%
\begin{equation}
\begin{array}{c}
\lambda _{\pm }=\mu \pm \zeta , \\ 
\\ 
\mu =-\frac{\kappa }{2},\ \ \ \zeta =\frac{\sqrt{\kappa ^{2}-2ia_{1}\kappa
\left( k+l\right) }}{2}.%
\end{array}
\label{Eq7.152}
\end{equation}%
Eq. (\ref{Eq7.149}) yields:%
\begin{equation}
\begin{array}{c}
\beta \left( t,\bar{t}\right) =\frac{2i\Omega ^{\prime }\left( t,\bar{t}%
\right) }{a_{1}\Omega \left( t,\bar{t}\right) }+k+l,%
\end{array}
\label{Eq7.156}
\end{equation}%
and%
\begin{equation}
\begin{array}{c}
\gamma ^{2}\left( t,\bar{t}\right) =\frac{2i}{a_{1}}\left( \gamma _{t}\left(
t,\bar{t}\right) +\kappa \beta \left( t,\bar{t}\right) \right) \\ 
\\ 
=\frac{2i}{a_{1}}\left( \gamma _{t}\left( t,\bar{t}\right) -\kappa \gamma
\left( t,\bar{t}\right) +\kappa \left( \kappa +k\right) \right) .%
\end{array}
\label{Eq7.157}
\end{equation}%
Thus,%
\begin{equation}
\begin{array}{c}
\alpha _{t}\left( t,\bar{t}\right) =\frac{ia_{0}}{a_{1}}\left( \gamma
_{t}\left( t,\bar{t}\right) -\kappa \gamma \left( t,\bar{t}\right) +\kappa
\left( k+l\right) \right) ,\ \ \ \alpha \left( \bar{t},\bar{t}\right) =0.%
\end{array}
\label{Eq7.158}
\end{equation}%
Accordingly,%
\begin{equation}
\begin{array}{c}
\alpha \left( t,\bar{t}\right) =\frac{ia_{0}}{a_{1}}\left( \gamma \left( t,%
\bar{t}\right) -l\right) -\frac{2a_{0}\kappa }{a_{1}^{2}}\ln \left( \Omega
\left( t,\bar{t}\right) \right) +\frac{a_{0}\kappa T}{a_{1}}i\left(
k+l\right) .%
\end{array}
\label{Eq7.159}
\end{equation}%
Finally,%
\begin{equation}
\begin{array}{c}
\mathcal{K}=\exp \left( \alpha +i\beta \left( t,\bar{t}\right) x-ik\bar{x}%
+i\gamma \left( t,\bar{t}\right) y-il\bar{y}\right) \\ 
\\ 
=\exp \left( -\frac{2a_{0}\kappa }{a_{1}^{2}}\ln \left( \Omega \left( t,\bar{%
t}\right) \right) +i\gamma \left( t,\bar{t}\right) \left( y-x+\frac{a_{0}}{%
a_{1}}\right) \right. \\ 
\\ 
\left. +il\left( x-\bar{y}+\frac{a_{0}}{a_{1}}\left( \kappa T-1\right)
\right) +ik\left( x-\bar{x}+\frac{a_{0}\kappa T}{a_{1}}\right) \right) .%
\end{array}
\label{Eq7.160}
\end{equation}%
To make sure that $\hat{\sigma}_{t}$ given by Eq. (\ref{Eq7.142}) and the
integrand (\ref{Eq7.160}) are well defined, it is assumed that%
\begin{equation}
\begin{array}{c}
a_{0}+a_{1}\left( y-x\right) >0,\ \ a_{0}+a_{1}\left( \ \bar{y}-\bar{x}%
\right) >0.%
\end{array}
\label{Eq7.161}
\end{equation}

\subsection{Example: OU-Like Process\label{Sec76}}

This Section considers several instances when an OU-inspired process becomes
non-Gaussian. This can happen for a variety of reasons, such as effects of
anomalous diffusion, the presence of jumps, effects of augmentation, and the
likes.

\subsubsection{Anomalous OU Process}

This section considers a mean-reverting process driven by a non-Gaussian
anomalous diffusion. For brevity, it is assumed that coefficient are
time-independent. The fractional forward Fokker-Planck and backward
Kolmogorov problems can be written as follows:%
\begin{equation}
\begin{array}{c}
\varpi _{\bar{t}}+a\left( -\frac{\partial ^{2}}{\partial \bar{y}^{2}}\right)
^{1/2}\varpi +\left( \left( \chi -\kappa \bar{y}\right) \varpi \right) _{%
\bar{y}}=0, \\ 
\\ 
\varpi \left( t,y,t,\bar{y}\right) =\delta \left( \bar{y}-y\right) ,%
\end{array}
\label{Eq7.162}
\end{equation}%
\begin{equation}
\begin{array}{c}
\varpi _{t}-a\left( -\frac{\partial ^{2}}{\partial y^{2}}\right)
^{1/2}\varpi +\left( \chi -\kappa y\right) \varpi _{y}=0, \\ 
\\ 
\varpi \left( \bar{t},y,\bar{t},\bar{y}\right) =\delta \left( y-\bar{y}%
\right) ,%
\end{array}
\label{Eq7.163}
\end{equation}%
respectively. Here $a>0$ is the anomalous diffusion coefficient.

As before, one can use Kelvin waves to solve Eq. (\ref{Eq7.163}) by choosing
a particular solution of the form (\ref{Eq7.47}). The corresponding $\left(
\alpha ,\gamma \right) $ satisfy the following ODEs:%
\begin{equation}
\begin{array}{c}
\alpha _{t}\left( t,\bar{t}\right) -a\left\vert \gamma \left( t,\bar{t}%
\right) \right\vert +i\chi \gamma \left( t,\bar{t}\right) =0,\ \ \ \alpha
\left( \bar{t},\bar{t}\right) =0, \\ 
\\ 
\gamma _{t}\left( t,\bar{t}\right) -\kappa \gamma \left( t,\bar{t}\right)
=0,\ \ \ \gamma \left( \bar{t},\bar{t}\right) =l,%
\end{array}
\label{Eq7.164}
\end{equation}%
so that%
\begin{equation}
\begin{array}{c}
\alpha \left( t,\bar{t}\right) =-\mathsf{\bar{B}}_{\kappa }\left( T\right)
\left( a\left\vert l\right\vert -i\chi l\right) , \\ 
\\ 
\gamma \left( t,\bar{t}\right) =e^{-\kappa T}l.%
\end{array}
\label{Eq7.165}
\end{equation}%
Accordingly,%
\begin{equation}
\begin{array}{c}
\varpi \left( t,y,\bar{t},\bar{y}\right) =\frac{1}{2\pi }\int\limits_{-%
\infty }^{\infty }\exp \left( -\mathsf{\bar{B}}_{\kappa }\left( T\right)
a\left\vert l\right\vert +\left( \mathsf{\bar{B}}_{\kappa }\left( T\right)
\chi +e^{-\kappa T}y-\bar{y}\right) il\right) dl \\ 
\\ 
=\frac{1}{\pi }\frac{\mathsf{\bar{B}}_{\kappa }\left( T\right) a}{\left(
\left( \mathsf{\bar{B}}_{\kappa }\left( T\right) a\right) ^{2}+\left(
e^{-\kappa T}\left( y-\frac{\chi }{\kappa }\right) -\left( \bar{y}-\frac{%
\chi }{\kappa }\right) \right) ^{2}\right) }.%
\end{array}
\label{Eq7.166}
\end{equation}%
Thus, in sharp contrast to the classical OU process, which is described by a
Gaussian distribution, the fractional OU process is described by a Cauchy
distribution. This distribution has fat tails and no first and second
moments.

\subsubsection{Non-Gaussian Augmented OU Process, I}

On occasion, problems seemingly not of the type given by Eq. (\ref{Eq7.1})
can be cast in the proper form via a suitable trick. Consider, for example,
the following system of SDEs:%
\begin{equation}
\begin{array}{c}
d\hat{x}_{t}=\hat{y}_{t}^{2}dt,\ \ \ \hat{x}_{t}=x, \\ 
\\ 
d\hat{y}_{t}=\left( \chi -\kappa \hat{y}_{t}\right) dt+\varepsilon d\hat{Z}%
_{t},\ \ \ \hat{y}_{t}=y.%
\end{array}
\label{Eq7.167}
\end{equation}%
Superficially, it does not belong to the class of processes studied above.
However, by introducing new variables $z_{1}=x$, $z_{2}=y^{2}$, $z_{3}=y$,
one can augment Eqs (\ref{Eq7.115}) as follows:%
\begin{equation}
\begin{array}{c}
d\hat{z}_{1,t}=\hat{z}_{2,t}dt,\ \ \ \hat{z}_{1,t}=x\equiv z_{1}, \\ 
\\ 
d\hat{z}_{2,t}=\left( \varepsilon ^{2}-2\kappa \hat{z}_{2,t}+2\chi \hat{z}%
_{3,t}\right) +2\varepsilon \hat{z}_{3,t}d\hat{Z}_{t},\ \ \ \hat{z}%
_{2,t}=y^{2}\equiv z_{2}, \\ 
\\ 
d\hat{z}_{3,t}=\left( \chi -\kappa \hat{z}_{3,t}\right) dt+\varepsilon d\hat{%
Z}_{t},\ \ \ \hat{z}_{3,t}=z_{3}\equiv y.%
\end{array}
\label{Eq7.168}
\end{equation}%
These equations are \textquotedblleft almost\textquotedblright\ in the
suitable form. The only snag is that one cannot claim that $\hat{z}_{3,t}=%
\sqrt{\hat{z}_{2,t}}$ since $\hat{z}_{3,t}$ is not always positive.

The corresponding Fokker-Planck and Kolmogorov problems can be written as
follows:%
\begin{equation}
\begin{array}{c}
\varpi _{\bar{t}}-2\varepsilon ^{2}\left( \bar{z}_{2}\varpi \right) _{\bar{z}%
_{2}\bar{z}_{2}}-2\varepsilon ^{2}\left( \bar{z}_{3}\varpi \right) _{\bar{z}%
_{2}\bar{z}_{3}}-\frac{1}{2}\varepsilon ^{2}\varpi _{\bar{z}_{3}\bar{z}_{3}}
\\ 
\\ 
+\bar{z}_{2}\varpi _{\bar{z}_{1}}+\left( \left( \varepsilon ^{2}-2\kappa 
\bar{z}_{2}+2\chi \bar{z}_{3}\right) \varpi \right) _{\bar{z}_{2}}+\left(
\left( \chi -\kappa \bar{z}_{3}\right) \varpi \right) _{\bar{z}_{3}}=0, \\ 
\\ 
\varpi \left( t,x,y^{2},y,t,\bar{z}_{1},\bar{z}_{2},\bar{z}_{3}\right)
=\delta \left( \bar{z}_{1}-x\right) \delta \left( \bar{z}_{2}-y^{2}\right)
\delta \left( \bar{z}_{3}-y\right) ,%
\end{array}
\label{Eq7.169}
\end{equation}%
\begin{equation}
\begin{array}{c}
\varpi _{t}+2\varepsilon ^{2}z_{2}\varpi _{z_{2}z_{2}}+2\varepsilon
^{2}z_{3}\varpi _{z_{2}z_{3}}+\frac{1}{2}\varepsilon ^{2}\varpi _{z_{3}z_{3}}
\\ 
\\ 
+z_{2}\varpi _{z_{1}}+\left( \varepsilon ^{2}-2\kappa z_{2}+2\chi
z_{3}\right) \varpi _{z_{2}}+\left( \chi -\kappa z_{3}\right) \varpi
_{z_{3}}=0, \\ 
\\ 
\varpi \left( \bar{t},z_{1},z_{2},z_{3},\bar{t},\bar{z}_{1},\bar{z}_{3}^{2},%
\bar{z}_{3}\right) =\delta \left( z_{1}-\bar{z}_{1}\right) \delta \left(
z_{2}-\bar{z}_{3}^{2}\right) \delta \left( z_{3}-\bar{z}_{3}\right) .%
\end{array}
\label{Eq7.170}
\end{equation}%
As usual, $\mathcal{K}$ has the form:%
\begin{equation}
\begin{array}{c}
\mathcal{K}\left( t,\bar{t},\mathbf{z},\mathbf{m}\right) \\ 
\\ 
=\exp \left( \alpha \left( t,\bar{t}\right) +im_{1}\left( z_{1}-\bar{z}%
_{1}\right) +i\delta _{2}\left( t,\bar{t}\right) z_{2}-im_{2}\bar{z}%
_{3}^{2}+i\delta _{3}\left( t,\bar{t}\right) z_{3}-im_{3}\bar{z}_{3}\right) .%
\end{array}
\label{Eq7.171}
\end{equation}%
The corresponding set of ODEs for $\alpha ,\delta _{2},\delta _{3}$ is as
follows:%
\begin{equation}
\begin{array}{c}
\alpha _{t}\left( t,\bar{t}\right) -\frac{\varepsilon ^{2}}{2}\delta
_{3}^{2}\left( t,\bar{t}\right) +i\varepsilon ^{2}\delta _{2}\left( t,\bar{t}%
\right) +i\chi \delta _{3}\left( t,\bar{t}\right) =0,\ \ \ \alpha \left( 
\bar{t},\bar{t}\right) =0, \\ 
\\ 
i\delta _{2}^{\prime }\left( t,\bar{t}\right) -2\varepsilon ^{2}\delta
_{2}^{2}\left( t,\bar{t}\right) -2i\kappa \delta _{2}\left( t,\bar{t}\right)
+im_{1}=0,\ \ \ \delta _{2}\left( \bar{t},\bar{t}\right) =m_{2}, \\ 
\\ 
i\delta _{3}^{\prime }\left( t,\bar{t}\right) -2\varepsilon ^{2}\delta
_{2}\left( t,\bar{t}\right) \delta _{3}\left( t,\bar{t}\right) +2i\chi
\delta _{2}v-i\kappa \delta _{3}\left( t,\bar{t}\right) =0,\ \ \ \delta
_{3}\left( \bar{t},\bar{t}\right) =m_{3}.%
\end{array}
\label{Eq7.172}
\end{equation}%
These are matrix Riccati equations.

Once again, the corresponding matrix Riccati equation can be solved
explicitly. Since the second equation is separable, and hence can be viewed
as a scalar Riccati equation, one can start with a familiar ansatz, and use
Eqs (\ref{Eq7.99})-(\ref{Eq7.101}) with $\left( \gamma ,l\right) $ replaced
by $\left( \delta _{2},m_{2}\right) $, and The corresponding characteristic
equation and its solutions are as follows:%
\begin{equation}
\begin{array}{c}
\lambda ^{2}+2\kappa \lambda +2i\varepsilon ^{2}m_{1}=0,%
\end{array}
\label{Eq7.173}
\end{equation}%
its solutions have the familiar form:%
\begin{equation}
\begin{array}{c}
\lambda _{\pm }=\mu \pm \zeta , \\ 
\\ 
\mu =-\kappa ,\ \ \ \zeta =\sqrt{\kappa ^{2}-2i\varepsilon ^{2}m_{1}}.%
\end{array}
\label{Eq7.174}
\end{equation}%
To linearize Eqs. (\ref{Eq7.172}) as a whole, use the following ansatz:%
\begin{equation}
\begin{array}{c}
\Omega =\mathsf{E}_{0}\left( \omega _{+}\mathsf{E}_{+}+\omega _{-}\mathsf{E}%
_{-}\right) ,\ \ \alpha =-\frac{1}{2}\ln \left( \Omega \right) +\frac{%
\mathsf{E}_{0}\left( a_{0}+a_{+}\mathsf{E}_{+}+a_{-}\mathsf{E}_{-}\right) }{%
\Omega }+g\left( \bar{t}-t\right) ,\  \\ 
\\ 
\delta _{2}=\frac{\mathsf{E}_{0}\left( b_{+}\mathsf{E}_{+}+b_{-}\mathsf{E}%
_{-}\right) }{\Omega },\ \ \ \delta _{3}=\frac{\mathsf{E}_{0}\left(
c_{0}+c_{+}\mathsf{E}_{+}+c_{-}\mathsf{E}_{-}\right) }{\Omega },%
\end{array}
\label{Eq7.178}
\end{equation}%
Where $a_{0}$, $a_{\pm }$, $b_{0}$, $b_{\pm }$, and $g$ are constants to be
determined. This ansatz is useful since terms proportional to $\sim \mathsf{E%
}_{0},\mathsf{E}_{+},\mathsf{E}_{-}$ balance each other, which allows us to
find the coefficients explicitly. Initial conditions complete the picture.
The actual calculation is omitted for brevity. The result is as follows:%
\begin{equation}
\begin{array}{c}
\omega _{\pm }=\mp \frac{\left( \lambda _{\mp }+2i\varepsilon
^{2}m_{2}\right) }{2\zeta },\ \ \ b_{\pm }=\frac{i\lambda _{\pm }\omega
_{\pm }}{2\varepsilon ^{2}}, \\ 
\\ 
c_{\pm }=\pm \frac{i\chi \lambda _{\pm }\omega _{\pm }}{\varepsilon
^{2}\zeta },\ \ \ c_{0}=m_{3}-c_{+}-c_{-},\ \ \ \ g=\frac{\chi ^{2}\lambda
_{+}\lambda _{-}}{2\varepsilon ^{2}\zeta ^{2}},\ \ \  \\ 
\\ 
a_{0}=-\frac{i\kappa \chi c_{0}}{\zeta ^{2}},\ \ \ a_{\pm }=-\omega _{\pm
}a_{0}\mp \left( \frac{\varepsilon ^{2}c_{0}^{2}}{4\zeta }+\frac{\chi
^{2}\kappa ^{2}\omega _{+}\omega _{-}}{\varepsilon ^{2}\zeta ^{3}}\right) ,%
\end{array}
\label{Eq7.180}
\end{equation}%
where $\lambda _{\pm }$ are given by Eqs (\ref{Eq7.174}). These expressions
can be substituted in the function $\mathcal{K}$ to obtain the corresponding
t.p.d.f.

\subsubsection{Non-Gaussian Augmented OU Process, II}

This Section studies an affine process of the form:%
\begin{equation}
\begin{array}{c}
d\hat{x}_{t}=\hat{y}_{t}d\hat{W}_{t},\ \ \ \hat{x}_{t}=x, \\ 
\\ 
d\hat{y}_{t}=\left( \chi -\kappa \hat{y}_{t}\right) dt+\varepsilon d\hat{Z}%
_{t},\ \ \ \hat{y}_{t}=y,%
\end{array}
\label{Eq7.181}
\end{equation}%
The killed process is studied in Chapter \ref{Chap8} in the context of the
Stein-Stein model.

Precisely as before, one can introduce a new variables $z_{1}=x$, $%
z_{2}=y^{2}$, $z_{3}=y$, and expand Eqs (\ref{Eq7.181}) as follows:%
\begin{equation}
\begin{array}{c}
d\hat{z}_{1,t}=\hat{z}_{3,t}d\hat{W}_{t},\ \ \ \hat{z}_{1,t}=x\equiv z_{1},
\\ 
\\ 
d\hat{z}_{2,t}=\left( \varepsilon ^{2}-2\kappa \hat{z}_{2,t}+2\chi \hat{z}%
_{3,t}\right) +2\varepsilon \hat{z}_{3,t}d\hat{Z}_{t},\ \ \ \hat{z}%
_{2,t}=y^{2}\equiv z_{2}, \\ 
\\ 
d\hat{z}_{3,t}=\left( \chi -\kappa \hat{z}_{3,t}\right) dt+\varepsilon d\hat{%
Z}_{t},\ \ \ \hat{z}_{3,t}=z_{3}\equiv y.%
\end{array}
\label{Eq7.182}
\end{equation}%
It is clear that Eqs (\ref{Eq7.182}) are affine.

The corresponding Fokker-Planck and Kolmogorov problems can be written as
follows:%
\begin{equation}
\begin{array}{c}
\varpi _{\bar{t}}-\frac{1}{2}\bar{z}_{2}\varpi _{\bar{z}_{1}\bar{z}%
_{1}}-2\rho \varepsilon \left( \bar{z}_{2}\varpi \right) _{\bar{z}_{1}\bar{z}%
_{2}}-\rho \varepsilon \left( \bar{z}_{3}\varpi \right) _{\bar{z}_{1}\bar{z}%
_{3}} \\ 
\\ 
-2\varepsilon ^{2}\left( \bar{z}_{2}\varpi \right) _{\bar{z}_{2}\bar{z}%
_{2}}-2\varepsilon ^{2}\left( \bar{z}_{3}\varpi \right) _{\bar{z}_{2}\bar{z}%
_{3}}-\frac{1}{2}\varepsilon ^{2}\varpi _{\bar{z}_{3}\bar{z}_{3}} \\ 
\\ 
+\left( \left( \varepsilon ^{2}-2\kappa \bar{z}_{2}+2\chi \bar{z}_{3}\right)
\varpi \right) _{\bar{z}_{2}}+\left( \left( \chi -\kappa \bar{z}_{3}\right)
\varpi \right) _{\bar{z}_{3}}=0, \\ 
\\ 
\varpi \left( t,x,y^{2},y,t,\bar{z}_{1},\bar{z}_{2},\bar{z}_{3}\right)
=\delta \left( \bar{z}_{1}-x\right) \delta \left( \bar{z}_{2}-y^{2}\right)
\delta \left( \bar{z}_{3}-y\right) ,%
\end{array}
\label{Eq7.183}
\end{equation}%
\begin{equation}
\begin{array}{c}
\varpi _{t}+\frac{1}{2}z_{2}\varpi _{z_{1}z_{1}}+2\rho \varepsilon
z_{2}\varpi _{z_{1}z_{2}}+\rho \varepsilon z_{3}\varpi _{z_{1}z_{3}} \\ 
\\ 
+2\varepsilon ^{2}z_{2}\varpi _{z_{2}z_{2}}+2\varepsilon ^{2}z_{3}\varpi
_{z_{2}z_{3}}+\frac{1}{2}\varepsilon ^{2}\varpi _{z_{3}z_{3}} \\ 
\\ 
+\left( \varepsilon ^{2}-2\kappa z_{2}+2\chi z_{3}\right) \varpi
_{z_{2}}+\left( \chi -\kappa z_{3}\right) \varpi _{z_{3}}=0, \\ 
\\ 
\varpi \left( \bar{t},z_{1},z_{2},z_{3},\bar{t},\bar{z}_{1},\bar{z}_{3}^{2},%
\bar{z}_{3}\right) =\delta \left( z_{1}-\bar{z}_{1}\right) \delta \left(
z_{2}-\bar{z}_{3}^{2}\right) \delta \left( z_{3}-\bar{z}_{3}\right) .%
\end{array}
\label{Eq7.184}
\end{equation}%
One can use $\mathcal{K}$ given by Eq. (\ref{Eq7.171}) and write the set of
ODEs for $\alpha ,\delta _{2},\delta _{3}$ is as follows:%
\begin{equation}
\begin{array}{c}
\alpha _{t}\left( t,\bar{t}\right) -\frac{\varepsilon ^{2}}{2}\delta
_{3}^{2}\left( t,\bar{t}\right) +i\varepsilon ^{2}\delta _{2}\left( t,\bar{t}%
\right) +i\chi \delta _{3}\left( t,\bar{t}\right) =0,\ \ \ \alpha \left( 
\bar{t},\bar{t}\right) =0, \\ 
\\ 
i\delta _{2}^{\prime }\left( t,\bar{t}\right) -2\varepsilon ^{2}\delta
_{2}^{2}\left( t,\bar{t}\right) -2i\left( \kappa -i\rho \varepsilon
m_{1}\right) \delta _{2}\left( t,\bar{t}\right) -\frac{1}{2}m_{1}^{2}=0,\ \
\ \delta _{2}\left( \bar{t},\bar{t}\right) =m_{2}, \\ 
\\ 
i\delta _{3}^{\prime }\left( t,\bar{t}\right) -2\varepsilon ^{2}\delta
_{2}\left( t,\bar{t}\right) \delta _{3}\left( t,\bar{t}\right) +2i\chi
\delta _{2}\left( t,\bar{t}\right) -i\left( \kappa -i\rho \varepsilon
m_{1}\right) \delta _{3}\left( t,\bar{t}\right) =0,\ \ \ \delta _{3}\left( 
\bar{t},\bar{t}\right) =m_{3}.%
\end{array}
\label{Eq7.185}
\end{equation}%
As before, this system can be linearized and solved analytically, which was
pointed out by\ \cite{Stein91, Schobel99}. One can repeat the result
obtained in the previous Section verbatim, except for Eq. (\ref{Eq7.174}).
The corresponding characteristic equation has the form: 
\begin{equation}
\begin{array}{c}
\lambda ^{2}+2\left( \kappa -i\rho \varepsilon m_{1}\right) \lambda
-\varepsilon ^{2}m_{1}^{2}=0,%
\end{array}
\label{Eq7.186a}
\end{equation}%
its solutions can be written as follows:%
\begin{equation}
\begin{array}{c}
\lambda _{\pm }=\mu \pm \zeta , \\ 
\\ 
\mu =-\left( \kappa -i\rho \varepsilon m_{1}\right) ,\ \ \ \zeta =\sqrt{\bar{%
\rho}^{2}\varepsilon ^{2}m_{1}^{2}-2i\rho \varepsilon \kappa m_{1}+\kappa
^{2}}.%
\end{array}
\label{Eq7.186b}
\end{equation}%
The rest of the formal analysis is the same.\ But the asymptotic behavior of
the t.p.d.f. is, of course, different.

\section{Pricing of Financial Instruments\label{Chap8}}

\subsection{Background\label{Sec81}}

Formulas derived in Chapters \ref{Chap6}, \ref{Chap7} can be used to solve
numerous problems of financial engineering within a consistent framework
based on Kelvin waves. Here are some representative examples.

Payoffs of European options, depend solely on the terminal value of $\bar{S}=%
\hat{S}_{\bar{t}}$ of the underlying at option's maturity. The most common
European options are calls and puts, but, on occasion, binary options and
other types are traded as well. Since the hedging and speculation needs of
market participants cannot be satisfied by European options alone, the whole
industry emerged to design, price, and hedge the so-called exotic options,
with payoffs depending on the entire price trajectory between inception and
maturity.

Prices of the fundamental financial instruments, such as forwards and
European calls and puts, depend on the underlying prices only at maturity.
However, the prices of many other instruments depend on the entire
underlying price history between the instrument's inception and maturity.
Typical examples are barrier, American, Asian, lookback, and passport
options; see, e.g., \cite{Lipton99a, Lipton01} and references therein.
Moreover, the prices of bonds also depend on the history of the interest
rates and credit spreads throughout their life. This section shows how to
price some path-dependent financial instruments using the methodology
developed in the previous sections.

\subsection{The Underlying Processes\label{Sec82}}

The original approach to modeling financial assets was developed by
Bachelier, who assumed that prices $\hat{S}_{t}$ of such instruments are
governed by an arithmetic Brownian motion; see \cite{Bachelier00}:%
\begin{equation}
\begin{array}{c}
d\hat{S}_{t}=r\hat{S}_{t}dt+\hat{\sigma}d\hat{W}_{t},\ \ \ \hat{S}_{t}=S.%
\end{array}
\label{Eq8.1}
\end{equation}%
Here, $r$ is the risk-neutralized drift, $\hat{\sigma}$ is the volatility,
and $\hat{W}_{t}$ is a Wiener process; $r$, $\hat{\sigma}$ are dimensional
quantities, $\left[ r\right] =T^{-1}$, $\left[ \sigma \right] =\$T^{-1/2}$.
The process for $\hat{S}_{t}$ given by Eq. (\ref{Eq8.1}) is affine; in fact,
it is an OU process with zero mean and mean-repulsion instead of
mean-reversion.

Subsequently, the academic community concluded that using a geometric
Brownian motion as a driver is more appropriate; see \cite{Boness64,
Samuelson65, Black73, Merton73}. At present, the basic assumption is that
the price $\hat{S}_{t}$ of an underlying financial instrument follows a
geometric Brownian motion process with constant coefficients:%
\begin{equation}
\begin{array}{c}
\frac{d\hat{S}_{t}}{\hat{S}_{t}}=rdt+\sigma d\hat{W}_{t},\ \ \ \hat{S}_{t}=S.%
\end{array}
\label{Eq8.2}
\end{equation}%
Here, $r$ is the risk-neutralized drift, and $\sigma $ is the volatility.
These are dimensional quantities, $\left[ r\right] =T^{-1}$, $\left[ \sigma %
\right] =T^{-1/2}$.

The choice between using the Bachelier and the Black-Scholes models often
depends on the nature of the underlying asset and the market's specific
characteristics. Since the Bachelier model assumes that the underlying asset
prices follow a normal distribution, it can be more appropriate for assets
whose price changes are additive and can theoretically go below zero, like
interest rates, some commodities, or certain types of bonds. Generally, the
price movements of the underlying asset are relatively small for short
periods, so the Bachelier model provides a good description of these
movements. The Bachelier model is often used for pricing commodities, some
interest-rate derivatives, and studying the optimal execution. In markets
with relatively low volatility, the Bachelier model's assumption of additive
price movements can provide a better fit for pricing and hedging derivatives
than the multiplicative approach of the Black-Scholes model.

It was realized very soon after the seminal paper by \cite{Black73} was
published, that in practice, it provides a rather poor description of
reality. Hence, considerable efforts were dedicated to developing more
adequate models. Such models include the jump-diffusion, local volatility,
path-dependent volatility, stochastic volatility, local-stochastic
volatility, rough volatility, and culminate in the universal volatility
model; see \cite{Merton76, Stein91, Bick93, Heston93, Derman94, Dupire94,
Rubinstein94, Hobson98, Jex99, Lewis00, Lipton00, Lipton01, Boyarchenko02,
Hagan02, Lipton02, Bergomi15, Reghai15, Gatheral18, Gershon22}, and
references therein.

Replacing constant volatility for a geometric Brownian motion with
stochastic volatility driven by a Feller process results in the popular
Heston model; see \cite{Heston93}. This model has numerous applications,
particularly for pricing equity and foreign exchange derivatives. The
governing SDEs:%
\begin{equation}
\begin{array}{c}
\frac{d\hat{S}_{t}}{\hat{S}_{t}}=rdt+\sqrt{\hat{v}_{t}}d\hat{W}_{t},\ \ \ 
\hat{S}_{t}=S, \\ 
\\ 
d\hat{v}_{t}=\left( \chi -\kappa \hat{v}_{t}\right) dt+\varepsilon \sqrt{%
\hat{v}_{t}}d\hat{Z}_{t},\ \ \ \hat{v}_{t}=\nu ,%
\end{array}
\label{Eq8.3}
\end{equation}%
where $d\hat{W}_{t}d\hat{Z}_{t}=\rho dt$. The logarithmic change of
variables, given by Eq. (\ref{Eq8.3}), yields Eqs (\ref{Eq7.124}).

Replacing constant volatility with stochastic volatility driven by an OU
process results in the (less popular) Stein-Stein model; see \cite{Stein91,
Schobel99}. The corresponding SDEs have the form:%
\begin{equation}
\begin{array}{c}
\frac{d\hat{S}_{t}}{\hat{S}_{t}}=rdt+\hat{\sigma}_{t}d\hat{W}_{t},\ \ \ \hat{%
S}_{t}=S, \\ 
\\ 
d\hat{\sigma}_{t}=\left( \chi -\kappa \hat{\sigma}_{t}\right) dt+\varepsilon
d\hat{Z}_{t},\ \ \ \hat{v}_{t}=\nu ,%
\end{array}
\label{Eq8.4}
\end{equation}%
\cite{Stein91} considered the special case of zero correlation, $d\hat{W}%
_{t}d\hat{Z}_{t}=0$, while \cite{Schobel99} studied the general case of
arbitrary correlation, $d\hat{W}_{t}d\hat{Z}_{t}=\rho dt$.

Now, it is shown how to use formulas derived in Chapters \ref{Chap6}, \ref%
{Chap7} in the context of financial engineering.

\subsection{European Derivatives\label{Sec83}}

\subsubsection{Forwards, Calls, Puts, and Covered Calls}

The most basic derivatives are forwards. Recall that a forward contract
obligates the buyer (seller) to buy (to sell) an underlying asset for an
agreed price at a specified future date. These contracts are not
standardized and are traded over-the-counter (OTC), not on exchanges.
Typical underlying assets are commodities, currencies, and financial
instruments. The choice of an asset depends on the needs of the contracting
parties. The price agreed upon in a forward contract is called the forward
price. This price is derived based on the spot price of the underlying
asset, adjusted for factors like time to maturity, interest rates, and
dividends. Forward contracts are primarily used for hedging price
fluctuations of the underlying asset or speculation. The payoff of a forward
contract with maturity $\bar{t}$ and strike $K$ has the form:%
\begin{equation}
\begin{array}{c}
U^{\left( F\right) }\left( \bar{S},K\right) =\bar{S}-K,%
\end{array}
\label{Eq8.5a}
\end{equation}%
where the strike is chosen in such a way that today's price of the forward
contract is equal to zero. This price can be found without knowing the
actual stochastic process $\hat{S}$. The hedging argument shows that the
only way to deliver the price of a non-dividend-paying stock at maturity $%
\bar{t}$ is to buy it outright at inception $t$. Similarly, to deliver the
strike $K$ at time $\bar{t}$, one has to buy a zero coupon bond at time $t$.
Let $\mathtt{Z}_{t,\bar{t}}$ be the price of a bond paying unity at maturity 
$\bar{t}$. Then%
\begin{equation}
\begin{array}{c}
F_{t,\bar{t}}\equiv K=\frac{S}{\mathtt{Z}_{t,\bar{t}}}.%
\end{array}
\label{Eq8.5b}
\end{equation}%
In contrast to forwards, a European call option grants the holder the right,
but impose no obligation, to buy an underlying asset at the option maturity
for a predetermined strike price. Similarly, a European put option grants
the holder the right to sell an underlying asset. Theoretically, buyers
utilize calls and puts to hedge future risks; however, they often buy
options for speculative purposes. American options can be exercised at any
time of the buyer's choice before the option's maturity. Bermudan options
are exercisable at fixed times between their inception and maturity. A call
option is a contract between two parties - a buyer and a seller. Typically,
the buyer takes the long position on the underlying (i.e., she expects that
at maturity, the underlying price will exceed the strike price) and does not
hedge her position. On the other hand, the seller or writer of the option
(typically a bank) does hedge and, hence, maintains a market-neutral
position. The seller receives cash up-front but incurs potential liabilities
at option maturity if the option is exercised. In contrast, the buyer pays
money up-front in exchange for the potential for future gains. For a put
option, the buyer takes a short position, while the seller is still
market-neutral.

Payoffs of call and put options with maturity $\bar{t}$ and strike $K$ have
the form:%
\begin{equation}
\begin{array}{c}
U^{\left( C\right) }\left( \bar{S},K\right) =\max \left\{ \bar{S}%
-K,0\right\} , \\ 
\\ 
U^{\left( P\right) }\left( \bar{S},K\right) =\min \left\{ K-\bar{S}%
,0\right\} , \\ 
\\ 
U^{\left( C,P\right) }\left( \bar{S},K\right) =\max \left\{ \phi \left( \bar{%
S}-K\right) ,0\right\} ,%
\end{array}
\label{Eq8.5c}
\end{equation}%
where $\phi =1$ for a call, and $\phi =-1$ for a put. Put-call parity
implies that their difference is linear in $\bar{S}$ and represents a
forward contract: 
\begin{equation}
\begin{array}{c}
U^{\left( C\right) }\left( \bar{S},K\right) -U^{\left( P\right) }\left( \bar{%
S},K\right) =\bar{S}-K.%
\end{array}
\label{Eq8.5d}
\end{equation}

Several popular models, including Bachelier, Black-Scholes, Heston, and
Stein-Stein, are considered below. While the Bachelier model is not scale
invariant, all the other models are. A general driver for a scale-invariant
model can be written as follows:%
\begin{equation}
\begin{array}{c}
\frac{d\hat{S}_{t}}{\hat{S}_{t}}=rdt+\sigma _{t}d\hat{W}_{t}+\upsilon d\hat{%
\Pi}_{t},\ \ \ \hat{S}_{t}=S,%
\end{array}
\label{Eq8.9}
\end{equation}%
where, potentially, the volatility $\hat{\sigma}_{t}$ and the intensity $%
\hat{\lambda}_{t}$ of the Poisson process $\hat{\Pi}_{t}$ are driven by SDEs
of their own. For such models, it is convenient to decompose call and put
payoffs (\ref{Eq8.7}) into parts, which are easier to study via Kevin waves;
see \cite{Lipton01, Lipton02}. To this end, introduce the covered call with
the payoff of the form 
\begin{equation}
\begin{array}{c}
U^{\left( CC\right) }\left( \bar{S},K\right) =\min \left( \bar{S},K\right) .%
\end{array}
\label{Eq8.10}
\end{equation}%
The call and put payoffs can be decomposed as follows:%
\begin{equation}
\begin{array}{c}
U^{\left( C\right) }\left( \bar{S},K\right) =\bar{S}-U^{\left( CC\right)
}\left( \bar{S},K\right) ,\ \ \ U^{\left( P\right) }\left( \bar{S},K\right)
=K-U^{\left( CC\right) }\left( \bar{S},K\right)%
\end{array}
\label{Eq8.11}
\end{equation}%
Thus, the call price is the difference between the forward price and the
covered call price, while the put price is the difference between the bond
price and the covered call price. In both cases, the covered call is the
source of optionality.

\subsubsection{Black-Scholes Model}

For the standard log-normal process, the backward pricing problem for
covered calls can be written as follows:%
\begin{equation}
\begin{array}{c}
U_{t}+\frac{1}{2}\sigma ^{2}S^{2}U_{SS}+rU_{S}-rU=0, \\ 
\\ 
U\left( \bar{t},S\right) =\min \left\{ S,K\right\} .%
\end{array}
\label{Eq8.12a}
\end{equation}%
It is helpful to rewrite it by using forward, rather that spot prices:%
\begin{equation}
\begin{array}{c}
\hat{U}_{t}+\frac{1}{2}\sigma ^{2}F^{2}\hat{U}_{FF}=0, \\ 
\\ 
\hat{U}\left( \bar{t},F\right) =\min \left\{ F,K\right\} ,%
\end{array}
\label{Eq8.12b}
\end{equation}%
where%
\begin{equation}
\begin{array}{c}
\hat{F}_{t,\bar{t}}=e^{r\left( \bar{t}-t\right) }\hat{S}_{t},\ \ \ \hat{U}%
\left( t,F\right) =e^{r\left( \bar{t}-t\right) }U\left( t,S\right) .%
\end{array}
\label{Eq8.12c}
\end{equation}%
Change of variables:%
\begin{equation}
\begin{array}{c}
\hat{F}_{t,\bar{t}}\rightarrow \hat{x}_{t,\bar{t}},\ \ \ \hat{F}_{t,\bar{t}%
}=Ke^{\hat{x}_{t,\bar{t}}},%
\end{array}
\label{Eq8.12d}
\end{equation}%
results in the following process for $\hat{x}_{t}$:%
\begin{equation}
\begin{array}{c}
d\hat{x}_{t,\bar{t}}=-\frac{1}{2}\sigma ^{2}dt+\sigma d\hat{W}_{t},\ \ \ 
\hat{x}_{t,\bar{t}}=x=\ln \left( \frac{F_{t,\bar{t}}}{K}\right) .%
\end{array}
\label{Eq8.13}
\end{equation}%
The t.p.d.f. for this process is Gaussian:%
\begin{equation}
\begin{array}{c}
\varpi \left( t,x,\bar{t},\bar{x}\right) =\frac{1}{\sqrt{2\pi \sigma ^{2}T}}%
\exp \left( -\frac{\left( \bar{x}-x+\sigma ^{2}/2T\right) ^{2}}{2\sigma ^{2}T%
}\right) .%
\end{array}
\label{Eq8.14}
\end{equation}%
Since the the non-dimensional payoff of the covered call has the form%
\begin{equation}
\begin{array}{c}
\tilde{U}^{\left( CC\right) }\left( x\right) =\min \left\{ e^{x},1\right\} ,%
\end{array}
\label{Eq8.15}
\end{equation}%
where $\tilde{U}=\hat{U}/K$, one obtains the following expression for $%
\tilde{U}^{\left( CC\right) }$:%
\begin{equation}
\begin{array}{c}
\tilde{U}^{\left( CC\right) }\left( t,x\right) =e^{x}\mathrm{N}\left( -\frac{%
x}{\sigma \sqrt{T}}-\frac{\sigma \sqrt{T}}{2}\right) +\mathrm{N}\left( \frac{%
x}{\sigma \sqrt{T}}-\frac{\sigma \sqrt{T}}{2}\right) ,%
\end{array}
\label{Eq8.16}
\end{equation}%
where $\mathrm{N}\left( .\right) $ is the cumulative normal function.

By using Eq. (\ref{Eq8.16}), one can represent call and put prices as
follows:%
\begin{equation}
\begin{array}{c}
\hat{U}^{\left( C,P\right) }\left( t,F_{T}\right) =\phi \left( F_{T}\mathrm{N%
}\left( \phi d_{+}\right) -K\mathrm{N}\left( \phi d_{-}\right) \right) , \\ 
\\ 
d_{\pm }=\frac{\ln \left( F_{T}/K\right) }{\sigma \sqrt{T}}\pm \frac{\sigma 
\sqrt{T}}{2},%
\end{array}
\label{Eq8.17}
\end{equation}%
see \cite{Black76}.

Returning to the original variables, write the classical \cite{Black73}
closed-form formula for the time $t$ prices of calls and puts in its
original form:%
\begin{equation}
\begin{array}{c}
U^{\left( C,P\right) }\left( t,S\right) =\phi \left( S\mathrm{N}\left( \phi
d_{+}\right) -e^{-rT}K\mathrm{N}\left( \phi d_{-}\right) \right) , \\ 
\\ 
d_{\pm }=\frac{\ln \left( e^{rT}S/K\right) }{\sigma \sqrt{T}}\pm \frac{%
\sigma \sqrt{T}}{2}.%
\end{array}
\label{Eq8.18}
\end{equation}%
Further transform%
\begin{equation}
\begin{array}{c}
\tilde{U}^{\left( CC\right) }\left( t,x\right) =e^{x/2}V^{\left( CC\right)
}\left( t,x\right) ,%
\end{array}
\label{Eq8.19}
\end{equation}%
yields the following backward problem%
\begin{equation}
\begin{array}{c}
V_{t}^{\left( CC\right) }+\frac{1}{2}\sigma ^{2}V_{xx}^{\left( CC\right) }-%
\frac{1}{8}\sigma ^{2}V^{\left( CC\right) }=0, \\ 
\\ 
V^{\left( CC\right) }\left( \bar{t},x\right) =e^{-\left\vert x\right\vert
/2},%
\end{array}
\label{Eq8.18a}
\end{equation}%
with symmetric \textquotedblleft peakon\textquotedblright\ payoff, which is
proportional to the Laplace distribution density. This transform removes the
drift in the x direction at the expense of adding killing with intensity $%
\sigma ^{2}/8$. Eq. (\ref{Eq8.16}) implies:%
\begin{equation}
\begin{array}{c}
V\left( t,x\right) =e^{x/2}\mathrm{N}\left( -\frac{x}{\sigma \sqrt{T}}-\frac{%
\sigma \sqrt{T}}{2}\right) +e^{-x/2}\mathrm{N}\left( \frac{x}{\sigma \sqrt{T}%
}-\frac{\sigma \sqrt{T}}{2}\right) .%
\end{array}
\label{Eq8.18b}
\end{equation}%
The Fourier transform of the \textquotedblleft peakon\textquotedblright\
payoff yield%
\begin{equation}
\begin{array}{c}
\int\limits_{-\infty }^{\infty }e^{-\left\vert x\right\vert /2-ikx}dx=\frac{1%
}{k^{2}+1/4}.%
\end{array}
\label{Eq8.20}
\end{equation}%
By using this formula, one can derive an alternative expression for $%
U^{\left( C,P\right) }$ based on Kelvin waves; see \cite{Lipton02}. It is
clear that Kelvin waves associated with the killed arithmetic Brownian
motion described by Eq. (\ref{Eq8.13}) are the standard Fourier waves of the
form:%
\begin{equation}
\begin{array}{c}
\mathcal{K}\left( t,x,k\right) =e^{-\left( k^{2}+1/4\right) \sigma
^{2}T/2+ikx}.%
\end{array}
\label{Eq8.21}
\end{equation}%
Eqs (\ref{Eq8.20}), (\ref{Eq8.21}) yield the following alternative
expression for the price of covered calls given by Eq. (\ref{Eq8.18b}):%
\begin{equation}
\begin{array}{c}
V^{\left( CC\right) }\left( t,x\right) =\frac{1}{2\pi }\int\limits_{-\infty
}^{\infty }\frac{e^{-\left( k^{2}+1/4\right) \sigma ^{2}T/2+ikx}}{k^{2}+1/4}%
dk,%
\end{array}
\label{Eq8.22}
\end{equation}%
see \cite{Lipton02}. Eq. (\ref{Eq8.22}) is central for the subsequent
developments. For a single strike, this formula is less efficient than its
classical counterpart , however, for a set of strikes, it is faster, because
all the prices can be computed in one go, via the Fast Fourier Transform.

As one shall see shortly, these formulas help to handle affine pricing
models very naturally.

\subsubsection{Heston Model}

The transformed forward pricing problem for the Heston model with the
\textquotedblleft peakon\textquotedblright\ payoff has the form:%
\begin{equation}
\begin{array}{c}
\begin{array}{c}
V_{t}^{\left( CC\right) }+\frac{1}{2}y\left( V_{xx}^{\left( CC\right)
}+2\rho \varepsilon V_{xy}^{\left( CC\right) }+\varepsilon
^{2}V_{yy}^{\left( CC\right) }\right) \\ 
\\ 
+\left( \chi -\hat{\kappa}y\right) V_{y}^{\left( CC\right) }-\frac{y}{8}%
V^{\left( CC\right) }=0, \\ 
\\ 
V^{\left( CC\right) }\left( \bar{t},x,y\right) =e^{-\left\vert x\right\vert
/2},%
\end{array}%
\end{array}
\label{Eq8.23}
\end{equation}%
where $\hat{\kappa}=\kappa -\rho \varepsilon /2$. Thus, one is dealing with
the killed stochastic process given by Eqs (\ref{Eq7.124}). Adapting the
corresponding equations to accommodate the updated mean-reversion rate and
the presence of the killing term, one gets the following system of ODEs for
the corresponding Kelvin wave parameters:%
\begin{equation}
\begin{array}{c}
\alpha _{t}\left( t,\bar{t}\right) +i\chi \gamma \left( t,\bar{t}\right)
=0,\ \ \ \alpha \left( \bar{t},\bar{t}\right) =0, \\ 
\\ 
i\gamma _{t}\left( t,\bar{t}\right) -\frac{1}{2}\varepsilon ^{2}\gamma
^{2}\left( t,\bar{t}\right) -\left( \hat{\kappa}-i\rho \varepsilon k\right)
i\gamma \left( t,\bar{t}\right) -\frac{1}{2}\left( k^{2}+\frac{1}{4}\right)
=0,\ \ \ \gamma \left( \bar{t},\bar{t}\right) =0.%
\end{array}
\label{Eq8.24a}
\end{equation}%
Formulas (\ref{Eq7.99})-(\ref{Eq7.102}) are still applicable. However, the
corresponding characteristic equation and its solution are:%
\begin{equation}
\begin{array}{c}
\lambda ^{2}+\left( \hat{\kappa}-i\rho \varepsilon k\right) \lambda -\frac{%
\varepsilon ^{2}}{4}\left( k^{2}+\frac{1}{4}\right) =0,%
\end{array}
\label{Eq8.24b}
\end{equation}%
\begin{equation}
\begin{array}{c}
\lambda _{\pm }=\mu \pm \zeta , \\ 
\\ 
\mu =-\frac{\left( \hat{\kappa}-i\rho \varepsilon k\right) }{2},\ \ \ \zeta =%
\frac{\sqrt{\bar{\rho}^{2}\varepsilon ^{2}k^{2}-2i\rho \varepsilon k+\hat{%
\kappa}^{2}+\varepsilon ^{2}/4}}{2}.%
\end{array}
\label{Eq8.24c}
\end{equation}%
It is convenient to write $\left( \alpha ,\gamma \right) $ as follows:%
\begin{equation}
\begin{array}{c}
\alpha \left( T,k\right) =-\frac{2\chi }{\varepsilon ^{2}}\left( \left( \mu
+\zeta \right) T+\ln \left( \frac{-\mu +\zeta +\left( \mu +\zeta \right)
e^{-2\zeta T}}{2\zeta }\right) \right) ,%
\end{array}
\label{Eq8.24d}
\end{equation}%
\begin{equation}
\begin{array}{c}
\gamma \left( T,k\right) =\left( k^{2}+\frac{1}{4}\right) \frac{i\left(
1-e^{-2\zeta T}\right) }{2\left( -\mu +\zeta +\left( \mu +\zeta \right)
e^{-2\zeta T}\right) }\equiv \left( k^{2}+\frac{1}{4}\right) i\varsigma
\left( T,k\right) .%
\end{array}
\label{Eq8.24e}
\end{equation}%
Hence, the price of the \textquotedblleft peakon\textquotedblright\ has the
form:%
\begin{equation}
\begin{array}{c}
V^{\left( CC\right) }\left( t,x,y\right) =\frac{1}{2\pi }\int\limits_{-%
\infty }^{\infty }\frac{e^{\alpha \left( T,k\right) -\left( k^{2}+1/4\right)
\varsigma \left( T,k\right) y+ikx}}{k^{2}+1/4}dk.%
\end{array}
\label{Eq8.24f}
\end{equation}%
Eq (\ref{Eq8.24f}) is frequently called the Lewis-Lipton formula; see, e.g., 
\cite{Lewis00, Lipton00, Lewis01, Lipton01, Lipton02, Schmelzle10, Janek11}.

\begin{figure}[tbp]
\begin{center}
\includegraphics[width=0.8\textwidth]
{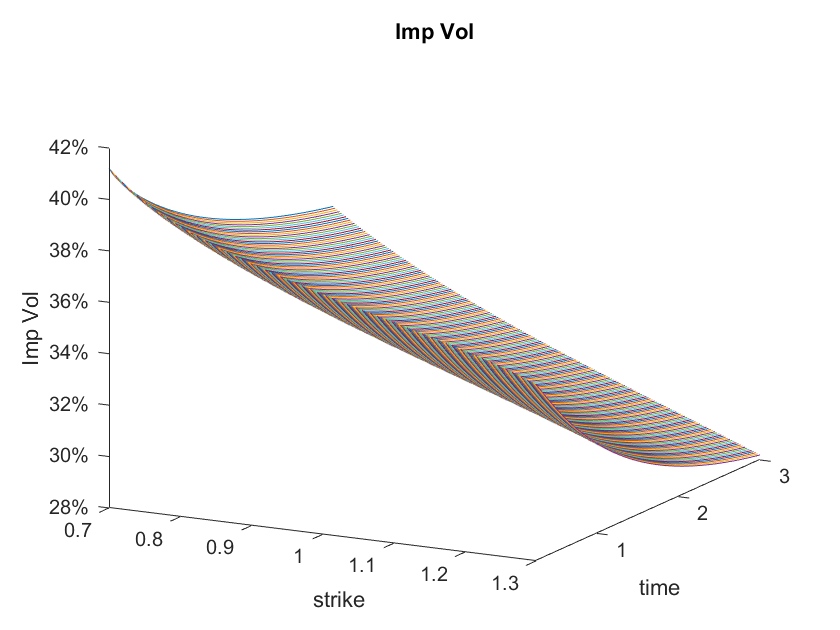}
\end{center}
\par
\vspace{-10pt}
\caption{A representative implied volatility surface generated by the Heston
model. Parameters are the same as in Figure \protect\ref{Fig7.4}. Own
graphics.}
\label{Fig8.1}
\end{figure}

The implied volatility surface generated by a representative Heston model is
shown in Figure \ref{Fig8.1}. 
Recall that the implied volatility $\Sigma \left( T,K\right) $ is the
volatility one must substitute into the Black-Scholes formula to reproduce
the market price of a call (or put) option with maturity $T$ and strike $K$.
Thus, the deviation of the volatility surface from the flat surface $\Sigma
\left( T,K\right) =\Sigma _{0}$ shows how far is a given market (or model)
from the idealized Black-Scholes framework.

\subsubsection{Stein-Stein Model}

The transformed forward pricing problem for the Stein-Stein model with the
\textquotedblleft peakon\textquotedblright\ payoff has the form:%
\begin{equation}
\begin{array}{c}
V_{t}^{\left( CC\right) }+\frac{1}{2}z_{2}V_{z_{1}z_{1}}^{\left( CC\right)
}+2\rho \varepsilon z_{2}V_{z_{1}z_{2}}^{\left( CC\right) }+\rho \varepsilon
z_{3}V_{z_{1}z_{3}}^{\left( CC\right) } \\ 
\\ 
+2\varepsilon ^{2}z_{2}V_{z_{2}z_{2}}^{\left( CC\right) }+2\varepsilon
^{2}z_{3}V_{z_{2}z_{3}}^{\left( CC\right) }+\frac{1}{2}\varepsilon
^{2}V_{z_{3}z_{3}}^{\left( CC\right) } \\ 
\\ 
+\left( \varepsilon ^{2}-2\hat{\kappa}z_{2}+2\chi z_{3}\right)
V_{z_{2}}^{\left( CC\right) }+\left( \chi -\hat{\kappa}z_{3}\right)
V_{z_{3}}^{\left( CC\right) }-\frac{z_{2}}{8}V^{\left( CC\right) }=0, \\ 
\\ 
V^{\left( CC\right) }\left( \bar{t},z_{1},z_{2},z_{3}\right) =e^{-\left\vert
z_{1}\right\vert /2},%
\end{array}
\label{Eq8.25a}
\end{equation}%
which corresponds to the killed stochastic process described by Eqs (\ref%
{Eq7.181}). By incorporating the killing term, one gets the following set of
ODEs for the Kelvin wave parameters%
\begin{equation}
\begin{array}{c}
\alpha _{t}\left( t,\bar{t}\right) -\frac{\varepsilon ^{2}}{2}\delta
_{3}^{2}\left( t,\bar{t}\right) +i\varepsilon ^{2}\delta _{2}\left( t,\bar{t}%
\right) +i\chi \delta _{3}\left( t,\bar{t}\right) =0,\ \ \ \alpha \left( 
\bar{t},\bar{t}\right) =0, \\ 
\\ 
i\delta _{2}^{\prime }\left( t,\bar{t}\right) -2\varepsilon ^{2}\delta
_{2}^{2}\left( t,\bar{t}\right) -2i\left( \hat{\kappa}-i\rho \varepsilon
m_{1}\right) \delta _{2}\left( t,\bar{t}\right) -\frac{1}{2}\left( m_{1}^{2}+%
\frac{1}{4}\right) =0,\ \ \ \delta _{2}\left( \bar{t},\bar{t}\right) =0, \\ 
\\ 
i\delta _{3}^{\prime }\left( t,\bar{t}\right) -2\varepsilon ^{2}\delta
_{2}\left( t,\bar{t}\right) \delta _{3}\left( t,\bar{t}\right) +2i\chi
\delta _{2}\left( t,\bar{t}\right) -i\left( \hat{\kappa}-i\rho \varepsilon
m_{1}\right) \delta _{3}\left( t,\bar{t}\right) =0,\ \ \ \delta _{3}\left( 
\bar{t},\bar{t}\right) =0.%
\end{array}
\label{Eq8.25b}
\end{equation}%
The corresponding solution has the form given by Eqs (\ref{Eq7.178}) with:%
\begin{equation}
\begin{array}{c}
\lambda _{\pm }=\mu \pm \zeta , \\ 
\\ 
\mu =-\left( \hat{\kappa}-i\rho \varepsilon m_{1}\right) ,\ \ \ \zeta =\sqrt{%
\bar{\rho}^{2}\varepsilon ^{2}m_{1}^{2}-2i\rho \varepsilon \kappa
m_{1}+\kappa ^{2}+\varepsilon ^{2}/4}, \\ 
\\ 
\omega _{\pm }=\mp \frac{\lambda _{\mp }}{2\zeta },\ \ \ b_{\pm }=\frac{%
i\lambda _{\pm }\omega _{\pm }}{2\varepsilon ^{2}}, \\ 
\\ 
c_{\pm }=\pm \frac{i\chi \lambda _{\pm }\omega _{\pm }}{\varepsilon
^{2}\zeta },\ \ \ c_{0}=-c_{+}-c_{-},\ \ \ \ g=\frac{\chi ^{2}\lambda
_{+}\lambda _{-}}{2\varepsilon ^{2}\zeta ^{2}},\ \ \  \\ 
\\ 
a_{0}=-\frac{i\kappa \chi c_{0}}{\zeta ^{2}},\ \ \ a_{\pm }=-\omega _{\pm
}a_{0}\mp \left( \frac{\varepsilon ^{2}c_{0}^{2}}{4\zeta }+\frac{\chi
^{2}\kappa ^{2}\omega _{+}\omega _{-}}{\varepsilon ^{2}\zeta ^{3}}\right) ,%
\end{array}
\label{Eq8.25c}
\end{equation}%
The generic expression for the price of the \textquotedblleft
peakon\textquotedblright\ has the form:%
\begin{equation}
\begin{array}{c}
V^{\left( CC\right) }\left( t,z_{1},z_{3}^{2},z_{3}\right) =\frac{1}{2\pi }%
\int\limits_{-\infty }^{\infty }\frac{e^{\alpha \left( T,m_{1}\right)
+i\delta _{2}\left( T,m_{1}\right) z_{3}^{2}+i\delta _{3}\left(
T,m_{1}\right) z_{3}+im_{1}z_{1}}}{m_{1}^{2}+1/4}dm_{1}.%
\end{array}
\label{Eq8.26}
\end{equation}%
It is clear that this price is a function of $\bar{t}$, $z_{1}$, $z_{3}$.

\subsubsection{Path-Dependent Volatility Model}

\cite{Hobson98} initially proposed path-dependent volatility models;
subsequently, they were studied by many authors; see \cite{Davis04,
DiFrancesco04, DiFrancesco05, Lipton23b} among others. They present a viable
alternative to the more popular local volatility models developed by \cite%
{Bick93, Derman94, Dupire94, Rubinstein94}.

The main advantage of path-dependent volatility models compared to their
local volatility brethren is that the former deal with volatility functions
depending on a non-dimensional argument, such as $\hat{S}_{t}/\hat{A}_{t}$,
where $\hat{S}_{t}$ is the stock price, and $\hat{A}_{t}$ is its average,
say, $\sigma =\sigma \left( \hat{S}_{t}/\hat{A}_{t}\right) $, while the
latter use volatilities depending on a dimensional argument $\hat{S}_{t}$, $%
\sigma =\sigma \left( \hat{S}_{t}\right) $, which is conceptually unsound
and results in model dynamics deviating from the one observed in the market.
The problem with path-dependent models is that building an analytically
tractable path-dependent model is exceedingly tricky, so gaining the
necessary intuition or benchmarking numerical solutions is complicated.
However, this Section develops such a model using results derived in\
Section \ref{Sec73}.

Here, an original path-dependent model with a semi-analytical solution is
presented for the first time. The dynamics is adapted from Section \ref%
{Sec73}, Eq. (\ref{Eq7.143}) as follows:%
\begin{equation}
\begin{array}{c}
\hat{A}_{t}=\exp \left( \kappa \int_{-\infty }^{t}e^{-\kappa \left(
t-t^{\prime }\right) }\ln \hat{S}_{t^{\prime }}dt^{\prime }\right) ,\ \ \ 
\hat{A}_{t}=A, \\ 
\\ 
\frac{d\hat{S}_{t}}{\hat{S}_{t}}=\sqrt{c_{0}+c_{1}\ln \left( \frac{\hat{S}%
_{t}}{A_{t}}\right) }d\hat{W}_{t},\ \ \ \hat{S}_{t}=S.%
\end{array}
\label{Eq8.27a}
\end{equation}%
It is not necessary to describe in detail how $\hat{S}_{\tilde{t}}$, and,
hence, $\hat{A}_{\tilde{t}}$, behave when $\tilde{t}<t$, since it becomes
unimportant provided that $\kappa T$ is sufficiently large. For instance,
one can assume that $\hat{S}_{\tilde{t}}\equiv A$, when $\tilde{t}<t$, then $%
A=S$. Additionally, it is assumed that $r=0$, so that spot and forward
prices coincide, $\hat{S}_{t}=\hat{F}_{t,\bar{t}}$.

In logarithmic variables $\hat{x}_{t}=\ln \left( \hat{A}_{t}\right) $, $\hat{%
y}_{t}=\ln \left( \hat{S}_{t}\right) $, Eqs (\ref{Eq8.27a}) assume the form
given by Eqs (\ref{Eq7.143}). Accordingly, the pricing equation for the
path-dependent model with the symmetric \textquotedblleft
peakon\textquotedblright\ payoff can be written as follows:%
\begin{equation}
\begin{array}{c}
V_{t}^{\left( CC\right) }+\frac{1}{2}\left( a_{0}+a_{1}\left( y-x\right)
\right) \left( V_{yy}^{\left( CC\right) }-\frac{1}{4}V^{\left( CC\right)
}\right) +\kappa \left( y-x\right) V_{x}^{\left( CC\right) }=0, \\ 
\\ 
V^{\left( CC\right) }\left( \bar{t},x,y\right) =e^{-\left\vert y\right\vert
/2}.%
\end{array}
\label{Eq8.27b}
\end{equation}%
The Kelvin wave parameters are governed by the equations of the form:%
\begin{equation}
\begin{array}{c}
\alpha _{t}\left( t,\bar{t}\right) -\frac{a_{0}}{2}\left( \gamma ^{2}\left(
t,\bar{t}\right) +\frac{1}{4}\right) =0,\ \ \ \alpha \left( \bar{t},\bar{t}%
\right) =0, \\ 
\\ 
i\beta _{t}\left( t,\bar{t}\right) +\frac{a_{1}}{2}\left( \gamma ^{2}\left(
t,\bar{t}\right) +\frac{1}{4}\right) -i\kappa \beta \left( t,\bar{t}\right)
=0,\ \ \ \beta \left( \bar{t},\bar{t}\right) =0, \\ 
\\ 
i\gamma _{t}\left( t,\bar{t}\right) -\frac{a_{1}}{2}\left( \gamma ^{2}\left(
t,\bar{t}\right) +\frac{1}{4}\right) +i\kappa \beta \left( t,\bar{t}\right)
=0,\ \ \ \gamma \left( \bar{t},\bar{t}\right) =l.%
\end{array}
\label{Eq8.27c}
\end{equation}

\subsubsection{Bachelier Model}

In the Bachelier model, the corresponding discounted t.p.d.f. is given by a
modified Eqs (\ref{Eq6.96}):%
\begin{equation}
\begin{array}{c}
\varpi \left( t,S,\bar{t},\bar{S}\right) =\frac{1}{\sqrt{2\pi \Sigma
^{2}\left( t,\bar{t}\right) }}\exp \left( -\frac{\left( \bar{S}-F_{T}\right)
^{2}}{2\Sigma ^{2}\left( t,\bar{t}\right) }\right) ,%
\end{array}
\label{Eq8.6}
\end{equation}%
where%
\begin{equation}
\begin{array}{c}
\Sigma ^{2}\left( t,\bar{t}\right) =\frac{\hat{\sigma}^{2}\left(
e^{2rT}-1\right) }{2r}.%
\end{array}
\label{Eq8.7}
\end{equation}
By virtue of Eq. (\ref{Eq8.5c}), one can price European calls and puts as
follows:%
\begin{equation}
\begin{array}{c}
V\left( t,F_{T}\right) =e^{-rT}\left( \phi \left( F_{T}-K\right) \mathrm{N}%
\left( \phi \frac{F_{T}-K}{\Sigma \left( T\right) }\right) +\Sigma \left(
T\right) \mathrm{n}\left( \frac{F_{T}-K}{\Sigma \left( T\right) }\right)
\right) ,%
\end{array}
\label{Eq8.8a}
\end{equation}%
or, in spot terms: 
\begin{equation}
\begin{array}{c}
V\left( t,S\right) =\phi \left( S-e^{-rT}K\right) \mathrm{N}\left( \phi 
\frac{S-e^{-rT}K}{\tilde{\Sigma}\left( T\right) }\right) +\tilde{\Sigma}%
\left( T\right) \mathrm{n}\left( \frac{S-e^{-rT}K}{\tilde{\Sigma}\left(
T\right) }\right) ,%
\end{array}
\label{Eq8.8b}
\end{equation}%
where%
\begin{equation}
\begin{array}{c}
\tilde{\Sigma}^{2}\left( T\right) =\frac{\hat{\sigma}^{2}\left(
1-e^{-2rT}\right) }{2r},%
\end{array}
\label{Eq8.8c}
\end{equation}%
see \cite{Bachelier00, Schachermayer08, Terakado19} for further details.

\subsection{Asian Options with Arithmetic and Geometric Averaging\label%
{Sec84}}

The most basic path-dependent options are fixed strike Asian calls and puts,
whose payoff depends on the underlying value averaged between the inception
and maturity. Such options are popular for commodity and energy trading and
in many other circumstances. The average $\hat{A}_{t,\bar{t}}$ on the
interval $\left[ t,\bar{t}\right] $ can be defined in several ways. The
simplest and, as a result, the most popular is an arithmetic average:%
\begin{equation}
\begin{array}{c}
\hat{A}_{t,\bar{t}}=\frac{1}{T}\dint\limits_{t}^{\bar{t}}\hat{S}_{s}ds.%
\end{array}
\label{Eq8.28}
\end{equation}%
A less frequent, but technically easier to deal with, alternative is a
geometric average:%
\begin{equation}
\begin{array}{c}
\hat{A}_{t,\bar{t}}=\exp \left( \frac{1}{T}\dint\limits_{t}^{\bar{t}}\ln
\left( \hat{S}_{s}\right) ds\right) .%
\end{array}
\label{Eq8.29}
\end{equation}%
The payoff of an Asian option with maturity $\bar{t}$ and fixed strike $K$ is%
\begin{equation}
\begin{array}{c}
U\left( \bar{A}_{t,\bar{t}}\right) =\max \left\{ \phi \left( \bar{A}_{t,\bar{%
t}}-K\right) ,0\right\} ,%
\end{array}
\label{Eq8.30}
\end{equation}%
as before, $\phi =1$ for a call, and $\phi =-1$ for a put. For the floating
strike, the payoff is%
\begin{equation}
\begin{array}{c}
U\left( \bar{S}_{\bar{t}},\bar{A}_{t,\bar{t}}\right) =\max \left\{ \phi
\left( \bar{S}_{\bar{t}}-k\bar{A}_{t,\bar{t}}\right) ,0\right\} ,%
\end{array}
\label{Eq8.31}
\end{equation}%
where the non-dimensional parameter $k$ is called weighting; typically, $k=1$%
.

Start with the Bachelier model. Equations for pricing Asian Options with an
arithmetic average are as follows:%
\begin{equation}
\begin{array}{c}
d\hat{A}_{t}=\hat{S}_{t}dt,\ \ \ \hat{A}_{t}=0, \\ 
\\ 
d\hat{S}_{t}=r\hat{S}_{t}dt+\sigma d\hat{W}_{t},\ \ \ \hat{S}_{t}=S.%
\end{array}
\label{Eq8.32}
\end{equation}%
Thus, Eqs (\ref{Eq6.114}), (\ref{Eq6.115}) are applicable. All one needs is
the marginal distribution for $\bar{A}_{t,\bar{t}}$, which is Gaussian:%
\begin{equation}
\begin{array}{c}
\varpi \left( \bar{A}\right) \sim \mathrm{N}\left( R,\Sigma ^{2}\right) ,%
\end{array}
\label{Eq8.33}
\end{equation}%
where%
\begin{equation}
\begin{array}{c}
R=\mathsf{B}_{-r}\left( T\right) S,\ \ \ \Sigma ^{2}=\frac{\sigma ^{2}}{r}%
\left( \mathsf{B}_{0}\left( T\right) -2\mathsf{B}_{-r}\left( T\right) +%
\mathsf{B}_{-2r}\left( T\right) \right) .%
\end{array}
\label{Eq8.34}
\end{equation}%
Consider the discounted payoff of the Asian call option (say):%
\begin{equation}
\begin{array}{c}
U\left( t,\bar{A}\right) =\left( \frac{\bar{A}}{T}-K\right) _{+}.%
\end{array}
\label{Eq8.35}
\end{equation}%
The corresponding calculation is straightforward:%
\begin{equation}
\begin{array}{c}
U\left( t,S\right) =e^{-rT}\int\limits_{TK}^{\infty }\frac{\left( \frac{\bar{%
A}}{T}-K\right) e^{-\frac{\left( \bar{A}-R\right) ^{2}}{2\Sigma ^{2}}}}{%
\sqrt{2\pi \Sigma ^{2}}}d\bar{A} \\ 
\\ 
=\frac{e^{-rT}\Sigma }{T}\int\limits_{\frac{\left( TK-R\right) }{\Sigma }%
}^{\infty }\frac{\eta e^{-\frac{\eta ^{2}}{2}}}{\sqrt{2\pi }}d\eta -\frac{%
e^{-rT}\left( TK-R\right) }{T}\int\limits_{\frac{\left( TK-R\right) }{\Sigma 
}}^{\infty }\frac{e^{-\frac{\eta ^{2}}{2}}}{\sqrt{2\pi }}d\eta \\ 
\\ 
=\frac{e^{-rT}\Sigma }{T}\mathrm{n}\left( \frac{R-TK}{\Sigma }\right) -\frac{%
e^{-rT}\left( TK-R\right) }{T}\mathrm{N}\left( \frac{R-TK}{\Sigma }\right) .%
\end{array}
\label{Eq8.36}
\end{equation}%
Analytical pricing of Asian options with arithmetic averaging for the
Black-Scholes model is notoriously tricky; see \cite{Geman95, Rogers95,
Lipton99b, Lipton01}. At the same time, pricing Asian options with geometric
averaging can be done quickly; see \cite{Barrucci01, Lipton01, DiFrancesco05}%
, and references therein. Such options can be priced using formula (\ref%
{Eq6.46}) derived in Chapter \ref{Chap6}. An alternative approach based on
the path integral method is discussed in \cite{Devreese10}. Define%
\begin{equation}
\begin{array}{c}
\hat{x}_{t}=\dint\limits_{t}^{t}\hat{y}_{s}ds,\ \ \ \hat{y}_{t}=\ln \left( 
\hat{S}_{t}\right) .%
\end{array}
\label{Eq8.37}
\end{equation}%
Then%
\begin{equation}
\begin{array}{c}
d\hat{x}_{t}=\hat{y}_{t}dt,\ \ \ \hat{x}_{t}=0, \\ 
\\ 
d\hat{y}_{t}=\left( r-\frac{\sigma ^{2}}{2}\right) dt+\sigma d\hat{W}_{t},\
\ \ \hat{y}_{t}=\ln \left( \hat{S}_{t}\right) \equiv y.%
\end{array}
\label{Eq8.38}
\end{equation}%
The value of the option can be written as follows:%
\begin{equation}
\begin{array}{c}
U\left( t,S\right) =e^{-rT}\dint\limits_{x^{\ast }}^{\phi \infty }\varpi
\left( \bar{x}\right) \left( \exp \left( \frac{\bar{x}}{T}\right) -\exp
\left( \ln K\right) \right) d\bar{x},%
\end{array}
\label{Eq8.39}
\end{equation}%
where%
\begin{equation}
\begin{array}{c}
x^{\ast }=T\ln K.%
\end{array}
\label{Eq8.40}
\end{equation}%
Since Eq. (\ref{Eq8.38}) is a special case of Eq. (\ref{Eq6.79}), one can
use Eqs (\ref{Eq6.86}) to obtain the marginal distribution for $\bar{x}$,
which is a Gaussian distribution of the form:%
\begin{equation}
\begin{array}{c}
\varpi \left( \bar{x}\right) =\frac{\exp \left( -\frac{\left( \bar{x}%
-p\right) ^{2}}{2\sigma _{x}^{2}}\right) }{\sqrt{2\pi \sigma _{x}^{2}}}, \\ 
\\ 
\sigma _{x}^{2}=\frac{\sigma ^{2}T^{3}}{3},\ \ \ p=\ln \left( S\right) T+%
\frac{1}{2}\left( r-\frac{\sigma ^{2}}{2}\right) T^{2}.%
\end{array}
\label{Eq8.41}
\end{equation}%
Thus,%
\begin{equation}
\begin{array}{c}
U\left( t,S\right) =\mathcal{J}_{1}\left( t,S\right) -\mathcal{J}_{2}\left(
t,S\right) ,%
\end{array}
\label{Eq8.42}
\end{equation}%
where%
\begin{equation}
\begin{array}{c}
\mathcal{J}_{1}\left( t,S\right) =e^{-rT}\dint\limits_{x^{\ast }}^{\phi
\infty }\frac{\exp \left( -\frac{\left( \bar{x}-p\right) ^{2}}{2\sigma
_{x}^{2}}+\frac{\bar{x}}{T}\right) }{\sqrt{2\pi \sigma _{x}^{2}}}d\bar{x}%
=\phi e^{-\frac{1}{2}\left( r+\frac{\sigma ^{2}}{6}\right) T}S\mathrm{N}%
\left( \phi d_{+}\right) , \\ 
\\ 
\mathcal{J}_{2}\left( t,S\right) =e^{-rT}\dint\limits_{x^{\ast }}^{\phi
\infty }\frac{\exp \left( -\frac{\left( \bar{x}-p\right) ^{2}}{2\sigma
_{x}^{2}}+\ln \left( K\right) \right) }{\sqrt{2\pi \sigma _{x}^{2}}}d\bar{x}%
=\phi e^{-rT}K\mathrm{N}\left( \phi d_{-}\right) ,%
\end{array}
\label{Eq8.43}
\end{equation}%
where%
\begin{equation}
\begin{array}{c}
d_{\pm }=\frac{\ln \left( S/K\right) +\frac{1}{2}\left( r-\frac{\sigma ^{2}}{%
6}\pm \frac{\sigma ^{2}}{3}\right) T}{\sqrt{\sigma ^{2}T/3}}.%
\end{array}
\label{Eq8.44}
\end{equation}%
Finally, one obtains a well-known formula for the price of a fixed strike
Asian option with geometric averaging:%
\begin{equation}
\begin{array}{c}
U\left( t,S\right) =\phi \left( e^{-\frac{1}{2}\left( r+\frac{\sigma ^{2}}{6}%
\right) T}S\mathrm{N}\left( \phi d_{+}\right) -e^{-rT}K\mathrm{N}\left( \phi
d_{-}\right) \right) .%
\end{array}
\label{Eq8.45}
\end{equation}%
Of course, a similar formula holds when $r,\sigma $ are time-dependent. The
derivation, although very simple, seems to be new.

\subsection{Volatility and Variance Swaps and Swaptions\label{Sec85}}

\subsubsection{Volatility Swaps and Swaptions}

Recall that the Stein-Stein stochastic volatility model assumes that the
volatility is driven by an OU process; see \cite{Stein91}. One needs to find
Green's function associated with the following augmented SDEs:%
\begin{equation}
\begin{array}{c}
d\hat{x}_{t}=\hat{y}_{t}dt,\ \ \ \hat{x}_{t}=0, \\ 
\\ 
d\hat{y}_{t}=\left( \chi ^{\left( Vol\right) }-\kappa ^{\left( Vol\right) }%
\hat{y}_{t}\right) dt+\varepsilon ^{\left( Vol\right) }d\hat{W}_{t},\ \ \ 
\hat{y}_{t}=y^{\left( Vol\right) },%
\end{array}
\label{Eq8.46}
\end{equation}%
or, equivalently,%
\begin{equation}
\begin{array}{c}
d\hat{x}_{t}=\hat{y}_{t}dt,\ \ \ \hat{x}_{t}=0, \\ 
\\ 
d\hat{y}_{t}=\kappa ^{\left( Vol\right) }\left( \theta ^{\left( Vol\right) }-%
\hat{y}_{t}\right) dt+\varepsilon ^{\left( Vol\right) }d\hat{W}_{t},\ \ \ 
\hat{y}_{t}=y^{\left( Vol\right) },%
\end{array}
\label{Eq8.46a}
\end{equation}%
which describe the evolution of the volatility $\hat{\sigma}_{t}\equiv \hat{y%
}_{t}$ and its integral $\hat{x}_{t}$; Eqs (\ref{Eq8.46}) are identical to
Eqs (\ref{Eq6.98}).

It can be shown that the pair $\left( \bar{x},\bar{y}\right) $ has the
bivariate Gaussian distribution with the covariance matrix $\mathfrak{H}$
given by Eq. (\ref{Eq6.113}), and mean $\left( p,q\right) $ given by Eq. (%
\ref{Eq6.114}): 
\begin{equation}
\begin{array}{c}
\left( 
\begin{array}{c}
p \\ 
q%
\end{array}%
\right) =\left( 
\begin{array}{c}
T\theta ^{\left( Vol\right) }+\mathsf{\bar{B}}_{\kappa ^{\left( Vol\right)
}}\left( T\right) \left( y^{\left( Vol\right) }-\theta ^{\left( Vol\right)
}\right) \\ 
\theta ^{\left( Vol\right) }+\mathsf{A}_{\kappa ^{\left( Vol\right) }}\left(
T\right) \left( y^{\left( Vol\right) }-\theta ^{\left( Vol\right) }\right)%
\end{array}%
\right) ,%
\end{array}
\label{Eq8.47}
\end{equation}%
Since the marginal distribution of $\hat{x}_{t}$ given by Eq. (\ref{Eq6.115}%
) is Gaussian, the fair strike of a volatility swap with maturity $t$ is
simply the expected value of $\hat{x}_{t}/T$:%
\begin{equation}
\begin{array}{c}
VolSwap=\theta ^{\left( Vol\right) }+\left( y^{\left( Vol\right) }-\theta
^{\left( Vol\right) }\right) \frac{\mathsf{\bar{B}}_{\kappa ^{\left(
Vol\right) }}\left( T\right) }{T}.%
\end{array}
\label{Eq8.48}
\end{equation}%
Here 
\begin{equation}
\begin{array}{c}
\left[ VolSwap\right] =\left[ \frac{\chi ^{\left( Vol\right) }}{\kappa
^{\left( Vol\right) }}\right] =\left[ \theta ^{\left( Vol\right) }\right] =%
\left[ y\right] =\frac{1}{T^{1/2}}.%
\end{array}
\label{Eq8.84b}
\end{equation}%
Of course, one can calculate the expected value of $\hat{x}_{t}/T$ via more
straightforward means. To this end, Eq. (\ref{Eq8.48}) can be derived
directly by taking expectations of SDE (\ref{Eq8.46}). However, as one shall
see in the following subsection, Eq. (\ref{Eq6.115}) for the marginal
distribution $\varpi ^{\left( x\right) }\left( t,y^{\left( Vol\right) },\bar{%
t},\bar{x}\right) $ allows one to solve more interesting problems, such as
calculating prices of bonds and bond options, see below.

Moreover, by using this equation, one can price volatility swaptions with
payoffs of the form:%
\begin{equation}
\begin{array}{c}
U\left( \bar{t},\bar{x}\right) =\max \left\{ \phi \left( \bar{x}-x^{\ast
}\right) ,0\right\} .%
\end{array}
\label{Eq8.50}
\end{equation}%
The price $U\left( t,y^{\left( Vol\right) }\right) $ becomes: 
\begin{equation}
\begin{array}{c}
U\left( t,y^{\left( Vol\right) }\right) =e^{-rT}\phi \dint\limits_{x^{\ast
}}^{\phi \infty }\left( \bar{x}-x^{\ast }\right) \varpi ^{\left( x\right)
}\left( t,y^{\left( Vol\right) },\bar{t},\bar{x}\right) d\bar{x} \\ 
\\ 
=\frac{e^{-rT}\phi }{\sqrt{2\pi h_{0}\left( t,\bar{t}\right) }}%
\dint\limits_{x^{\ast }}^{\phi \infty }\left( \bar{x}-x^{\ast }\right) \exp
\left( \frac{\left( \bar{x}-p\right) ^{2}}{2h_{0}}\right) d\bar{x} \\ 
\\ 
=e^{-rT}\left( \phi \left( p-x^{\ast }\right) \mathrm{N}\left( \phi \frac{%
\left( p-x^{\ast }\right) }{\sqrt{h_{0}}}\right) +\sqrt{h_{0}}\mathrm{n}%
\left( \frac{\left( p-x^{\ast }\right) }{\sqrt{h_{0}}}\right) \right) .%
\end{array}
\label{Eq8.51}
\end{equation}%
It is clear that formula (\ref{Eq8.51}) is a variant of the Bachelier
formula (\ref{Eq8.8a}).

\subsubsection{Variance Swaps and Swaptions}

In contrast to volatility, which, despite common misconceptions, can be
negative, variance must be nonnegative since it is a square of a real-valued
quantity. Accordingly, the easiest way to model it is by using the augmented
Feller process with $\vartheta >0$; see Eq. (\ref{Eq7.87}).

Using Eq. (\ref{Eq7.115}), one can immediately obtain the following
expression for the fair value of a variance swap for the Feller process:%
\begin{equation}
\begin{array}{c}
VarSwap=\theta ^{\left( Var\right) }+\left( y^{\left( Var\right) }-\theta
^{\left( Var\right) }\right) \frac{\mathsf{\bar{B}}_{\kappa ^{\left(
Var\right) }}\left( T\right) }{T},%
\end{array}
\label{Eq8.52}
\end{equation}%
where $\theta ^{\left( Var\right) }=\chi ^{\left( Var\right) }/\kappa
^{\left( Var\right) }$. Here 
\begin{equation}
\begin{array}{c}
\left[ VarSwap\right] =\left[ \theta ^{\left( Var\right) }\right] =\left[
y^{\left( Var\right) }\right] =\frac{1}{T}.%
\end{array}
\label{Eq8.52a}
\end{equation}%
While formulas (\ref{Eq8.48}), (\ref{Eq8.52}) look the same but deal with
the volatility and variance, respectively, the corresponding parameters have
different meanings.

Alternatively, one can use the degenerate augmented OU process, see Eqs (\ref%
{Eq7.167}). Averaging away stochastic terms, one gets the following formula
for the fair price of the variance swap:%
\begin{equation}
\begin{array}{c}
VarSwap=\left( \theta ^{\left( Vol\right) }\right) ^{2}+\left( \left(
y^{\left( Vol\right) }\right) ^{2}-\left( \theta ^{\left( Vol\right)
}\right) ^{2}\right) \frac{\mathsf{\bar{B}}_{\kappa ^{\left( Vol\right)
}}\left( T\right) }{T}.%
\end{array}
\label{Eq8.53}
\end{equation}%
It is clear that Eqs (\ref{Eq8.52}) and (\ref{Eq8.53}) provide different
fair values for a variance swap, although these values asymptotically agree.
This fact reflects the so-called model risk - by using different models, one
gets different answers to the same question.

Eq. (\ref{Eq7.111}) can be used to calculate the price of a variance
swaption:%
\begin{equation}
\begin{array}{c}
U\left( t,y^{\left( Var\right) }\right) =\frac{1}{2\pi }\dint\limits_{x^{%
\ast }}^{\phi \infty }\dint\limits_{-\infty }^{\infty }\phi \left( \bar{x}%
-x^{\ast }\right) \digamma \left( \bar{t},k\right) e^{ik\bar{x}}dkd\bar{x}
\\ 
\\ 
=\frac{1}{2\pi }\dint\limits_{-\infty }^{\infty }\digamma \left( \bar{t}%
,k\right) \left( \phi \dint\limits_{x^{\ast }}^{\phi \infty }\left( \bar{x}%
-x^{\ast }\right) e^{ik\bar{x}}d\bar{x}\right) dk \\ 
\\ 
=\frac{1}{2\pi }\underset{\epsilon \rightarrow 0}{\lim }\dint\limits_{-%
\infty }^{\infty }\digamma \left( \bar{t},k\right) e^{ikx^{\ast }}\left( -%
\frac{\partial }{\partial \epsilon }\dint\limits_{x^{\ast }}^{\phi \infty
}e^{\left( ik-\phi \epsilon \right) \bar{x}}d\bar{x}\right) \\ 
\\ 
=\frac{1}{2\pi }\underset{\epsilon \rightarrow 0}{\lim }\dint\limits_{-%
\infty }^{\infty }\frac{\digamma \left( \bar{t},k\right) e^{ikx^{\ast }}}{%
\left( ik-\phi \epsilon \right) ^{2}}dk,%
\end{array}
\label{Eq8.55}
\end{equation}%
where $\digamma \left( \bar{t},k\right) $ is given by Eq. (\ref{Eq7.112}).

\subsection{Automated Market Makers\label{Sec86}}

Variance and volatility swaps had long occupied a specific niche within the
financial product landscape. Recently, they experienced an unexpected surge
in interest due to the influence of cryptocurrency trading. These swaps have
proven effective in hedging impermanent loss, a phenomenon generated by
automated market makers; see \cite{Lipton21b, Lipton22, Cartea23, Fukusawa23}
and others. This section closely follows \cite{Lipton21b}.

Let us consider a smart contract (SC), called an automated market maker
(AMM) designed to facilitate exchanges of two tokens, $TN_{1}$ and $TN_{2}$.
The analytical formula for the price of the second token in terms of the
first, defines the nature of the contract. AMMs have gained significant
traction in recent years. Initially, anyone can participate as a market
maker and liquidity provider by depositing $TN_{1}$ and $TN_{2}$
simultaneously and in the correct ratio into the collateral pool.
Subsequently, participants can withdraw one token from the pool by
delivering the other token according to the rules established by the
underlying SC. While AMMs excel in facilitating stablecoin swaps, they can
easily accommodate the exchange of various tokens, such as swapping a
stablecoin, say USDT, for ethereum (ETH).

The actual exchange rate is determined by rules that rely on prior
agreement. The available options are the constant sum, constant product, and
mixture rules. Sources including \cite%
{Angeris19,Egorov19,Zhang18,Lipton21b,Lipton22} and references therein offer
detailed coverage of AMMs and comprehensive insights into their mechanisms.

Assuming that initially tokens $TN_{1}$, $TN_{2}$ are equal in value, one
can define a constant sum AMM:%
\begin{equation}
\begin{array}{c}
X+Y=\Sigma _{0},\ \ \ X_{0}=Y_{0}=N,\ \ \ \Sigma _{0}=2N.%
\end{array}
\label{Eq8.56}
\end{equation}%
Here $X,Y$ are the quantities of $TN_{1}$, $TN_{2}$ in the pool. Eq. (\ref%
{Eq8.56}) yields:%
\begin{equation}
\begin{array}{c}
Y=\Sigma _{0}-X,\ \ \ \left\vert \frac{dY}{dX}\right\vert =1.%
\end{array}
\label{Eq8.57}
\end{equation}%
As per Eq. (\ref{Eq8.57}), the pool reaches depletion at $X=\Sigma _{{0}}$,
as it becomes advantageous for an arbitrageur to increase $X$ from $N$ to $%
2N $ when $TN_{{2}}$ surpasses $TN_{1}$ in value. The marginal price of $%
TN_{2}$ relative to $TN_{1}$, as expressed in the second equation (\ref%
{Eq8.57}), remains consistent and equal to one. A constant price is optimal
for a constant sum AMM, particularly when dealing with stablecoins like $%
TN_{1}$ and $TN_{2}$, whose prices fluctuate mildly around their equilibrium
values. Depleting the pool is rational in scenarios where transaction fees
are nonexistent, even with a minimal deviation from equilibrium. However,
under more realistic conditions with non-zero transaction fees, arbitrage
becomes profitable only if the deviation surpasses a certain threshold.

The constant product rule defines more intricate and, importantly, practical
AMMs:%
\begin{equation}
\begin{array}{c}
XY=\Pi _{0},\ \ \ X_{0}=Y_{0}=N,\ \ \ \Pi _{0}=N^{2}.%
\end{array}
\label{Eq8.58}
\end{equation}%
It is clear that%
\begin{equation}
\begin{array}{c}
Y=\frac{\Pi _{0}}{X},\ \ \ \left\vert \frac{dY}{dX}\right\vert =\frac{\Pi
_{0}}{X^{2}}.%
\end{array}
\label{Eq8.59}
\end{equation}%
Consequently, an arbitrageur is unable to deplete such a pool, allowing it
to persist indefinitely. In this scenario, it becomes evident that the price
of $TN_{2}$ relative to $TN_{1}$ is no longer steady; instead, it rises (or
falls) as $X$ decreases (or increases).

\begin{figure}[tbp]
\begin{center}
\includegraphics[width=0.8\textwidth]
{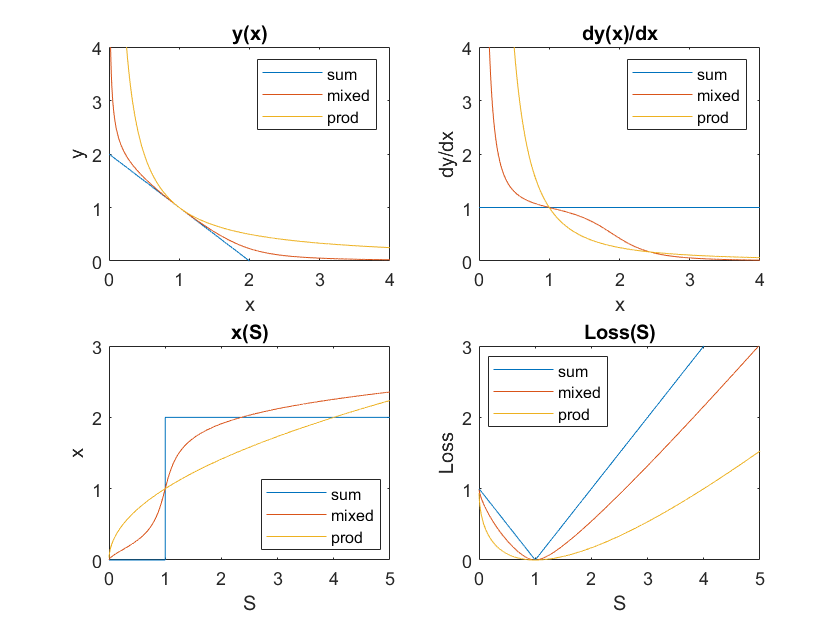}
\end{center}
\par
\vspace{-10pt}
\caption{The constant sum, constant product, and mixed rule curves, along
with the relative prices of $TN_{2}$ in terms of $TN_{1}$ and the associated
impermanent losses; $\protect\alpha =10$. Own graphics.}
\label{Fig8.2}
\end{figure}

To make liquidity provision more attractive to potential market makers, one
can generalize the constant sum and constant product rules. Expressions (\ref%
{Eq8.56}), (\ref{Eq8.58}) representing these rules can be formulated as
follows:%
\begin{equation}
\begin{array}{c}
\left( \frac{\Sigma }{\Sigma _{0}}-1\right) =0,\ \ \ X_{0}=Y_{0}=N,\ \ \
\Sigma _{0}=2N, \\ 
\\ 
\left( \frac{\Pi _{0}}{\Pi }-1\right) =0,\ \ \ X_{0}=Y_{0}=N,\ \ \ \Pi
_{0}=N^{2}.%
\end{array}
\label{Eq8.60}
\end{equation}%
where $\Sigma =X+Y$, $\Pi =XY$ are the current sum and product,
respectively. These rules can be combines as follows: 
\begin{equation}
\begin{array}{c}
\left( \frac{\Pi _{0}}{\Pi }-1\right) +\alpha \left( \frac{\Sigma }{\Sigma
_{0}}-1\right) =0, \\ 
\\ 
X_{0}=Y_{0}=N,\ \ \ \Sigma _{0}=2N,\ \ \ \Pi _{0}=N^{2}.%
\end{array}
\label{Eq8.61}
\end{equation}%
Here, $\alpha >0$ is an adaptive parameter, characterizing the transition
from the constant product to the constant sum rule. The product $\Pi $ is in
the denominator to avoid the possibility of exhausting the entire pool and
ensuring that:%
\begin{equation}
\begin{array}{c}
Y\left( X\right) \underset{X\rightarrow 0}{\rightarrow }\infty ,\ \ \ \ \
X\left( Y\right) \underset{Y\rightarrow 0}{\rightarrow }\infty .%
\end{array}
\label{Eq8.62}
\end{equation}%
Certainly, when AMM liquidity providers are exposed to arbitragers, they
face potential losses stemming from a decline in collateral value below its
buy-and-hold threshold. In financial terms, an AMM liquidity provider is an
option seller experiencing negative convexity, so that they must impose
transaction fees to offset these losses. The losses incurred by AMMs are
(somewhat misleadingly) termed "impermanent" because they tend to vanish
under the assumption of mean reversion. However, the validity of the
mean-reversion assumption in real-world scenarios can vary. Introducing
variables $x$ and $y$ where $X=Nx$ and $Y=Ny$, one can express the constant
sum rule described by Eqs (\ref{Eq8.56}), (\ref{Eq8.57}) as follows:%
\begin{equation}
\begin{array}{c}
x+y=2,\ \ \ x_{0}=y_{0}=1,%
\end{array}
\label{Eq8.63}
\end{equation}%
\begin{equation}
\begin{array}{c}
y\left( x\right) =2-x,\ \ \left\vert \frac{dy}{dx}\right\vert =1.%
\end{array}
\label{Eq8.64}
\end{equation}%
In terms of $x,y$, the constant product rule given by Eqs (\ref{Eq8.58}), (%
\ref{Eq8.59}) can be written in the form:%
\begin{equation}
\begin{array}{c}
xy=1,\ \ \ \ x_{0}=y_{0}=1,%
\end{array}
\label{Eq8.65}
\end{equation}%
\begin{equation}
\begin{array}{c}
y\left( x\right) =\frac{1}{x},\ \ \left\vert \frac{dy}{dx}\right\vert =\frac{%
1}{x^{2}}.%
\end{array}
\label{Eq8.66}
\end{equation}%
Finally, mixed rule Eqs (\ref{Eq8.61}) written in terms of $x,y$ become:%
\begin{equation}
\begin{array}{c}
\left( \frac{1}{xy}-1\right) +\alpha \left( \frac{x+y}{2}-1\right) =0,\ \ \
x_{0}=y_{0}=1.%
\end{array}
\label{Eq8.67}
\end{equation}%
Straightforward algebra yields: 
\begin{equation}
\begin{array}{c}
y_{\alpha }=\frac{1}{2\alpha }\left( -\left( 2\left( 1-\alpha \right)
+\alpha x\right) +\left( \left( 2\left( 1-\alpha \right) +\alpha x\right)
^{2}+\frac{8\alpha }{x}\right) ^{1/2}\right) , \\ 
\\ 
\frac{dy_{\alpha }}{dx}=\frac{1}{2}\left( -1+\frac{2\left( 1-\alpha \right)
+\alpha x-\left. 4\right/ x^{2}}{\left( \left( 2\left( 1-\alpha \right)
+\alpha x\right) ^{2}+\frac{8\alpha }{x}\right) ^{1/2}}\right) , \\ 
\\ 
\frac{d^{2}y_{\alpha }}{dx^{2}}=\frac{1}{2}\left( \frac{\alpha +\left.
8\right/ x^{3}}{\left( \left( 2\left( 1-\alpha \right) +\alpha x\right) ^{2}+%
\frac{8\alpha }{x}\right) ^{1/2}}-\frac{\alpha \left( 2\left( 1-\alpha
\right) +\alpha x-\left. 4\right/ x^{2}\right) }{\left( \left( 2\left(
1-\alpha \right) +\alpha x\right) ^{2}+\frac{8\alpha }{x}\right) ^{3/2}}%
\right) .%
\end{array}
\label{Eq8.68}
\end{equation}%
Assume that the external exchange price $S$ of $TN_{2}$ expressed in terms
of $TN_{1}$ moves away from its equilibrium value $S_{0}=1$. Let $S>1$. For
the constant sum contract, an arbitrageur can choose a number $x$, $1<x\leq
2 $, and deliver $\left( x-1\right) $ of $TN_{1}$ tokens to the pool in
exchange for getting $\left( x-1\right) $ of $TN_{2}$ tokens. The profit or
loss ($P\&L$) is given by%
\begin{equation}
\begin{array}{c}
\Omega \left( x\right) =\left( S-1\right) \left( x-1\right) .%
\end{array}
\label{Eq8.69}
\end{equation}%
Since $\Omega $ is a linear function of $x$, it is rational to exhaust the
entire pool, by choosing the following optimal values $\left( x^{\ast
},y^{\ast },\Omega ^{\ast }\right) $:%
\begin{equation}
\begin{array}{c}
x^{\ast }=2,\ \ \ y^{\ast }=0,\ \ \ \Omega ^{\ast }=\left( S-1\right) .%
\end{array}
\label{Eq8.70}
\end{equation}%
Similarly, when $S<1$:%
\begin{equation}
\begin{array}{c}
x^{\ast }=0,\ \ \ y^{\ast }=2,\ \ \ \Omega ^{\ast }=-\left( S-1\right) .%
\end{array}
\label{Eq8.71}
\end{equation}%
The arbitraged portfolio's value is $\pi ^{\ast }\left( S\right) $, where%
\begin{equation}
\begin{array}{c}
\pi ^{\ast }\left( S\right) =\left\{ 
\begin{array}{c}
2,\ \ \ S\geq 1, \\ 
2S,\ \ \ S<1.%
\end{array}%
\right. ,%
\end{array}
\label{Eq8.72}
\end{equation}%
while the buy and hold portfolio's value is $\left( S+1\right) $. The
difference $\omega $ has the form 
\begin{equation}
\begin{array}{c}
\omega =\left( S+1\right) -\pi ^{\ast }\left( S\right) =\left\vert
S-1\right\vert .%
\end{array}
\label{Eq8.73}
\end{equation}%
In the DeFi parlance, $\omega $ is termed as impermanent loss. However, this
description can be misleading as the loss can swiftly become permanent when $%
S$ moves away from its assumed "equilibrium" value of one. The percentage
loss in the actual portfolio compared to the buy-and-hold portfolio is
structured as follows:%
\begin{equation}
\begin{array}{c}
\lambda =1-\frac{\left\vert S-1\right\vert }{S+1}.%
\end{array}
\label{Eq8.74}
\end{equation}%
A similar calculation can be performed for the constant product contract.
When $S$ deviates from one, an arbitrageur can choose a number $x>1$ and
deliver $\left( x-1\right) $ tokens $TN_{1}$ to the pool, while taking $%
\left( 1-y\right) $ tokens $TN_{2}$ from the pool, where $y=1/x$. The $P\&L$
has the form: 
\begin{equation}
\begin{array}{c}
\Omega \left( x\right) =\left( S\left( 1-\frac{1}{x}\right) -\left(
x-1\right) \right) .%
\end{array}
\label{Eq8.75}
\end{equation}%
The optimality condition has the form 
\begin{equation}
\begin{array}{c}
\Omega ^{\prime }\left( x\right) =\left( \frac{S}{x^{2}}-1\right) =0,%
\end{array}
\label{Eq8.76}
\end{equation}%
so that the corresponding optimal values $\left( x^{\ast },y^{\ast },\Omega
^{\ast }\right) $ are%
\begin{equation}
\begin{array}{c}
x^{\ast }=\sqrt{S},\ \ \ y^{\ast }=\frac{1}{\sqrt{S}},\ \ \ \Omega ^{\ast
}=\left( \sqrt{S}-1\right) ^{2}.%
\end{array}
\label{Eq8.77}
\end{equation}%
Hence, a constant product collateral pool remains inexhaustible. Throughout
each phase, the ideal quantities of $TN_{1}$ and $TN_{2}$ maintained in the
portfolio are both $\sqrt{S}$. As both tokens' values within the portfolio
must equate, the suggested optimal value of $TN_{2}$ in terms of $TN_{1}$ is 
$S^{\ast }=x^{\ast }/y^{\ast }=S$. The value of the arbitrage-driven
portfolio stands at $\pi ^{\ast }=2\sqrt{S}$, whereas the value of the
buy-and-hold portfolio amounts to $\left( S+1\right) $. The difference is
given by: 
\begin{equation}
\begin{array}{c}
\omega =\left( S+1\right) -2\sqrt{S}=\left( \sqrt{S}-1\right) ^{2}.%
\end{array}
\label{Eq8.78}
\end{equation}%
The corresponding percentage loss is%
\begin{equation}
\begin{array}{c}
\lambda =1-\frac{2\sqrt{S}}{\left( S+1\right) }=\frac{\left( \sqrt{S}%
-1\right) ^{2}}{\left( S+1\right) }.%
\end{array}
\label{Eq8.79}
\end{equation}%
For the mixed rule AMM, the arbitrageur's profit for $S>1$ has the form:%
\begin{equation}
\begin{array}{c}
\Omega \left( x\right) =\left( S\left( 1-y_{\alpha }\left( x\right) \right)
-\left( x-1\right) \right) ,%
\end{array}
\label{Eq8.80}
\end{equation}%
with the optimum achieved at $x_{\alpha }^{\ast },y_{\alpha }^{\ast },\Omega
_{\alpha }^{\ast }$ of the form: 
\begin{equation}
\begin{array}{c}
y_{\alpha }^{\prime }\left( x_{\alpha }^{\ast }\right) =-\frac{1}{S},\ \ \
y_{\alpha }^{\ast }=y_{\alpha }\left( x_{\alpha }^{\ast }\right) ,\ \ \
\Omega _{\alpha }^{\ast }=\left( S\left( 1-y_{\alpha }^{\ast }\right)
-\left( x_{\alpha }^{\ast }-1\right) \right) ,%
\end{array}
\label{Eq8.81}
\end{equation}%
with the optimal $x_{\alpha }^{\ast }$ via the Newton-Raphson method
starting with a suitable $x_{\alpha }^{\left( 0\right) }$: 
\begin{equation}
\begin{array}{c}
x_{\alpha }^{\left( n+1\right) }=x_{\alpha }^{\left( n\right) }-\frac{%
y_{\alpha }^{\prime }\left( x_{\alpha }^{\left( n\right) }\right) +\frac{1}{S%
}}{y_{\alpha }^{\prime \prime }\left( x_{\alpha }^{\left( n\right) }\right) }%
.%
\end{array}
\label{Eq8.82}
\end{equation}%
Here $y_{\alpha }^{\prime }$, $y_{\alpha }^{\prime \prime }$ are given by
Eqs (\ref{Eq8.68}). Due to quadratic convergence of the Newton-Raphson
method, ten iterations provide machine accuracy, so that one can set $%
x_{\alpha }^{\ast }=x_{\alpha }^{\left( 10\right) }$. The value of the
arbitraged portfolio is%
\begin{equation}
\begin{array}{c}
\pi ^{\ast }=x_{\alpha }^{\ast }+Sy_{\alpha }\left( x_{\alpha }^{\ast
}\right) .%
\end{array}
\label{Eq8.83}
\end{equation}%
Figure \ref{Fig8.2} shows the constant sum, constant product, and mixed rule
curves, along with the relative prices of $TN_{2}$ in terms of $TN_{1}$ and
the associated impermanent losses. It demonstrates that deviations from the
tokens' equilibrium values result in losses for the market maker.
Impermanent loss is relatively minor for the constant product rule, moderate
for the mixed rule, and notably high for the constant sum rule. Even when
the price $S$ sways by a factor of five from its equilibrium, the
impermanent loss within the constant product rule remains manageable,
especially compared to the mixed rule. 

One can use variance swaps to hedge impermanent loss. For brevity, consider
the constant product rule. The corresponding impermanent loss, shown in
Figure \ref{Fig8.2}, is given by Eq. (\ref{Eq8.78}). It can be viewed as a
payoff of a non-standard European option. The hedging approach is
straightforward - one approximates this payoff with payoffs of options,
which can be priced explicitly. Specifically, one can use two such options:
the log and entropy contracts. The corresponding payoffs are as follows:

\begin{equation}
\begin{array}{c}
U^{LC}\left( S\right) =c^{LC}\left( S-1-\ln \left( S\right) \right) ,%
\end{array}
\label{Eq8.84}
\end{equation}%
\begin{equation}
\begin{array}{c}
U^{EC}\left( S\right) =c^{EC}\left( S\ln \left( S\right) -\left( S-1\right)
\right) .%
\end{array}
\label{Eq8.85}
\end{equation}%
The prefactors $c^{LC}$, $c^{EC}$ are chosen in such a way that the value of
the impermanent loss (\ref{Eq8.78}) and the hypothetical payoffs (\ref%
{Eq8.84}) and ((\ref{Eq8.85})) agree at the point $S=1$ up to the third
derivative, so that%
\begin{equation}
\begin{array}{c}
c^{LC}=c^{EC}=\frac{1}{2}.%
\end{array}
\label{Eq8.86}
\end{equation}%
Assuming that $S$ is driven by the geometric Brownian motion with stochastic
volatility, one can find the value of the log and entropy contracts at time $%
t$ at the point $S=1$, by solving the following problems: 
\begin{equation}
\begin{array}{c}
U_{t}+\frac{1}{2}v\left( S^{2}U_{SS}+2\varepsilon \rho SU_{Sv}+\varepsilon
^{2}U_{vv}\right) +\left( \chi -\kappa v\right) U_{v}=0,%
\end{array}
\label{Eq8.87}
\end{equation}%
supplied with terminal conditions of the form%
\begin{equation}
\begin{array}{c}
U^{LC}\left( \bar{t},S,v\right) =\left( S-1-\ln \left( S\right) \right) ,%
\end{array}
\label{Eq8.88}
\end{equation}%
and 
\begin{equation}
\begin{array}{c}
U^{EC}\left( \bar{t},S,v\right) =\left( S\ln \left( S\right) -\left(
S-1\right) \right) ,%
\end{array}
\label{Eq8.89}
\end{equation}%
respectively.

The corresponding solutions are well-known and easy to find. One can present 
$U^{LC}\left( t,S\right) $ as follows%
\begin{equation}
\begin{array}{c}
U^{LC}\left( t,S,v,\bar{t}\right) =\Phi ^{LC}\left( t,v,\bar{t}\right)
+\left( S-1-\ln \left( S\right) \right) ,%
\end{array}
\label{Eq8.90}
\end{equation}%
where%
\begin{equation}
\begin{array}{c}
\Phi _{t}^{LC}+\frac{1}{2}v\left( 1+\varepsilon ^{2}\Phi _{vv}^{LC}\right)
+\left( \chi -\kappa v\right) \Phi _{v}^{LC}=0, \\ 
\\ 
\Phi ^{LC}\left( \bar{t},v,\bar{t}\right) =0.%
\end{array}
\label{Eq8.91}
\end{equation}%
Accordingly,%
\begin{equation}
\begin{array}{c}
\Phi ^{LC}\left( t,v,\bar{t}\right) =\alpha ^{LC}\left( t,\bar{t}\right)
+\beta ^{LC}\left( t,\bar{t}\right) v,%
\end{array}
\label{Eq8.92}
\end{equation}%
where%
\begin{equation}
\begin{array}{c}
\alpha _{t}^{LC}\left( t,\bar{t}\right) +\chi \beta ^{LC}\left( t,\bar{t}%
\right) =0,\ \ \ \alpha ^{LC}\left( \bar{t},\bar{t}\right) =0, \\ 
\\ 
\beta _{t}^{LC}\left( t,\bar{t}\right) -\kappa \beta ^{LC}\left( t,\bar{t}%
\right) +\frac{1}{2}=0,\ \ \ \beta ^{LC}\left( \bar{t},\bar{t}\right) =0.%
\end{array}
\label{Eq8.93}
\end{equation}%
Thus,%
\begin{equation}
\begin{array}{c}
\alpha ^{LC}\left( t,\bar{t}\right) =\frac{\chi }{2\kappa }\left( T-\mathsf{%
\bar{B}}_{\kappa }\left( T\right) \right) , \\ 
\\ 
\beta ^{LC}\left( t,\bar{t}\right) =\frac{\mathsf{\bar{B}}_{\kappa }\left(
T\right) }{2},%
\end{array}
\label{Eq8.94}
\end{equation}%
so that%
\begin{equation}
\begin{array}{c}
U^{LC}\left( t,S,v,\bar{t}\right) =\frac{1}{2}\left( \frac{\chi T}{\kappa }%
+\left( v-\frac{\chi }{2\kappa }\right) \mathsf{\bar{B}}_{\kappa }\left(
T\right) \right) +\left( S-1-\ln \left( S\right) \right) .%
\end{array}
\label{Eq8.94a}
\end{equation}%
It is clear that $U^{LC}\left( t,1,v,\bar{t}\right) $ is in agreement with
Eq. (\ref{Eq8.52}).

One can calculate $U^{EC}\left( t,S,v\right) $ in a similar fashion by
representing it in the form:%
\begin{equation}
\begin{array}{c}
U^{EC}\left( t,S,v,\bar{t}\right) =\Phi ^{EC}\left( t,v,\bar{t}\right)
S+\left( S\ln \left( S\right) -\left( S-1\right) \right) ,%
\end{array}
\label{Eq8.95}
\end{equation}%
where, once the common factor $S$ is omitted,%
\begin{equation}
\begin{array}{c}
\Phi _{t}^{EC}+\frac{1}{2}v\left( 1+2\varepsilon \rho \Phi
_{v}^{EC}+\varepsilon ^{2}\Phi _{vv}^{EC}\right) +\left( \chi -\kappa
v\right) \Phi _{v}^{EC}=0, \\ 
\\ 
\Phi ^{LC}\left( \bar{t},v,\bar{t}\right) =0.%
\end{array}
\label{Eq8.96}
\end{equation}%
As before,%
\begin{equation}
\begin{array}{c}
\Phi ^{EC}\left( t,v,\bar{t}\right) =\alpha ^{EC}\left( t,\bar{t}\right)
+\beta ^{EC}\left( t,\bar{t}\right) v,%
\end{array}
\label{Eq8.97}
\end{equation}%
where%
\begin{equation}
\begin{array}{c}
\alpha _{t}^{EC}\left( t,\bar{t}\right) +\chi \beta ^{EC}\left( t,\bar{t}%
\right) =0,\ \ \ \alpha ^{EC}\left( \bar{t},\bar{t}\right) =0, \\ 
\\ 
\beta _{t}^{EC}\left( t,\bar{t}\right) -\left( \kappa -\varepsilon \rho
\right) \beta ^{EC}\left( t,\bar{t}\right) +\frac{1}{2}=0,\ \ \ \beta
^{EC}\left( \bar{t},\bar{t}\right) =0.%
\end{array}
\label{Eq8.98}
\end{equation}%
Thus,%
\begin{equation}
\begin{array}{c}
\alpha ^{EC}\left( t,\bar{t}\right) =\frac{\chi }{2\left( \kappa
-\varepsilon \rho \right) }\left( T-\mathsf{\bar{B}}_{\kappa -\varepsilon
\rho }\left( T\right) \right) , \\ 
\\ 
\beta ^{EC}\left( t,\bar{t}\right) =\frac{\mathsf{\bar{B}}_{\kappa
-\varepsilon \rho }\left( T\right) }{2},%
\end{array}
\label{Eq8.99}
\end{equation}%
\begin{equation}
\begin{array}{c}
U^{EC}\left( t,S,v,\bar{t}\right) =\frac{1}{2}\left( \frac{\chi T}{\kappa
_{1}}+\left( v-\frac{\chi }{2\kappa _{1}}\right) \mathsf{\bar{B}}_{\kappa
_{1}}\left( T\right) \right) S+\left( S\ln \left( S\right) -\left(
S-1\right) \right) .%
\end{array}
\label{Eq8.100}
\end{equation}%
where $\kappa _{1}=\kappa -\varepsilon \rho $. Eqs (\ref{Eq8.94a}), (\ref%
{Eq8.100}) allow us to estimate the amount a liquidity provider needs to
collect to cover the expected impermanent loss.

However, it turns out (which comes a surprise, at least to the present
author), that one can solve the pricing problem (\ref{Eq8.87}) with the
exact terminal condition (\ref{Eq8.78}) \emph{explicitly}, since the
impermantent loss does not have any optionality and is a linear combination
of the so-called power contracts with payoffs of the form $S,\sqrt{S},1$.%
\footnote{%
This fact is true even for nonzero interest rates.}

Thus, by using an appropriate Kelvin wave, one can solve the problem (\ref%
{Eq8.87}) with the power terminal condition:%
\begin{equation}
\begin{array}{c}
U^{\left( \nu \right) }\left( S\right) =S^{\nu }.%
\end{array}
\label{Eq8.101}
\end{equation}%
Of course, for $\nu =0,1$, the solution is trivial, for other values of $\nu 
$, additional efforts are needed. To be concrete, it is assumed that $0<\nu
<1$; for other values of $\nu $, the solution can blow-up in real time. The
price of the power contract with the payoff $S^{\nu }$ (even when the
interest rate $r\neq 0$) is given by a Kelvin wave:%
\begin{equation}
\begin{array}{c}
V\left( t,S,\bar{t}\right) =e^{\alpha \left( t,\bar{t}\right) +\beta \left(
t,\bar{t}\right) v}S^{\nu },%
\end{array}
\label{Eq8.102}
\end{equation}%
where $\alpha \left( t\right) ,\beta \left( t\right) $ solve the following
system of ODEs:%
\begin{equation}
\begin{array}{c}
\alpha _{t}\left( t,\bar{t}\right) +\chi \beta \left( t,\bar{t}\right) =0,\
\ \ \alpha \left( \bar{t},\bar{t}\right) =0, \\ 
\\ 
\beta _{t}\left( t,\bar{t}\right) +\frac{\varepsilon ^{2}}{2}\beta
^{2}\left( t,\bar{t}\right) -\left( \kappa -\nu \rho \varepsilon \right)
\beta \left( t,\bar{t}\right) +\frac{\nu \left( \nu -1\right) }{2}=0,\ \ \
\beta \left( \bar{t},\bar{t}\right) =0,%
\end{array}
\label{Eq8.103}
\end{equation}%
which has an explicit solution given by Eqs (\ref{Eq7.99})-(\ref{Eq7.102})
with%
\begin{equation}
\begin{array}{c}
\lambda _{\pm }^{2}+\left( \kappa -\nu \rho \varepsilon \right) \lambda
_{\pm }+\frac{\varepsilon ^{2}\nu \left( \nu -1\right) }{4}=0,%
\end{array}
\label{Eq8.104a}
\end{equation}%
\begin{equation}
\begin{array}{c}
\lambda _{\pm }=\mu \pm \zeta , \\ 
\\ 
\mu =-\frac{\left( \kappa -\nu \rho \varepsilon \right) }{2},\ \ \ \zeta =%
\frac{\sqrt{\left( \kappa -\nu \rho \varepsilon \right) ^{2}-\varepsilon
^{2}\nu \left( \nu -1\right) }}{2}.%
\end{array}
\label{Eq8.104b}
\end{equation}%
Thus, both $\mu $ and $\zeta $ are real. Accordingly, one can represent $%
\alpha $ and $\beta $ as follows:%
\begin{equation}
\begin{array}{c}
\alpha \left( T\right) =-\frac{2\chi }{\varepsilon ^{2}}\left( -\frac{\left(
\kappa -\nu \rho \varepsilon \right) T}{2}+\ln \left( \frac{\zeta \cosh
\left( T\right) -\mu \sinh \left( T\right) }{\zeta }\right) \right) , \\ 
\\ 
\beta \left( T\right) =\frac{\nu \left( \nu -1\right) \left( \sinh \left(
T\right) \right) }{2\left( \zeta \cosh \left( T\right) -\mu \sinh \left(
T\right) \right) }.%
\end{array}
\label{Eq8.104c}
\end{equation}%
The exact impermanent loss and its approximations are shown in Figure \ref%
{Fig8.3}. 
\begin{figure}[tbp]
\begin{center}
\includegraphics[width=0.8\textwidth]
{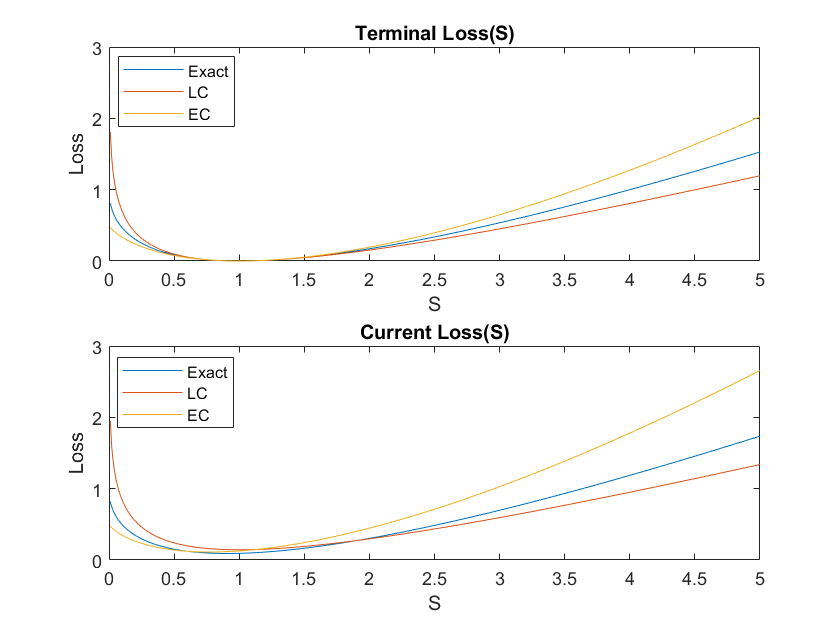}
\end{center}
\par
\vspace{-10pt}
\caption{The exact impermanent loss and its approximations via log and
enthropy contracts. The corresponding parameters are $\protect\alpha =10$, $%
T=1$. Own graphics.}
\label{Fig8.3}
\end{figure}
This Figure shows that $\max \left( U^{LC},U^{EC}\right) $ strictly
dominates the exact solution $U^{EX}$, but, as time of liquidity provision
grows, the corresponding upper bound becomes inaccurate.

The calculation of the mixed rule impermanent loss and its approximations is
left to the reader as a difficult exercise.

In $P\&L$ modeling for AMMs, the primary aim is to ensure that the liquidity
provider makes a profit or, at least, does not incur a loss. This profit
stems from transaction fees charged by the pool, which must exceed the
impermanent loss caused by collateral value dropping below its buy-and-hold
threshold. These fees must exceed the impermanent loss. An arbitrageur needs
to add more tokens to the pool than the rule dictates to account for
transaction fees. In the presence of non-zero transaction costs the actual
composition of the pool is time- and path-dependent. Given the stochastic
nature of the log price, the analysis of $\overline{P\&L}$ can only be
conducted probabilistically through Monte Carlo simulations; see \cite%
{Lipton21a, Lipton22}. For the parameter selection used by these authors,
automated liquidity provision is profitable on average. This profitability
arises because the AMM accumulates more tokens by the process's conclusion
than initially possessed.

\subsection{Bonds and Bond Options\label{Sec87}}

\subsubsection{Background}

We now use the machinery developed in\ Chapters \ref{Chap6}, \ref{Chap7} for
pricing bonds and bond options in some popular fixed-income models,
including Vasicek-Hull-White and Cox-Ingersoll-Ross.

\subsubsection{Vasicek Model}

One can use formulas derived in the previous subsection to price bonds and
bond options in the popular Vasicek and Hull-White models; see \cite%
{Vasicek77, Hull90}. Recall that Vasicek postulated the following dynamics
for the short interest rate $\hat{y}_{t}$:%
\begin{equation}
\begin{array}{c}
d\hat{y}_{t}=\left( \chi -\kappa \hat{y}_{t}\right) dt+\varepsilon d\hat{W}%
_{t},\ \ \ \hat{y}_{t}=y,%
\end{array}
\label{Eq8.105}
\end{equation}%
or, alternatively,%
\begin{equation}
\begin{array}{c}
d\hat{y}_{t}=\kappa \left( \theta -\hat{y}_{t}\right) dt+\varepsilon d\hat{W}%
_{t},\ \ \ \hat{y}_{t}=y,%
\end{array}
\label{Eq8.106}
\end{equation}%
where $\kappa \theta =\chi $.

At time $t$, the price of a bond maturing at time $\bar{t}$, which is
denoted by $\mathtt{Z}\left( t,y,\bar{t}\right) $, boils down to solving the
following classical backward problem:%
\begin{equation}
\begin{array}{c}
\mathtt{Z}_{t}\left( t,y,\bar{t}\right) +\left( \chi -\kappa y\right) 
\mathtt{Z}_{y}\left( t,y,\bar{t}\right) +\frac{1}{2}\varepsilon ^{2}\mathtt{Z%
}_{yy}\left( t,y,\bar{t}\right) -y\mathtt{Z}\left( t,y,\bar{t}\right) =0, \\ 
\\ 
\mathtt{Z}\left( \bar{t},y,\bar{t}\right) =1.%
\end{array}
\label{Eq8.107}
\end{equation}%
The standard affine ansatz yields%
\begin{equation}
\begin{array}{c}
\mathtt{Z}\left( t,y,\bar{t}\right) =\exp \left( \mathsf{C}-\mathsf{B}%
_{\kappa }y\right) \\ 
\\ 
\mathsf{C}=\left( \theta -\frac{\varepsilon ^{2}}{2\kappa ^{2}}\right)
\left( \mathsf{B}_{\kappa }-T\right) -\frac{\varepsilon ^{2}}{4\kappa }%
\mathsf{B}_{\kappa }^{2} \\ 
\\ 
=\left( \mathsf{B}_{\kappa }-T\right) \theta +\frac{h_{0}}{2},%
\end{array}
\label{Eq8.108}
\end{equation}%
where $h_{0}$ is given by Eq. (\ref{Eq6.114}).

One can use formulae derived in the previous section to come up with an
alternative derivation. Introduce $\hat{x}_{t}=\int_{t}^{t}\hat{y}_{s}ds$.
The distribution of $\left( \hat{x}_{t},\hat{y}_{t}\right) $ is given by Eq.
(\ref{Eq6.46}) with the covariance matrix $\mathfrak{H}$, given by Eq. (\ref%
{Eq6.114}) and the expected value $\mathbf{r}$ given by Eq. (\ref{Eq6.115}).
Accordingly, the price of a bond can be written as follows:%
\begin{equation}
\begin{array}{c}
\mathtt{Z}\left( t,y,\bar{t}\right) =\mathbb{E}\left\{ e^{-\bar{x}}\right\} =%
\frac{1}{\sqrt{2\pi h_{0}}}\dint\limits_{-\infty }^{\infty }e^{-\bar{x}-%
\frac{\left( \bar{x}-p\right) ^{2}}{2h_{0}}}d\bar{x} \\ 
\\ 
=e^{-p+\frac{h_{0}}{2}}=\exp \left( \mathsf{C}\left( T\right) -\mathsf{B}%
_{\kappa }\left( T\right) y\right) ,%
\end{array}
\label{Eq8.109}
\end{equation}%
so that Eqs. (\ref{Eq8.108}) and (\ref{Eq8.109}) are in agreement.

Knowing the joint Gaussian distribution for $\left( \hat{x}_{t},\hat{y}%
_{t}\right) $ one can price an option on zero coupon bond maturing at time $%
\breve{t}>t$, $\breve{t}-t=\breve{T}$. The payoff of a European option with
strike $K$ has the form:%
\begin{equation}
\begin{array}{c}
U\left( \bar{t},\bar{y}\right) =\max \left\{ \phi \left( \exp \left( \mathsf{%
C}\left( \breve{T}\right) -\mathsf{B}_{\kappa }\left( \breve{T}\right) \bar{y%
}\right) -\exp \left( \ln K\right) \right) ,0\right\} .%
\end{array}
\label{Eq8.110}
\end{equation}%
At maturity $\bar{t}$, the payoff is independent on $\bar{x}$, however, at
inception it does depend on the realized value of $\bar{x}$. By using Eqs (%
\ref{Eq6.46}), (\ref{Eq6.114}), (\ref{Eq6.115}), one can write $U\left(
t,y\right) $ (recall that here $x=0$) as follows:%
\begin{equation}
\begin{array}{c}
U\left( t,y\right) =\mathcal{J}_{1}\left( t,y\right) -\mathcal{J}_{2}\left(
t,y\right) ,%
\end{array}
\label{Eq8.111}
\end{equation}%
where%
\begin{equation}
\begin{array}{c}
\mathcal{J}_{1}\left( t,y\right) =\frac{1}{2\pi \det \left( \mathfrak{H}%
\right) ^{1/2}}\dint\limits_{-\infty }^{\infty }\dint\limits_{-\phi \infty
}^{\mathsf{y}^{\ast }}\exp \left( -\Lambda \left( \bar{x},\bar{y}\right) -%
\bar{x}+\mathsf{C}\left( \breve{T}\right) -\mathsf{B}_{\kappa }\left( \breve{%
T}\right) \bar{y}\right) d\bar{x}d\bar{y},%
\end{array}
\label{Eq8.112}
\end{equation}%
\begin{equation}
\begin{array}{c}
\mathcal{J}_{2}\left( t,y\right) =\frac{1}{2\pi \det \left( \mathfrak{H}%
\right) ^{1/2}}\dint\limits_{-\infty }^{\infty }\dint\limits_{-\phi \infty
}^{\mathsf{y}^{\ast }}\exp \left( -\Lambda \left( \bar{x},\bar{y}\right) -%
\bar{x}+\ln K\right) d\bar{x}d\bar{y},%
\end{array}
\label{Eq8.113}
\end{equation}%
\begin{equation}
\begin{array}{c}
\Lambda \left( \bar{x},\bar{y}\right) =\frac{\left( h_{2}\left( \bar{x}%
-p\right) ^{2}-2h_{1}\left( \bar{x}-p\right) \left( \bar{y}-q\right)
+h_{0}\left( \bar{y}-q\right) ^{2}\right) }{2\det \left( \mathfrak{H}\right) 
},%
\end{array}
\label{Eq8.113a}
\end{equation}%
with $h_{i}$ given by Eq. (\ref{Eq6.114}), $\det \left( \mathfrak{H}\right)
=h_{0}h_{2}-h_{1}^{2}$. Here $\mathsf{y}^{\ast }$ is defined as follows:%
\begin{equation}
\begin{array}{c}
\mathsf{y}^{\ast }=\frac{\mathsf{C}\left( \breve{T}\right) -\ln K}{\mathsf{B}%
_{\kappa }\left( \breve{T}\right) }.%
\end{array}
\label{Eq8.114}
\end{equation}%
First, consider $\mathcal{J}_{1}$. Completing the square, one gets:%
\begin{equation}
\begin{array}{c}
-\Lambda \left( \bar{x},\bar{y}\right) -\bar{x}+\mathsf{C}\left( \breve{T}%
\right) -\mathsf{B}_{\kappa }\left( \breve{T}\right) \bar{y} \\ 
\\ 
=-\frac{\left( h_{2}\left( \left( \bar{x}-p\right) -\frac{\Xi \left(
y\right) }{\sqrt{h_{2}}}\right) ^{2}-\Xi ^{2}\left( \bar{y}\right)
+h_{0}\left( \bar{y}-q\right) ^{2}\right) }{2\det \left( \mathfrak{H}\right) 
}-\mathsf{B}_{\kappa }\left( \breve{T}\right) \left( \bar{y}-q\right) -p+%
\mathsf{C}\left( \breve{T}\right) -\mathsf{B}_{\kappa }\left( \breve{T}%
\right) q,%
\end{array}
\label{Eq8.115}
\end{equation}%
where%
\begin{equation}
\begin{array}{c}
\Xi \left( \bar{y}\right) =\frac{\left( h_{1}\left( \bar{y}-q\right) -\det
\left( \mathfrak{H}\right) \right) }{\sqrt{h_{2}}}.%
\end{array}
\label{Eq8.116}
\end{equation}%
Integrating over $\bar{x}$, one obtains the following expression for $%
\mathcal{J}_{1}$:%
\begin{equation}
\begin{array}{c}
\mathcal{J}_{1}\left( t,y\right) =\frac{e^{-p+\mathsf{C}\left( \breve{T}%
\right) -\mathsf{B}_{\kappa }\left( \breve{T}\right) q}}{\sqrt{2\pi h_{2}}}%
\dint\limits_{-\phi \infty }^{\mathsf{y}^{\ast }}\exp \left( -\frac{\left(
-\Xi ^{2}\left( \bar{y}\right) +h_{0}\left( \bar{y}-q\right) ^{2}+2\det
\left( \mathfrak{H}\right) \mathsf{B}_{\kappa }\left( \breve{T}\right)
\left( \bar{y}-q\right) \right) }{2\det \left( \mathfrak{H}\right) }\right) d%
\bar{y}.%
\end{array}
\label{Eq8.117}
\end{equation}%
Completing the square one more time, one gets:%
\begin{equation}
\begin{array}{c}
-\frac{-\Xi ^{2}+h_{0}\left( \bar{y}-q\right) ^{2}+2\det \left( \mathfrak{H}%
\right) \mathsf{B}_{\kappa }\left( \breve{T}\right) \left( \bar{y}-q\right) 
}{2\det \left( \mathfrak{H}\right) } \\ 
\\ 
=-\frac{\left( \bar{y}-q+h_{1}+\mathsf{B}_{\kappa }\left( \breve{T}\right)
h_{2}\right) ^{2}}{2h_{2}}+\frac{h_{0}}{2}+\mathsf{B}_{\kappa }\left( \breve{%
T}\right) h_{1}+\frac{\mathsf{B}_{\kappa }^{2}\left( \breve{T}\right) h_{2}}{%
2},%
\end{array}
\label{Eq8.118}
\end{equation}%
so that%
\begin{equation}
\begin{array}{c}
\mathcal{J}_{1}\left( t,y\right) \\ 
\\ 
=\frac{e^{-p+\mathsf{C}\left( \breve{T}\right) \mathsf{-B}_{\kappa }\left( 
\breve{T}\right) q+\frac{h_{0}}{2}+\mathsf{B}_{\kappa }\left( \breve{T}%
\right) h_{1}+\frac{\mathsf{B}_{\kappa }^{2}\left( \breve{T}\right) h_{2}}{2}%
}}{\sqrt{2\pi h_{2}\left( t,t\right) }}\dint\limits_{-\phi \infty }^{\mathsf{%
y}^{\ast }}\exp \left( -\frac{\left( \bar{y}-q+h_{1}+\mathsf{B}_{\kappa
}\left( \breve{T}\right) h_{2}\right) ^{2}}{2h_{2}}\right) d\bar{y} \\ 
\\ 
=\phi e^{-p+\mathsf{C}\left( \breve{T}\right) \mathsf{-B}_{\kappa }\left( 
\breve{T}\right) q+\frac{h_{0}}{2}+\mathsf{B}_{\kappa }\left( \breve{T}%
\right) h_{1}+\frac{\mathsf{B}_{\kappa }^{2}\left( \breve{T}\right) h_{2}}{2}%
}\mathrm{N}\left( \frac{\phi \left( \mathsf{y}^{\ast }-q+h_{1}+\mathsf{B}%
_{\kappa }\left( \breve{T}\right) h_{2}\right) }{\sqrt{h_{2}}}\right) .%
\end{array}
\label{Eq8.119}
\end{equation}%
It is easy to see that $\mathtt{Z}\left( t,y,\breve{t}\right) $ is given by
Eq. (\ref{Eq8.119}) with $\phi =1$ and $\mathsf{y}^{\ast }=\infty $, so that 
\begin{equation}
\begin{array}{c}
\mathtt{Z}\left( t,y,\breve{t}\right) =e^{-p+\mathsf{C}\left( \breve{T}%
\right) \mathsf{-B}_{\kappa }\left( \breve{T}\right) q+\frac{h_{0}}{2}+%
\mathsf{B}_{\kappa }\left( \breve{T}\right) h_{1}+\frac{\mathsf{B}_{\kappa
}^{2}\left( \breve{T}\right) h_{2}}{2}}.%
\end{array}
\label{Eq8.120}
\end{equation}%
Thus,%
\begin{equation}
\begin{array}{c}
\mathcal{J}_{1}\left( t,y\right) =\phi \mathtt{Z}\left( t,y,\breve{t}\right) 
\mathrm{N}\left( \frac{\phi \left( \mathsf{C}\left( \breve{T}\right) -\ln K-%
\mathsf{B}_{\kappa }\left( \breve{T}\right) q+\mathsf{B}_{\kappa }\left( 
\breve{T}\right) h_{1}+\mathsf{B}_{\kappa }^{2}\left( \breve{T}\right)
h_{2}\right) }{\sqrt{h_{2}}\mathsf{B}_{\kappa }\left( \breve{T}\right) }%
\right) .%
\end{array}
\label{Eq8.121}
\end{equation}%
Direct verification of Eq. (\ref{Eq8.119}) is left to the reader as a useful
exercise. By using this equation, it is easy but tedious to show that%
\begin{equation}
\begin{array}{c}
\mathcal{J}_{1}\left( t,y\right) =\phi \mathtt{Z}\left( t,y,\breve{t}\right) 
\mathrm{N}\left( \phi d_{+}\right) , \\ 
\\ 
d_{+}=\frac{\ln \left( \frac{\mathtt{Z}\left( t,y,\breve{t}\right) }{\mathtt{%
Z}\left( t,y,\bar{t}\right) K}\right) }{\Sigma \left( t,\bar{t},\breve{t}%
\right) }+\frac{\Sigma \left( t,\bar{t},\breve{t}\right) }{2},%
\end{array}
\label{Eq8.122}
\end{equation}%
where%
\begin{equation}
\begin{array}{c}
\Sigma \left( t,\bar{t},\breve{t}\right) =\sqrt{h_{2}}\mathsf{B}_{\kappa
}\left( \breve{T}\right) .%
\end{array}
\label{Eq8.123}
\end{equation}%
Second, consider $\mathcal{J}_{2}$, proceed in the same way as before, and
represent $\mathcal{J}_{2}\left( t,y\right) $ in the form: 
\begin{equation}
\begin{array}{c}
\mathcal{J}_{2}\left( t,y\right) =\phi \mathtt{Z}\left( t,y,\bar{t}\right) 
\mathrm{N}\left( \phi d_{-}\right) , \\ 
\\ 
d_{-}=\frac{\ln \left( \frac{\mathtt{Z}\left( t,y,\breve{t}\right) }{\mathtt{%
Z}\left( t,y,\bar{t}\right) K}\right) }{\Sigma \left( t,\bar{t},\breve{t}%
\right) }-\frac{\Sigma \left( t,\bar{t},\breve{t}\right) }{2}.%
\end{array}
\label{Eq8.128}
\end{equation}%
Finally, one arrives at the following familiar expression for the bond
option price: 
\begin{equation}
\begin{array}{c}
U\left( t,y\right) =\phi \left( \mathtt{Z}\left( t,y,\breve{t}\right) 
\mathrm{N}\left( \phi d_{+}\right) -\mathtt{Z}\left( t,y,\bar{t}\right) K%
\mathrm{N}\left( \phi d_{-}\right) \right) .%
\end{array}
\label{Eq8.129}
\end{equation}

\subsubsection{CIR\ Model}

The CIR model postulates that the short rate follows the Feller process; see 
\cite{Cox85}. Accordingly, the bond price can be calculated by using Eq. (%
\ref{Eq7.111}) with $x=0$, and $k=-i$:%
\begin{equation}
\begin{array}{c}
\mathtt{Z}\left( t,y,\bar{t}\right) =\int_{-\infty }^{\infty }\varpi
^{\left( x\right) }\left( t,y,\bar{t},\bar{x}\right) e^{-\bar{x}}d\bar{x} \\ 
\\ 
=\frac{1}{2\pi }\int_{-\infty }^{\infty }\int_{-\infty }^{\infty }\digamma
\left( t,y,\bar{t},k\right) e^{\left( ik-1\right) \bar{x}}dkd\bar{x}%
=\digamma \left( t,y,\bar{t},-i\right) ,%
\end{array}
\label{Eq8.130}
\end{equation}%
where%
\begin{equation}
\begin{array}{c}
\digamma \left( t,y,\bar{t},-i\right) =\exp \left( \frac{2\chi \mu T}{%
\varepsilon ^{2}}-\frac{2\chi }{\varepsilon ^{2}}\ln \left( \frac{-\lambda
_{-}\mathsf{E}_{+}+\lambda _{+}\mathsf{E}_{-}}{2\zeta }\right) +\frac{%
2\lambda _{+}\lambda _{-}\left( \mathsf{E}_{+}-\mathsf{E}_{-}\right) y}{%
\varepsilon ^{2}\left( -\lambda _{-}\mathsf{E}_{+}+\lambda _{+}\mathsf{E}%
_{-}\right) }\right) ,%
\end{array}
\label{Eq8.131}
\end{equation}%
with%
\begin{equation}
\begin{array}{c}
\lambda _{\pm }=\mu \pm \zeta , \\ 
\\ 
\mu =-\frac{\kappa }{2},\ \ \ \zeta =\frac{\sqrt{\kappa ^{2}+2\varepsilon
^{2}}}{2}.%
\end{array}
\label{Eq8.132}
\end{equation}%
Thus,%
\begin{equation}
\begin{array}{c}
\mathtt{Z}\left( t,y,\bar{t}\right) =e^{\mathsf{\tilde{C}}-\mathsf{\tilde{B}}%
y},%
\end{array}
\label{Eq8.133}
\end{equation}%
where%
\begin{equation}
\begin{array}{c}
\mathsf{\tilde{C}}=\frac{\chi \kappa T}{\varepsilon ^{2}}-\frac{2\chi }{%
\varepsilon ^{2}}\ln \left( \frac{-\lambda _{-}\mathsf{E}_{+}+\lambda _{+}%
\mathsf{E}_{-}}{2\zeta }\right) , \\ 
\\ 
\mathsf{\tilde{B}}=\frac{\left( \mathsf{E}_{+}-\mathsf{E}_{-}\right) }{%
\left( -\lambda _{-}\mathsf{E}_{+}+\lambda _{+}\mathsf{E}_{-}\right) },%
\end{array}
\label{Eq8.134}
\end{equation}%
which coincides with the standard expressions given by \cite{Cox85}.

\subsection{European Options with Stochastic Interest Rates\label{Sec88}}

This section shows how to price equity options with stochastic interest
rates. While the formulation of this problem may appear straightforward, its
solution proves to be tedious. It is assumed that interest rate is governed
by the Ornstein-Uhlenbeck-Vasicek processes.

\begin{equation}
\begin{array}{c}
d\hat{y}_{t}=\left( \chi -\kappa \hat{y}_{t}\right) dt+\varepsilon d\hat{Z}%
_{t},\ \ \ \hat{y}_{t}=y,%
\end{array}
\label{Eq8.135}
\end{equation}%
where $\hat{Z}_{t}$ is the standard Wiener processes. The risk-neutral
evolution of the foreign exchange is governed by the following equation: 
\begin{equation}
\begin{array}{c}
\frac{d\hat{S}_{t}}{\hat{S}_{t}}=\hat{y}_{t}dt+\sigma d\hat{W}_{t},\ \ \ 
\hat{S}_{t}=S,%
\end{array}
\label{Eq8.136}
\end{equation}%
or, equivalently,%
\begin{equation}
\begin{array}{c}
d\hat{x}_{t}=\left( \hat{y}_{t}-\frac{1}{2}\sigma ^{2}\right) dt+\sigma d%
\hat{W}_{t},\ \ \ \hat{x}_{t}=x,%
\end{array}
\label{Eq8.137}
\end{equation}%
where $\hat{x}=\ln \left( \hat{S}/K\right) $. In general, $d\hat{Z}_{t}$ and 
$d\hat{W}_{t}$ are correlated, so that $d\hat{Z}_{t}d\hat{W}_{t}=\rho dt$.

Consider the familiar backward Kolmogorov problem for European calls and
puts:%
\begin{equation}
\begin{array}{c}
U_{t}+\frac{1}{2}\varepsilon ^{2}U_{rr}+\rho \varepsilon \sigma U_{rx}+\frac{%
1}{2}\sigma ^{2}U_{xx} \\ 
\\ 
+\left( \chi -\kappa r\right) U_{y}+\left( y-\frac{1}{2}\sigma ^{2}\right)
U_{x}-rU=0, \\ 
\\ 
U\left( \bar{t},y,x\right) =K\left( \phi \left( e^{x}-1\right) \right) _{+}.%
\end{array}
\label{Eq8.138}
\end{equation}%
As usual, start with the change of the dependent variable:%
\begin{equation}
\begin{array}{c}
U=KB_{1}V.%
\end{array}
\label{Eq8.139}
\end{equation}%
where $B=\exp \left( \alpha _{1}-\beta _{1}y\right) $ is the domestic bond
price, given by Eq. (\ref{Eq8.108}), so that%
\begin{equation}
\begin{array}{c}
B_{t}+\frac{1}{2}\varepsilon ^{2}B_{rr}+\left( \chi -\kappa r\right)
B_{y}-rB=0.%
\end{array}
\label{Eq8.140}
\end{equation}%
Hence,%
\begin{equation}
\begin{array}{c}
V_{t}+\frac{1}{2}\varepsilon ^{2}V_{rr}+\rho \varepsilon \sigma V_{rx}+\frac{%
1}{2}\sigma ^{2}V_{xx} \\ 
\\ 
+\left( x-\frac{1}{2}\varepsilon ^{2}\beta _{1}-\kappa r\right) V_{y}+\left(
y-\frac{1}{2}\sigma ^{2}-\rho \varepsilon \sigma \beta _{1}\right) V_{x}=0,
\\ 
\\ 
V\left( \bar{t},y,x\right) =\left( \phi \left( e^{x}-1\right) \right) _{+}.%
\end{array}
\label{Eq8.141}
\end{equation}%
Now, change independent variables $\left( t,y,x\right) \rightarrow \left(
t,\eta _{1},\eta _{2}\right) $, where%
\begin{equation}
\begin{array}{c}
\eta _{1}=y,\ \ \ \eta _{2}=-\alpha _{1}+\beta _{1}y+x.%
\end{array}
\label{Eq8.142}
\end{equation}%
Thus,%
\begin{equation}
\begin{array}{c}
\frac{\partial }{\partial t}=\frac{\partial }{\partial t}+\left( -\alpha
_{1}^{\prime }+\beta _{1}^{\prime }\eta _{1}\right) \frac{\partial }{%
\partial \eta _{2}}, \\ 
\\ 
\frac{\partial }{\partial y}=\frac{\partial }{\partial \eta _{1}}+\beta _{1}%
\frac{\partial }{\partial \eta _{2}},\ \ \ \frac{\partial }{\partial x}=%
\frac{\partial }{\partial \eta _{2}}.%
\end{array}
\label{Eq8.143}
\end{equation}%
so that%
\begin{equation}
\begin{array}{c}
V_{t}+\left( -\alpha _{1}^{\prime }+\beta _{1}^{\prime }\eta _{1}\right)
V_{\eta _{2}} \\ 
\\ 
+\frac{1}{2}\varepsilon ^{2}\left( V_{\eta _{1}\eta _{1}}+2\beta _{1}V_{\eta
_{1}\eta _{2}}+\beta _{1}^{2}V_{\eta _{2}\eta _{2}}\right) +\rho \varepsilon
\sigma \left( V_{\eta _{1}\eta _{2}}+\beta _{1}V_{\eta _{2}\eta _{2}}\right)
+\frac{1}{2}\sigma ^{2}V_{\eta _{2}\eta _{2}} \\ 
\\ 
+\left( \chi -\frac{1}{2}\varepsilon ^{2}\beta _{1}-\kappa \eta _{1}\right)
\left( V_{\eta _{1}}+\beta _{1}V_{\eta _{2}}\right) +\left( \eta _{1}-\frac{1%
}{2}\sigma ^{2}-\rho \varepsilon \sigma \beta _{1}\right) V_{\eta _{2}}=0,
\\ 
\\ 
V\left( \bar{t},\eta _{1},\eta _{2}\right) =\left( \phi \left( e^{\eta
_{2}}-1\right) \right) _{+}.%
\end{array}
\label{Eq8.144}
\end{equation}%
Assume that $V\left( t,\eta _{1},\eta _{2}\right) $ only depends on $t,\eta
_{2}$, $V\left( t,\eta _{1},\eta _{2}\right) =V\left( t,\eta _{2}\right) $,
which is consistent with the terminal condition. Thus,%
\begin{equation}
\begin{array}{c}
V_{t}+\left( \frac{1}{2}\varepsilon ^{2}\beta _{1}^{2}+\rho \varepsilon
\sigma \beta _{1}+\frac{1}{2}\sigma ^{2}\right) V_{\eta _{2}\eta _{2}} \\ 
\\ 
+\left( -\alpha _{1}^{\prime }+\beta _{1}^{\prime }\eta _{1}+\left( b_{1}-%
\frac{1}{2}\varepsilon ^{2}\beta _{1}-\kappa \eta _{1}\right) \beta _{1}-%
\frac{1}{2}\sigma ^{2}-\rho \varepsilon \sigma \beta _{1}+\eta _{1}\right)
V_{\eta _{2}}=0 \\ 
\\ 
V\left( \bar{t},\eta _{1},\eta _{2},\eta _{2}\right) =\left( \phi \left(
e^{\eta _{2}}-1\right) \right) _{+}.%
\end{array}
\label{Eq8.145}
\end{equation}%
But%
\begin{equation}
\begin{array}{c}
\alpha _{1}^{\prime }-\beta _{1}^{\prime }\eta _{1}+\frac{1}{2}\varepsilon
^{2}\beta _{1}^{2}-\left( \chi -\kappa \eta _{1}\right) \beta _{1}-\eta
_{1}=0,%
\end{array}
\label{Eq8.146}
\end{equation}%
so that%
\begin{equation}
\begin{array}{c}
V_{t}+\left( \frac{1}{2}\varepsilon ^{2}\beta _{1}^{2}+\rho \varepsilon
\sigma \beta _{1}+\frac{1}{2}\sigma ^{2}\right) \left( V_{\eta _{2}\eta
_{2}}-V_{\eta _{2}}\right) =0, \\ 
\\ 
V\left( \bar{t},\eta _{1},\eta _{2}\right) =\left( \phi \left( e^{\eta
_{2}}-1\right) \right) _{+}.%
\end{array}
\label{Eq8.147}
\end{equation}%
This is the classical Black-Scholes problem with time-dependent volatility: 
\begin{equation}
\begin{array}{c}
V_{t}+\frac{1}{2}\Sigma ^{2}\left( V_{\eta _{2}\eta _{2}}-V_{\eta
_{2}}\right) =0, \\ 
\\ 
V\left( \bar{t},\eta _{2}\right) =\left( \phi \left( e^{\eta _{2}}-1\right)
\right) _{+},%
\end{array}
\label{Eq8.148}
\end{equation}%
where%
\begin{equation}
\begin{array}{c}
\Sigma ^{2}=\varepsilon ^{2}\mathsf{B}_{\kappa }^{2}+2\rho \varepsilon
\sigma \mathsf{B}_{\kappa }+\sigma ^{2}.%
\end{array}
\label{Eq8.149}
\end{equation}%
Thus, the price is%
\begin{equation}
\begin{array}{c}
U=B_{1}U^{\left( C,P\right) }\left( \frac{B_{2}S}{B_{1}};T,K,\sqrt{\frac{%
\int \Sigma ^{2}ds}{T}}\right) ,%
\end{array}
\label{Eq8.150}
\end{equation}%
where $U^{\left( C,P\right) }$ are given by Eq. (\ref{Eq8.17}).

A similar technique can be used for the Heston model and the Stein-Stein
model with stochastic interest rates. However, there is one significant
difference between these two models - the former model works only when
volatility and rate innovations are uncorrelated, while the latter model can
handle arbitrary correlations.

\section{Conclusions\label{Chap9}}

Due to the space constraints, the discussion must be concluded here. It is
left to the reader to explore further the application of mathematical tools
and techniques based on Kelvin waves in financial engineering. Three
particularly compelling problems are

\begin{itemize}
\item the pricing and risk management of credit derivatives;

\item the exploration of mean-reverting trading strategies, such as pairs
trading;

\item the examination of affine jump-diffusion and pseudo-differential
processes.
\end{itemize}

References such as \cite{Lipton12, Lipton20} and others provide additional
insights into these problems.

This book has established a unified methodology for determining t.p.d.fs and
expectations for affine processes through integral representations based on
Kelvin waves. This approach has bridged various disciplines, uncovering
profound connections between hydrodynamics, molecular physics, stochastic
processes, and financial engineering. Both degenerate problems, which
possess more independent variables than sources of uncertainty, and their
non-degenerate counterparts are covered, showcasing the versatility of the
method.

A surprising link is established between the Langevin equation for
underdamped Brownian motion and the vorticity equation for two-dimensional
flows in viscous incompressible fluids. Utilizing Kelvin wave expansions,
the book solves several relevant financial problems, including the deriving
convenient formulas for t.p.d.fs and expectations for processes with
stochastic volatility, developing an analytically solvable model for
path-dependent volatility, pricing of Asian options with geometric
averaging, and pricing bonds and bond options by augmenting the short-rate
process with its integral process.

The methodology introduced in this book can address a wide spectrum of
complex problems, significantly enhancing the comprehension and modeling of
stochastic systems across diverse fields.

\begin{acknowledgement}
I am grateful to my ADIA\ colleagues Majed Alromaithi, Marcos Lopez de
Prado, Koushik Balasubramanian, Andrey Itkin, Oleksiy Kondratiev, Arthur
Maghakian, Dmitry Muravey, Adil Reghai, other Q-team colleagues, my ADIA Lab
colleague Horst Simon, and a former Bank of America colleague Artur Sepp for
their encouragement and council. The kind invitation by Riccardo Rebonato to
contribute to Cambridge Elements in Quantitative Finance is much
appreciated. I am grateful to Drs Nicola Ghazi and Piergiorgio Neri from
Cleveland Clinic Adu Dhabi for saving vision in my left eye, thus allowing
me to finish this book. Last but not least, the help of my wife, Marsha
Lipton, especially her editorial suggestions and financial insights, has
been critical in producing this book.
\end{acknowledgement}


\begin{thebibliography}{Friedlander \& Lipton-Lifschitz (2003)}
\bibitem[Abramowitz \& Stegun (1964)]{Abramowitz64} Abramowitz, M.\& Stegun,
I.A. eds. 1968 \textit{Handbook of mathematical functions with formulas,
graphs, and mathematical tables}. Washington, D.C.: US Government Printing
Office.

\bibitem[Aksenov (1995)]{Aksenov95} Aksenov, A.V. 1995 Symmetries of linear
partial differential equations and fundamental solutions. \textit{Dokl. Math.%
} \textbf{51 (3)}, 329--331.

\bibitem[Andersen \& Piterbarg (2007)]{Andersen07} Andersen, L.B.G. \&
Piterbarg, V. 2007 Moment explosions in stochastic volatility models. 
\textit{Finance Stoch} \textbf{11}, 29--50.

\bibitem[Angeris \textit{et al.} (2019)]{Angeris19} Angeris, G., Kao, H.T.,
Chiang, R., Noyes, C. \& Chitra, T. 2019 An analysis of Uniswap markets. 
\textit{Cryptoeconomic Systems Journal}.

\bibitem[Arnold (1974)]{Arnold74} Arnold, L. 1974 \textit{Stochastic
Equations: Theory and Applications}. New York, NY: John Wiley.

\bibitem[Bachelier (1900)]{Bachelier00} Bachelier, L. 1900 Th\'{e}orie de la
sp\'{e}culation. \textit{Annales de l'Ecole Normale Sup\'{e}rieure }\textbf{%
17}, 21-86.

\bibitem[Barrucci \textit{et al.} (2001)]{Barrucci01} Barucci, E., Polidoro,
S. \& Vespri, V. 2001 Some results on partial differential equations and
Asian options. \textit{Math. Models Methods Appl. Sci.} \textbf{11(3)},
475--497.

\bibitem[Bayly (1986)]{Bayly86} Bayly, B. J. 1986 Three-dimensional
instability of elliptical flow.\ \textit{Phys. Rev. Lett.} \textbf{57},
2160-2163.

\bibitem[Bayly \textit{et al.} (1996)]{Bayly96} Bayly, B.J., Holm, D.D. \&
Lifschitz, A., 1996. Three-dimensional stability of elliptical vortex
columns in external strain flows. \textit{Phil. Trans. R. Soc. Lond. Ser. A} 
\textbf{354(1709)}, 895-926.

\bibitem[Berest (1993)]{Berest93} Berest, Yu.Yu. 1993 Group analysis of
linear differential equations in distributions and the construction of
fundamental solutions. \textit{Differ. Equations} \textbf{29 (11)},
1700--1711.

\bibitem[Bergomi (2015)]{Bergomi15} Bergomi, L. 2015 \textit{Stochastic
volatility modeling.} Boca Raton: CRC press.

\bibitem[Bharucha-Reid (1960)]{Bharucha60} Bharucha-Reid, A. T. 1960 \textit{%
Elements of the Theory of Markov Processes and their Applications}. New
York, NY: McGraw-Hill Book Co.

\bibitem[Bick \& Reisman (1993)]{Bick93} Bick, A. \& Reisman, H. 1993
Generalized implied volatility. \textit{Preprint.}

\bibitem[Black (1976)]{Black76} Black, F. 1976 The pricing of commodity
contracts. \textit{Journal of Financial Economics }\textbf{3}, 167-179.

\bibitem[Black \& Scholes (1973)]{Black73} Black, F. \& Scholes, M. 1973 The
pricing of options and corporate liabilities. \textit{Journal of Political
Economy }\textbf{81}, no.3, 637-659.

\bibitem[Bluman \& Kumei (1989).]{Bluman1989} Bluman, G. \& Kumei, S. 1989 
\textit{Symmetries and Differential Equations.} Berlin: Springer.

\bibitem[Boness (1964)]{Boness64} Boness, A. J. 1964 Elements of a theory of
a stock option value. \textit{Journal of Political Economy }\textbf{72(2)},
163-175.

\bibitem[Boyarchenko \& Levendorsky (2002)]{Boyarchenko02} Boyarchenko, S.
\& Levendorskii, S. 2002 \textit{Non-Gaussian Merton-Black-Scholes Theory}.
River Edge, NJ: World Scientific.

\bibitem[Carr \textit{et al.} (2002)]{Carr02} Carr, P., Lipton, A. \& Madan,
D. 2002 The reduction method for valuing derivative securities. \textit{%
Working Paper, New York University.}

\bibitem[Cartea \textit{et al.} (2023)]{Cartea23} Cartea, \'{A}., Drissi, F.
\& Monga, M. 2023 Predictable Losses of Liquidity Provision in Constant
Function Markets and Concentrated Liquidity Markets. Available at SSRN
4541034.

\bibitem[Chandresekhar (1943)]{Chandrasekhar43} Chandresekhar, S. 1943
Stochastic problems in physics and astronomy. \textit{Rev. Modern Phys.} 
\textbf{15}, 1--89.

\bibitem[Chandrasekhar (1961)]{Chandrasekhar61} Chandrasekhar, S. 1961 
\textit{Hydrodynamic and hydromagnetic stability.} Oxford: Clarendon.

\bibitem[Chapman (1928)]{Chapman28} Chapman, S. 1928 On the Brownian
displacements and thermal diffusion of grains suspended in a nonuniform
fluid. \textit{Proc R Soc Lond Ser A} \textbf{119(781)}, 34--54

\bibitem[Cordes (1995)]{Cordes95} Cordes, H.O. 1995 \textit{The technique of
pseudodifferential operators}. Cambridge: Cambridge University Press.

\bibitem[Cox \textit{et al.} (1985)]{Cox85} Cox, J. C. , Ingersoll Jr., J.
E. \& Ross, S. A. 1985 A theory of the term structure of interest rates. 
\textit{Econometrica} \textbf{53}, 385-408.

\bibitem[Craddock (2012)]{Craddock12} Craddock, M. 2012 Lie symmetry methods
for multi-dimensional parabolic PDEs and diffusions. \textit{J. Differential
Equations} \textbf{252(1)}, 56--90.

\bibitem[Craddock \& Platen (2004)]{Craddock04} Craddock, M. \& Platen, E.
2004 Symmetry group methods for fundamental solutions. \textit{J.
Differential Equations} \textbf{207(2)}, 285--302.

\bibitem[Craik \& Criminale (1986)]{Craik86} Craik, A. D. D. \& Criminale,
W. O. 1986 Evolution of wavelike disturbances in shear flows: A class of
exact solutions of the Navier--Stokes equations.\ \textit{Proc. R. Soc.
London, Ser. A} \textbf{406}, 13-26.

\bibitem[Dai \& Singleton (2000)]{Dai00} Dai, Q. \& Singleton, K. J. 2000
Specification analysis of affine term structure models. \textit{Journal of
Finance} \textbf{55}, 1943-1978.

\bibitem[Davis (2004)]{Davis04} Davis, M. H. A. 2004 Complete-market models
of stochastic volatility. \textit{Proc. R. Soc. Lond. Ser. A} \textbf{460},
11--26.

\bibitem[Derman \& Kani (1994)]{Derman94} Derman, E. \& Kani, I. 1994 Riding
on a smile. \textit{Risk Magazine }\textbf{7(2)}, 32-39.

\bibitem[Devreese \textit{et al.} (2010)]{Devreese10} Devreese, J.P.A.,
Lemmens, D. \& Tempere, J. 2010 Path integral approach to Asian options in
the Black-Scholes model. \textit{Physica A} \textbf{389 (4)}, 780--788.

\bibitem[Di Francesco \& Pascucci (2004)]{DiFrancesco04} Di Francesco, M. \&
Pascucci, A. 2004 On the complete model with stochastic volatility by Hobson
and Rogers. \textit{Proc. R. Soc. Lond. Ser. A} \textbf{460(2051)},
3327--3338.

\bibitem[Di Francesco \& Pascucci (2005)]{DiFrancesco05} Di Francesco, M. \&
Pascucci, A. 2005 On a class of degenerate parabolic equations of Kolmogorov
type. \textit{Applied Mathematics Research eXpress} \textbf{3}, 77-116.

\bibitem[Dragulescu \& Yakovenko (2002)]{Dragulescu02} Dragulescu, A.A. \&
Yakovenko, V.M. 2002 Probability distribution of returns in the Heston model
with stochastic volatility. \textit{Quantitative finance} \textbf{2(6)}, 443.

\bibitem[Duffie \textit{et al.} (2003)]{Duffie03} Duffie, D., Filipovic, D.
\& Schachermayer, W. 2003 Affine processes and applications in finance. 
\textit{Ann. Appl. Probab.} \textbf{13(3)}, 984--1053.

\bibitem[Duffie \& Kan (1996)]{Duffie96} Duffie, J.D. \& Kan, R. 1996 A
yield-factor model of interest rates. \textit{Mathematical Finance} \textbf{6%
}, 379-406.

\bibitem[Duffie \textit{et al.} (2000)]{Duffie00} Duffie, D., Pan, J. \&
Singleton, K. 2000 Transform analysis and asset pricing for affine
jump-diffusions. \textit{Econometrica} \textbf{68(6)}, 1343-1376.

\bibitem[Duong \& Tran (2018)]{Duong18} Duong, M.H. \& Tran, H.M. 2018 On
the fundamental solution and a variational formulation for a degenerate
diffusion of Kolmogorov type. \textit{Discrete \& Continuous Dynamical
Systems: Series A} \textbf{38(7)}.

\bibitem[Dupire (1994)]{Dupire94} Dupire, B. 1994 Pricing with a smile. 
\textit{Risk Magazine }\textbf{7(1)}, 18-20.

\bibitem[Ebeling \textit{et al.} (2008)]{Ebeling08} Ebeling, W.,
Gudowska-Nowak, E. \& Sokolov, I.M. 2008 On stochastic dynamics in physics
-- remarks on history and terminology. \textit{Acta Physica Polonica B} 
\textbf{39(5)}, 1003-1017.

\bibitem[Egorov (2019)]{Egorov19} Egorov, M. 2019 StableSwap - efficient
mechanism for Stablecoin liquidity. White paper.

\bibitem[Fabijonas \textit{et al.} (1997)]{Fabijonas97} Fabijonas, B., Holm,
D.D. \& Lifschitz, A. 1997 Secondary instabilities of flows with elliptic
streamlines. \textit{Physical Review Letters} \textbf{78(10)}, 1900-1903.

\bibitem[Feller (1951)]{Feller51} Feller, W. 1951 Two singular diffusion
problems. \textit{Ann. Math.} \textbf{54(1)}, 173--182.

\bibitem[Feller (1952)]{Feller52} Feller, W. 1952 The parabolic differential
equations and the associated semi-groups of transformations. \textit{Ann.
Math.} \textbf{55}, 468--518.

\bibitem[Feller (1971)]{Feller71} Feller, W. 1971 \textit{An Introduction to
Probability Theory and Its Application}, vol. 2, 2nd ed. New York, NY: John
Wiley.

\bibitem[Filipovic (2009)]{Filipovic09} Filipovic, D. 2009 \textit{%
Term-structure models.} Berlin: Springer-Verlag.

\bibitem[Fokker (1914)]{Fokker14} Fokker, A.D. 1914 Die mittlere Energie
rotierender elektrischer Dipole im Strahlungsfeld. \textit{Annalen der Physik%
} \textbf{348(5)}, 810--820.

\bibitem[Fourier (1822)]{Fourier22} Fourier, J.B. 1822 \textit{Th\'{e}orie
Analytique de la Chaleur}. Paris: Firmin Didot, P\`{e}re et Fils.

\bibitem[Friedlander \& Vishik (1991)]{Friedlander91} Friedlander, S. \&
Vishik, M. 1991 Instability criteria for the flow of an inviscid
incompressible fluid. \textit{Phys. Rev. Lett.} \textbf{66}, 2204-2206

\bibitem[Friedlander \& Lipton-Lifschitz (2003)]{Friedlander03} Friedlander,
S. \& Lipton-Lifschitz, A. 2003 Localized instabilities in fluids. In 
\textit{Handbook of Mathematical Fluid Dynamics} \textbf{2}, 289-354.
North-Holland: Amsterdam.

\bibitem[Friz \& Keller-Ressel (2010)]{Friz10} Friz, P. \& Keller-Ressel, M.
2010. Moment explosions in stochastic volatility models. \textit{%
Encyclopedia of quantitative finance}, 1247-1253.

\bibitem[Fukusawa \textit{et al.} (2023)]{Fukusawa23} Fukasawa, M., Maire,
B. \& Wunsch, M. 2023 Weighted variance swaps hedge against impermanent
loss. \textit{Quantitative Finance} \textbf{23(6)}, 1-11.

\bibitem[Gatheral \textit{et al.} (2018)]{Gatheral18} Gatheral, J., Jaisson,
T. \& Rosenbaum, M. 2018 Volatility is rough. \textit{Quant. Finance }%
\textbf{18(6)}, 933-949.

\bibitem[Geman \& Eydeland (1995)]{Geman95} Geman, H. \& Eydeland, A. 1995
Asian options revisited: inverting the Laplace transform. \textit{Risk
Magazine }\textbf{8(4)}, 65-67.

\bibitem[Gershon \textit{et al.} (2022)]{Gershon22} Gershon, D., Lipton, A.,
Rosenbaum, M. \& Wiener, Z. eds. 2022 \textit{Options-45 Years Since the
Publication of the Black-Scholes-Merton Model: The Gershon Fintech Center
Conference}. Singapore: World Scientific.

\bibitem[Gihman \& Skorohod (1972)]{Gihman72} Gihman, I. I. \& Skorohod, A.
V. 1972 \textit{Stochastic Differential Equations}. New York, NY:
Springer-Verlag.

\bibitem[Giorno \& Nobile (2021)]{Giorno21} Giorno, V. \& Nobile, A.G. 2021
Time-inhomogeneous Feller-type diffusion process in population dynamics. 
\textit{Mathematics} \textbf{9(16)}, 1879.

\bibitem[Guyon (2014)]{Guyon14} Guyon, J. 2014 Path-Dependent volatility. 
\textit{Risk Magazine} \textbf{27(10)}.

\bibitem[Hagan \textit{et al.} (2002)]{Hagan02} Hagan, P., Kumar, D.,
Lesniewski, A. \& Woodward, D. 2002 Managing smile risk. \textit{Wilmott
Magazine} \textbf{(9)}, 84--108.

\bibitem[H\"{a}nggi \textit{et al.} (1990)]{Hanggi90} H\"{a}nggi, P.,
Talkner, P. \& Borkovec, M. 1990 Reaction-rate theory: fifty years after
Kramers. \textit{Reviews of modern physics} \textbf{62(2)}, 251-341.

\bibitem[Hanson (2007)]{Hanson07} Hanson, F.B. 2007. \textit{Applied
stochastic processes and control for jump-diffusions: modeling, analysis and
computation}. Philadelphia: Society for Industrial and Applied Mathematics.

\bibitem[He \textit{et al.} (2021)]{He21} He, C., Chen, J., Fang, H. \& He,
H. 2021 Fundamental solution of fractional Kolmogorov--Fokker--Planck
equation. \textit{Examples and Counterexamples} (\textbf{1}), 100031.

\bibitem[Heston (1993)]{Heston93} Heston, S. L. 1993 A closed-form solution
for options with stochastic volatility with applications to bond and
currency options. \textit{Review of Financial Studies }\textbf{6}, 327-343.

\bibitem[Hobson \& Rogers (1998)]{Hobson98} Hobson, D. G. \& Rogers, L. C.
G. 1998 Complete models with stochastic volatility. \textit{Math. Finance} 
\textbf{8(1)}, 27--48.

\bibitem[H\"{o}rmander (1967)]{Hormander67} H\"{o}rmander, L. 1967
Hypoelliptic second order differential equations. \textbf{Acta Math.} 
\textit{119}, 147--171.

\bibitem[Hull \& White (1990)]{Hull90} Hull, J. \& A. White, A. 1990 Pricing
interest rate derivative securities. \textit{Review of Financial Studies }%
\textbf{3}, 573-592.

\bibitem[Ibragimov (1985)]{Ibragimov85} Ibragimov, N.H. 1985 \textit{%
Transformation Groups Applied to Mathematical Physics. }Dordrecht: D. Reidel.

\bibitem[Ivasishen \& Medynsky (2010)]{Ivasishen10} Ivasishen, S.D. \&
Medynsky, I.P. 2010 The Fokker Planck Kolmogorov equations for some
degenerate diffusion processes. \textit{Theory of Stochastic Processes} 
\textbf{16(1)}, 57-66.

\bibitem[Jacob \& Schilling (2001)]{Jacob01} Jacob, N. \& Schilling, R.L.
2001 L\'{e}vy-type processes and pseudodifferential operators. In \textit{L%
\'{e}vy Processes: Theory and Applications, }O. E. Barndorff-Nielsen \textit{%
et al.} (eds.), (pp. 139-168). Boston, MA: Birkh\"{a}user Boston.

\bibitem[Janek \textit{et al.} (2011)]{Janek11} Janek, A., Kluge, T., Weron,
R. and Wystup, U. 2011 FX smile in the Heston model. \textit{In Statistical
tools for finance and insurance} (pp. 133-162). Berlin, Heidelberg: Springer
Berlin Heidelberg.

\bibitem[Jex \textit{et al.} (1999)]{Jex99} Jex, M., Henderson, R. and Wang,
D. 1999 Pricing exotics under the smile. \textit{Risk Magazine} \textbf{%
12(11)}, 72--75.

\bibitem[Kelvin (1887)]{Kelvin87} Kelvin, Lord 1887 Stability of fluid
motion: rectilinear motion of viscous fluid parallel plates. \textit{Phil.
Mag.} \textbf{24}, 188-196.

\bibitem[Klein (1921)]{Klein21} Klein, O. 1921 Zur statistischen Theorie der
Suspension und L\"{o}sungen. Inaugural-Dissertation. Uppsala: Almqvist \&
Wiksells.

\bibitem[Kolmogoroff (1931)]{Kolmogorov31} Kolmogoroff, A. 1931 \"{U}ber die
analytischen Methoden in der Wahrscheinlichkeitsrechnung. \textit{Math. Ann.}
\textbf{104(1)}, 415--458.

\bibitem[Kolmogoroff (1933)]{Kolmogorov33} Kolmogoroff, A. 1933 Zur Theorie
der stetigen zuf\"{a}lligen Prozesse. \textit{Math. Ann.} \textbf{108},
149--160

\bibitem[Kolmogoroff (1934)]{Kolmogorov34} Kolmogoroff, A. 1934 Zufallige
Bewegungen (Zur Theorie der Brownschen Bewegung), \textit{Ann. Math.} 
\textbf{35(1)}, 116--117.

\bibitem[Kovalenko \textit{et al.} (2014)]{Kovalenko14} Kovalenko, S.,
Stogniy, V. \& Tertychnyi, M. 2014 Lie symmetries of fundamental solutions
of one (2+ 1)-dimensional ultra-parabolic Fokker--Planck--Kolmogorov
equation. \textit{arXiv preprint} arXiv:1408.0166.

\bibitem[Kramers (1940)]{Kramers40} Kramers, H.A. 1940 Brownian motion in a
field of force and the diffusion model of chemical reactions. \textit{%
Physica. Elsevier BV.} \textbf{7(4)}, 284--304.

\bibitem[Kuptsov (1972)]{Kuptsov72} Kuptsov, L. P. 1972 The fundamental
solutions of a certain class of elliptic-parabolic second order equations. 
\textit{Differential Equations} \textbf{8}, 1649--1660.

\bibitem[Lanconelli \textit{et al.} (2002)]{Lanconelli02} Lanconelli, E.,
Pascucci, A. \& Polidoro, S. 2002 Linear and nonlinear ultraparabolic
equations of Kolmogorov type arising in diffusion theory and in finance. In: 
\textit{Nonlinear Problems in Mathematical Physics and Related Topics}, vol.
II, 243-265. New York: Kluwer/Plenum.

\bibitem[Langevin (1908)]{Langevin08} Langevin, P. 1908 Sur la th\'{e}orie
du mouvement brownien. \textit{C. R. Acad. Sci. Paris.} \textbf{146},
530--533.

\bibitem[Lewis (2000)]{Lewis00} Lewis, A. L. 2000 \textit{Option Valuation
under Stochastic Volatility with Mathematica Code. }Newport Beach: Finance
Press.

\bibitem[Lewis (2001)]{Lewis01} Lewis, A. 2001 Asimple option formula for
general jump-diffusion and other exponential L\'{e}vy processes, Envision
Financial Systems and OptionCity.net, California.
http://optioncity.net/pubs/ExpLevy.pdf.

\bibitem[Lifschitz (1991)]{Lifschitz91} Lifschitz, A. 1991 Short wavelength
instabilities of incompressible three-dimensional flows and generation of
vorticity. \textit{Physics Letters A} \textbf{157(8-9)}, 481-487.

\bibitem[Lifschitz (1995)]{Lifschitz95} Lifschitz, A. 1995 Exact description
of the spectrum of elliptical vortices in hydrodynamics and
magnetohydrodynamics.\ \textit{Phys. Fluids A} \textbf{7}, 1626-1636.

\bibitem[Lifschitz \& Hameiri (1991a)]{Lifschitz91a} Lifschitz, A. \&
Hameiri, E. 1991 A universal instability in fluid dynamics. In 1991
International Sherwood Fusion Theory Conference.

\bibitem[Lifschitz \& Hameiri (1991b)]{Lifschitz91b} Lifschitz, A. \&
Hameiri, E. 1991 Local stability conditions in fluid dynamics.\ \textit{%
Phys. Fluids A} \textbf{3}, 2644-2651.

\bibitem[Lipton-Lifschitz (1999)]{Lipton99a} Lipton-Lifschitz, A. 1999
Predictability and unpredictability in financial markets. \textit{Physica D:
Nonlinear Phenomena} \textbf{133(1-4)}, 321-347.

\bibitem[Lipton (1999)]{Lipton99b} Lipton, A. 1999 Similarities via
self-similarities. \textit{Risk Magazine} \textbf{12(9)}, 101--105.

\bibitem[Lipton (2000)]{Lipton00} Lipton,\ A. 2000 \textit{Pricing and
Risk-Managing Exotics on Assets with Stochastic Volatility}. Presentation,
Risk Minds, Geneva.

\bibitem[Lipton (2001)]{Lipton01} Lipton, A. 2001 \textit{Mathematical
Methods For Foreign Exchange: A Financial Engineer's Approach.} Singapore:
World Scientific.

\bibitem[Lipton (2002)]{Lipton02} Lipton, A. 2002 The volatility smile
problem. \textit{Risk Magazine} \textbf{15(2)}, 61--65.

\bibitem[Lipton (2018)]{Lipton18} Lipton, A. 2018 \textit{Financial
Engineering: Selected Works of Alexander Lipton.} Singapore: World
Scientific.

\bibitem[Lipton (2023)]{Lipton23a} Lipton, A. 2023 \textit{Kelvin Waves,
Klein-Kramers and Kolmogorov Equations, Path-Dependent Financial
Instruments: Survey and New Results}. arXiv preprint arXiv:2309.04547.

\bibitem[Lipton \textit{et al.} (2014)]{Lipton14} Lipton, A., Gal, A. \&
Lasis, A. 2014 Pricing of vanilla and first-generation exotic options in the
local stochastic volatility framework: survey and new results. \textit{%
Quantitative Finance} \textbf{14(11)}, 1899-1922.

\bibitem[Lipton \& Hardjono (2021)]{Lipton21a} Lipton, A. \& Hardjono, T.
2021 Blockchain intra-and interoperability. In \textit{Innovative Technology
at the Interface of Finance and Operations: Volume II} (pp. 1-30). Cham:
Springer International Publishing.

\bibitem[Lipton \& Lopez de Prado (2020)]{Lipton20} Lipton, A. \& Lopez de
Prado, M. 2020 A closed-form solution for optimal ornstein--uhlenbeck driven
trading strategies. \textit{International Journal of Theoretical and Applied
Finance} \textbf{23(08)}, 2050056.

\bibitem[Lipton \& Reghai (2023)]{Lipton23b} Lipton, A. \& Reghai, A. 2023
SPX, VIX and scale-invariant LSV. \textit{Wilmott Magazine} \textbf{2023(126)%
}, 78-84.

\bibitem[Lipton \& Sepp (2008)]{Lipton08} Lipton, A. \& Sepp, A. 2008
Stochastic volatility models and Kelvin waves.\textit{\ J. Phys. A: Math.
Theor.} \textbf{41}, 344012 (23pp)

\bibitem[Lipton \& Sepp (2022)]{Lipton22} Lipton, A. \& Sepp, A. 2022
Automated market-making for fiat currencies. \textit{Risk Magazine} \textbf{%
35(5)}.

\bibitem[Lipton \& Shelton (2012)]{Lipton12} Lipton, A. \& Shelton, D. 2012
Credit default swaps with and without counterparty and collateral
adjustments. \textit{Stochastics An International Journal of Probability and
Stochastic Processes} \textbf{84(5-6)}, 603-624.

\bibitem[Lipton \& Treccani (2021)]{Lipton21b} Lipton, A. \& Treccani, A.
2021 \textit{Blockchain and distributed ledgers: Mathematics, technology,
and economics.} Singapore: World Scientific.

\bibitem[Masoliver (2016)]{Masoliver16} Masoliver, J. 2016 Nonstationary
Feller process with time-varying coefficients. \textit{Phys. Rev. E} \textbf{%
93}, 012122.

\bibitem[Merton (1973)]{Merton73} Merton, R. C. 1973 Theory of rational
option pricing. \textit{Bell Journal of Economics and Management Science }%
\textbf{4}, 141-183.

\bibitem[Merton (1976)]{Merton76} Merton, R. C. 1976 Option pricing when
underlying stock returns are discontinuous. \textit{Journal of Financial
Economics }\textbf{3}, 125-144.

\bibitem[Morse \& Feschbach (1953)]{Morse53} Morse, P. M. \& Feshbach, H.
1953 \textit{Methods of Theoretical Physics, Part I.} New York: McGraw-Hill.

\bibitem[Olver (1986)]{Olver86} Olver, P. J. 1986 Applications of Lie Groups
to Differential Equations, 1st ed. New York: Springer--Verlag.

\bibitem[Orr (1907)]{Orr07} Orr, W. McF. 1907 The stability or instability
of the steady motions of a perfect fluid. \textit{Irish Acad. A} \textbf{27}%
, 9-69.

\bibitem[Ovsiannikov (1982)]{Ovsiannikov82} Ovsiannikov, L.V. 1982 \textit{%
Group Analysis of Differential Equations.} New York: Academic Press.

\bibitem[Pascucci (2005)]{Pascucci05} Pascucci, A. 2005 Kolmogorov Equations
in Physics and in Finance. In: \textit{Progress in Nonlinear Differential
Equations and their Applications}, \textbf{63}, 313-324. Basel: Birkh\"{a}%
user.

\bibitem[Piessens (2000)]{Piessens00} Piessens, R. 2000 The Hankel
Transform. In: \textit{The Transforms and Applications Handbook: Second
Edition}. Ed. Poularikas, A.D. Boca Raton: CRC Press LLC.

\bibitem[Planck (1917)]{Planck17} Planck, M. 1917 \"{U}ber einen Satz der
statistischen Dynamik und seine Erweiterung in der Quantentheorie. \textit{%
Sitzungsberichte der K\"{o}niglich Preussischen Akademie der Wissenschaften}%
, 324--341.

\bibitem[Reghai (2015)]{Reghai15} Reghai, A. 2015 \textit{Quantitative
Finance, Back to Basics}. New York: Palgrave MacMillan.

\bibitem[Risken (1989)]{Risken89} Risken, H. 1989 \textit{The Fokker--Planck
Equation: Method of Solution and Applications.} New York: Springer-Verlag.

\bibitem[Rogers \& Shi (1995)]{Rogers95} Rogers, L.C.G. \& Shi, Z. 1995 The
value of an Asian option. \textit{Journal of Applied Probability} \textbf{%
32(4)}, 1077--1088.

\bibitem[Rubinstein (1994)]{Rubinstein94} Rubinstein, M. 1994 Implied
binomial trees. \textit{Journal of Finance }\textbf{49}, 771-818.

\bibitem[Samuelson (1965)]{Samuelson65} Samuelson, P. A. 1965 Rational
theory of warrant pricing. \textit{Industrial Management Review }\textbf{6},
13-32.

\bibitem[Schachermayer \& Teichmann (2008)]{Schachermayer08} Schachermayer,
W. \& Teichmann, J. 2008 How close are the option pricing formulas of
Bachelier and Black--Merton--Scholes? \textit{Mathematical Finance: an
International Journal of Mathematics, Statistics and Financial Economics}
18(1), 155-170.

\bibitem[Schmelzle (2010)]{Schmelzle10} Schmelzle, M. 2010 Option pricing
formulae using Fourier transform: Theory and application. Preprint,
http://pfadintegral. com.

\bibitem[Sch\"{o}bel \& Zhu (1999)]{Schobel99} Sch\"{o}bel, R. \& Zhu, J.
1999 Stochastic volatility with an Ornstein--Uhlenbeck process: an
extension. \textit{Review of Finance} \textbf{3(1)}, 23-46.

\bibitem[Sepp (2007)]{Sepp07} Sepp, A., 2007. \textit{Affine models in
mathematical finance: an analytical approach.} Tartu: University Press.

\bibitem[Stein \& Stein (1991)]{Stein91} Stein, E. M. \& Stein, J.\ C. 1991
Stock price distributions with stochastic volatility: an analytic approach. 
\textit{Review of Financial Studies }\textbf{4}, 727-752.

\bibitem[Terakado (2019)]{Terakado19} Terakado, S., 2019 On the option
pricing formula based on the Bachelier model. Available at SSRN 3428994.

\bibitem[Uhlenbeck \& Ornstein (1930)]{Uhlenbeck30} Uhlenbeck, G. E. \&
Ornstein, L. S. 1930 On the theory of Brownian motion. \textit{Physical
Review }\textbf{36}, 823-841.

\bibitem[Vasicek (1977)]{Vasicek77} Vasicek, O.\ A. 1977 An equilibrium
characterization of the term structure. \textit{Journal of Financial
Economics }\textbf{5}, 177-188.

\bibitem[Weber (1951)]{Weber51} Weber, M. 1951 The fundamental solution of a
degenerate partial differential equation of parabolic type. \textit{Trans.
Amer. Math. Soc.} \textbf{71}, 24--37.

\bibitem[Wong (2014)]{Wong14} Wong, M.W. 2014 \textit{Introduction to
pseudo-differential Operators}. Singapore: World Scientific Publishing
Company.

\bibitem[Zhang \textit{et al.} (2018)]{Zhang18} Zhang, Y., Chen, X. \& Park,
D. 2018 Formal specification of constant product (xy= k) market maker model
and implementation. White paper.
\end{thebibliography}
\end{document}